\documentclass[aps,prb,twocolumn,superscriptaddress,floatfix,showpacs,10pt,letterpaper]{revtex4-1}

\usepackage{amsthm}

\pdfoutput=1

\newcommand{\be}{\begin{equation}}
\newcommand{\ee}{\end{equation}}

\begin{document}

\begin{titlepage}

\title{Symmetry protected topological orders \\
and the group cohomology of their symmetry group}

\author{Xie Chen}
\affiliation{Department of Physics, University of California, Berkeley, CA 94720, USA}
\affiliation{Department of Physics, Massachusetts Institute of Technology, Cambridge, Massachusetts 02139, USA}

\author{Zheng-Cheng Gu}
\affiliation{Institute for Quantum Information and Matter, California Institute of Technology, Pasadena, CA 91125, USA}
\affiliation{Department of Physics, California Institute of Technology, Pasadena, CA 91125, USA}

\author{Zheng-Xin Liu}
\affiliation{Institute for Advanced Study, Tsinghua University, Beijing, 100084, P. R. China}
\affiliation{Department of Physics, Massachusetts Institute of Technology, Cambridge, Massachusetts 02139, USA}

\author{Xiao-Gang Wen}
\affiliation{Perimeter Institute for Theoretical Physics, Waterloo, Ontario, N2L 2Y5 Canada}
\affiliation{Department of Physics, Massachusetts Institute of
Technology, Cambridge, Massachusetts 02139, USA}
\affiliation{Institute for Advanced Study, Tsinghua University,
Beijing, 100084, P. R. China}

\begin{abstract}
Symmetry protected topological (SPT) phases are gapped short-range-entangled
quantum phases with a symmetry $G$. They can all be smoothly connected to the
same trivial product state if we break the symmetry.  The Haldane phase of
spin-1 chain is the first example of SPT phases which is protected by $SO(3)$
spin rotation symmetry.  The topological insulator is another example of SPT
phases which is protected by $U(1)$ and time reversal symmetries.  
%
%It has been shown that free fermion SPT phases can be systematically described
%by the K-theory.  But how to label different SPT phases that have the same
%symmetry and cannot be distinguished by local order parameter?  We have shown
%that all bosonic SPT phases with on-site symmetry group $G$ are classified by
%$\cH^2[G,U(1)]$ (the projective representations of $G$) in  $1$ spatial
%dimension, and by $\cH^1[G,U(1)]$ (the 1D representations of $G$) in $0$
%spatial dimension.  
%
In this paper, we show that interacting bosonic SPT phases can be
systematically described by  group cohomology theory: distinct  $d$-dimensional
bosonic SPT phases with on-site symmetry $G$ (which may contain anti-unitary
time reversal symmetry) can be labeled by the elements in $\cH^{1+d}[G,U_T(1)]$
-- the Borel $(1+d)$-group-cohomology classes of $G$ over the $G$-module
$U_T(1)$.  Our theory, which leads to explicit ground state wave functions and
commuting projector Hamiltonians, is based on a new type of topological term
that generalizes the topological $\th$-term in continuous non-linear
$\si$-models to lattice non-linear $\si$-models.  The boundary excitations of
the non-trivial SPT phases are described by lattice non-linear $\si$-models
with a non-local Lagrangian term that generalizes the Wess-Zumino-Witten term
for continuous non-linear $\si$-models.  As a result, the symmetry $G$ must be
realized as a non-on-site symmetry for the low energy boundary excitations, and
those boundary states must be gapless or degenerate.
% if the symmetry is not explicitly broken on the boundary. 
As an application of our result, we can use
$\cH^{1+d}[U(1)\rtimes Z_2^T,U_T(1)]$ to obtain interacting bosonic topological
insulators (protected by time reversal $Z_2^T$ and boson number conservation),
which contain one non-trivial phases in 1D or 2D, and three in 3D.  We also
obtain interacting bosonic topological superconductors (protected by time
reversal symmetry only), in term of $\cH^{1+d}[Z_2^T,U_T(1)]$, which contain
one non-trivial phase in odd spatial dimensions and none for even.  Our result
is much more general than the above two examples, since it is for any symmetry
group.  For example, we can use $\cH^{1+d}[U(1)\times Z_2^T,U_T(1)]$ to
construct the SPT phases of integer spin systems with time reversal and $U(1)$
spin rotation symmetry, which contain three non-trivial SPT phases in 1D, none
in 2D, and seven in 3D.  Even more generally, we find that the different
bosonic symmetry breaking short-range-entangled phases are labeled by the
following three mathematical objects:
$\Big(G_H,G_\Psi,\cH^{1+d}[G_\Psi,U_T(1)]\Big)$, where $G_H$ is the symmetry
group of the Hamiltonian and $G_\Psi$ the symmetry group of the ground states.

\end{abstract}


\maketitle

\vspace{2mm}

\end{titlepage}

{\small \setcounter{tocdepth}{1} \tableofcontents }

\section{Introduction}

\subsection{Background}

Understanding phases of matter is one of the central issues in condensed matter
physics. For a long time, we believed that all the phases and phases
transitions were described by Landau symmetry breaking
theory.\cite{L3726,GL5064,LanL58} In 1989, it was realized that many quantum
phases can contain a new kind of orders which are beyond the Landau symmetry
breaking theory.\cite{Wtop}  A quantitative theory of the new orders was
developed based on robust ground state degeneracy and the robust non-Abelian
Berry's phases of the degenerate ground states, which can be viewed as new
``topological non-local order parameters''.\cite{WNtop,Wrig} The new orders
were named topological order.  Topologically ordered states contain gapless
edge excitations and/or degenerate sectors that encode all the information of
bulk topological orders.\cite{H8285,Wedgerev} The nontrivial edge states provide
us a practical way to experimentally probe topological order and illustrate the
holographic principle which was introduced later.\cite{H9326,S9577} The
excitations in those topologically ordered states in general carry fractional
charges\cite{JR7698} and obey fractional
statistics.\cite{LM7701,W8257,H8483,ASW8422}

\begin{figure}[b]
\begin{center}
\includegraphics[scale=0.47]{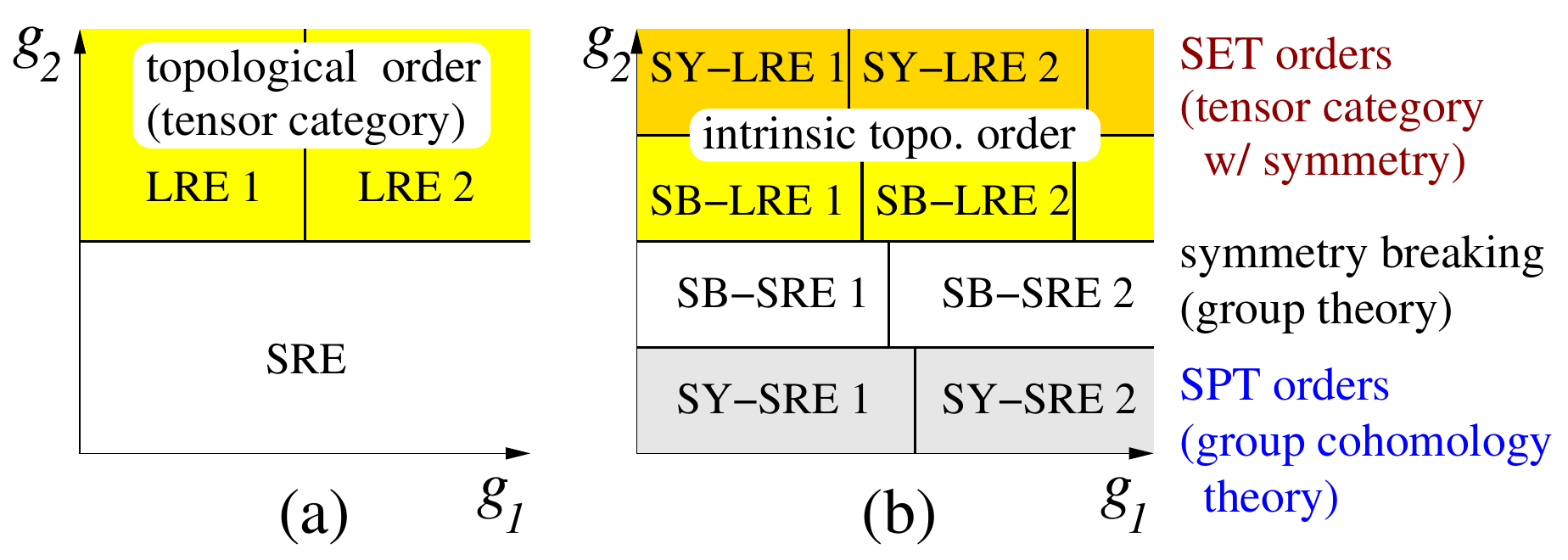}
%Fig. 1
\end{center}
\caption{
(Color online)
(a) The possible gapped phases for a class of Hamiltonians $H(g_1,g_2)$ without
any symmetry restriction.  (b) The possible gapped phases for the class of
Hamiltonians $H_\text{symm}(g_1,g_2)$ with symmetry.  Each phase is labeled by
its entanglement properties and symmetry breaking properties.  SRE stands for
short range entanglement, LRE for long range entanglement, SB for symmetry
breaking, SY for no symmetry breaking.  SB-SRE phases are the Landau symmetry
breaking phases, which are understood by introducing group theory.  The SY-SRE
phases are the SPT phases, and we will show that they can be understood by
introducing group cohomology theory.
The  SY-LRE phases are the SET phases.
}
\label{topsymm}
\end{figure}

Since its discovery, we have been trying to obtain a systematic and deeper
understanding of topological orders.  The studies of entanglement entropy show
signs that topological orders are related to long-range
entanglements.\cite{KP0604,LW0605} Recently, we found that topological orders
actually can be regarded as patterns of long range entanglements\cite{CGW1038}
defined through local unitary (LU) transformations.\cite{LW0510,VCL0501,V0705}

The notion of topological orders and long range entanglements leads to the
following more general and more systematic picture of phases and phase
transitions (see Fig.  \ref{topsymm}).\cite{CGW1038} For gapped quantum systems
without any symmetry, their quantum phases can be divided into two classes:
short range entangled (SRE) states and long range entangled (LRE) states.

SRE states are states that can be transformed into direct product states via LU
transformations. All SRE states can be transformed into each other via  LU
transformations. So all SRE states belong to the same phase (see Fig.
\ref{topsymm}a).

LRE states are states that cannot be transformed into  direct product
states via LU transformations.  It turns out that, many LRE states
also cannot be transformed into each other. The LRE
states that are not connected via LU transformations belong to different
classes and represent different quantum phases.  Those different quantum phases
are nothing but the topologically ordered phases.  Fractional quantum Hall
states\cite{TSG8259,L8395}, chiral spin liquids,\cite{KL8795,WWZ8913} $Z_2$
spin liquids,\cite{RS9173,W9164,MS0181} non-Abelian fractional quantum Hall
states,\cite{MR9162,W9102,WES8776,RMM0899} \etc are examples of topologically
ordered phases.  The mathematical foundation of topological orders is closely
related to tensor category theory\cite{FNS0428,LW0510,CGW1038,GWW1017} and
simple current algebra.\cite{MR9162,LWW1024}

For gapped quantum systems with symmetry, the structure of phase diagram is
even richer (see Fig. \ref{topsymm}b).  Even SRE states now can belong to
different phases. The Landau symmetry breaking states belong to this class of
phases.  However, there are more interesting examples in this class. Even SRE
states that do not break any symmetry and have the same symmetry can belong to
different phases.  The 1D Haldane phases for spin-1 chain\cite{H8364,AKL8877}
and topological insulators\cite{KM0501,BZ0602,KM0502,MB0706,FKM0703,QHZ0824}
are examples of non-trivial SRE phases that do not break any symmetry.  Those
phases are beyond Landau symmetry breaking theory since they do not break any
symmetry.  We will call those phases Symmetry Protected Topological (SPT)
phases.

For gapped quantum systems with symmetry, the corresponding LRE phases will be
much richer than those without symmetry. We may call those phases Symmetry
Enriched Topological (SET) phases.  Projective symmetry group (PSG) was
introduced to study the SET phases.\cite{W0213,W0303a} Many examples of this
kind of states can be found in \Ref{W0213,KLW0834,KW0906,LS0903,YFQ1070}, but a
systematic understanding is still lacking.

\subsection{Motivation}

The notion of topological order and long range entanglements deepens our
understanding of quantum phases and guides our research strategy. This allows
us to make significant progress.  

For example, there is no long range entanglement in gapped 1D
states.\cite{VCL0501,CGW1107} So, without symmetry, all gapped 1D quantum
states belong to the same phase.  For systems with a certain symmetry, all
gapped 1D phases are either SPT phases protected by symmetry or symmetry
breaking states.  Since both SPT phases and symmetry breaking states are short
range entangled, it is easy to understand them.  As a result, a complete
classification of all 1D gapped bosonic/fermionic quantum phases for any
symmetry can be obtained.\cite{CGW1107,SPC1032,CGW1123,PBT1039} (A special case
of the above result, a classification of 1D fermionic systems with $T^2=1$ time
reversal symmetry can also be found in \Ref{TPB1102,FK1103,CGW1123}). Using the
idea of LU transformations, we also developed a systematic and quantitative
theory for non-chiral topological orders in 2D interacting boson and fermion
systems.\cite{LW0510,CGW1038,GWW1017} We would like to mention that symmetry
protected Berry phases have been used to study various topological
phases.\cite{H1,H2}

Motivated by the 1D classification result, in this paper and in \Ref{CLW1152}
we would like to study SPT phases in higher dimensions.  Since SPT phases are
short range entangled, it is relatively easy to obtain a systematic
understanding.  (Another way to make the classification problem easier is to
consider only free fermion systems which are classified by
K-theory.\cite{K0886,SRF0825}) In \Ref{CLW1152}, we study some simple but
highly non-trivial examples.  Those non-trivial examples lead to the generic
and systematic results discussed in this paper.  Some other examples of 2D
gapped SPT phases are given in \Ref{YK1065,CGW1107,LS0903,SPC1032,ML}.

\subsection{Summary of results}

Using group theory, we can obtain a systematic understanding of symmetry
breaking phases (or more precisely, short-range entangled symmetry breaking
phases).  In this paper, we will show that, using group cohomology theory, we
can obtain a systematic understanding of short-range entangled symmetric phases
of bosons/qubits, even with strong interactions.  Those phases are called
bosonic SPT phases.  In particular, we have obtained the following results for
bosonic systems:
\begin{enumerate}
\item
{}From each element in $(1+d)$-Borel-cohomology group $\cH^{1+d}[G,U_T(1)]$, we
can construct a distinct SPT phase that respects the on-site symmetry $G$ in
$d$-spatial dimensions.  Here $G$ may contain anti-unitary time reversal
transformation and $\cH^{1+d}[G,U_T(1)]$ is introduced in appendix \ref{Gcoh}.
Note that $\cH^{1+d}[G,U_T(1)]$ is an Abelian group that can be calculated from
the symmetry group $G$.  The identity element in $\cH^{1+d}[G,U_T(1)]$
correspond to trivial SPT phases while other elements corresponds to non-trivial
SPT phases.  For example, $\cH^{2}[SO(3),U(1)]=Z_2$.  So a 1D integer spin
chain with $SO(3)$ spin rotation symmetry (but no translation symmetry) has
two kinds of SPT phases: one is the trivial $S=0$ phase and the other is the
Haldane phase.\cite{H8364,AKL8877}

\item
The low energy effective theory of a SPT phase with symmetry $G$ is given by a
topological non-linear $\si$-model that contains only a $2\pi$-quantized
topological $\th$-term.  The $2\pi$-quantized topological $\th$-term in
$(d+1)$D is classified by $\cH^{1+d}[G,U_T(1)]$ which generalizes the topological
term for non-linear $\si$-model with continuous symmetry.

\item
We argue that a $d$-dimensional non-trivial SPT phase has gapless boundary
excitations or degenerate boundary states. 
The  boundary  degeneracy may come from spontaneous symmetry breaking
or topological orders. This is
similar to the holographic principle for intrinsic topological
orders.\cite{H8285,Wedgerev} The boundary excitations of the SPT phase are
described by a non-linear $\si$-model with a non-local Lagrangian (NLL) term
that generalizes the Wess-Zumino-Witten (WZW) term\cite{WZ7195,W8322} for
continuous non-linear $\si$-models.  In (1+1)D  and for continuous symmetry
group, it is shown that non-linear $\si$-model with  WZW term is gapless which
is described by Kac-Moody current algebra.\cite{W8322}  

\item
The SPT phases that respect on-site symmetry $G$ and translation symmetry can
be obtained in the following way: In $1$-dimension, those phases are
labeled by 
$\cH^{1}[G,U_T(1)]\times\cH^{2}[G,U_T(1)]$.\cite{CGW1107,SPC1032,CGW1123} In
$2$-dimension, they are labeled by
$\cH^{1}[G,U_T(1)]\times\Big\{\cH^{2}[G,U_T(1)]\Big\}^2\times\cH^{3}[G,U_T(1)]$.
A partial result
$\cH^{1}[G,U_T(1)]\times\Big\{\cH^{2}[G,U_T(1)]\Big\}^2$ was obtained in
\Ref{CGW1107}.  In $3$-dimension, they are labeled by
$\cH^{1}[G,U_T(1)]\times\Big\{\cH^{2}[G,U_T(1)]\Big\}^3\times\Big\{\cH^{3}[G,U_T(1)]\Big\}^3\times\cH^{4}[G,U_T(1)]$.

\end{enumerate}

This paper is organized as the following: In section \ref{example}, we list the
bosonic SPT phases for many symmetry groups in dimension 0,1,2,3, and discuss
some examples of those SPT phases.  In section \ref{LUt}, we give a brief
review of local unitary transformations.  In section \ref{Cfm}, we discuss
canonical form of the ground state wave function for SPT phases.  In section
\ref{clsymm}, we study on-site symmetry transformations that leave the
canonical ground state wave function unchanged.  In section \ref{cnsymm}, we
construct the on-site symmetry transformations through the cocycles of the
symmetry group.  In section \ref{clnlm}, we introduced topological non-linear
$\si$-model and discuss their SPT phases.  We also argue that the boundary
states of the topological non-linear $\si$-model are gapless or degenerate if
the symmetry is not explicitly broken.  In section \ref{const}, we construct
and classify topological non-linear $\si$-model through the cocycles of the
symmetry group.  In section \ref{tritop} and \ref{gwvclass}, we show that the
ground states of the topological non-linear $\si$-model all have trivial
intrinsic topological orders, and the same SPT order if constructed from
equivalent cocycles.  In section \ref{AppBerry}, we discuss the relation
between the cocycles in the topological non-linear $\si$-model and the Berry's
phase.  In section \ref{SPTtrans}, we study SPT phases with both on-site and
translation symmetries.

\begin{table*}[tb]
 \centering
 \begin{tabular}{ |c||c|c|c|c| }
 \hline
 Symm. group & $d=0$ & $d=1$ & $d=2$ & $d=3$  \\
\hline
\hline
\color{blue}{$Z_2^T$}  & \color{blue}{$\Z_1$} & \color{blue}{$\Z_2$} & \color{blue}{$\Z_1$} & \color{blue}{$\Z_2$}   \\
\color{blue}{$Z_2^T\times \text{trn}$}  & \color{blue}{$\Z_1$} & \color{blue}{$\Z_2$} & \color{blue}{$\Z_2^2$} & \color{blue}{$\Z_2^4$}   \\
\hline
$Z_n$ & $\Z_n$  & $\Z_1$ & $\Z_n$ & $\Z_1$    \\
$Z_n\times \text{trn}$ & $\Z_n$  & $\Z_n$ & $\Z_n^2$ & $\Z_n^4$    \\
\hline
\color{dgrn}{$U(1)$} & \color{dgrn}{$\Z$}  & \color{dgrn}{$\Z_1$} & \color{dgrn}{$\Z$} & \color{dgrn}{$\Z_1$}    \\
\color{dgrn}{$U(1)\times \text{trn}$} & \color{dgrn}{$\Z$}  & \color{dgrn}{$\Z$} & \color{dgrn}{$\Z^2$} & \color{dgrn}{$\Z^4$}    \\
\hline
\color{red}{$U(1)\rtimes Z_2^T$}  & \color{red}{$\Z$} & \color{red}{$\Z_2$} & \color{red}{  $\Z_2$ } & \color{red}{ $\Z^2_2$ }   \\
\color{red}{$U(1)\rtimes Z_2^T\times \text{trn}$}  & \color{red}{$\Z$} & \color{red}{$\Z\times \Z_2$} & \color{red}{$\Z\times \Z_2^3$ } & \color{red}{$ \Z\times\Z_2^8$ }   \\
\hline
$U(1)\times Z_2^T$  & $\Z_1$ & $\Z^2_2$ & $\Z_1$ & $\Z^3_2$   \\
$U(1)\times Z_2^T\times \text{trn}$  & $\Z_1$ & $\Z^2_2$ & $\Z_2^4$ & $\Z_2^9$   \\
\hline
$U(1) \rtimes Z_2$
& $\Z_2$ & $\Z_2$ & $\Z\times \Z_2$ & $\Z_2$   \\
$U(1) \times Z_2$
& $\Z\times \Z_2$ & $\Z_1$ & $\Z\times \Z^2_2$ & $\Z_1$   \\
\hline
%$S_3$ & $\Z_2$  & $\Z_1$ & $\Z_6$ & $\Z_1$    \\
%$S_3\times \text{trn}$ & $\Z_2$  & $\Z_2$ & $\Z_2\times\Z_6$ & $\Z_2\times\Z_6^3$    \\
%\hline
$Z_n \rtimes Z_2^T$
& $\Z_n$ & $\Z_2\times \Z_{(2,n)}$ & $\Z^2_{(2,n)}$ & $\Z_2\times \Z^2_{(2,n)}$   \\
$Z_n \times Z_2^T$
& $\Z_{(2,n)}$ & $\Z_2\times \Z_{(2,n)}$ & $\Z^2_{(2,n)}$ & $\Z_2\times \Z^2_{(2,n)}$   \\
%$Z_2^T\times  Z_n\times \text{trn}$
%& $\Z_{(2,n)}$ & $\Z_2\times \Z^2_{(2,n)}$ & $\Z^2_2\times \Z^5_{(2,n)}$ & $\Z^4_2\times \Z^{12}_{(2,n)}$   \\
%\hline
$Z_n \rtimes Z_2$
& $\Z_2\times \Z_{(2,n)}$ & $\Z_{(2,n)}$ & $\Z_n\times \Z_2\times \Z_{(2,n)}$ & $\Z^2_{(2,n)}$   \\
$Z_m \times Z_n$
& $\Z_m\times \Z_n$ & $\Z_{(m,n)}$ & $\Z_m\times \Z_n\times \Z_{(m,n)}$ & $ \Z^2_{(m,n)}$   \\
\hline
$D_2 \times Z_2^T =D_{2h}$ & $\Z^2_2$  & $\Z^4_2$ & $\Z^6_2$ & $\Z^9_2$    \\
$ Z_m \times Z_n\times Z_2^T$
& $\Z_{(2,m)} \times \Z_{(2,n)}$ &
$\Z_2\times\Z_{(2,m)}\times  \Z_{(2,n)}\times\Z_{(m,n)} $ & 
$\Z^2_{(2,m,n)}\times  \Z^2_{(2,m)}\times\Z^2_{(2,n)} $ & 
$\Z_2\times\Z^4_{(2,m,n)}\times  \Z^2_{(2,m)}\times\Z^2_{(2,n)} $   \\
$SU(2)$ & $\Z_1$  & $\Z_1$ & $\Z$ & $\Z_1$    \\
\hline
$SO(3)$ & $\Z_1$  & $\Z_2$ & $\Z$ & $\Z_1$    \\
$SO(3)\times \text{trn}$ & $\Z_1$  & $\Z_2$ & $\Z\times \Z_2^2$ & $\Z^3\times\Z_2^3$    \\
\hline
$SO(3)\times Z_2^T$ & $\Z_1$  & $\Z_2^2$ & $\Z_2$ & $\Z_2^3$    \\
$SO(3)\times Z_2^T\times \text{trn}$ & $\Z_1$  & $\Z_2^2$ & $\Z_2^5$ & $\Z_2^{12}$    \\
\hline
 \end{tabular}
 \caption{
(Color online)
SPT phases of interacting bosonic systems in $d$-spatial dimensions protected
by on-site symmetry $G$.  In absence of translation symmetry,
the above table lists $\cH^{1+d}[G,U_T(1)]$ whose elements
label the SPT phases.  Here $\Z_1$ means that our construction only gives rise
to the trivial phase.  $\Z_n$ means that the constructed non-trivial SPT phases
plus the trivial phase are labeled by the elements in $\Z_n$.  $Z_2^T$
represents time reversal symmetry, ``$\text{trn}$'' represents translation
symmetry, $U(1)$ represents $U(1)$ symmetry, $Z_n$ represents cyclic symmetry,
\etc.  Also $(m,n)$ is the greatest common divisor of $m$ and $n$.  The red
rows are for bosonic topological insulators and the blue rows bosonic
topological superconductors.  The red/blue rows without translation symmetry
correspond to strong bosonic topological insulators/superconductors and the
red/blue rows with translation symmetry also contain weak bosonic topological
insulators/superconductors.
}
 \label{tb}
\end{table*}

\section{Examples of bosonic SPT phases}
\label{example}

In table \ref{tb}, we list the SPT phases for some simple symmetry groups.  In
the following, we discuss some of those phases in detail.  We also give some
simple examples for some of the listed SPT states.

\subsection{$SO(3)$ SPT states}

For integer spin systems with the full $SO(3)$ spin rotation symmetries, the
symmetry group is $SO(3)$.  From $\cH^{1+d}[SO(3),U(1)]$, we find one
non-trivial SPT phase in 1D and infinite many in 2D.  Those 2D SPT phases
labeled by $k\in \Z$ have a special property that they can be described by
continuous non-linear $\si$-model with $2\pi$-quantized topological $\th$-term:
\begin{align}
\label{SO3S}
 S &=\int \dd\tau\dd^2 x\; \big(
\frac{1}{2\rho } \Tr(\prt_\mu g^\dag \prt_\mu g)
\\
&\ \ \ \ \ \ \ \
+\imth \frac{\th}{2\pi^2}\frac{\eps^{\mu\nu\la}}{6}\frac{1}{8}\Tr[
(g^{-1}\prt_\mu g)
(g^{-1}\prt_\nu g)
(g^{-1}\prt_\la g)]\Big),
\nonumber
\end{align}
where $g(\v x,t)$ is a $3\times 3$ matrix in $SO(3)$ and $\th=2\pi k$, $k\in
\Z$.  This is because the topological term, when $k=0$ mod 4,\cite{LWwzw}
\begin{align}
\int \dd\tau\dd^2 x\;
\frac{k}{2\pi^2}
\frac{\eps^{\mu\nu\la}}{6}\frac{1}{8}\Tr[ (g^{-1}\prt_\mu g)
(g^{-1}\prt_\nu g) (g^{-1}\prt_\la g)]
\end{align}
corresponds to $\cH^{3}[SO(3),U(1)]$ whose elements are labeled by $k/4 \in
\Z$.  

At the boundary, the topological term reduces to the well known WZW term
which gives rise to gapless edge excitations.\cite{W8322} We would like to
point out that $g^{-1}(\prt_\mu g)$ and $(\prt_\mu g)g^{-1}$ create excitations
on the edge that move in the opposite directions.  Since the $SO(3)$ symmetry
acts as $g(x,t) \to h g(x,t)$, $h\in SO(3)$, only $(\prt_\mu g)g^{-1}$ carries
non trivial $SO(3)$ quantum numbers while  $g^{-1}(\prt_\mu g)$ is a $SO(3)$
singlet.  So on the edge, the excitations with non-trivial $SO(3)$ quantum
numbers all move in one direction, which breaks the time reversal and parity
symmetry.  In fact, under time reversal or parity transformations, $\th\to
-\th$ and $k\to -k$.

In the above example, we see that a $2\pi$-quantized topological $\th$-terms in
a non-linear $\si$-model gives rise to a non-trivial SPT phase.  However, the
topological $\th$-terms in \emph{continuous} non-linear $\si$-model do not
always correspond to non-trivial SPT phases.  For example, $\pi_2(SO(3))=0$ and
the continuous $SO(3)$ non-linear $\si$-model has no topological $\th$-terms in
(1+1)D. But we do have a 1D non-trivial SPT phase protected by $SO(3)$
symmetry.  Also, $\pi_4(SO(3))=\Z_2$  and the continuous $SO(3)$ non-linear
$\si$-model has non-trivial topological $\th$-term in (3+1)D.  But such
topological $\th$-term cannot produce non-trivial SPT phase protected by
$SO(3)$ symmetry, since the topological term becomes trivial once we include
the cut-off.  In this paper, we will show that non-trivial $2\pi$-quantized
topological $\th$-terms can even be defined for  \emph{lattice} non-linear
$\si$-models on discretized space-time.  It is the $2\pi$-quantized topological
$\th$-terms in \emph{lattice} non-linear $\si$-models that give rise to
non-trivial SPT phases.

It is also possible that the above $SO(3)$ SPT phases labeled by $k$ and $k+1$
are connected by a continuous phase transition that do not break any symmetry.
The gapless critical point is likely to be described by \eqn{SO3S} with
$\th=2\pi(k+\frac12)$.  When $\th< 2\pi(k+\frac12)$, it may flow to $2\pi k$
at low energies and when $\th> 2\pi(k+\frac12)$, it may flow to $2\pi (k+1)$.
Since all the SPT phases are described by $2\pi$-quantized topological
$\th$-terms in \emph{lattice} non-linear $\si$-models, the above picture about
the transitions between SPT phases may be valid for generic SPT phases.

For integer spin systems with time reversal and the full $SO(3)$ spin rotation
symmetries, the symmetry group is $SO(3)\times Z_2^T$.  From
$\cH^{1+d}[SO(3)\times Z_2^T,U_T(1)]$, we find one non-trivial SPT phase in 2D
and seven in 3D.  Note that on systems with boundary, the topological
$\th$-term in \eqn{SO3S} breaks the time reversal symmetry.  So we cannot use
those $\Z$ classified topological $\th$-terms to produce 2D SPT phases with
time reversal and $SO(3)$ spin rotation symmetries.  As a result, the 2D SPT
phases with time reversal and $SO(3)$ spin rotation symmetries are only
described by $\Z_2$.

\subsection{$SU(2)$ SPT states}

For bosonic systems with $SU(2)$ symmetry, the SPT phases are labeled by
$\cH^{1+d}[SU(2),U(1)]$. We find infinite many non-trivial $SU(2)$ SPT phases
in $(2+4n)$ spatial dimension.  Those $SU(2)$ SPT phases labeled by $k\in \Z$.
There is no non-trivial $SU(2)$ SPT phase in other dimensions.  Similarly,
those  $SU(2)$ SPT phases in 2-dimensions can be described by continuous
non-linear $\si$-model with $2\pi$-quantized topological $\th$-term:
\begin{align}
\label{SU2S}
 S &=\int \dd\tau\dd^2 x\; \big(
\frac{1}{2\rho } \Tr(\prt_\mu g^\dag \prt_\mu g)
\\
&\ \ \ \ \ \ \ \
+\imth \frac{\th}{2\pi^2}\frac{\eps^{\mu\nu\la}}{6}\frac{1}{2}\Tr[
(g^{-1}\prt_\mu g)
(g^{-1}\prt_\nu g)
(g^{-1}\prt_\la g)]\Big),
\nonumber
\end{align}
where $g(\v x,t)$ is a $2\times 2$ matrix in $SU(2)$ and $\th=2\pi k$, $k\in
\Z$.

\subsection{$U(1)$ SPT states}

{}From $\cH^{1+d}[U(1),U(1)]=\Z$ for even $d$ and $\cH^{1+d}[U(1),U(1)]=\Z_1$
for odd $d$, we find that spin/boson systems with $U(1)$ on-site symmetry have
infinite non-trivial SPT phases labeled by non-zero integer in $d$ = even
dimensions.  This generalizes a result obtained by Levin for $d=2$.\cite{ML} We
note that $\cH^3[SU(2),U(1)]=\cH^3[U(1),U(1)]=\Z$.  The SPT states with $SU(2)$
symmetry can also be viewed as SPT states with $U(1)$ symmetry.  We know that
an $SU(2)$ SPT state labeled by $k\in \Z$ is described by \eqn{SU2S} with
$\th=2\pi k$.  Such an $SU(2)$ SPT state is also an
non-trivial  $U(1)$ SPT state  labeled by $k\in \Z$.

We like to point out that it is believed that all 2D gapped phases with Abelian
statistics are classified by $K$-matrix and the related $U(1)$ Chern-Simons
theory.\cite{BW9045,R9002,WZ9290} All the quasiparticles in the 2D SPT phases
are bosons. So the SPT phases are also described by $K$-matrices.  We just need
to find a way to include symmetry in the $K$-matrix approach, which is done in
\Ref{LS0903}.  In particular, Michael Levin\cite{LU1} pointed out that a 2D
$U(1)$ SPT phase can be described by a $U(1)\times U(1)$ Chern-Simons theory
(or a double-layer quantum Hall state) (see also \Ref{LS,LV})
\begin{align}
\cL=
\frac{1}{4\pi} K_{IJ} a_{I\mu}\prt_\nu a_{J\la}\eps^{\mu\nu\la}
+
\frac{1}{2\pi} q_{I} A_{\mu}\prt_\nu a_{I\la}\eps^{\mu\nu\la}
+...
\end{align}
with the $K$-matrix and the charge vector $\v q$:\cite{BW9045,R9002,WZ9290}
\begin{align}
\label{Kq}
K=\bpm
0 & 1 \\
1 & 2k \\
\epm,\ \ \ \ 
\v q=\bpm
1  \\
1  \\
\epm.
\end{align}
We note that such a $K$-matrix has two  null vectors
$\v n_1=\bpm
1  \\
k  \\
\epm,
\v n_2=\bpm
0  \\
1  \\
\epm
$
that satisfy $\v n_i^T K^{-1} \v n_i=0$.  The null vectors correspond to
quasiparticles with Bose statistics.  Such null vectors would destabilize the
state if we did not have the $U(1)$ symmetry, since we could include \emph{one}
of the corresponding quasiparticle operators in the Hamiltonian which would gap
the edge excitations.\cite{Hsta}  In the presence of $U(1)$ symmetry, the
quasiparticles carry $U(1)$ charges $\v q^TK^{-1}\v n_1=1-k$ and $\v q^TK^{-1}\v
n_2=1$.  We see that when $k\neq 1$, both quasiparticles that correspond to the
null vectors carry non-zero $U(1)$ charges. Thus, the quasiparticle operators
cannot be included in the Hamiltonian, and they do not gap the gapless edge
excitations.  The correspond state will have $U(1)$ protected gapless
excitation and correspond to a non-trivial $U(1)$ SPT state.  We see that the
$K$-matrix and the charge vector $\v q$ describe a  non-trivial $U(1)$ SPT
state when $k\neq 1$ and a trivial state when $k=1$.  The 2D $U(1)$ SPT states
are labeled by an integer.

\subsection{Bosonic topological insulators/superconductors}

The $U(1)\rtimes Z_2^T$ line in Table \ref{tb} describes the SPT phases for
interacting bosons with time reversal symmetry $Z_2^T$ and boson number
conservation (symmetry group = $U(1)\rtimes Z_2^T$, where time reversal $T$ and
$U(1)$ transformations $U_\th$ satisfy $TU_\th=U_{-\th} T$).  Those phases are
bosonic analogues of free fermion topological insulators protected by the same
symmetry.  From $\cH^{1+d}[U(1)\rtimes Z_2^T,U(1)]$, we find one kind of
non-trivial bosonic topological insulators in 1D or 2D, and three kinds in 3D.
The only non-trivial topological insulator in 1D is the same as the Haldane
phase.

The $Z_2^T$ line in Table \ref{tb} describes interacting bosonic analogues of
free fermion topological
superconductors\cite{SMF9945,RG0067,R0664,QHR0901,SF0904} with only time
reversal symmetry, $Z_2^T$ .  Since $\cH^{1+d}[Z_2^T,U(1)] =\Z_2$ for odd $d$
and $\cH^{1+d}[Z_2^T,U(1)] =\Z_1$ for even $d$, we find one kind of ``bosonic
topological superconductors'' or non-trivial SPT phases in every odd dimensions
(for the spin/boson systems with only time reversal symmetry).

\subsection{Other SPT states}

The $U(1)\times Z_2^T$ line describes the SPT phases for integer spin systems
with time reversal and $U(1)$ spin rotation symmetries (symmetry group =
$U(1)\times Z_2^T$, where time reversal $T$ and $U(1)$ transformations
$U_\th$ satisfy $TU_\th=U_{\th} T$).  From $\cH^{1+d}[U(1)\times
Z_2^T,U(1)]$, we find three non-trivial SPT phases in 1D, none in 2D, and seven in 3D.

We also find that
$\cH^{1+d}[Z_n,U(1)]=\Z_n$ for even $d$ and $\cH^{1+d}[Z_n,U(1)]=\Z_1$ for
odd $d$. So spin/boson systems with $Z_n$ on-site symmetry have $n-1$ kinds of
non-trivial SPT phases in $d$ = even dimensions.

For integer spin systems with $D_{2h}$ symmetry but no translation symmetry, we
discover $15$ new SPT phases in 1D,\cite{CGW1123,LCW1121} $63$ new SPT phases in
2D, and $511$ new SPT phases in 3D.

\begin{figure}[tb]
\begin{center}
\includegraphics[scale=0.5]{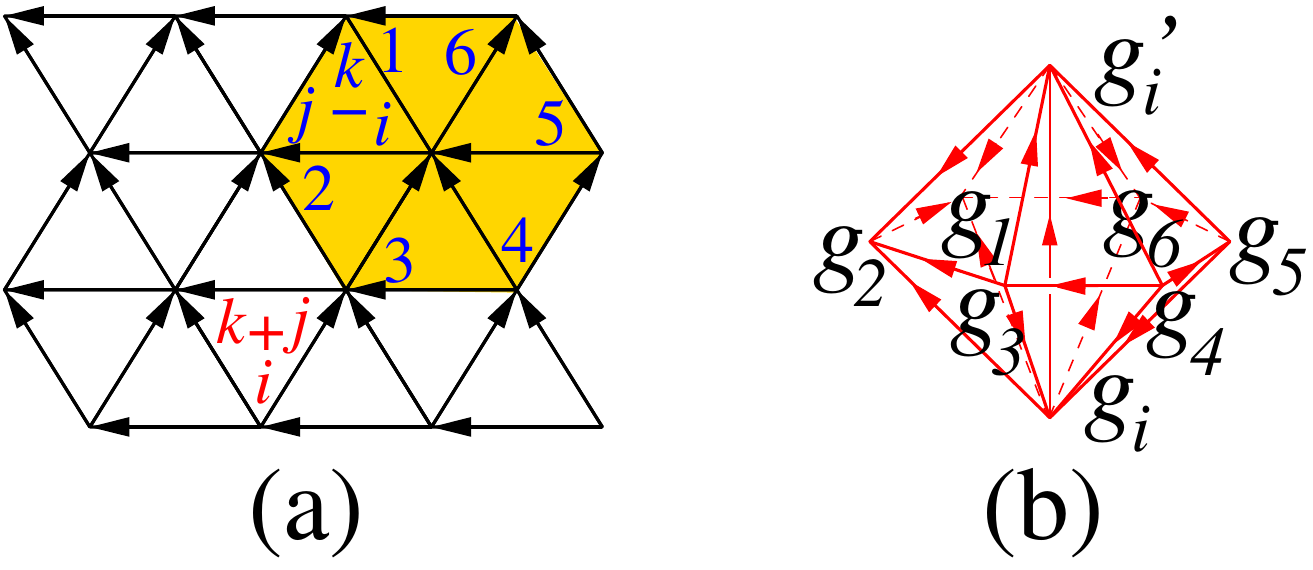}
%Fig. 2
\end{center}
\caption{
%(Color online)
(a) 
A triangular lattice.
The Hamiltonian term \eq{Hi}
acts on the seven sites in the shaded area.
(b) A geometric representation of the
the phase factors in \eqn{Hi}.
}
\label{discreteS}
\end{figure}

\subsection{Ideal ground state wave functions and exactly soluble 
Hamiltonians for SPT phases}

We can construct the idea ground state wave functions and
exactly soluble Hamiltonians for all the
SPT phases described by $\cH^{1+d}[G,U_T(1)]$.
The elements
in $\cH^{1+d}[G,U_T(1)]$ are complex functions of $d+2$ variables
$\nu_{d+1}(g_0,...,g_{d+1})$, $g_i \in G$.
$\nu_{d+1}(g_0,...,g_{d+1})$ is a pure phase $|\nu_{d+1}(g_0,...,g_{d+1})|=1$
that satisfy certain cocycle conditions \eq{cchcnd} and \eq{dnud}.
From each element $\nu_{d+1}(g_0,...,g_{d+1})$ we can construct
the $d$-dimensional ground state wave function for the corresponding SPT phase.
In 2D, we can start with a triangle lattice model where the physical states on
site-$i$ are given by $|g_i\>$, $g_i\in G$ (see Fig.  \ref{discreteS}a).  
%The
%ground state wave function can be obtained by viewing the 2D lattice (which is
%a torus) as the surface of a solid torus. The evaluation of the path integral
%of the topological non-linear $\si$-model (which is given by
%$\nu_3(g_0,g_1,g_2,g_3)$, an element in $\cH^3[G,U(1)]$) on the solid torus
%gives rise to the following 
The ideal ground state wave function is then given by $ \Phi(\{ g_i \}) =
\prod_{\bigtriangleup} \nu_3(1,g_i, g_j, g_k) \prod_{\bigtriangledown}
\nu_3^{-1}(1,g_i, g_j, g_k) $, where $ \prod_{\bigtriangleup} $ and $
\prod_{\bigtriangledown} $ multiply over all up- and down-triangles, and the
order of $ijk$ is clockwise for up-triangles and anti-clockwise for
down-triangles (see Fig.  \ref{discreteS}a). 

To construct exactly soluble Hamiltonian $H$ that realizes the above wave
function as the ground state, we start with an exactly soluble Hamiltonian
$H_0=-\sum_i |\phi_i\>\<\phi_i|$, $|\phi_i\>=\sum_{g_i\in G} |g_i\>$, whose
ground state is $\Phi_0(\{ g_i \}) =1$.  Then, using the local unitary
transformation $U=\prod_{\bigtriangleup} \nu_3(1,g_i, g_j, g_k)
\prod_{\bigtriangledown} \nu_3^{-1}(1,g_i, g_j, g_k)$, we find that the above
ideal ground state wave function is given by $\Phi=U\Phi_0$ and the
corresponding  exactly soluble Hamiltonian is given by $H=\sum_i H_i$, where
$H_i=U |\phi_i\>\<\phi_i| U^\dag$. $H_i$ acts on a seven-spin cluster labeled
by $i$, 1 -- 6 in shaded area in Fig. \ref{discreteS}a
\begin{align}
\label{Hi}
&
 H_i|g_i,g_1g_2g_3g_4g_5g_6\> =\sum_{g_i'}
|g_i',g_1g_2g_3g_4g_5g_6\> \times
\nonumber\\
&\ \ \ \ \ \ \
\frac{
\nu_3(g_4,g_5,g_i,g_i')
\nu_3(g_5,g_i,g_i',g_6)
\nu_3(g_i,g_i',g_6,g_1)
}{
\nu_3( g_i,g_i',g_2,g_1)
\nu_3(g_3,g_i,g_i',g_2)
\nu_3(g_4,g_3,g_i,g_i')
} 
\end{align}
The above phase factor has a graphic representation as in Fig.
\ref{discreteS}b.  (For a detailed explanation of the graphic representation
see Fig. \ref{d1d2}.) $H$ has a short ranged interaction and has the symmetry
$G$: $|\{g_i\}\> \to |\{g g_i\}\>, \ g\in G$, if $\nu_3(g_0,...,g_3)$ satisfies
the 3-cocycle conditions \eqn{cchcnd} and \eqn{dnu3}.

For symmetry $G=Z_2$ and using the 3-cocycle calculated in section \ref{Z2gc},
we find that the Hamiltonian that realize the non-trivial
$Z_2$ SPT state in two dimensions is given by
\begin{align}
 H_i=
 \si^+_i\eta^+_{21}\eta^+_{32}\eta^+_{43}\eta^+_{45}\eta^+_{56} \eta^+_{61}
+\si^-_i\eta^-_{21}\eta^-_{32}\eta^-_{43}\eta^-_{45}\eta^-_{56} \eta^-_{61}
\end{align}
where
\begin{align}
 \si^+_i=
\bpm 
0 & 0\\
1 & 0\\
\epm, \ \ \ \ \
 \si^-_i=
\bpm 
0 & 1\\
0 & 0\\
\epm,
\end{align}
which act on site-$i$.
Also $\eta^\pm_{ij}$ are operators acting on site-$i$ and site-$j$:
\begin{align}
& \eta^+_{ij}|0\>_i\otimes|1\>_j=- |0\>_i\otimes|1\>_j,
\nonumber\\
&
\eta^+_{ij}|\al\>_i\otimes|\bt\>_j= |\al\>_i\otimes|\bt\>_j,\ \ \
(\al,\bt)\neq (0,1)
\nonumber\\
& \eta^-_{ij}|1\>_i\otimes|0\>_j=- |1\>_i\otimes|0\>_j,
\nonumber\\
&
\eta^-_{ij}|\al\>_i\otimes|\bt\>_j= |\al\>_i\otimes|\bt\>_j,\ \ \
(\al,\bt)\neq (1,0) .
\end{align}

\subsection{A classification of
short-range-entangled states with or without symmetry
breaking}

The above results are for bosonic states that do not break any symmetry of the
Hamiltonian.  Combining group theory (that describe the symmetry break states)
and group cohomology theory (that describes the SPT states), we can obtain a
theory for more general short-range-entangled states that may break the
symmetry $G_H$ of the Hamiltonian down to the symmetry $G_\Psi$ of the ground
states.  We find that the different symmetry breaking short-range-entangled
phases are described/labeled by the following three mathematical objects:
$\Big(G_H,G_\Psi,\cH^{1+d}[G_\Psi,U_T(1)]\Big)$.

Landau symmetry breaking theory tries to use $(G_H,G_\Psi)$ to describe/label
all the symmetry breaking short-range-entangled phases.  We see that Landau
symmetry breaking theory misses the  third label $\cH^{1+d}[G_\Psi,U_T(1)]$.
The SPT phases do not break any symmetry and are described by
$\Big(G_H,G_H,\cH^{1+d}[G_H,U_T(1)]\Big)\sim \cH^{1+d}[G_H,U_T(1)]$.

\section{Local unitary transformations}

\label{LUt}

In the rest of the paper, we will explain the ideas, the way of thinking, and
the detailed calculations that allow us to obtain the results described in
the above section. We will start with a short review of local unitary (LU)
transformation.\cite{LW0510,VCL0501,V0705,CGW1038}

LU transformation is an important concept which is directly related to the
definition of quantum phases.\cite{CGW1038}  In this section, we will explain
what it is.  Let us first introduce local unitary evolution.  A LU evolution is
defined as the following unitary operator that act on the degrees of freedom in
a quantum system:
\begin{align}
\label{LUdef}
  \cT[e^{-i\int_0^1 dg\, \t H(g)}]
\end{align}
where $\cT$ is the path-ordering operator and $\t H(g)=\sum_{\v i} O_{\v i}(g)$
is a sum of local Hermitian operators.  Two gapped quantum states belong to the
same phase if and only if they are related by a LU
evolution.\cite{HW0541,BHM1044,CGW1038}

\begin{figure}[tb]
\begin{center}
\includegraphics[scale=0.5]{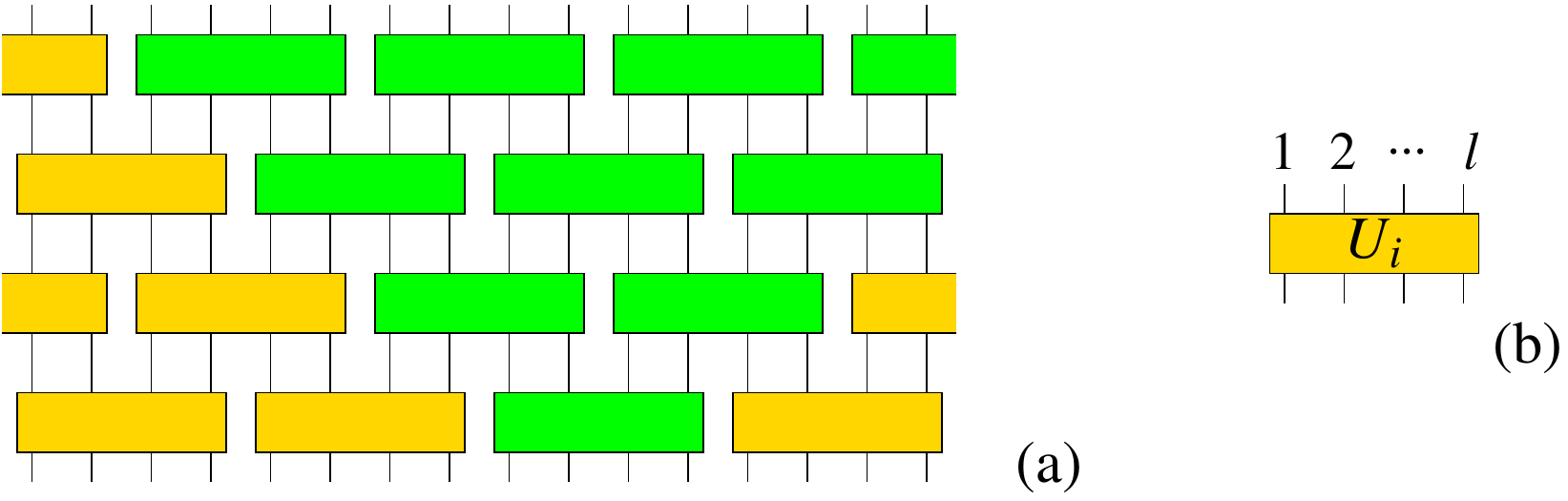}
%Fig. 3
\end{center}
\caption{
(Color online)
(a) A graphic representation of a quantum circuit, which is form by (b) unitary
operations on blocks of finite size $l$. The green shading represents a causal
structure.
}
\label{qc}
\end{figure}

The LU evolutions is closely related to \emph{quantum circuits with finite
depth}.  To define quantum circuits, let us introduce  piecewise local unitary
operators.  A piecewise local unitary operator has a form
\begin{equation*}
 U_{pwl}= \prod_{i} U^i
\end{equation*}
where $\{ U^i \}$ is a set of unitary operators that act on non overlapping
regions. The size of each region is less than some finite number $l$. The
unitary operator $U_{pwl}$ defined in this way is called a piecewise local
unitary operator with range $l$.  A quantum circuit with depth $M$ is given by
the product of $M$ piecewise local unitary operators:
\begin{equation*}
 U^M_{circ}= U_{pwl}^{(1)} U_{pwl}^{(2)} \cdots U_{pwl}^{(M)}
\end{equation*}
We will call $U^M_{circ}$ a LU transformation.  In quantum information theory,
it is known that finite time unitary evolution with local Hamiltonian (LU
evolution defined above) can be simulated with constant depth quantum circuit
(\ie a  LU transformation) and vice-verse:
\begin{align}
  \cT[e^{-i\int_0^1 dg\, \t H(g)}] =U^M_{circ}.
\end{align}
So two gapped quantum states belong to the same phase if and only if they are
related by a LU transformation.

In this paper, we will use the LU transformations to simplify gapped quantum
states within the same phase.  This allows us to gain a deeper understanding and
even to classify gapped quantum phases.

\section{Canonical form of many-body states with short range entanglements}

\label{Cfm}

A generic many-body wave function $\Phi( m_1,...,m_N)$ is very complicated. It
is hard to see and identify the quantum phase represented by a many-body wave
function. In this section, we will use LU transformations to simplify many-body
wave functions in order to understand the structure of quantum phases.

Such an approach is very effective in 1D\cite{CGW1107,SPC1032,CGW1123} which
leads to a complete classification of gapped 1D phases. In two dimensions, the
approach allows us to classify non-chiral topological
orders.\cite{LW0510,CGW1038,GWW1017}
In this paper, we will study another problem where such an approach is
effective. We will use LU transformations to study SRE quantum phases with
symmetries and study SPT phases that do not break any symmetry.

\subsection{Cases without any symmetry}

\begin{figure}[tb]
\begin{center}
\includegraphics[scale=0.6]{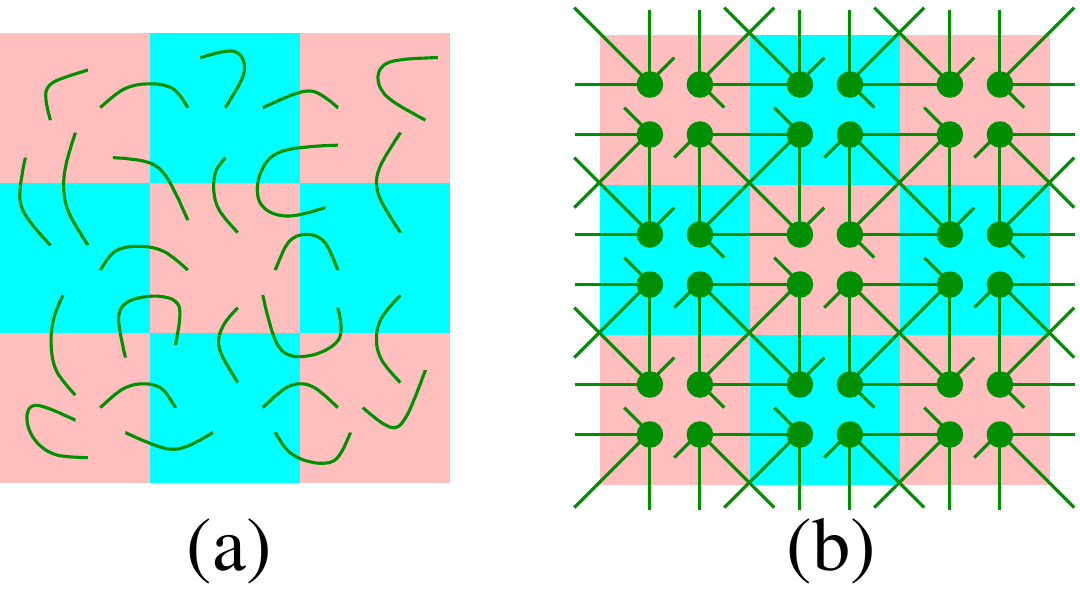}
%Fig. 4
\end{center}
\caption{
(Color online)
Transforming a SRE state to a tensor-network state which take simple canonical
form.  (a) A SRE state. (b) Using the unitary transformations that act within
each block, we can transform the  SRE state to a tensor-network state.
Entanglements exist only between the degrees of freedom on the connected
tensors.
}
\label{cForm}
\end{figure}

\begin{figure}[tb]
\begin{center}
\includegraphics[scale=0.6]{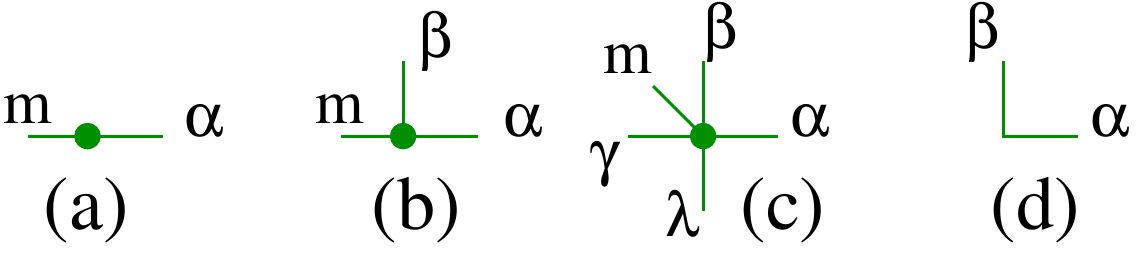}
%Fig. 5
\end{center}
\caption{
(Color online)
Graphic representations of tensors:
(a) $A^m_{\al}$, (b) $A^m_{\al\bt}$, and (c) $A^m_{\al\bt\ga\la}$.
(d) A corner represents a special rank-2 tensor $A_{\al\bt}=\del_{\al\bt}$.
}
\label{AA}
\end{figure}

Without any symmetry, we can always use LU transformations to transform a SRE
wave function into a product state.  In the following, we will describe how to
choose such LU transformation and what is the form of the resulting product
state.

We first divide our system into patches of size $l$ as in Fig.  \ref{cForm}a.
If $l$ is large enough, entanglement only exists between regions that share an
edge or a corner.  In this case, we can use LU transformation to transform the
state in Fig.  \ref{cForm}a into a state with many unentangled regions (see
Fig.  \ref{cForm}b).  For example, some degrees of freedom in the middle square
in Fig.  \ref{cForm}a may be entangled with the degrees of freedom in the three
squares below, to the right, and to the lower-right of the middle square.  We
can use the LU transformation inside the middle square to move all those
degrees of freedom to the  lower-right corner of the middle square.  Similarly,
we can use the LU transformation to  move all the degrees of freedom that are
entangled with the three squares below, to the left, and to the lower-left  of
the middle square to the  lower-left corner of the middle square, etc.  Repeat
such operation to every square and we obtain a state described by Fig.
\ref{cForm}b.  For stabilizer states, such reduction procedure has been
established explicitly.\cite{SB}

Fig.  \ref{cForm}b is a graphic representation of a tensor-network description
of the state.\cite{GMN0391,NMG0415,M0460,VC0466,JOV0802,GLWtergV,JWX0803}  In
the graphic representation, a dot with $n$ legs represents a rank $n$ tensor
(see Fig. \ref{AA}).  If two legs are connected, the indexes on those legs will
take the same value and are summed over.  In the  tensor-network representation
of states, we can see the entanglement structure. The disconnected parts of
tensor-network are not entangled.  In particular, the tensor-network state Fig.
\ref{cForm}b is a direct product state.

If there is no symmetry, we can transform any  direct product state to any
other  direct product state via LU transformations.  So all SRE states belong
to one phase.

\subsection{Cases with an on-site symmetry}

However, when we study phases of systems with certain symmetry,
we can only use the LU transformations that respect the symmetry to connect
states within the same phase.  In this case, even SRE states with the same
symmetry can belong to different phases.

Let us consider $d$-dimensional systems of $N$ sites that have only an
on-site symmetry group $G$. We also assume that the states $|m\>$
on each site form a linear representation $U_{mm'}(g),\ g\in G$ of the
group $G$.

To understand the structure of quantum phases of the symmetric states that do
not break the symmetry $G$, we can only use symmetric LU transformation that
respects the on-site symmetry $G$ to define phases.  Two gapped symmetric states
are in the same phase if and only if they can be connected by a {\it symmetric}
LU transformation.\cite{CGW1038}

\begin{figure}[tb]
\begin{center}
\includegraphics[scale=0.4]{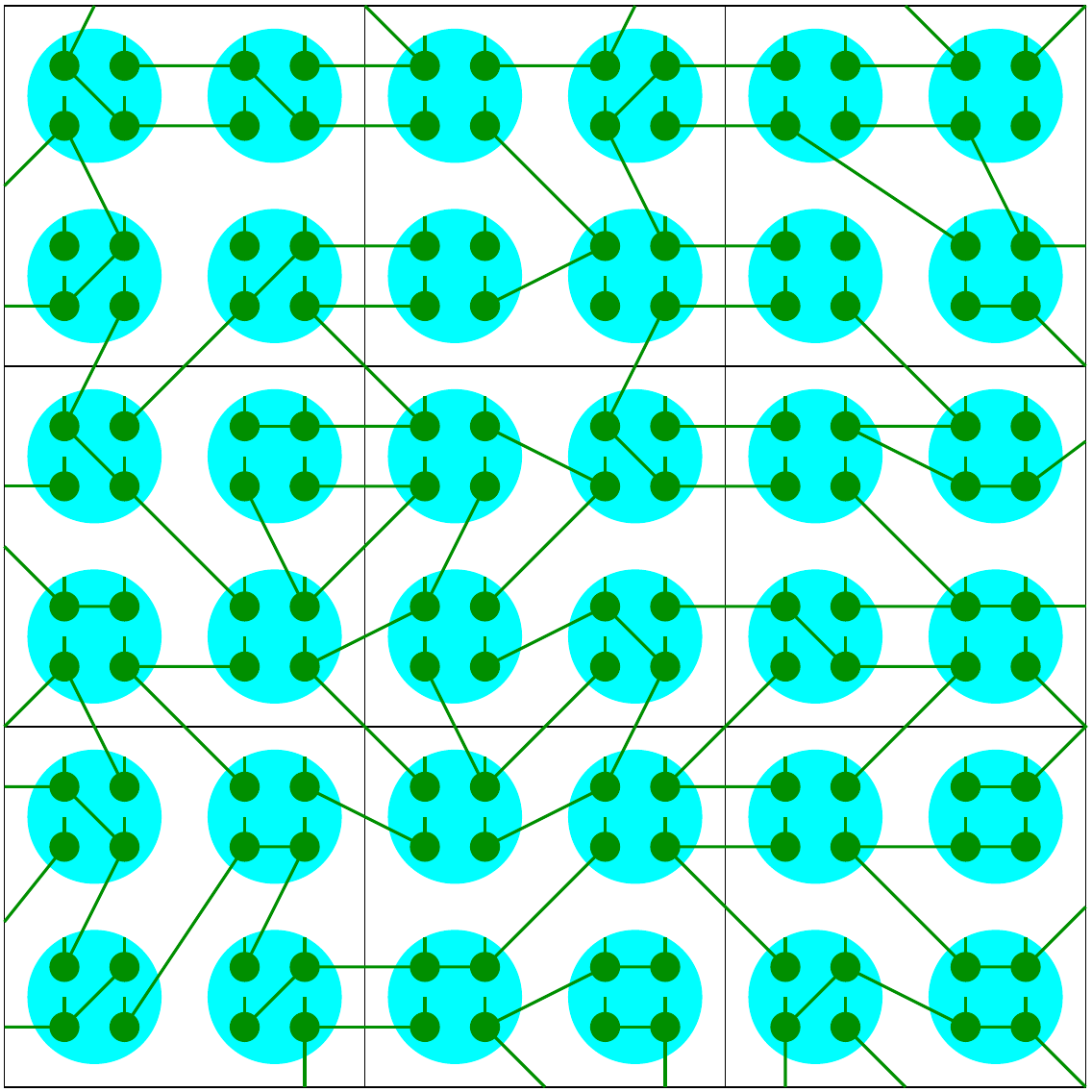}
%Fig. 6
\end{center}
\caption{
(Color online)
A tensor network representation of a SRE state with on-site symmetry $G$.  The
all the dots in each shaded circle form a site. The degrees of freedom on each
site (\ie in each  shaded circle) form a linear representation of $G$.
However, the degrees of freedom on each dot may not form a linear
representation of $G$.
}
\label{SREsymm}
\end{figure}

We have argued that generic LU transformations can change a SRE state in Fig.
\ref{cForm}a to a tensor-network state in Fig. \ref{cForm}b.  The  LU
transformations rearrange the spatial distributions of the entanglements which
should not be affected by the on-site symmetry $G$.  So, in the following, we
would like to argue that symmetric LU transformations can still change a SPT
state in Fig.  \ref{cForm}a to a symmetric tensor-network state in Fig.
\ref{cForm}b (although a generic proof is missing).

\begin{figure}[tb]
\begin{center}
\includegraphics[scale=1.3]{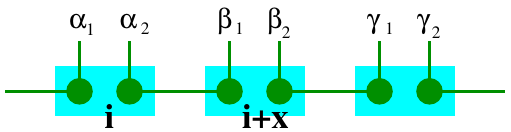}
%Fig. 7
\end{center}
\caption{
(Color online)
The canonical tensor network representation for 1D SRE state
$|\Psi_\text{pSRE}\>$.  The two dots in each rectangle represent a physical
site.
}
\label{csymm1D}
\end{figure}

We first assume that symmetric SRE states have tensor network
representation as shown in Fig. \ref{SREsymm}.  The linked dots
represent the entangled degrees of freedom.  The dots in each shaded
circle represent a site, which forms a linear representation of the
on-site symmetry group $G$.  We then divide the systems into large
squares (see Fig. \ref{SREsymm}).  The size of the square is large
enough such that entanglement only appears between squares that
share an edge or a vertex.  Now we view the degrees of freedom in
each square as a large effective site.  The  degrees of freedom on
each  effective site form a linear representation of $G$.  Now, we
can use an unitary transformation in each square to rearrange the
degrees of freedom in that square (which corresponds to change basis
in the large effective site).  This way, we can transform the SPT
state in Fig. \ref{SREsymm} into the canonical form in Fig.
\ref{cForm}b, where the degrees of freedom on each shaded square
form a linear representation of $G$.  So Fig. \ref{cForm}b is a
symmetric tensor-network state.  We would like to point out that
although in  Fig.  \ref{cForm}b, we only present a 2D tensor-network
state in canonical form,  the similar reduction can be done  in any
dimensions.

\section{Classify symmetry transformations of SPT states}
\label{clsymm}

\begin{figure}[tb]
\begin{center}
\includegraphics[scale=1.3]{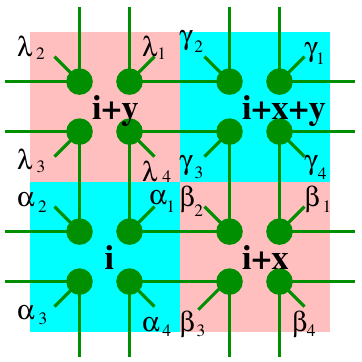}
%Fig. 8
\end{center}
\caption{
(Color online)
The canonical tensor network representation for 2D SRE state
$|\Psi_\text{pSRE}\>$.  The four dots in each square represent a physical site.
}
\label{csymm}
\end{figure}

After the symmetric state being reduced to the canonical form in Fig.
\ref{cForm}b, the on-site symmetry transformation is generated by the following
matrix on the effective site-$\v i$: $U^{\v i}_{\al_1 \al_2 \al_3
\al_4,\al_1'\al_2'\al_3'\al_4'}$ which forms a linear representation of the
on-site symmetry group $G$.  The symmetry transformation $U^{\v i}_{\al_1
\al_2 \al_3 \al_4,\al_1'\al_2'\al_3'\al_4'}$ keeps the SRE state
$|\Psi_\text{pSRE}\>$ in Fig.  \ref{csymm1D} or Fig.  \ref{csymm} invariant:
\begin{align}
\label{UPsi}
\otimes_{\v i}  U^{\v i} |\Psi_\text{pSRE}\>=|\Psi_\text{pSRE}\>
\end{align}
for any lattice size.

Eqn. (\ref{UPsi}) is one of the key equations.  It describes the condition that
the on-site symmetry transformations $U^{\v i}$ must satisfy so that those
on-site symmetry transformations can represent the symmetry of a SRE state. So
to classify all possible symmetry transformations of SPT states, we
need to find all the pairs $( U^{\v i}, |\Psi_\text{pSRE}\>)$ that satisfy
\eqn{UPsi}.  Those different solutions can correspond to different SRE
symmetric phases.

However, two different solutions $U^{\v i}$ may not correspond to different
phases.  They may be ``equivalent'' and can correspond to the same phase. So to
understand the structure of SRE symmetric phases, we also need to find out
those  ``equivalent'' relations.  Clearly one ``equivalent'' relation is
generated by unitary transformations
$
W^{\v i}_{\al_1 \al_2 \al_3 \al_4,\bt_1\bt_2\bt_3\bt_4}
$
on each effective physical site:
\begin{align}
\label{WUW}
&\ \ \ U^{\v i}_{\al_1 \al_2 \al_3 \al_4,\al_1'\al_2'\al_3'\al_4'}
\sim
\t U^{\v i}_{\al_1 \al_2 \al_3 \al_4,\al_1'\al_2'\al_3'\al_4'}
\\
& =
W^{\v i}_{\al_1 \al_2 \al_3 \al_4,\bt_1\bt_2\bt_3\bt_4}
 U^{\v i}_{\bt_1 \bt_2 \bt_3 \bt_4,\bt_1'\bt_2'\bt_3'\bt_4'}
W^{\v i \dag}_{\bt_1'\bt_2'\bt_3'\bt_4',\al_1'\al_2'\al_3'\al_4'}
\nonumber
\end{align}
where the repeated indices are summed over.  Here $W^{\v i}_{\al_1 \al_2 \al_3
\al_4,\bt_1\bt_2\bt_3\bt_4}$ are not the most general on-site unitary
transformations.  They are the on-site unitary transformations that map
$|\Psi_\text{pSRE}\>$ to another state $|\Psi'_\text{pSRE}\>$ having the same
form as described by Fig.  \ref{csymm1D} or Fig.  \ref{csymm}.

\begin{figure}[tb]
\begin{center}
\includegraphics[scale=1.0]{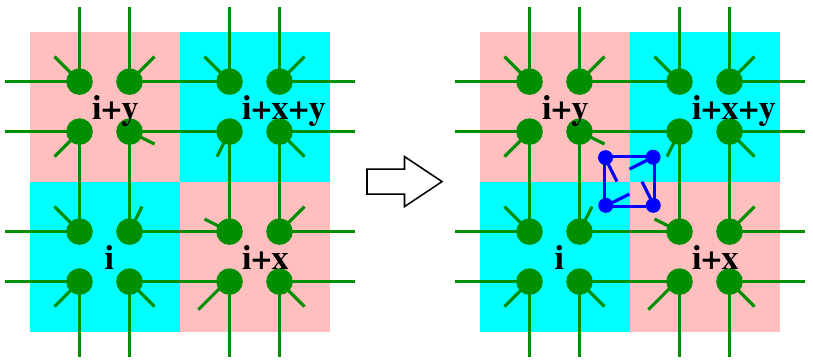}
%Fig. 9
\end{center}
\caption{
(Color online)
Adding four local degrees of freedom that form a 1D representation does not
change the phase of state.
}
\label{addsite}
\end{figure}

The second ``equivalent'' relation is given by
\begin{align}
\label{UWWWW}
&\ \ \ U^{\v i}_{\al_1 \al_2 \al_3 \al_4,\al_1'\al_2'\al_3'\al_4'}
\sim
\t U^{\v i}_{
\al_1\bt_1 \al_2\bt_2 \al_3\bt_3 \al_4\bt_4,
\al_1'\bt_1' \al_2'\bt_2' \al_3'\bt_3' \al_4'\bt_4'}
\nonumber\\
&=
U^{\v i}_{\al_1 \al_2 \al_3 \al_4,\al_1'\al_2'\al_3'\al_4'}
W^{1,\v i}_{\bt_1,\bt_1'}
W^{2,\v i}_{\bt_2,\bt_2'}
W^{3, \v i}_{\bt_3,\bt_3'}
W^{4,\v i}_{\bt_4,\bt_4'}
\end{align}
where $W^{a, \v i}_{\bt_a,\bt_a'}$, $a=1,2,3,4$
are linear representations of the on-site symmetry group $G$
which satisfy a condition that
the direct product representation
$
W^{1,\v i}\otimes
W^{2,\v i+\v x}\otimes
W^{3,\v i+\v x+\v y}\otimes
W^{4,\v i+\v y}
$
contains a trivial 1D representation:
\begin{align}
W^{1,\v i}\otimes
W^{2,\v i+\v x}\otimes
W^{3,\v i+\v x+\v y}\otimes
W^{4,\v i+\v y}
=1\oplus ...
\end{align}
Such an ``equivalent'' relation arises from the fact that adding local degrees
of freedom that form a 1D representation does not change the phase of state (see
Fig. \ref{addsite}) It is clear that if the transformations $U^{\v i}$ satisfy
\eqn{UPsi},  $\t U^{\v i}$ from the second ``equivalent'' relation also
satisfy \eqn{UPsi}.

The solutions of \eqn{UPsi} can be grouped into classes using the equivalence
relations \eqn{WUW} and \eqn{UWWWW}.  Those classes should correspond different
SPT states.

We note that the condition \eqn{UPsi} involves the whole many-body wave
function.  In appendix \ref{local}, we will show that the condition \eqn{UPsi}
can be rewritten as a local condition where only a local region of the
many-body wave function is used.
Although we only discuss the 2D case in the above,
similar result can be obtained in any dimensions.

The discussions in the last a few sections outline some ideas that may lead to
a classification of SPT phases.  In this paper, we will not attempt to directly
find all the solutions of  \eqn{UPsi} and to directly classify all the SPT
phases.  Instead, we will try to explicitly construct, as general as possible,
the solutions of  \eqn{UPsi}.  Our goal is to find a general construction that
produces all the possible solutions.

\section{Constructing SPT phases through group cocycles}
\label{cnsymm}

In this section, we will construct solutions of \eqn{UPsi} through the cocycles
of the symmetry group $G$.  The different solutions will correspond to
different SPT phases.

\subsection{Group cocycles}

The cocycles, cohomology group, and their graphic representations
on simplex with branching structure are discussed in appendix
\ref{Gcoh} and \ref{branchapp}.  Here we just briefly introduce those concepts.
A $d$-cochain of group $G$ is a complex function $\nu_d(g_0,g_1,...,g_d)$ of
$1+d$ variables in $G$ that satisfy
\begin{align}
\label{cchcnd}
 |\nu_d(g_0,g_1,...,g_d)|&=1,
\\
 \nu_d^{s(g)}(g_0,g_1,...,g_d)&=
 \nu_d(gg_0,gg_1,...,gg_d), \ \ g\in G
\nonumber
\end{align}
where $s(g)=1$ if $g$ contains no anti-unitary time reversal transformation $T$
and $s(g)=-1$ if $g$ contains one anti-unitary time reversal transformation
$T$.
[When $G$ is continuous, we do not require the cochain
$\nu_d(g_0,g_1,...,g_d)$ to be a continuous function of $g_i$.  Rather, we only
require $\nu_d(g_0,g_1,...,g_d)$ to be a so called \emph{measurable}
function of $g_i$.\cite{MFun} A measurable function
is not continuous only on a measure zero space.]

The $d$-cocycles are special $d$-cochains that satisfy
\begin{align}
\label{dnud}
&\ \ \
\prod_{i=0}^{d+1} \nu_d^{(-1)^i}( g_0,.., g_{i-1}, g_{i+1},...,
g_{d+1}) =1  .
\end{align}
For $d=1$,
the 1-cocycles satisfy
\begin{align}
\label{dnu1}
 \nu_1 ( g_1,  g_2)  \nu_1 ( g_0,  g_1)
/  \nu_1 ( g_0, g_2)=1
\end{align}
The 2-cocycles satisfy
\begin{align}
\label{dnu2}
\frac{ \nu_2 ( g_1,  g_2, g_3) \nu_2
( g_0, g_1, g_3)}{
 \nu_2 ( g_0,  g_2, g_3)  \nu_2 ( g_0, g_1, g_2) }=1
\end{align}
and the 3-cocycles satisfy
\begin{align}
\label{dnu3} \frac{
\nu_3(g_1,g_2,g_3,g_4)\nu_3(g_0,g_1,g_3,g_4)\nu_3(g_0,g_1,g_2,g_3)}{
 \nu_3(g_0,g_2,g_3,g_4)\nu_3( g_0,g_1,g_2,g_4) }=1
\end{align}

The $d$-coboundaries $\la_d$ are special $d$-cocycles that
can be constructed from the $(d-1)$-cochains $\mu_{d-1}$:
\begin{align}
&\ \ \
\la_d(g_0,...,g_d)=
\prod_{i=0}^{d} \mu_{d-1}^{(-1)^i}( g_0,.., g_{i-1}, g_{i+1},...,
g_{d})   .
\end{align}
For $d=1$, the 1-coboundaries are given by
\begin{align}
\label{la1}
\la_1(g_0,g_1)= \mu_0 ( g_1)  /  \mu_0 ( g_0)
\end{align}
The 2-coboundaries are given by
\begin{align}
\label{la2}
\la_2(g_0,g_1,g_2)= \mu_1 ( g_1,  g_2)  \mu_1 ( g_0,  g_1)
/  \mu_1 ( g_0, g_2) ,
\end{align}
and the 3-coboundaries by
\begin{align}
\label{la3}
\la_3(g_0,g_1,g_2,g_3)=
\frac{ \mu_2 ( g_1,  g_2, g_3) \mu_2
( g_0, g_1, g_3)}{
 \mu_2 ( g_0,  g_2, g_3)  \mu_2 ( g_0, g_1, g_2) }.
\end{align}

Two $d$-cocycles, $\nu_d$ and $\nu'_d$, are said to be equivalent iff
they differ by a coboundary $\la_d$:
$\nu_d=\nu'_d\la_d$.  The equivalence classes of
cocycles give rise to the $d$-cohomology group $\cH^d[G,U_T(1)]$.

\begin{figure}[tb]
\begin{center}
\includegraphics[scale=0.6]{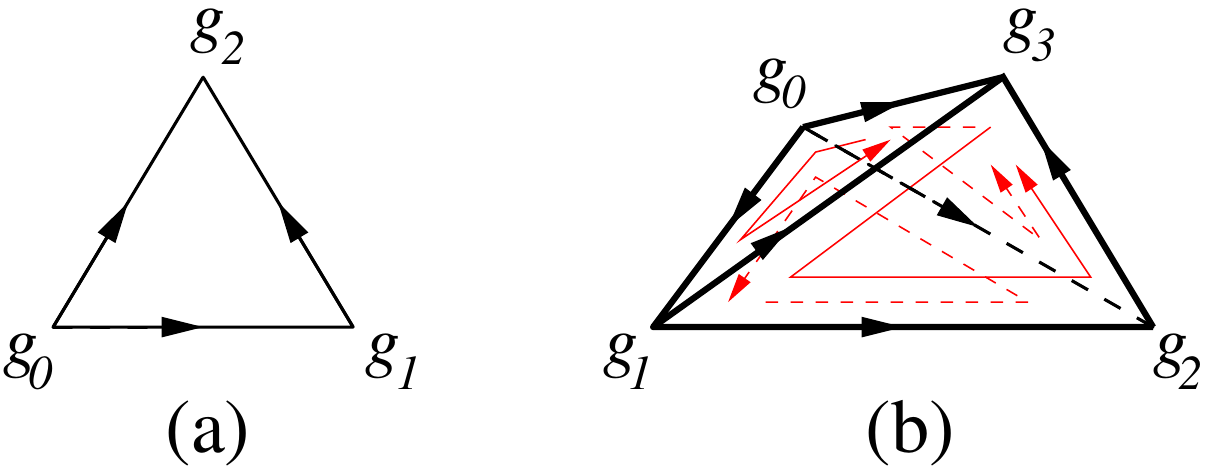}
%Fig. 10
\end{center}
\caption{
(Color online)
(a) The line from $g_0$ to $g_1$ is a graphic representation of $  \nu_1 ( g_0,
g_1)$.  The triangle $(g_0,g_1,g_2)$ with a branching structure (see appendix
\ref{branchapp}) is a graphic representation of $  \nu_2 (g_0,g_1,g_2)$.  Note
that, for the first variable, the $g_0$-vertex is connected to two outgoing
edges, and  for the last variable, the $g_2$-vertex is connected to two
incoming edges.
(a) can also be viewed
as the graphic representation of \eqn{dnu1} and \eqn{d1}.  The triangle
corresponds to $ (\dd_1 \nu_1) ( g_0,  g_1, g_2) $ in \eqn{d1} and the three
edges correspond to $\nu_1 ( g_1, g_2)$, $ \nu_1 ( g_0,  g_1)$ and $\nu_1^{-1}
( g_0, g_2)$.  (b) The tetrahedron $(g_0,g_1,g_2,g_3)$ with a branching
structure is a graphic representation of $ \nu_3 ( g_0,  g_1, g_2, g_3)$.  (b)
can also be viewed as the graphic representation of \eqn{dnu2} and \eqn{d2}.
The tetrahedron corresponds to $(\dd_2  \nu_2) ( g_0,g_1,g_2,g_3)$ in \eqn{d2},
and the four faces correspond to $\nu_2 ( g_1,  g_2, g_3)$, $\nu_2 ( g_0, g_1,
g_3)$, $ \nu_2^{-1} ( g_0, g_2, g_3)$, and $ \nu_2^{-1} ( g_0, g_1, g_2)$.
}
\label{d1d2}
\end{figure}

A $d$-cochain can be represented by a $d$-dimensional simplex with a branching
structure (see Fig.  \ref{d1d2}).  A branching structure (see appendix
\ref{branchapp}) is represented by arrows on the edges of the simplex that
never form an oriented loop on any triangles. We note that the first variable
$g_0$ in $\nu_d(g_0,g_1,...,g_{d})$ corresponds to the vertex with no incoming
edge, the second variable $g_1$ to the vertex with one incoming edge, and the
third variable $g_2$ to the vertex with two incoming edges, \etc.  The
conditions \eqn{dnu1} and \eqn{dnu2} can also be represented as in Fig.
\ref{d1d2}.  For example, Fig.  \ref{d1d2}a has three edges which correspond to
$\nu_1 ( g_1, g_2)$, $ \nu_1 ( g_0,  g_1)$ and $\nu_1^{-1} ( g_0, g_2)$.  The
evaluation of a $1$-cochain $\nu_1$ on the complex Fig.  \ref{d1d2}a is given
by the product of the factors $\nu_1 ( g_1, g_2)$, $ \nu_1 ( g_0,  g_1)$ and
$\nu_1^{-1} ( g_0, g_2)$.  Such an evaluation will be 1 if $\nu_1$ is a
cocycle.  In general, the evaluations of cocycles on any complex without
boundary are 1.

Such a geometric picture will help us to obtain most of the results in this paper.

\subsection{(1+1)D case}

Let us discuss the 1D case first.  We will choose the 1D SPT wave function to
have a fixed form of a ``dimer crystal'' (see Fig. \ref{csymm1D}):
\begin{align}
\label{pSRE1D}
 |\Psi_\text{pSRE}\> &=
...
\otimes (\sum_{g\in G} |\al_2=g,\bt_1=g\>)
\otimes
\nonumber\\
&
(\sum_{g\in G} |\bt_2=g,\ga_1=g\>)\otimes ...
\end{align}
where we have assumed that physical states on each dot in Fig. \ref{csymm1D}
are labeled by the elements of the symmetry group $G$: $\al_i,\bt_i \in G$.
The dimmer in  Fig. \ref{csymm1D} corresponds to a maximally entangled state
$\sum_{g\in G} |\al_2=g,\bt_1=g\>$.

\begin{figure}[tb]
\begin{center}
\includegraphics[scale=1.6]{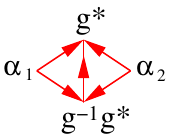}
%Fig. 11
\end{center}
\caption{
(Color online)
The graphic representation of \eqn{fom2om2}.  $f_2(\al_1,\al_2,g,g^*)$
is represented by the polygon with a branching structure as represented by the
arrows on the edge which never form a oriented loop on any triangle.
$\nu_2(\al_1,g^{-1}g^*,  g^* ) $ and
$ \nu_2(\al_2,g^{-1}g^*,  g^*) $
are represented by the two triangles as in Fig.
\ref{d1d2}a.  The value of the cocycle $\nu_2$ on a triangle (say
$\nu_2(\al_1,g^{-1}g^*,  g^*)$) can be viewed as flux going through the
corresponding triangle.
}
\label{fom2}
\end{figure}

Next, we need to choose an on-site symmetry transformation \eq{UPsi} such that
the state $|\Psi_\text{pSRE}\>$ is invariant (where the two dots in each shaded
box represent a site).  We note that $U^{\v i}(g)$ acts on the states on the
$\v i$ site which are linear combinations of $|\al_1,\al_2\>$ in Fig.
\ref{csymm1D}.  Note that $\al_1,\al_2 \in G$.  So we can choose the action of
$U^{\v i}(g)$ to be (see Fig. \ref{fom2})
\begin{align}
\label{alalU}
U^{\v i}(g) |\al_1,\al_2\>
=
f_2(\al_1,\al_2,g,g^*)
|g\al_1,g\al_2\>
\end{align}
where $f_2(\al_1,\al_2,g,g^*)$ is a phase factor $|f_2(\al_1,\al_2,g,g^*)|=1$.
We will use a 2-cocycle $\nu_2 \in \cH^2[G,U_T(1)]$ for the symmetry group $G$
to construct the phase factor $f_2$.  (A discussion of the group cocycles is
given in the appendix \ref{Gcoh}.)

Using a 2-cocycle $\nu_2$, we construct the phase factor $f_2$ as the follows
(see Fig. \ref{fom2}):
\begin{align}
\label{fom2om2}
 f_2(\al_1,\al_2,g,g^*)&=
\frac{\nu_2(\al_1,g^{-1}g^*,  g^*)}{ \nu_2( \al_2,g^{-1}g^*,  g^*)
}.
\end{align}
Here $g^*$ is a fixed element in $G$.  For example we may choose $g^*=1$.  In
appendix \ref{11D}, we will show that $U^{\v i}(g)$ defined above is indeed a
linear representation of $G$ that satisfies \eqn{UPsi}.  In this way, we obtain a
SPT phase described by $|\Psi_\text{pSRE}\>$ that transforms as $U^{\v i}(g)$.

Note that here we only discussed a fixed SRE wave function. If
we choose different cocycles in \eqn{fom2om2}, the same wave function
\eq{pSRE1D} can indeed represent different phases. One may wonder how
a fixed SRE wave function can represent different quantum phases.

To see this, let us examine how the state varies under the symmetry group.
Notice that the phase factor $ f_2(\al_1,\al_2,g,g^*)$ is factorized, the
basis $|\al_1\rangle$ varies as
\begin{align*}
M(g)|\al_1\rangle&= \nu_2(\al_1,g^{-1}g^*,  g^*)|g\al_1\rangle.%,\\
\end{align*}
The states $|\al_1\rangle$ form a representation of $G$ itself, and the
operator $g$ transforms a state into another.  The representation matrix
element is given as $M(g)_{\al_1,g\al_1}=\nu_2(g^{-1}g^*,  g^*,\al_1)$, and
eqn.(\ref{fom2om2}) can be rewritten as $
f_2(\al_1,\al_2,g,g^*)=M(g)_{\al_1,g\al_1}[M(g)_{\al_2,g\al_2}]^\dag$.  From
eqn.(\ref{alalU}) we have $U^i(g)=M(g)\otimes [M(g)]^\dag$.  Actually, this
matrix $M(g)$ is a projective representation of the group $G$, corresponding to
the 2-cocycle $\nu_2$.

Different classes of cocycles $\nu_2$ correspond to different projective
representations. In the trivial case, where $\nu_2(\al_1,g^{-1}g^*, g^*)=1$,
$M(g)$ can be reduced into linear representations, and the corresponding SPT
phase is a trivial phase.

We will also show, in appendix \ref{11D}, that on a finite segment of chain,
the state $|\Psi_\text{pSRE}\>$ has low energy excitations on the chain end.
The excitations on one end of the chain form a projective representation
described by the same cocycle $\nu_2$ that is used to construct the solution
$U^{\v i}(g)$.  The end states and their projective representation describe
the universal properties of bulk SPT phase.

The different solutions of \eqn{UPsi}
constructed from different 2-cocycles do not always
represent different SPT phases.
If $\nu_2( g_0,  g_1, g_2)$ satisfies \eqn{cchcnd} and \eqn{dnu2}, then
\begin{align}
 \nu_2'( g_0,  g_1, g_2)=
 \nu_2( g_0,  g_1, g_2)
\frac{\mu_1(g_1,g_2) \mu_1(g_0,g_1)}{\mu_1(g_0,g_2)}
\end{align}
also satisfies \eqn{cchcnd} and \eqn{dnu2}, for any $\mu_1(g_0,g_1)$ satisfying
$\mu_1(gg_0,gg_1)=\mu_1^{s(g)}(g_0,g_1)$, $g\in G$.  So $\nu_2'( g_0,  g_2,
g_3)$ also gives rise to a solution of \eqn{UPsi}.  But the two solutions
constructed from $ \nu_2( g_0,  g_1, g_2)$ and $ \nu_2'( g_0,  g_1, g_2)$ are
related by a symmetric LU transformations
(for details, see discussion near the end of
appendix \ref{MPUO}).  They are also smoothly
connected since we can smoothly deform $\mu_1(g_0,g_1)$ to $\mu_1(g_0,g_1)=1$.
So we say that the two solutions obtained from $\nu_2(
g_0, g_2, g_3)$ and  $\nu_2'( g_0,  g_2, g_3)$ are equivalent.  We note that
$\nu_2( g_0,  g_2, g_3)$ and  $\nu_2'( g_0,  g_2, g_3)$ differ by a
2-coboundary $\frac{\mu_1(g_1,g_2) \mu_1(g_0,g_1)}{\mu_1(g_0,g_2)}$.  So the set of
equivalence classes of $\nu_2( g_0,  g_2, g_3)$ is nothing but the cohomology
group $\cH^2[G,U_T(1)]$.  Therefore, the different SPT phases are classified by
$\cH^2[G,U_T(1)]$.

We see that, in our approach here, the different SPT phases are not encoded in
the different wave functions, but encoded in the different methods of fractionalizing the
symmetry transformations $U^{\v i}(g)$.

\subsection{(2+1)D case}
\label{2+1Dcase}

\begin{figure}[tb]
\begin{center}
\includegraphics[scale=1.4]{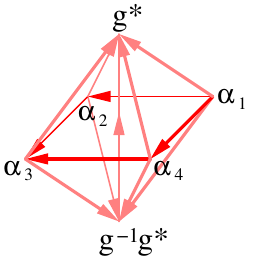}
%Fig. 12
\end{center}
\caption{ (Color online) The graphic representation of the phase
factor $ f_3(\al_1,\al_2,\al_3,\al_4, g,g^*) $ in \eqn{fom3om3}.
The arrows on the edges that never form a oriented loop on any
triangle represent the branching structure on the complex. The four
tetrahedrons give rise to $\nu_3( \al_1,\al_2,g^{-1}g^*,g^*)$,
$\nu_3( \al_2,\al_3,g^{-1}g^*,g^*)$, $\nu_3^{-1}(
\al_4,\al_3,g^{-1}g^*,g^*)$, and $\nu_3^{-1}(
\al_1,\al_4,g^{-1}g^*,g^*)$ } \label{fom3}
\end{figure}

The above discussion and result can be generalized to higher dimensions.
Here we will discuss 2D SPT state as an example.
We choose the 2D SPT state to be  a ``plaquette state'' (see Fig. \ref{csymm})
\begin{align}
\label{pSRE2D}
 |\Psi_\text{pSRE}\> &=
\otimes_\text{squares} (\sum_{g\in G} | \al_1=g,\bt_2=g,
\ga_3=g,\la_4=g \>)
\end{align}
where we have assumed that physical states on each dot in Fig.
\ref{csymm} are labeled by the elements of the symmetry group $G$:
$\al_i,\bt_i,... \in G$. The four dots in a linked square in  Fig.
\ref{csymm} form a maximally entangled state $\sum_{g\in G}
|\al_1=g,\bt_2=g, \ga_3=g,\la_4=g \>$.  We require that the state
$|\Psi_\text{pSRE}\>$ is invariant under an on-site symmetry
transformation \eq{UPsi} (where the four dots in each shaded square
represent a site).

To construct an on-site symmetry transformation
\eq{UPsi},
in 2 dimensions, the action of $U^{\v i}$ is chosen to be
\begin{align}
\label{fom3om3}
& \ \ \ \
U^{\v i}(g)
|\al_1,\al_2,\al_3,\al_4\>
\\
&=
f_3(\al_1,\al_2,\al_3,\al_4,
g,g^*)
|g\al_1,g\al_2,g\al_3,g\al_4\>.
\nonumber
\end{align}
Here $f_3(\al_1,\al_2,\al_3,\al_4, g,g^*)$ is a phase factor that
corresponds to the value of a 3-cocycle $\nu_3\in \cH^3[G,U_T(1)]$
evaluated on the complex with a branching structure in Fig.
\ref{fom3}:
\begin{align}\label{eq: sym2D}
&\ \ \ \
f_3(\al_1,\al_2,\al_3,\al_4, g,g^*)
\nonumber\\
&=
\frac{
\nu_3( \al_1,\al_2,g^{-1}g^*,g^*)
\nu_3( \al_2,\al_3,g^{-1}g^*,g^*)}{
\nu_3( \al_4,\al_3,g^{-1}g^*,g^*)
\nu_3( \al_1,\al_4,g^{-1}g^*,g^*)
} .
\end{align}
In appendix \ref{12D}, we will show that $U^{\v i}(g)$ defined above
is indeed a linear representation of $G$ that satisfies \eqn{UPsi}.
We will also show that (see appendix \ref{MPUO} and \Ref{CLW1152})
\emph{
in a basis where the many-body ground state is a simple product state, although
$\otimes_{\v i} U^{\v i}(g)$ is an on-site symmetry transformation when acting
on the bulk state, it cannot be an on-site symmetry transformation when viewed
as a symmetry transformation acting on the effective low energy degrees of
freedom on the boundary when the 3-cocycle $\nu_3$ is non-trivial.
}

\section{SPT phases and
topological non-linear $\si$-models}

\label{clnlm}

\subsection{The fixed-point action that does not depend on the
space-time metrics}

In the above,  we have constructed SPT states and their symmetry
transformations using the cocycles of the symmetry group.  We can easily find
the Hamiltonians such that the constructed SPT states are the exact
ground states.  In the following, we are going to discuss a Lagrangian
formulation of the construction.  We will systematically construct models in
$d+1$ space-time dimensions that contain SPT orders characterized by elements
in $\cH^{1+d}[G,U_T(1)]$.  It turns out that the  Lagrangian formulation is
simpler than the Hamiltonian formulation.

A SPT phase can be described by a non-linear $\si$-model
of a field $\v n(\v x,\tau)$,
whose imaginary-time path integral is given by
\begin{align}
Z=\int D\v n\ \e^{-\int \dd^d \v x \dd \tau\; \cL[\v n(\v x,\tau)]}.
\end{align}
We will call the term $\e^{- \int \dd^d \v x \dd \tau\; \cL[\v n(\v x,\tau)]}$
the action-amplitude.  The imaginary-time evolution operator from $\tau_1$ to
$\tau_2$, $U[ \v n_2(\v x), \v n_1(\v x), \tau_2,\tau_1]$, can also be expressed
as a path integral
\begin{align}
\label{Upath}
U[ \v n_2(\v x), \v n_1(\v x), \tau_2,\tau_1]
=\int D\v n\
\e^{-\int \dd^d \v x \int_{\tau_1}^{\tau_2} \dd \tau\; \cL[\v n(\v x,\tau)]}
\end{align}
with the boundary condition
$\v n(\v x,\tau_1)=\v n_1(\v x)$ and
$\v n(\v x,\tau_2)=\v n_2(\v x)$.

If the model has a symmetry, the field $\v n$ transforms as $\v n \to
g\cdot \v n$ under the symmetry transformation $g\in G$.  The action-amplitude
has the $G$ symmetry
\begin{align}
 \e^{ - \int \dd^d \v x \dd \tau\; \cL[\v n(\v x,\tau)]}
=\e^{ - \int \dd^d \v x \dd \tau\; \cL[g\cdot \v n(\v x,\tau)]} .
\end{align}
To understand the low energy physics, we concentrate on the ``orbit'' generated
by $G$ from a fixed
$\v n_0$: $\{g\cdot \v n_0; g\in G\}$.
Such an ``orbit'' is a symmetric space $G/H$ where $H$ is the subgroup of $G$ that keeps
$\v n_0$ invariant: $H=\{h; h\cdot \v n_0=\v n_0, h\in G\}$. We can always add degrees
of freedom to expand the symmetric space $G/H$ to the maximal symmetric space,
which is the whole space of the group $G$. So to study SPT phase, we can always
start with a  non-linear $\si$-model whose field takes value in the symmetry
group $G$, the maximal symmetric space. Such a non-linear $\si$-model is
described by a path integral
\begin{align}
Z=\int Dg\ \e^{- \int \dd^d \v x \dd \tau\; \cL[g(\v x,\tau)]},\ \ \ \
g\in G.
\end{align}

We would like to consider non-linear $\si$-models that describe a SRE phase
with finite energy gap and finite correlations. So a low energy fixed point
action-amplitude $\e^{- \int \dd^d \v x \dd \tau\; \cL[g(\v x,\tau)]}$ must not
depend on the space-time metrics.  In other words, the fixed-point non-linear
$\si$-model must be a topological quantum field theory.\cite{W8951}  We will
call such non-linear $\si$-model a topological non-linear $\si$-model.  A
trivial topological non-linear $\si$-model is given by the following
fixed-point Lagrangian $\cL_\text{fix}[g(\v x,\tau)]=0$ which describes
the trivial SPT phase.

A non-trivial topological non-linear $\si$-model has a non-zero Lagrangian
$\cL_\text{fix}[g(\v x,t)]\neq 0$.  However, the corresponding fixed-point
action-amplitude $\e^{- \int \dd^d \v x \dd \tau\; \cL[g(\v x,\tau)]}$ does not
depend on the space-time metrics.  One possible form of the fixed-point
Lagrangian $\cL_\text{fix}[g(\v x,\tau)]$ is given by a pure topological
$\th$-term.  As stated in section \ref{AppBerry}, the origin of the topological
$\th$-term may be the Berry phase in coherent state path integral.  For a
continuous non-linear $\si$-model whose field takes values in a continuous
group $G$, the topological $\th$-term is described by the action-amplitude
$\e^{ \int \dd^d \v x \dd \tau\; \cL_\text{topo}[g(\v x,\tau)]}$  that only
depends on the mapping class from the space-time manifold $M$ to the group
manifold $G$.  Such kind of topological term is given by a closed $(1+d)$-form
$\om_{1+d}$ on the group manifold $G$ which is classified by $H^{1+d}(G,\R)$.
The corresponding action is given by $ \int \dd^d \v x \dd \tau\;
\cL_\text{topo}[g(\v x,\tau)]=\int \om_{1+d}$.

The other possible form of the fixed-point Lagrangian $\cL_\text{fix}[g(\v
x,\tau)]$ is given by a WZW term.\cite{WZ7195,W8322} The
WZW term is described by the action $S_\text{WZW}$ that cannot be
expressed as a local integral on the space-time manifold $M$. That is to say, we cannot
express  $ S_\text{WZW}$ as $S_\text{WZW}=\int \dd^d \v x \dd \tau\;
\cL_\text{WZW}[g(\v x,\tau)]$. We have to view the space-time manifold $M$ as a
boundary of another manifold $M_\text{ext}$ in one higher dimensions, $M=\prt
M_\text{ext}$, and extend the field on $M$ to a field on $M_\text{ext}$.  Then
the WZW term $ S_\text{WZW}$ can be expressed as a local
integral on the extended  manifold $M_\text{ext}$:
\begin{align}
 S_\text{WZW}=\int_{M_\text{ext}}
\dd^{1+d} \v x \dd \tau\; \cL_\text{WZW}[g(\v x,\tau)]
\end{align}
such that  $S_\text{WZW} $ mod $2\pi \imth$ does not depend on how we extend
$M$ to $M_\text{ext}$.  A WZW term is given by a quantized
closed $(d+2)$-form $\om_{d+2}$ on the group manifold $G$:
\begin{align}
 S_\text{WZW}=\int_{M_\text{ext}} \om_{d+2} ,
\end{align}
which clearly does not depend on the space-time metrics.
Later, we will show that WZW terms in $(d+1)$-dimension space-time and for
group $G$ are classified by the elements in $\cH^{d+2}[G,U_T(1)]$.

We see that both the  topological $\th$-term and the WZW term do not
depend on the space-time metrics.  So the fixed-point Lagrangian may be given
by a pure topological $\th$-term and/or a pure WZW term.

\subsection{Lattice non-linear $\si$-model}

We would like to stress that the topological $\th$-term and the WZW
term discussed above require both continuous group manifold and continuous
space-time manifold.

On the other hand, in this paper, we are considering quantum disordered states
that do not break any symmetry.  So the field $g(\v x,\tau)$ fluctuates strongly
at all length scale.  The low energy effective theory has no smooth limit.
Therefore, the low energy effective theory must be one defined on
discrete space-time.

For discrete space-time, we no longer have non-trivial mapping class from
space-time to the group $G$, and we no longer have topological $\th$-term and
WZW term.  In this section we will show that although a generic topological
$\th$-term cannot be defined for discrete space-time, we can construct a new
topological term on discrete space-time that corresponds to a quantized
topological $\th$-term in the limit of continuous space-time.  Here a quantized
topological $\th$-term is defined as a topological $\th$-term that always gives
rise to an action-amplitude $\e^{ \int \dd^d \v x \dd \tau\;
\cL_\text{fix}[g(\v x,\tau)]} = 1$ on closed space-time.
We will also call the new topological term on discrete space-time a
quantized topological $\th$-term.

\begin{figure}[tb]
\begin{center}
\includegraphics[scale=0.6]{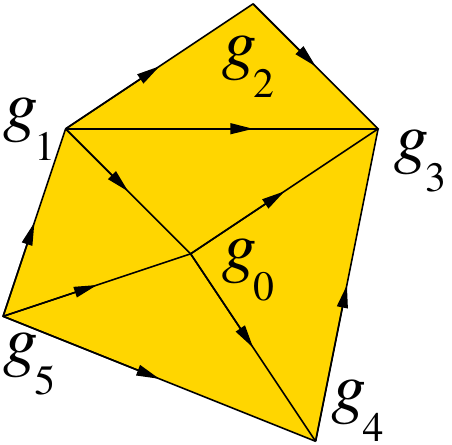}
%Fig. 13
\end{center}
\caption{ (Color online) The graphic representation of the action-amplitude
$\e^{-S (\{g_i\}) }$ on a complex with a branching structure represented by the
arrows on the edge.  The vertices of the complex are labeled by $i$.  Note that
the arrows never for a loop on any triangle.
} \label{Lg}
\end{figure}

To understand  the new topological term on discrete space-time, let us start
with a continuous non-linear $\si$-model whose field takes values in a group
$G$: $g(\v x,t)$. The imaginary-time path integral of the model is given by
\begin{align}
Z=\int Dg\ \e^{- \int \dd^d \v x \dd \tau\; \cL[g(\v x,\tau)]}
\end{align}
with a symmetry described by $G$:
\begin{align}
\e^{ - \int \dd^d \v x \dd \tau\; \cL[g(\v x,\tau)]}
=
\e^{ - \int \dd^d \v x \dd \tau\; \cL[g g(\v x,\tau)]},
\ \ \ \ g\in G .
\end{align}
If we discretize the space-time into a complex with a branching
structure (such as the complex obtained by a triangularization of
the space-time manifold), the path integral can be rewritten as (see
Fig. \ref{Lg})
\begin{align}
\label{AAdis}
Z &=|G|^{-N_v}\sum_{\{g_i\}} \e^{ - S(\{g_i\}) }
\nonumber\\
 \e^{ - S(\{g_i\}) }&=\prod_{\{ij...k\}} \nu_{1+d}^{s_{ij...k}}(g_i,g_j,...,g_k)
\end{align}
where $ \e^{ - S(\{g_i\}) }$ is the action-amplitude on the discretized
space-time that corresponds to $\e^{- \int \dd^d \v x \dd \tau\; \cL[g(\v
x,\tau)]} $ of the continuous non-linear $\si$-model, and
$\nu_{1+d}^{s_{ij...k}}(g_i,g_j,...,g_k)$ corresponds to the action-amplitude
$\e^{- \int_{(i,j,...,k)} \dd^d \v x \dd \tau\; \cL[g(\v x,\tau)]}$ on a single
simplex $(i,j,...,k)$.
Also, $s_{ij...k}=\pm 1$ depending on the orientation
of the simplex (which will be explained in detail later).

Here on each vertex of the space-time complex, we have a $g_i \in G$. $g_i$
corresponds to the field $g(\v x,t)$ and  $\sum_{\{g_i\}}$ corresponds to the
path integral $\int Dg$ in the continuous non-linear $\si$-model.
$|G|$ is the number of elements in $G$, $N_v$ is the number of
vertices in the complex.

We note that, on discrete space-time, the model can be defined for both
continuous group and discrete group.  When $G$ is a continuous group, 
$|G|^{-1}\sum_{g_i}$ should be interpreted as an integral over the group
manifold $\int \dd g_i$.

We see that a non-linear $\si$-model on $(d+1)$D discrete space-time is
described by a complex function $\nu_{1+d}(g_i,g_j,...,g_k)$.  Different choices
of $\nu_{1+d}(g_i,g_j,...,g_k)$ give different theories/models.

So we would like to ask: what
$\nu_{1+d}(g_i,g_j,...,g_k)$ will give rise to a quantized topological $\th$-term
on discrete space-time?  Very simply, we need to choose
$\nu_{1+d}(g_i,g_j,...,g_k)$ so that
\begin{align}
 \prod_{\{ij...k\}} \nu_{1+d}^{s_{ij...k}}(g_i,g_j,...,g_k)
=1
\end{align}
on every closed  space-time complex without boundary.

There are uncountably many choices of $\nu_{1+d}(g_i,g_j,...,g_k)$ that
satisfy the above condition and give rise to quantized topological $\th$-terms.
However, we can group them into equivalent classes, and each class corresponds
to a type of quantized topological $\th$-terms.  We will show later that the
types of quantized topological $\th$-terms are classified by
$\cH^{1+d}[G,U_T(1)]$.  So we can have non-trivial quantized topological
$\th$-terms discrete space-time only when $\cH^{1+d}[G,U_T(1)]$ is non trivial.
The number of equivalence classes of non-trivial quantized topological $\th$-terms is given by the
number of the non-trivial elements in $\cH^{1+d}[G,U_T(1)]$.

From the above discussion, it is also clear that we cannot generalize
un-quantized topological $\th$-terms to discrete space-time.  So on discretized
space-time complex, the only possible topological $\th$-terms are the quantized
ones.

After generalizing quantized topological $\th$-terms to discrete space-time, we
can now generalize WZW term to discrete space-time. We will call the
generalized WZW term a non-local Lagrangian (NLL) term.  To construct a NLL
term on a closed $(d+1)$D space-time complex $M_{d+1}$, we first view $M_{d+1}$
as a boundary of a  $(d+2)$D space-time complex $M_{d+2}$.  We then choose a
function $\nu_{2+d}(g_i,g_j,...,g_k)$ that defines a quantized topological
$\th$-term on the $(d+2)$D space-time complex $M_{d+2}$.  Then the
action-amplitude of $\nu_{2+d}(g_i,g_j,...,g_k)$ on $M_{d+2}$ only depends on
the $g_i$ on $M_{d+1}=\prt M_{d+2}$, the boundary of $M_{d+2}$. Thus such an
action-amplitude actually defines a theory on the $(d+1)$D space-time complex
$M_{d+1}$. Such an action-amplitude is the NLL term on the space-time complex
$M_{d+1}$.  We will see that the types of NLL terms on $(d+1)$D space-time
complex and for group $G$ are classified by $\cH^{2+d}[G,U_T(1)]$.

We would like to stress that the proper topological non-linear $\si$-models are
for disordered phases, and they must be defined on discrete space-time.  Only
quantized topological $\th$-terms can be defined on discrete space-time.  On
the other hand, the WZW term can always be generalized to discrete space-time,
which is called NLL term.  Both quantized topological $\th$-terms and NLL terms
on discrete space-time can be defined for \emph{discrete} groups.

\subsection{Quantized topological $\th$-terms lead to gapped SPT phases}

We know that the action-amplitude defines a physical model, in particular,
defines imaginary-time evolution operator $U(\tau_1,\tau_2)$.  For a SPT phase,
its fixed point action-amplitude must have the following properties (on a
closed spatial complex):\\ (a) The singular values of the imaginary-time
evolution operator $U[ g_i(\tau_1), g_i(\tau_2),\tau_1,\tau_2]$ are 1's or
0's.\\ (b) The singular values of the imaginary-time evolution operator $U[
g_i(\tau_1), g_i(\tau_2),\tau_1,\tau_2]$ contain only one 1.

Usually, the imaginary-time evolution operator is given by
$U(\tau_1,\tau_2)=\e^{-(\tau_2-\tau_1) H}$.  One expects that the log of the
eigenvalues of $U(\tau_1,\tau_2)$ correspond to the negative energies.
However, in general, the basis of the Hilbert space at different time $\tau$
can be chosen to be different.  Such a time dependent choice of the basis
corresponds to adding a total time derivative term to the Lagrangian $\cL\to
\cL+\frac{\dd  F}{\dd \tau}$.  It is well known that adding a total time
derivative term to the Lagrangian does not change any physical properties.  For
such more general cases, the log of the eigenvalues of $U(\tau_1,\tau_2)$ do
not correspond to the negative energies, since the eigenvalues of
$U(\tau_1,\tau_2)$ may be complex numbers.  In those cases, the log of the
singular values of $U(\tau_1,\tau_2)$ correspond to the negative energies.
This is why we use the singular values of $U(\tau_1,\tau_2)$ instead of the
eigenvalues of $U(\tau_1,\tau_2)$.

At the low energy fixed point of a gapped system, the fixed-point energies are
either 0 or infinite. Thus the singular values of the imaginary-time evolution
operator are either 1 or 0.  For a SPT phase without any intrinsic topological
order and without any symmetry breaking, the ground state degeneracy on a
closed spatial complex is always one.  Thus the singular values of the
imaginary-time evolution operator contain only one 1.

For the action-amplitude given by a quantized topological $\th$-term, its
corresponding imaginary-time evolution operator does have a property that its
singular values contain only one 1 and the rest are 0's.  This is due to the
fact that the action-amplitude for each closed path is always equal to 1.  So a
quantized topological $\th$-term indeed describes a SPT state.

\subsection{NLL terms lead to gapless excitations or degenerate boundary
states}

On the other hand, if the fixed-point action-amplitude in $(d+1)$ space-time
dimension is given by a pure NLL term, its corresponding imaginary-time
evolution operator, we believe, does not have the property that its singular
values contain only one 1 and the rest are 0's, since the action-amplitude for
different closed paths can be different.

In addition, if the pure NLL term corresponds to a non-trivial cocycle
$\nu_{d+2}$ in $\cH^{d+2}[G,U_T(1)]$, adding different coboundary to
$\nu_{d+2}$ will lead to different action-amplitude on closed paths.  There is
no coboundary that we can add to the cocycle $\nu_{d+2}$ to make the
action-amplitude for closed paths all equal to 1.  Further more, a
renormalization group flow only adds local Lagrangian term $\del \cL$ that is
well defined on the space-time complex. The renormalization group flow cannot
change the NLL term and cannot change the corresponding  cocycle $\nu_{d+2}$,
which is defined in one higher dimensions.  This leads us to conclude that an
action-amplitude with a NLL term cannot describe a SPT state.  Therefore
\[\frm{\emph{ An action-amplitude with a NLL term must have gapless
excitations, or degenerate boundary ground states due to symmetry breaking 
and/or topological order.
}} \]

The above is a highly non-trivial conjecture.  Let us examine its validity for
some simple cases.  Consider a $G$ symmetric non-linear $\si$-model in (1+0)
dimension which is described by an action-amplitude with a NLL term.  In (1+0)
dimension, the  NLL term is classified by 2-cocycles $\nu_2$ in
$\cH^2[G,U_T(1)]$, which correspond to the projective representations of the
symmetry group $G$.  So the ground states of the non-linear $\si$-model form a
projective representation of $G$ characterized by the same 2-cocycle $\nu_2$.
Since projective representations are always more than one dimension, (1+0)D
systems with  NLL terms cannot have a non-degenerate ground state.  In
(1+1)-dimension, continuous non-linear $\si$-models with the  WZW term are
shown to be described by the current algebra of the continuous symmetry group
and are gapless.\cite{W8322} In \Ref{CLW1152}, we further show that lattice
non-linear $\si$-models with the NLL term in (1+1)D must be gapless if the
symmetry is not broken, for both continuous and discrete symmetry. The above
conjecture generalize such a result to higher dimensions.

We note that the boundary excitations of the SPT phases characterized by
$(1+d)$-cocycle $\nu_{1+d}$ are described by an effective boundary non-linear
$\si$-model that contains a NLL term characterized by the same $(1+d)$-cocycle
$\nu_{1+d}$.

As discussed before, a non-linear $\si$-model with a non-trivial NLL term
cannot describe a SPT state.  Thus the boundary state must be gapless, or break
the symmetry, or have degeneracy due to non-trivial topological order.
However, the SPT state is a direct product state.  The degrees of freedom on
the boundary also form a product state.  Therefore the boundary state must be
gapless, or break the symmetry.  Thus, \[\frm{\emph{ A non-trivial SPT state
described by a non-trivial $(1+d)$-cocycle must have gapless 
excitations or degenerate ground states at the boundary.}} \]

We would like to stress that the symmetry plays a very important role in the
above discussion. It is the reason why the non-linear $\si$-model field $g_i$
takes many different values.  If there was no symmetry, at low energies, the
non-linear $\si$-model field $g_i$ would only take a single value that
minimizes the local potential energy.  In this case, there were no non-trivial
topological terms.

\section{Constructing symmetric fixed-point path integral through the cocycles
of the symmetry group} \label{const}

In the last section, we argue that SPT phases in $d$-dimension with on-site
symmetry $G$ are described by quantized topological $\th$-terms.  In this
section, we are going to explicitly construct quantized topological $\th$-terms
that realize those SPT orders in each space-time dimension.  We will also show
that the quantized topological $\th$-terms are classified by
$\cH^{1+d}[G,U_T(1)]$.

\subsection{(1+1)D symmetric fixed-point action-amplitude}

Let us first discuss (1+1)D fixed-point action-amplitude with a symmetry group
$G$. For a $(1+1)$D system on a complex with a branching structure, a
fixed-point action-amplitude (\ie a quantized topological $\th$-term) has a
form (see Fig.  \ref{Lg})
\begin{align}
\label{Lnu2}
&\ \ \ \ \e^{-S(\{g_i\}) }=\prod_{\{ijk\}} \nu_2^{s_{ijk}}(g_i,g_j,g_k)
\nonumber\\
&=
\nu_2^{-1} (g_1,g_2,g_3)
\nu_2 (g_0,g_4,g_3)
\nu_2^{-1} (g_5,g_0,g_1,) \times
\nonumber\\
&\ \ \ \ \ \ \ \
\nu_2 (g_1,g_0,g_3)
\nu_2^{-1} (g_5,g_0,g_4)
\end{align}
where each triangle contributes to a phase factor
$\nu_2^{s_{ijk}}(g_i,g_j,g_k)$, $\prod_{\{ijk\}}$ multiply over all the
triangles in the complex Fig.  \ref{Lg}.  Note that the first variable $g_i$ in
$\nu_2(g_i,g_j,g_k)$ corresponds to the vertex with two out going edges, the
second variable $g_j$ to the vertex with one out going edge, and the third
variable $g_k$ to the vertex with no out going edge.  $s_{ijk}=\pm 1$
depending on the orientation of $i\to j \to k$ to be anti-clock-wise or
clock-wise.

\begin{figure}[tb]
\begin{center}
\includegraphics[scale=0.7]{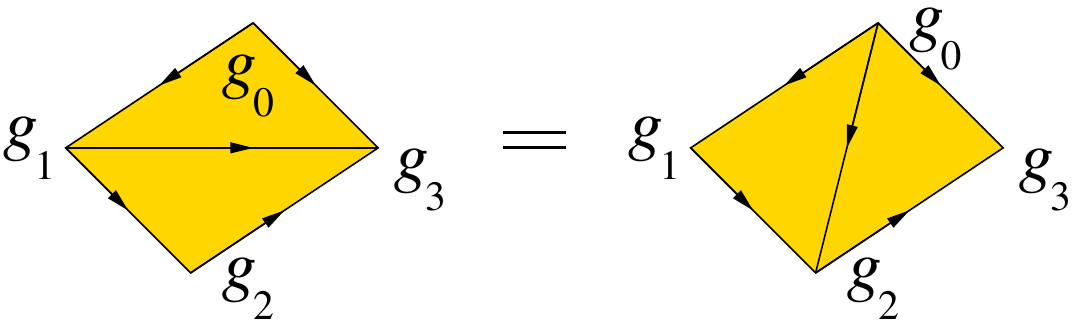}
%Fig. 14
\end{center}
\caption{
(Color online)
Graphic representation of
 $ \nu_2(g_0,g_1,g_2) \nu_2(g_0,g_2,g_3) = \nu_2(g_1,g_2,g_3)
\nu_2(g_0,g_1,g_3)$
The arrows on the edges represent the branching structure.
}
\label{Tflip}
\end{figure}

\begin{figure}[tb]
\begin{center}
\includegraphics[scale=0.7]{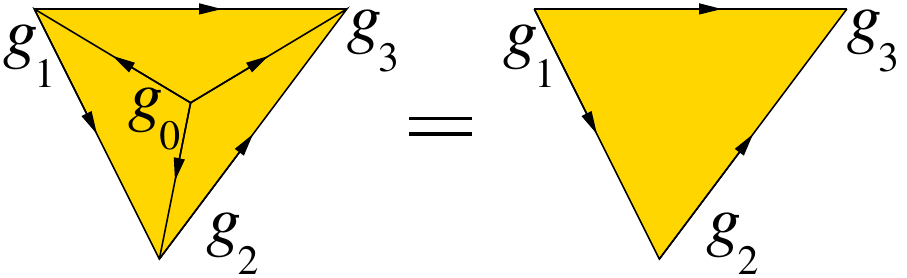}
%Fig. 15
\end{center}
\caption{
(Color online)
Graphic representation of
 $ \nu_2(g_1,g_2,g_3)= \nu_2(g_0,g_1,g_2) \nu_2(g_0,g_2,g_3) \nu_2^{-1}(g_0,g_1,g_3)$.
The arrows on the edges represent the branching structure.
}
\label{T1to3}
\end{figure}

In order for the action-amplitude to represent a quantized topological
$\th$-term, we must choose $\nu_2(g_i,g_j,g_k)$ such that
\begin{align}
\e^{-S(\{g_i\}) }=\prod_{\{ijk\}} \nu_2^{s_{ijk}}(g_i,g_j,g_k)=1
\end{align}
on closed space-time complex without boundary, in particular, on a tetrahedron
with four triangles (see Fig.  \ref{d1d2}):
\begin{align}
\e^{-S(\{g_i\}) } &=\prod_{\{ijk\}} \nu_2^{s_{ijk}}(g_i,g_j,g_k)
\nonumber\\
&=
\frac{
\nu_2(g_1,g_2,g_3) \nu_2(g_0,g_1,g_3)
}{
\nu_2(g_0,g_1,g_2) \nu_2(g_0,g_2,g_3)
}=1
\end{align}
Also, in order for our system to have the symmetry generated by the
group $G$, its action-amplitude must satisfy
\begin{align}
\label{nu2a}
 \e^{-S(\{g_i\}) } &= \e^{-S(\{gg_i\}) },
 \text{ if } g \text{ contains no T}
\nonumber\\
 \Big(\e^{-S(\{g_i\}) }\Big)^\dag &= \e^{-S(\{gg_i\}) },
 \text{ if } g \text{ contains one T}
\end{align}
where $T$ is the time-reversal transformation.
This requires
\begin{align}
\label{nu2b}
\nu_2^{s(g)}(g_i,g_j,g_k)=\nu_2(gg_i,gg_j,gg_k).
\end{align}
Eqn. \eq{nu2a} and \eqn{nu2b} happen to be the conditions of
2-cocycles $\nu_2(g_0,g_1,g_2)$ of $G$.
Thus the action-amplitude \eqn{Lnu2} constructed from a 2-cocycle
$\nu_2(g_0,g_1,g_2)$ is a quantized topological $\th$-term.

If $\nu_2( g_0,  g_2, g_3)$ satisfies \eqn{nu2a} and \eqn{nu2b}, then
\begin{align}
 \nu_2'( g_0,  g_2, g_3)=
 \nu_2( g_0,  g_2, g_3)
\frac{\mu_1(g_1,g_2) \mu_1(g_0,g_1)}{\mu_1(g_0,g_2)}
\end{align}
also satisfies \eqn{nu2a} and \eqn{nu2b}, for any $\mu_1(g_0,g_1)$ satisfying
$\mu_1(gg_0,gg_1)=\mu_1(g_0,g_1)$, $g\in G$.  So $\nu_2'( g_0,  g_2, g_3)$ also
gives rise to a quantized topological $\th$-term.  As we continuously deform
$\mu_1(g_0,g_1)$, the two quantized topological $\th$-terms can be smoothly
connected.  So we say that the two quantized topological $\th$-terms obtained
from $\nu_2( g_0,  g_2, g_3)$ and  $\nu_2'( g_0,  g_2, g_3)$ are equivalent.
We note that $\nu_2( g_0,  g_2, g_3)$ and  $\nu_2'( g_0,  g_2, g_3)$ differ by
a 2-coboundary $\frac{\mu_1(g_1,g_2) \mu_1(g_0,g_1)}{\mu_1(g_0,g_2)}$.  So the set of
equivalence classes of $\nu_2( g_0,  g_2, g_3)$ is nothing but the cohomology
group $\cH^2[G,U_T(1)]$.  Therefore, the quantized topological $\th$-terms are
classified by $\cH^2[G,U_T(1)]$.

We can also show that \eqn{Lnu2} is a fixed-point action-amplitude from the
cocycle conditions on $\nu_2(g_i,g_j,g_k)$.  From the geometrical picture of
the cocycles (see Fig. \ref{d1d2}), we have the following relations: $
\nu_2(g_0,g_1,g_2) \nu_2(g_0,g_2,g_3) = \nu_2(g_1,g_2,g_3) \nu_2(g_0,g_1,g_3)$
(see Fig. \ref{Tflip}) and $ \nu_2(g_1,g_2,g_3)= \nu_2(g_0,g_1,g_2)
\nu_2(g_0,g_2,g_3) \nu_2^{-1}(g_0,g_1,g_3)$.  (see Fig.  \ref{T1to3}).  We can
use those two basic moves to generate a renormalization flow that induces a coarse-grain transformation of 
the complex. The two relations Fig. \ref{Tflip} and Fig.  \ref{T1to3} imply
that the action-amplitude is invariant under the  renormalization flow.  So it
is a fixed-point action-amplitude.  Certainly, the above construction applies
to any dimensions.

\subsection{(1+1)D fixed-point ground state wave function}

\begin{figure}[tb]
\begin{center}
\includegraphics[scale=0.6]{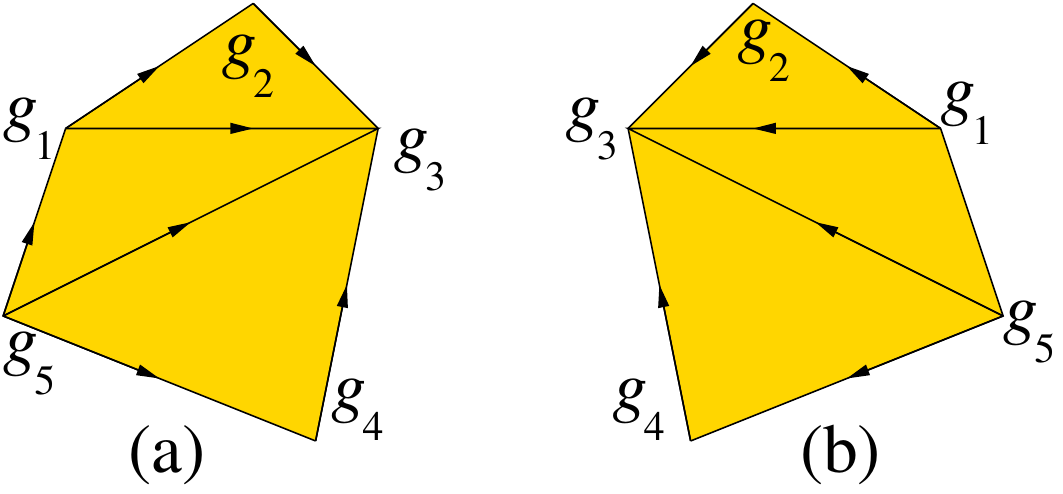}
%Fig. 16
\end{center}
\caption{
(Color online)
(a) The graphic representation of $\Psi(\{g_i\}_\text{edge}) =
\nu_2^{-1}(g_1,g_2,g_3) \nu_2^{-1}(g_1,g_3,g_5) \nu_2^{-1}(g_3,g_4,g_5) $
(b) The graphic representation of $\Psi^\dag(\{g_i\}_\text{edge}) =
\nu_2(g_1,g_2,g_3) \nu_2(g_1,g_3,g_5) \nu_2(g_3,g_4,g_5) $
The arrows on the edges represent the branching structure.
}
\label{PsiG}
\end{figure}

For our fixed-point theory described by a
quantized topological $\th$-term, its
ground state wave function $\Psi(\{g_i\})$ can be obtained by putting $g_i$ on
the edge of a disk and making a triangularization of the disk (see Fig. \ref{Lg}).
We sum the action-amplitude over the $g_i$ on the internal vertices while fixing the
$g_i$'s on the edge (see Fig. \ref{PsiG}a):
\begin{align}
\label{psigrnd2}
\Psi(\{g_i\}_\text{edge})
&=\frac{\sum_{g_i \in \text{internal} }}{|G|^{N^\text{internal}_v}}
\prod_{\{ijk\}} \nu_2^{s_{ijk}}(g_i,g_j,g_k)
\end{align}
where $\sum_{g_i \in \text{internal} }$ sums over $g_i$ on the internal
vertices and $N^\text{internal}_v$ is the number of internal vertices on the
disk.

Clearly the ideal wave function  $\Psi(\{g_i\}_\text{edge})$
satisfies
\begin{align}
\label{psipsi}
 \Psi^{s(g)}(\{g_i\}_\text{edge})=\Psi(\{gg_i\}_\text{edge}), \ \ \ \
 |\Psi(\{g_i\}_\text{edge})|=1 ,
\end{align}
which represents a symmetric state. We also note that $\Psi^\dag
(\{g_i\}_\text{edge})$ can be represented by Fig. \ref{PsiG}b, since the
product of the wave functions in Fig. \ref{PsiG}a and Fig. \ref{PsiG}b is the
value of the cocycle on a sphere which is equal to 1.

\subsection{(2+1)D symmetric fixed-point action-amplitude}

\begin{figure}[tb]
\begin{center}
\includegraphics[scale=0.7]{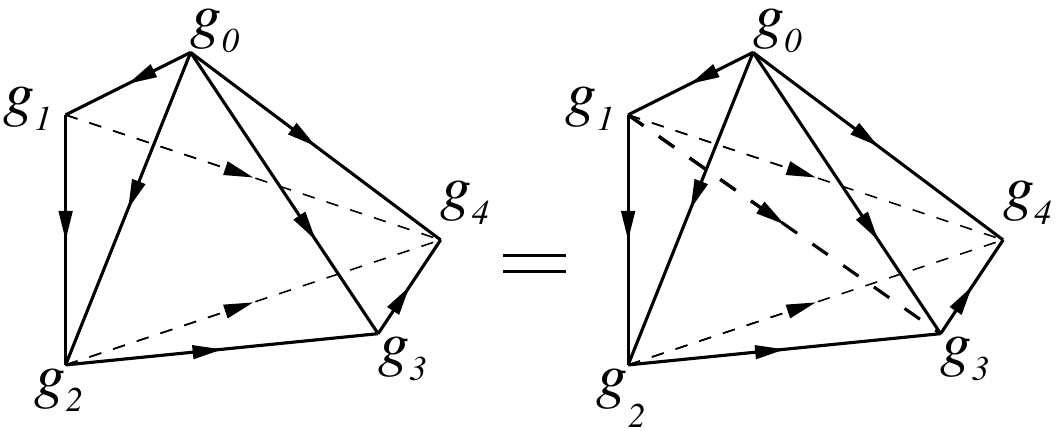}
%Fig. 17
\end{center}
\caption{
Two solid tetrahedrons
$g_0g_1g_2g_4$,
$g_0g_2g_3g_4$
and three solid tetrahedrons
$g_0g_1g_2g_3$,
$g_0g_1g_3g_4$,
$g_1g_2g_3g_4$
occupy the same volume, which leads to the
graphic representation of
$
\nu_3(g_0,g_1,g_2,g_4)
\nu_3(g_0,g_2,g_3,g_4)=
\nu_3(g_0,g_1,g_2,g_3)
\nu_3(g_0,g_1,g_3,g_4)
\nu_3(g_1,g_2,g_3,g_4)
$ (see \eqn{dnu3}).
The arrows on the edges represent the branching structure.
}
\label{2to3}
\end{figure}

In $(2+1)$D, our ideal model with on-site symmetry $G$ is defined
by the action-amplitude on a 3D complex with $g_i \in G$ on each
vertex:
\begin{align}
\e^{-S(\{g_i\}) } = \prod_{\{ijkl\}} \nu_3^{s_{ijkl}}(g_i,g_j,g_k,g_l)
\end{align}
where $\nu_3(g_i,g_j,g_k,g_l)$ is a three cocycle and $\prod_{\{ijkl\}}$
multiply over all the tetrahedrons in the complex Fig. \ref{2to3}. The 3D
complex has a branching structure. The first variable $g_i$ is on the vertex
with no incoming edge, the second variable $g_j$ is on the vertex with one
incoming edge, \etc.  Also $s_{ijkl}=\pm 1$ depending on the orientation of the
$ijkl$-tetrahedron. On a close space-time complex, the above action-amplitude
is always equal to 1 due to the cocycle condition on $\nu_3(g_i,g_j,g_k,g_l)$.
Thus the above action-amplitude is a quantized  topological $\th$-term.

\begin{figure}[tb]
\begin{center}
\includegraphics[scale=0.7]{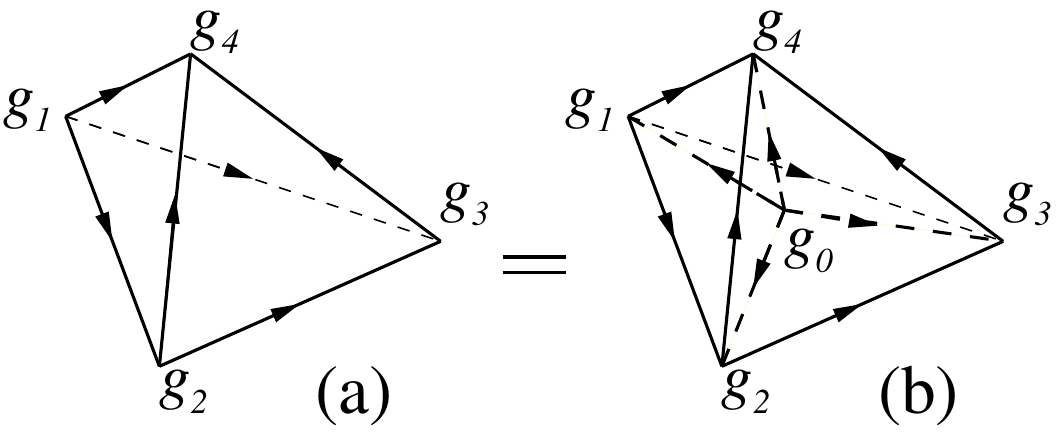}
%Fig. 18
\end{center}
\caption{
One solid tetrahedron $g_1g_2g_3g_4$, and four solid tetrahedrons
$g_0g_1g_2g_4$, $g_0g_2g_3g_4$, $g_0g_1g_3g_4$, $g_1g_1g_2g_3$ occupy the same
volume, which leads to the graphic representation of $ \nu_3(g_1,g_2,g_3,g_4)=
\nu_3(g_0,g_1,g_2,g_4) \nu_3(g_0,g_2,g_3,g_4) \nu_3^{-1}(g_0,g_1,g_3,g_4)
\nu_3^{-1}(g_0,g_1,g_2,g_3) $ (see \eqn{dnu3}).
The arrows on the edges represent the branching structure.
}
\label{1to4}
\end{figure}

The conditions of 3-cocycle lead to the two relations in Fig.
\ref{2to3} and Fig. \ref{1to4}.  These lead to a renormalization
flow of the complex in which the above action-amplitude is a
fixed-point action-amplitude.  The fixed-point action-amplitude
leads to an ideal short-range-entangled state (see section
\ref{tritop}) that has a symmetry $G$ and is characterized by $\nu_3
\in \cH^3[G,U_T(1)]$.

\subsection{$(d+1)$D symmetric fixed-point action-amplitude}

Through the above two examples in (1+1)D and (2+1)D,
we see that the $(d+1)$D symmetric fixed-point action-amplitude
is given by
\begin{align}
\label{Zd}
Z &=\frac{\sum_{\{g_i\} }}{|G|^{N_v}}
\prod_{\{ij...k\}} \nu_{1+d}^{s_{ij...k}}(g_i,g_j,...,g_k)
\end{align}
where $g_i$ is associated with each vertex on the space-time complex
and $N_v$ is the number of vertices.  $\sum_{\{g_i\}}$ sums over all
possible configurations of $\{g_i\}$ and
$\nu_{1+d}(g_i,g_j,...,g_k)$ is a $(1+d)$-cocycle in
$\cH^{1+d}[G,U_T(1)]$.

When the space-time complex is closed (\ie has no boundary), the
action-amplitude $\prod_{\{ij...k\}}
\nu_{1+d}^{s_{ij...k}}(g_i,g_j,...,g_k)$ is always equal to 1. Thus
the action-amplitude represents a  topological $\th$-term.

When the space-time complex has a boundary, the action-amplitude
will not always be equal to 1 and is not trivial.  We note that, due
to the cocycle condition on
$\nu_{1+d}^{s_{ij...k}}(g_i,g_j,...,g_k)$, such a action-amplitude
will only depend on $g_i$'s on the boundary of the space-time
complex. Thus such an action-amplitude can be viewed as an
action-amplitude of the boundary theory.

As an  action-amplitude of the boundary theory, $\prod_{\{ij...k\}}
\nu_{1+d}^{s_{ij...k}}(g_i,g_j,...,g_k)$ is actually a
NLL term, which is a generalization of the
WZW topological term for continuous non-linear
$\si$-models to lattice  non-linear $\si$-models.  So the boundary
excitations of our model defined by \eqn{Zd} are described by a
non-linear $\si$-model with a NLL term composed by
the same $\nu_{1+d} \in \cH^{1+d}[G,U_T(1)]$.  We see a close
relation between the  topological $\th$-term in $(d+1)$ space-time
dimensions and the NLL term in $d$ space-time
dimensions.  An example of such a relation has been discussed by Ng
for a (1+1)D model with $SO(3)$ symmetry.\cite{N9455}
When $\nu_{1+d}$ is non-trivial, we believe that the boundary
states are gapless or degenerate on the
boundary. 
%(To make sure that the symmetry is not broken on the
%boundary, we may need to add non-topological term $\del \cL$ to the
%boundary Lagrangian.)

In the following, we will show that the ground state wave function of our model
\eq{Zd} describes a SPT state.

\section{Trivial intrinsic topological order
in our fixed-point models}
\label{tritop}

\begin{figure}[tb]
\begin{center}
\includegraphics[scale=0.7]{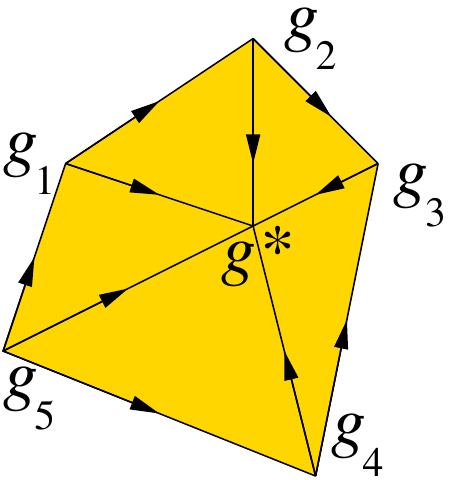}
%Fig. 19
\end{center}
\caption{ (Color online) The graphic representation of \eqn{PsiM}.
The boundary is the complex $M$, and the whole complex
$M_\text{ext}$ is an extension of $M$. } \label{Psi}
\end{figure}

The ground state of our $d$-dimensional model \eq{Zd} is a wave
function $\Psi_{M}$ on $M$,  a $d$-dimensional complex.  It
is given by (see Fig. \ref{Psi})
\begin{align}
\label{PsiM}
\Psi_{M}(\{g_i\}_M)
&=\frac{\sum_{g_i \in \text{internal} }}{|G|^{N^\text{internal}_v}}
\prod_{\{ij...k\}} \nu_{1+d}^{s_{ij...k}}(g_i,g_j,...,g_k)
\end{align}
which generalizes \eqn{psigrnd2} from (1+1)-D to $(d+1)$-D.  We use
$M_\text{ext}$ to denote a  $(d+1)$ dimensional complex whose
boundary is $M$. $\{g_i\}_M$ are on the vertices on $M$ and
$\sum_{g_i \in \text{internal} }$ sums over the $g_i$'s on the
vertices inside the complex $M_\text{ext}$ (not on its
boundary $M$). Also $ \prod_{\{ij...k\}}$ is product over all
simplices on $M_\text{ext}$.

Due to the cocycle condition satisfied by
$\nu_{1+d}(g_i,g_j,...,g_k)$, we see that, for fixed  $\{g_i\}_M$,
the product $\prod_{\{ij...k\}}
\nu_{1+d}^{s_{ij...k}}(g_i,g_j,...,g_k)$ does not depend on  $g_i$'s
on the vertices inside the complex $M_\text{ext}$. Thus
\begin{align}
\label{PsiMB}
\Psi_{M}(\{g_i\}_M)
&= \prod_{\{ij...*\}} \nu_{1+d}^{s_{ij...*}}(g_i,g_j,...,g^*).
\end{align}
if we choose $M_\text{ext}$ to be $M$ plus one more vertex with label $g^*$ (see
Fig. \ref{Psi}).  The state on $M$ (the boundary of Fig. \ref{Psi}) does not
depend on the choice of $g^*$.

Using the above expression, we can show that the ground state wave function of
our fixed-point model is SRE state with no intrinsic topological orders.
Let us first write the ground state of our fixed-point model in a form
\begin{align}\label{Psiold}
| \Psi_{M}\>
&= \sum_{\{g_i\}_M} \prod_{\{ij...*\}} \nu_{1+d}^{s_{ij...*}}(g_i,g_j,...,g^*)
|\{g_i\}_M\> ,
\end{align}
where $|\{g_i\}_M\>$ form a basis of our model on $d$-dimensional
complex $M$. The on-site symmetry acts in a simple way:
\begin{align}
g:\ |\{g_i\}_M\> \to |\{gg_i\}_M\>,\
g\in G
\label{eq:Lgsymm}
\end{align}

We note that if we choose the particular form of $M_\text{ext}$ in Fig.
\ref{Psi} to obtain state $\Phi_M$ on $M$, the phase factor
$\prod_{\{ij...*\}} \nu_{1+d}^{s_{ij...*}}(g_i,g_j,...,g^*)$ can be viewed as a
LU transformation. We can write $| \Psi_{M}\>$ in a new basis
$|\{g_i\}_M\>'=\prod_{\{ij...*\}} \nu_{1+d}^{s_{ij...*}}(g_i,g_j,...,g^*)
|\{g_i\}_M\>$:
\begin{align}\label{Psinew}
| \Psi_{M}\> &= \sum_{\{g_i\}_M} |\{g_i\}_M\>' .
\end{align}
Thus, on any complex $M$ that can be viewed as a boundary of another
complex $M_\text{ext}$, the state on $M$ can be transformed by an LU
transformation into a state that is the equal weight superposition
of all possible states $|\{g_i\}_M\>$ on $M$. The wave function in
the new bases is very simple, which is actually a product state. In appendix \ref{twosymm},
we will show that under a dual transformation, this product state is equivalent to the
canonical form of wave function discussed in Sec. \ref{Cfm} and \ref{clsymm}.

We have used the $(1+d)$-cocycles in $\cH^{1+d}[G,U_T(1)]$ to construct our
fixed-point models which have ground state wave functions that also depend on
the  $(1+d)$-cocycles.  In the above, we have shown that all those states can be
mapped to the same simple product state via LU transformations. Does this mean
that those states from different $(1+d)$-cocycles all belong to the same phase?
The answer depends on if symmetry is included or not.

If we do not include any symmetry, those states from different
$(1+d)$-cocycles indeed all belong to the same trivial phase.  Thus our
fixed-point states constructed from different $(1+d)$-cocycles all have trivial
intrinsic topological order.  This is consistent with the fact that the
fixed-point partition function on any space-time complex has the form
\begin{align}
 Z=e^{-S_0 V} ,
\end{align}
where $V$ is the volume of the space-time complex (say $V$ is the number of
simplices in the space-time complex).  We would like to stress that the above
expression is exact.  So after we remove the term that is proportional to the
space-time volume, we have $Z=1$.  This means that the ground state is not
degenerate on any closed space complex, which in turn implies that the ground
state contains no intrinsic topological order.

On the other hand, if we include the on-site symmetry $G$, states from
different $(1+d)$-cocycles belong to the different phases which correspond to
different SPT phases.  This is because the LU transformation represented by
$\prod_{\{ij...*\}} \nu_{1+d}^{s_{ij...*}}(g_i,g_j,...,g^*)$ is not a symmetric
LU transformation under the on-site symmetry $G$.  To see this,
we first note that the  LU
transformation $\prod_{\{ij...*\}} \nu_{1+d}^{s_{ij...*}}(g_i,g_j,...,g^*)$
contains several layers of non-overlapping terms.  For example, for the (1+1)D
system in Fig. \ref{Psi}, the  LU transformation has two layers
\begin{align}
&\prod_{\{ijk\}} \nu_{2}(g_i,g_j,g_k)
=
[
\nu_{2}(g_3,g_2,g^*)
\nu_{2}(g_5,g_4,g^*)
]\times
\nonumber\\
&\ \ \ \ \ \ \ \ \ \ \ \
[
\nu_{2}(g_2,g_1,g^*)
\nu_{2}(g_4,g_3,g^*)
\nu_{2}(g_1,g_5,g^*)
]
\end{align}
In order for the LU transformation to be a symmetric, each local term, such as
$\nu_{2}(g_2,g_1,g^*)$, must transform as
\begin{align}
 \nu_{2}^{s(g)}(g_2,g_1,g^*)=\nu_{2}(gg_2,gg_1,g^*)
\end{align}
under the on-site symmetry transformation generated by $g\in G$: Although
$\nu_{2}^{s(g)}(g_2,g_1,g^*)=\nu_{2}(gg_2,gg_1,gg^*)$, in general
$\nu_{2}^{s(g)}(g_2,g_1,g^*)\neq \nu_{2}(gg_2,gg_1,g^*)$.  In fact, only
trivial cocycle in $\cH^{1+d}[G,U_T(1)]$ can satisfy $
\nu_{1+d}^{s(g)}(g_1,g_2,..,g_{1+d}, g^*) =\nu_{1+d}(gg_1,gg_2,..,gg_{1+d},
g^*) $.  Thus the fixed-point states from different $(1+d)$-cocycles belong to
the different SPT phases.

We have seen that we can use two different basis $|\{g_i\}_M\>$ and
$|\{g_i\}_M\>'$ to expand the fixed-point wave function $|\Psi_M\>$.  The
old basis $|\{g_i\}_M\>$ transforms simply under the symmetry transformation:
$|\{g_i\}_M\> \to |\{gg_i\}_M\>$.  But the wave function $\Psi_M(\{g_i\})$ in
the old basis is complicated.  In the new basis, the wave function is very
simple $\Psi'_M(\{g_i\})=1$.  But the symmetry transformation is more
complicated in the new basis which will be discussed in appendix \ref{twosymm}.

In section \ref{cnsymm}, we discuss the SPT phase by starting with a
simple many-body wave function, and try to classify all the allowed on-site
symmetry transformations.  Such a formalism is closely related to the new
basis.

\section{Equivalent cocycles give rise to the same SPT phase}
\label{gwvclass}

The ground state wave function
$\Psi_M(\{g_i\}_M)$ of a SPT phase is constructed from a cocycle
$\nu_{1+d}$ as in \eqn{PsiMB}.
Let $\nu'_{1+d}$ be a cocycle that is equivalent to $\nu_{1+d}$.
That is $\nu_{1+d}$ and $\nu'_{1+d}$ only differ by a coboundary
\begin{align}
&\ \ \ \
 \nu'_{1+d}(g_0,...,g_{1+d})
\nonumber\\
&=
 \nu_{1+d}(g_0,...,g_{1+d})
\prod_{i=0}^{1+d} \mu_d^{(-)^i}(...,g_i,g_{i+1},...)
\end{align}
where $ \mu_d(g_0,...,g_d)$ is a $d$-cochain.  Then $\nu'_{1+d}$ will give rise
to a new ground state wave function $\Psi_M'(\{g_i\}_M)$ of a SPT phase.
One can show that
$\Psi_M(\{g_i\}_M)$ and
$\Psi_M'(\{g_i\}_M)$ are related:
\begin{align}
 \Psi'_M(\{g_i\}_M)=
 \Psi_M(\{g_i\}_M) \prod_{\{ij...\}} \mu_d^{s_{ij...}}(g_i,g_j,...)
\end{align}
where $\prod_{\{ij...\}}$ multiply over all the $d$-simplices in $M$.  Note
that, when we calculate $\Psi'_M(\{g_i\}_M)$, the terms
$\mu_d(g_i,g_j,...,g^*)$ containing $g^*$ all cancel out.  Due to the cochain
condition \eqn{cchcnd} satisfied by $\mu_d$, the factor $\prod_{\{ij...\}}
\mu_d^{s_{ij...}}(g_i,g_j,...)$ actually represents a \emph{symmetric} LU
transformation.  So the two wave functions $\Psi_M(\{g_i\}_M)$ and
$\Psi_M'(\{g_i\}_M)$ belong to the same SPT phase.  Hence equivalent cocycles
give rise to the same SPT phase, and different SPT phases are classified by the
equivalence classes of cocycles which form $\cH^{1+d}[G,U_T(1)]$.

\section{Relation between cocycles and Berry phase}
\label{AppBerry}

In this section, from path integral formalism, we will discuss some
relations between the Berry phase and the cocycles that we used to construct
topological non-linear $\si$-models.  The Berry phase is defined in continuum
parameter space, so we need to embed the discrete symmetry group $G$ into a
continuous group $\tilde G$, with $G\subset\tilde G$. For example, a discrete
rotation group can be embedded into the $SO(3)$ group. The coherent state path
integral is performed in forms of $\tilde G$. After obtaining the topological
$\th$-term, we will reduce the symmetry group back to $G$.

Suppose a rotation operator $g$ is a symmetry operation $g\in G$,
and $g|g\rangle_0\propto|g\rangle_0$ is its eigenstate. The spin
coherent state is defined as the following
\begin{eqnarray*}
|g(\v m)\rangle=R(\v m)|g\rangle_0,
\end{eqnarray*}
where $R(\v m)\in \tilde G$. We can write $|g(\v m)\rangle$ as
$|\v m\rangle$ for simplicity. They satisfy the complete relation
\begin{eqnarray*}
\int \dd \v m |\v m\rangle\langle \v m|\propto 1,
\end{eqnarray*}
where the integration is performed over the group space of $\tilde
G$.

The Berry phase in the spin coherent path
integral is very important in our discussion. For a non-symmetry
breaking system, the low energy effective theory can be written as
the following path integral
\begin{eqnarray}
Z&=&\int D \v{m}(\v x,\tau)\exp\{-\int \dd ^d\v x\dd \tau
\mathcal L_0(\v{m})+iS_{\mathrm {top}}\},\nonumber\\
S_{\mathrm {top}}&=&\int \dd ^d\v x\dd \tau\mathcal L_{\mathrm
{top}}(\v A,A_0),
\end{eqnarray}
where $\mathcal L_0$ is the dynamic part of the Lagrangian which
respects the symmetry group $G$ (it is not important at the fixed
point), and $S_{\mathrm {top}}$ is the topological $\th$-term of the
action, which respects the enlarged symmetry group $\tilde G$. The
`gauge' field is defined as $\v A=\langle{\v m}(\v
x,\tau)|\nabla| {\v m}(\v x,\tau)\rangle$ and $A_0=\langle{\v
m}(\v x,\tau)|\partial_\tau| {\v m}(\v x,\tau)\rangle$. The
following is a generalization of result of $O(3)$ nonlinear sigma
model discussed in Ref.~\onlinecite{N9455}.

At zero temperature, the partition function only contains the contribution from
the ground state. Under periodic boundary condition, the ground state is a
singlet, as a consequence, the Berry phase is trivial (integer times $2\pi$).
Under open boundary conditions, the Berry phase is contributed from the edge
states. The topological $\th$-term is dependent on dimension. We will study it
case by case.

\begin{figure}[tb]
\centering
\includegraphics[width=3in]{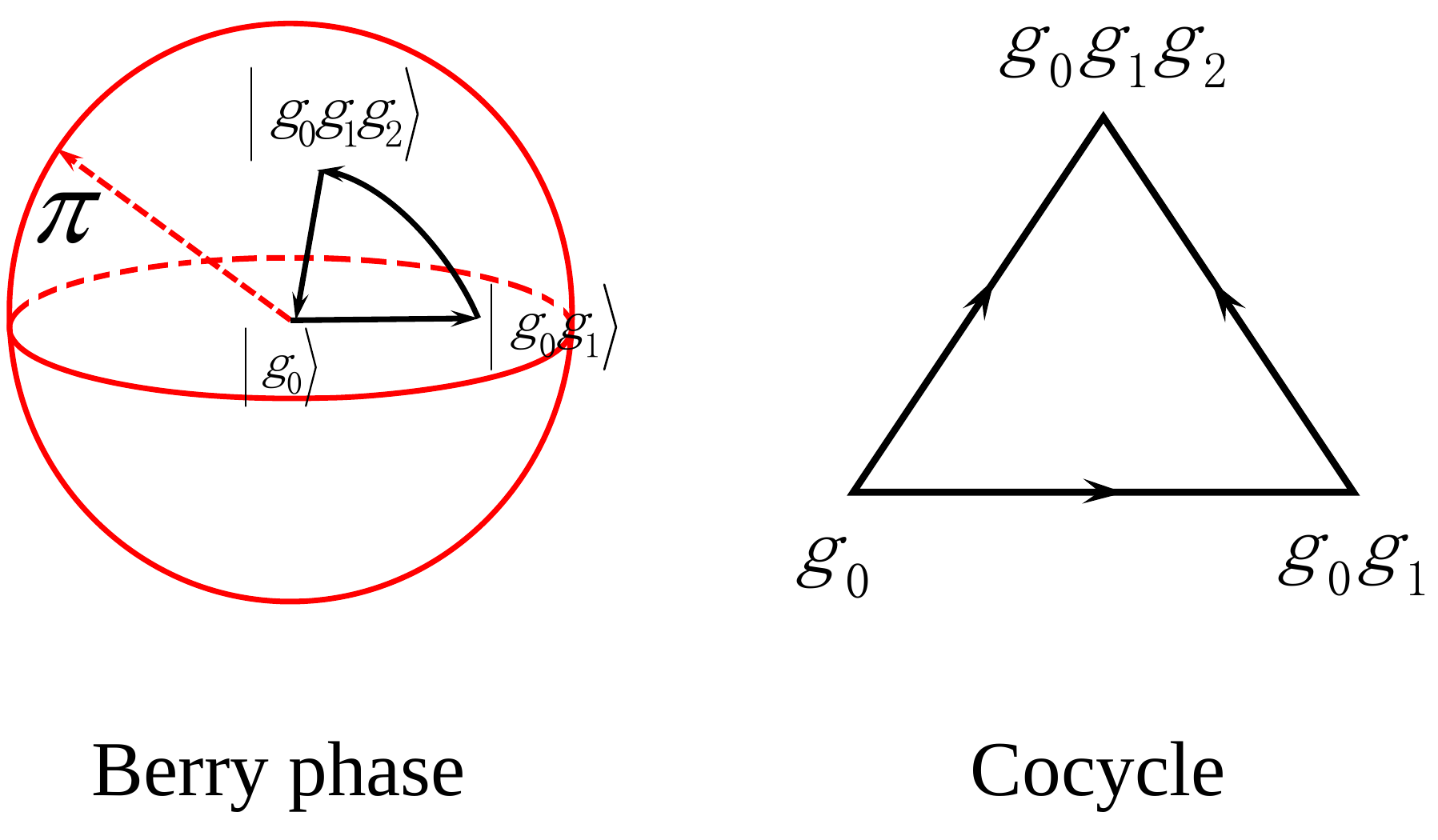}
%Fig.20 
\caption{(Color online) Relations between Berry phase (of the end spin) in the loop
($|g_0\rangle, |g_0g_1\rangle, |g_0g_1g_2\rangle$) and 2-cocycle $\nu_2(g_0,g_0g_1,g_0g_1g_2$). In the group
space of $SO(3)$, the two ends of a diameter stand for the same group element and can be seen as the same point. When $\theta=2\pi$, the Berry phase is equal to zero when the the loop intersect the shell even times and is equal to $\pi$ when the loop intersection the shell odd times. If $\theta$ is equal to even times of $2\pi$, the corresponding 2-cocycle is trivial. If $\theta$ is equal to odd times of $2\pi$, the corresponding 2-cocycle
is nontrivial.} \label{fig:cocycle}
\end{figure}

In $(1+1)$D, the topological $\th$-term is given as
\[
S_{\mathrm {top}}=\theta\oint_{S_1\times S_1} \dd x\dd \tau F,
\]
where $F=\partial_xA_0-\partial_0A_x$, and $S_1\times S_1$ is the
space-time manifold. $\theta$ is an important constant which
determines the topological properties of the system.
Under periodic boundary condition, the above integral is
quantized and is equal to an integer (Chern number) times $2\pi$
which results in a trivial phase $e^{iS_\mathrm {top}}=1$. However,
at open boundary condition (where the space-time manifold becomes a
cylinder), the integral is not quantized. From Stokes theorem, it is
determined by the boundaries,
\begin{eqnarray*}
S_{\mathrm{top}}=S_{\mathrm{L}}-S_{\mathrm{R}}
\end{eqnarray*}
with (similar expression for $S_{\mathrm{R}}$)
\begin{eqnarray}\label{2-cocycle}
S_{\mathrm{L}}&=&\theta\oint_{S_1}\dd \tau A_{0}(\v
x_{\mathrm{L}},\tau)\nonumber
\\&=&\theta\oint_{S'_1}\dd \lambda(\tau)\tilde A_{\lambda}[\lambda(\tau)]\nonumber
\\&=&\theta\int_{D_1}d^2\lambda \tilde F,
\end{eqnarray}
where $S'_1$ is a path in the parameter space (i.e., the group space
of $\tilde G$, which is parameterized by $\lambda$), $D_1$ is the
area enclose by $S'_1$, and $\tilde A_\lambda$, $\tilde F$ are the
Berry connection and Berry curvature in the parameter space,
respectively. The cyclic path $S'_1$ can be chosen as a sequence of
symmetry operators in the symmetry group $G$. A closed path contains
at least three points $|g_0\rangle, |g_0g_1\rangle$, and $|g_0g_1g_2\rangle$
(see figure.~\ref{fig:cocycle}). The above integral gives a 2-cocycle or
a product of 2-cocycles if we choose a proper gauge (\ie, multiply a proper coboundary)
\[
e^{iS_{\mathrm{L}}}%=\omega_2(g_1,g_2)
=\nu_2(g_0,g_0g_1,g_0g_1g_2),
\]
where $g_0$ is an arbitrary symmetry operator in $G$.

In $(2+1)$D, the possible topological $\th$-term is the Hopf term,
\begin{eqnarray}
S_{\mathrm {top}}=\theta\oint_{S_1\times S_1\times S_1} \dd ^2x\dd \tau
\varepsilon^{ijk}A_iF_{jk},
\end{eqnarray}
where $F_{ij}=\partial_iA_j-\partial_jA_i$, $i,j,k=x,y,\tau$. The space is compacted to $S_1\times S_1$, and the time is compacted to the last $S_1$. The Hopf term can be written as a total differential locally, $\varepsilon^{ijk}A_iF_{jk}=\varepsilon^{ijk}
\partial_k[f_{ij}(A_i,A_j)]$, where $f_{ij}(A_i,A_j)$ is a (nonlocal) function of
$A_i$ and $A_j$. Thus, at open boundary condition, the integral is
determined by the boundary values, $S_{\mathrm{top}}=S_{\mathrm{L}}
-S_{\mathrm{R}}$, here we have cut the space along $y$-direction.
$S_{\mathrm{L}}$ is given as
\begin{eqnarray}\label{3-cocycle}
S_{\mathrm L}&=&\theta\oint_{S_1\times S_1} \dd y\dd \tau
(f_{y\tau}(A_y,A_\tau)-f_{\tau y}(A_\tau,A_y))\nonumber\\
&=&\theta\oint_{S'_1\times S''_1} \dd \lambda_1\dd \lambda_2 (\tilde
f_{\lambda_1{\lambda_2}}(\tilde A_{\lambda_1},\tilde A_{\lambda_2})
-\tilde f_{\lambda_2 \lambda_1}(\tilde A_{\lambda_2},\tilde A_{\lambda_1}))\nonumber\\
&=&\theta\int_{S_1'\times D_1}d^3\lambda \varepsilon^{IJK}\tilde
A_{I}\tilde F_{JK}
\end{eqnarray}
where $I,J,K=\lambda_1,\lambda_2,\lambda_3$ are parameters of the group space
of $\tilde G$. $S_1'$ is the circle formed by parameter $\lambda_1(y)$, and
$S_1''$ is the circle formed by parameter $\lambda_2(\tau)$. $D_1$ is the area
enclosed by the $S_1''$. Above we have mapped the two-dimensional integral on
the boundary of space-time manifold into a three-dimensional integral on the
group space of $\tilde G$. Notice that the spatial dimension of the boundary is
1D, the above topological $\th$-term is actually an effective WZW term of the
boundary.

Since Eq.~(\ref{3-cocycle}) is a 3-dimensional integral over the group space of $\tilde G$, when reducing to the symmetry group $G$, we need at least four points to span the 3-d space $S_1'\times D_1$: $|g_0\rangle, |g_0g_1\rangle, |g_0g_1g_2\rangle, $ and $|g_0g_1g_2g_3\rangle$. Thus we can identify Eq.~(\ref{3-cocycle}) as a 3-cocycle or product of 3-cocycles under proper gauge choice
\begin{eqnarray*}
e^{iS_{L}}
&=&\nu_3(g_0,g_0g_1,g_0g_1g_2,g_0g_1g_2g_3),
\end{eqnarray*}
Here $g_0,g_1,g_2,g_3$ are group elements in the symmetry group $G$.

Above arguments can be generalized to arbitrary $d$-dimension.
For example, in (1+3)D, we may have
\begin{align}
S_{\text{top}}=\theta\oint_{S_1\times S_3} \dd ^3x\dd \tau
\varepsilon^{\mu\nu\ga\la} F_{\mu\nu}F_{\ga\la}.
\end{align}

The topological $\th$-term (or $\theta$ term in literature) plays important
roles in various many-body systems. In reference \onlinecite{YL1017}, the
authors came up with a new method to calculate the topological $\th$-term.

We have shown that the topological term (or the $\theta$-term originating from
Berry phase) reduce to cocycles if we discretize the space and time. The
discrete topological term even exist for discrete groups (they are related to
the $\theta$-term by the embedding argument mentioned previously). Although the
discrete topological nonlinear sigma models constructed from group cocycle are
formally close to $\theta$-terms in continuous nonlinear model, they actually
describe quite different physics. 
%Mathematically, the $\theta$-terms in continuous nonlinear models are
%classified by group homotopy while the discrete topological nonlinear sigma
%models constructed in this paper are classified by group cohomology. The
%physical reason why they are quite different is that
The $\theta$-terms in continuous nonlinear model ignore the physics at cut-off
length-scale, which should be very important in general, especially for those
gapped quantum systems, whose fixed point actions have zero correlation length
and quantum fluctuation can be non-smooth at arbitrary energy scale. Thus, the
discrete topological nonlinear sigma models can be regarded as the quantum
analogous of $\theta$-terms in continuous nonlinear sigma model and help us
understand the nature of symmetry protected topological order in interacting
systems.

\section{SPT orders with translation symmetry}
\label{SPTtrans}

In the above we have discussed bosonic SPT phases with on-site symmetry $G$ but
no other symmetries.  Here we would like to stress that when we say a SPT phase
have no  other symmetries, we do mean that the ground state wave function of
the SPT phase has no other symmetries.  In fact  the ground state wave function
of the SPT phase can have some other symmetries.  What we really mean is that
when we deform the Hamiltonian to construct phase diagram, the deformed
Hamiltonians can have no other symmetries.

In this section, we will discuss the SPT phases with both on-site symmetry $G$
and translation symmetry.  We will use the non-linear $\si$-model approach to
obtain our results.  We have argued that the $d$-dimensional SPT phases with
on-site symmetry $G$ are classified by fixed-point non-linear $\si$-models that
contain only a  topological $\th$-term constructed from a $(1+d)$-cocycle in
$\cH^{1+d}[G,U_T(1)]$.  The action-amplitude (in imaginary time) for such a
fixed-point non-linear $\si$-model is given by
\begin{align}
\e^{-\int \cL^{1+d}(\nu_{1+d})}=
\prod_{\{ij...k\}} \nu_{1+d}^{s_{ij...k}}(g_i,g_j,...,g_k) .
\end{align}

When the system has translation symmetry, we can include additional topological
$\th$-terms which lead to richer SPT phases.  Let us use (2+1)-dimensional systems as
examples to discuss those addition topological $\th$-terms.

\begin{figure}[tb]
\begin{center}
\includegraphics[scale=0.6]{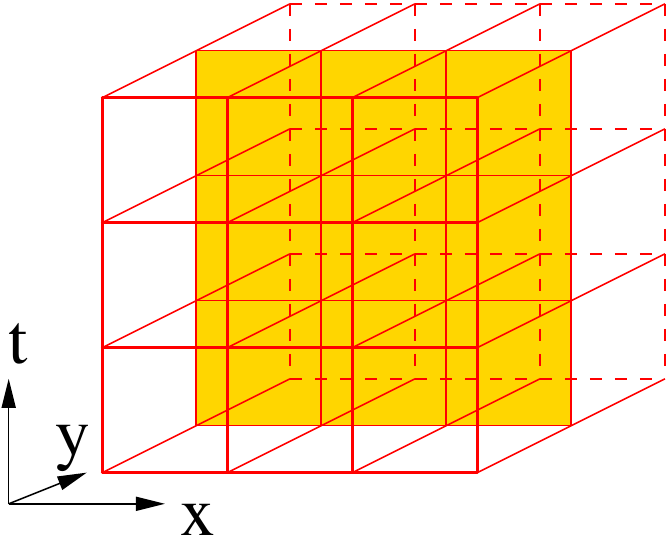}
%Fig. 21
\end{center}
\caption{
(Color online)
A triangularization of (2+1)D space-time where each cube represent five
tetrahedrons.  The shaded area represents a $[xt]$ plane, on which each square
represents two triangles.
}
\label{xt}
\end{figure}

When we say a (2+1)-dimensional system has a translation symmetry, we mean that
the system has a \emph{discrete} translation symmetry in the two spatial
directions.  We must choose the triangularization of the space-time in a way to
be consistent with the \emph{discrete} spatial translation symmetry.
In this case, we can include a new topological $\th$-term:
\begin{align}
& \e^{-\int \cL^{3}_\text{fix}}=
\e^{-\int \cL^{3}(\nu_{3})}
\e^{-\int \cL^{3}(\nu^{xt}_{2})}
\end{align}
where
\begin{align}
 \e^{-\int \cL^{3}(\nu^{xt}_{2})}=\prod_{[xt]}
\prod_{\{ijk\}\in [xt]} (\nu^{xt}_{2})^{s_{ijk}}(g_i,g_j,g_k) .
\end{align}
The translation invariant space-time complex can be viewed as formed by
many $2$-dimensional sheets, say, in $x$-$t$ directions (see Fig.
\ref{xt}).  We pick a sheet $[xt]$ in $x$-$t$ directions, then
$\prod_{\{ijk\}\in [xt]} (\nu^{xt}_{2})^{s_{ijk}}(g_i,g_j,g_k)$ is
simply a  topological $\th$-term on the $[xt]$ sheet constructed
from a $2$-cocycle $\nu^{xt}_2$ in $\cH^2[G,U_T(1)]$.  In the above
expression, $\prod_{\{ijk\}\in [xt]}$ multiply over all triangles in
the $[xt]$ sheet and $\prod_{[xt]}$ multiply over all the $[xt]$
sheets in the space-time complex.

We can include a similar topological $\th$-term $ \e^{-\int
\cL^{3}(\nu^{yt}_{2})}$ by considering the sheets in $y$-$t$
directions and using another 2-cocycle $\nu^{yt}_2$.  A third
topological $\th$-term can be added by viewing space-time complex as
formed by many $1$-dimensional lines in time direction:
\begin{align}
 \e^{-\int \cL^{3}(\nu^{t}_{1})}=\prod_{[t]}
\prod_{\{ij\}\in [t]} (\nu^{t}_{1})^{s_{ij}}(g_i,g_j) .
\end{align}
Here
 $\prod_{\{ij\}\in [t]}$ multiply
over all segments in the $[t]$ line and $\prod_{[t]}$ multiply over
all the $[t]$ lines in the space-time complex. In fact
$\prod_{\{ij\}\in [t]} (\nu^{t}_{1})^{s_{ij}}(g_i,g_j)$ is a
topological $\th$-term on a single $[t]$ line constructed from a 1-cocycle
$\nu^{t}_1$.

We can also try to include the fourth new topological $\th$-term
by considering the sheets in $x$-$y$ directions:
\begin{align}
 \e^{-\int \cL^{3}(\nu^{xy}_{2})}=\prod_{[xy]}
\prod_{\{ijk\}\in [xy]} (\nu^{xy}_{2})^{s_{ijk}}(g_i,g_j,g_k) .
\end{align}
But such a topological $\th$-term corresponds to a LU transformation with a few
layers. In fact $ \prod_{\{ijk\}\in [xy]}
(\nu^{xy}_{2})^{s_{ijk}}(g_i,g_j,g_k)$ is a LU transformation when viewed as a
time-evolution operator.  So there is no fourth new topological $\th$-term.

We see that SPT phases in (2+1)-dimensions with an on-site symmetry $G$ and
translation symmetry are characterized by one 1-cocycles $\nu^t_1\in
\cH^1[G,U_T(1)]$, two 2-cocycles $\nu^{xt}_2,\nu^{yt}_2\in \cH^2[G,U_T(1)]$,
and one 3-cocycle $\nu_3 \in \cH^3[G,U_T(1)]$.  If we believe that those are
all the possible topological $\th$-terms, we argue that the SPT phases in
(2+1)-dimensions with an on-site symmetry $G$ and translation symmetry are
classified by
$\cH^1[G,U_T(1)]\times\cH^2[G,U_T(1)]\times\cH^2[G,U_T(1)]\times\cH^3[G,U_T(1)]$.  A special
case of this result with $\nu_3=0$ is discussed in \Ref{CGW1107} where the
physical meaning of $\nu^t_1,\nu^{xt}_2,\nu^{yt}_2$ is explained in terms of
1D representations and projective representations of $G$.  Certainly, the
above construction can be generalized to any dimensions.

If we do not have translation symmetry, we can still add the new topological
$\th$-terms, such as $\e^{-\int \cL^{3}(\nu^{xt}_{2})}$.  But in this case, we can
combine $n$ $[xt]$ planes in to one.  If $\cH^{2}[G,U_T(1)]$ is finite,
the new topological $\th$-term on the combined plane can be trivial if we choose $n$
properly. So, we cannot have new topological $\th$-terms if we do not have
translation symmetry and if $\cH^{d}[G,U_T(1)]$ is finite.

\section{Summary}

Since the introduction of topological order in 1989, we have been trying to
gain a global and systematic understanding of topological order.  We have made
a lot of progress in understanding topological orders without symmetry in low
dimensions.  We have used the $K$-matrix to classify all Abelian fractional
quantum Hall states,\cite{BW9045,R9002,WZ9290} and have used string-net
condensation\cite{LW0510,CGW1038} to classify non-chiral topological orders in
two spatial dimensions, and have constructed a large class of topological
orders in higher dimensions.

The LU transformations deepen our understanding of topological order and link
topological orders to patterns of long range entanglements.\cite{CGW1038}  Such
a deeper understanding allows us to obtain a systematic description of
topological orders in 2D fermion systems.\cite{GWW1017} The LU transformations
also allow us to start to understand topological order with symmetries.  In
particular, it allows us to classify all gapped quantum phases in one spatial
dimension.  We find that all gapped 1D phases are SPT phases  (SPT phases are
gapped quantum phases with certain symmetry which can be smoothly connected to
the same trivial product state if we remove the symmetry).  In 1D, the SPT
phases can be classified by 2-cohomology classes of the symmetry group.

In this paper, we try to understand topological order with symmetry in higher
dimensions. In particular, we try to classify SPT phases in higher dimensions.
We find that distinct SPT phases with on-site symmetry $G$ in $d$ spatial
dimensions can be constructed from distinct elements in
$(1+d)$-Borel-cohomology classes of the symmetry group $G$.  We summarize our
results in table \ref{tb} for some simple symmetry groups.

We have used two approaches to obtain the above result: the LU transformations
and topological non-linear $\si$-models.  We generalized the usual topological
$\th$-term and the WZW term in continuous non-linear $\si$-model to the
topological $\th$-term and the NLL terms in lattice non-linear $\si$-models
(with both discrete space-time and discrete target space).

Our results demonstrate how many-body entanglements interact with symmetry in a
simple situation where there is no long range entanglements (\ie no intrinsic
topological orders).  This may prepare us to study the more important and
harder problem: how to classify quantum states with long range entanglements
(\ie with intrinsic topological orders) and symmetry.  Those phases with  long
range entanglements and symmetry are called symmetry enriched topological
orders.  Also, our approach can be modified and generalized to describe/classify
fermionic SPT phases, through generalizing the group
cohomology theory to  group
super-cohomology theory.\cite{GW}

\section{Acknowledgements}

X.G.W. would like to thank Michael  Levin for helpful discussions and for
sharing his result of bosonic SPT phases in $(2+1)$-dimensions.\cite{LU1}  This
motivated us to calculate $\cH^{1+d}[U(1),U(1)]$, which reproduced his results
for $d=2$.  We would like to thank Geoffrey Lee, Jian-Zhong Pan, and
Zhenghan Wang for many very helpful discussions on group cohomology for
discrete and continuous groups. Z.C.G.  would like to thank Dung-Hai Lee for
discussion on the possibility of discretized Berry phase term in $1+1$D. This
research is supported by NSF Grant No. DMR-1005541 and NSFC 11074140. Z.C.G. is
supported by NSF Grant No. PHY05-51164.

\appendix

\section{Making the condition \eqn{UPsi} a local condition} \label{local}

We can make the condition \eqn{UPsi} on $U^{\v i}$ a local condition.  Instead
of requiring  \eqn{UPsi}, we may require $U^{\v i}$ to satisfy
\begin{align}
\label{UUUUWV}
&\ \ \ \
(U^{\v i}\otimes U^{\v i+\v x}\otimes U^{\v i+\v y}\otimes U^{\v i+\v x+\v y})
(P^{\v i}\otimes p^{\v i})
\\
&=
(P^{\v i}\otimes p^{\v i})
(U^{\v i}\otimes U^{\v i+\v x}\otimes U^{\v i+\v y}\otimes U^{\v i+\v x+\v y})
(P^{\v i}\otimes p^{\v i})
\nonumber
\end{align}
for certain  projection operators $ P^{\v i}$ and $ p^{\v i}$ with Tr$p^{\v
i}=1$.  Here $U^{\v i}\otimes U^{\v i+\v x}\otimes U^{\v i+\v y}\otimes U^{\v
i+\v x+\v y}$ and $P^{\v i}\otimes p^{\v i}$ are matrices given by
\begin{align}
&\ \ \ \
(U^{\v i}\otimes U^{\v i+\v x}\otimes U^{\v i+\v y}\otimes U^{\v i+\v x+\v y})
\nonumber\\
& \to
 U^{\v i}_{\al_1 \al_2 \al_3 \al_4,\al_1'\al_2'\al_3'\al_4'}
 U^{\v i+\v x}_{\bt_1 \bt_2 \bt_3 \bt_4,\bt_1'\bt_2'\bt_3'\bt_4'}
\times
\nonumber\\
&\ \ \ \ \ \ \ \ \ \ \ \ \ \ \ \ \ \
 U^{\v i+\v y}_{\ga_1 \ga_2 \ga_3 \ga_4,\ga_1'\ga_2'\ga_3'\ga_4'}
 U^{\v i+\v x+\v y}_{\la_1 \la_2 \la_3 \la_4,\la_1'\la_2'\la_3'\la_4'}
\end{align}
and
\begin{align}
& (P^{\v i}\otimes p^{\v i})
  \to p^{\v i}_{\al_1 \bt_2 \ga_3 \la_4,\al_1'\bt_2'\ga_3'\la_4'}\times
\\
& \ \ \ \
 P^{\v i}_{
\al_2 \al_3 \al_4
\bt_1 \bt_3 \bt_4
\ga_1 \ga_2 \ga_4
\la_1 \la_2 \la_3,
\al_2'\al_3'\al_4'
\bt_1'\bt_3'\bt_4'
\ga_1'\ga_2'\ga_4'
\la_1'\la_2'\la_3'},
\nonumber
\end{align}
The condition \eqn{UPsi} implies the condition \eqn{UUUUWV}
because, in the canonical form, the states on the sites
$\al_1,\ \bt_2,\ \ga_3,\ \la_4$ and the states on the sites $\al_2,\ \al_3,\
\al_4,\ \bt_1,\ \bt_3,\ \bt_4,\ \ga_1,\ \ga_2,\ \ga_4,\ \la_1,\ \la_2,\ \la_3$
are unentangled (see Fig. \ref{csymm}).

\section{Representations and projective Representations}
\label{prorep}

Let us consider a group $G$ that may contain anti-unitary time reversal
transformation. We can divide the group elements into two classes:
\begin{align}
 s(g)=1 \text{ or } -1,\ \ \ \ g\in G.
\end{align}
The group elements that contain an odd number of time-reversal operations have
$s(g)=-1$ and the group elements that contain an even number of time-reversal
operations have $s(g)=1$.

Unitary matrices $u(g)$ form a representation of symmetry group $G$ if
\begin{align}
\label{usuu}
 u(g_1) u_{s(g_1)}(g_2) = u(g_1g_2).
\end{align}
where $u_{s(g_1)}(g_2)=u(g_2)$ if $s(g_1)=1$ and $u_{s(g_1)}(g_2)=[u(g_2)]^*$ if $s(g_1)=-1$.

The above relation is obtained from the following mapping
\begin{align}
 g \to u(g),  \text{ if } s(g)=1;\ \ \ \
 g \to u(g)K, \text{ if } s(g)=-1.
\end{align}
Here $K$ is the anti-unitary operator
\begin{align}
 K a = a^* K,
\end{align}
where $a$ is a complex number. For example, if $s(g_1)=s(g_2)=-1$
and $s(g_1g_2)=1$, we require that
\begin{align}
 u(g_1)K u(g_2)K = u(g_1g_2)
\end{align}
which leads to \eqn{usuu}.

Matrices $u(g)$ form a projective representation of symmetry group $G$ if
\begin{align}
 u(g_1)u_{s(g_1)}(g_2)=\om(g_1,g_2)u(g_1g_2),\ \ \ \ \
g_1,g_2\in G.
\end{align}
Here $\om(g_1,g_2) \in U(1)$ and $\om(g_1,g_2) \neq 0$, which is called the
factor system of the projective representation.
The associativity requires that
\begin{align}
 [ u(g_1)u_{s(g_1)}(g_2) ] u_{s(g_1g_2)}(g_3)
= u(g_1) [u(g_2) u_{s(g_2)}(g_3)]_{s(g_1)}
\end{align}
or
\begin{align}
&\ \ \ \
 \om(g_1,g_2) \om(g_1g_2,g_3)u(g_1g_2g_3)
\nonumber\\
&=\om^{s(g_1)}(g_2,g_3) \om(g_1,g_2g_3) u(g_1g_2g_3)
\end{align}
Thus, the factor system satisfies
\begin{align}
\label{pRepCond}
 \om^{s(g_1)} (g_2,g_3)\om(g_1,g_2g_3)&=
 \om(g_1,g_2)\om(g_1g_2,g_3),
\end{align}
for all $g_1,g_2,g_3\in G$.
If $\om(g_1,g_2)=1$, $u(g)$ reduces to the usual linear representation of
$G$.

A different choice of pre-factor for the representation matrices
$u'(g)= \bt(g) u(g)$ will lead to a different factor system
$\om'(g_1,g_2)$:
\begin{align}
\label{omom}
 \om'(g_1,g_2) =
\frac{\bt(g_1g_2)}{\bt(g_1)\bt^{s(g_1)}(g_2)}
 \om(g_1,g_2).
\end{align}
We regard $u'(g)$ and $u(g)$ that differ only by a pre-factor as equivalent
projective representations and the corresponding factor systems $\om'(g_1,g_2)$
and $\om(g_1,g_2)$ as belonging to the same class $\om$.

Suppose that we have one projective representation $u_1(g)$ with factor system
$\om_1(g_1,g_2)$ of class $\om_1$ and another $u_2(g)$ with factor system
$\om_2(g_1,g_2)$ of class $\om_2$, obviously $u_1(g)\otimes u_2(g)$ is a
projective presentation with factor group $\om_1(g_1,g_2)\om_2(g_1,g_2)$. The
corresponding class $\om$ can be written as a sum $\om_1+\om_2$. Under such an
addition rule, the equivalence classes of factor systems form an Abelian group,
which is called the second cohomology group of $G$ and denoted as
$\cH^2[G,U_T(1)]$.  The ``zero'' element $0 \in \cH^2[G,U_T(1)]$ is the class
that corresponds to the linear representation of the group.  The best known
example of projective representation is the spin-1/2 representation of $SO(3)$.
The integer spins correspond to the linear representations of $SO(3)$.

\section{1D representations and projective representations of
$U(1)\times Z_2$ and $U(1)\rtimes Z_2$ }

\label{1Dproj}

In this section, we are going to discuss the 1D representations and projective
representations of four groups $U(1)\times Z_2$, $U(1)\rtimes Z_2$, $U(1)\times
Z_2^T$, and $U(1)\rtimes Z_2^T$.  As group, $Z_2$ and $Z_2^T$ are actually the
same group.  However, the generator $t$ of $Z_2$ corresponds to a usual
symmetry transformation which has a unitary representation.  The generator $T$
of $Z_2^T$ corresponds to the time reversal transformation which has a
anti-unitary representation.

Let $U_\th$, $\th\in [0,2\pi)$, be an element in $U(1)$.
The four groups are defined by the following relations
\begin{align}
 U(1)\times Z_2 :\ \ & t U_\th=U_\th t;
\nonumber\\
 U(1)\times Z_2^T :\ \ & T U_\th=U_\th T;
\nonumber\\
 U(1)\rtimes Z_2 :\ \ & t U_\th=U_{-\th} t;
\nonumber\\
 U(1)\rtimes Z_2^T :\ \ & T U_\th=U_{-\th} T.
\end{align}
Their representations are given by matrix
functions $M(U_\th)$ and $M(t)$ (or $M(T)K$).

A 1D representation of $U(1)\times Z_2$ has a form
\begin{align}
\label{MU1a}
M(U_\th) = \e^{n \imth \th}, \ \ \ M(t)=\eta=\pm 1,
\end{align}
where $n \in \Z$.  One can check that  $M(T)M(U_\th)=M(U_\th)M(T)$, for any
$n$.  So the 1D representation of $U(1)\times Z_2$ is labeled by $n$ and
$\eta$ (or by $\Z\times \Z_2$).

A 1D representation of $U(1)\rtimes Z_2$ also has a form \eqn{MU1a} One can
check that  $M(T)M(U_\th)=M(U_{-\th})M(T)$ only when $n=0$.  So there are two 1D
representation of $U(1)\rtimes Z_2$ labeled by $\Z_2$ (corresponding to
$\eta=\pm 1$).

A 1D representation of $U(1)\times Z_2^T$ has a form
\begin{align}
\label{MU1}
M(U_\th) = \e^{n \imth \th}, \ \ \ M(T)K=\e^{\imth \phi}K,
\end{align}
where $n \in \Z$ and $\phi\in \R$.
Note that $M(T)KM(T)K=1$ for any $\phi$.
We find
\begin{eqnarray}
& \ \ \ \
M(T)K M(U_\th)
= M(T)K \e^{n \imth \th}
\nonumber\\
&=  \e^{-n \imth \th} M(T)K.
\end{eqnarray}
Thus $M(T)K M(U_\th) =  M(U_\th)M(T)K$ only when $n=0$.  Also
under an unitary transformation
$\e^{\imth \phi}$, $M(T)K$ transforms as
\begin{align}
\e^{\imth \phi} (\e^{\imth \phi}K) \e^{-\imth \phi}=
\e^{\imth (\phi+2\vphi)}K.
\end{align}
So different $\phi$ correspond to the same 1D representation.
Therefore, there is only one
1D representation for $U(1)\times Z_2^T$.

From the above calculation, we note that $M(T)K |n\> \propto |-n\>$ where
$|n\>$ is an eigenstate of $M(U_\th)$: $M(U_\th)|n\>=  \e^{n \imth \th}|n\>$.
So the group $ Z_2^T\times U(1)$ describes the symmetry group of a spin system
with time reversal and $S_z$ spin rotation symmetry (the $U(1)$ symmetry).

A 1D representation of $U(1)\rtimes Z_2^T$ also has a form
\eqn{MU1}.
We find
\begin{eqnarray}
\label{MU2}
& \ \ \ \
M(T)K M(U_\th)
= M(T)K \e^{n \imth \th}
\nonumber\\
&=  \e^{-n \imth \th} M(T)K
= M(U_{-\th}) M(T)K .
\end{eqnarray}
Thus $M(T)K M(U_\th) =  M(U_{-\th})M(T)K$
for any $n \in \Z$, and the
1D representations for $U(1)\rtimes Z_2^T$
are labeled by $\Z$.

The above relation \eq{MU2} also allows us to show $M(T)K |n\> \propto |n\>$,
where $|n\>$ is an eigenstate of $M(U_\th)$: $M(U_\th)|n\>=  \e^{n \imth
\th}|n\>$.  This is the expected transformation of time reversal for boson
systems, where $n$ is the boson number. Therefore $U(1) \rtimes Z_2^T$ is the
symmetry group of boson systems with time reversal symmetry and boson number
conservation.

Next, let us discuss the projective representations
of the four groups.  First, let us consider
$U(1)\times Z_2$, whose projective representations
may have a form
\begin{align}
M(U_\th)=\bpm \e^{n\imth \th} & 0  \\ 0 & \e^{m\imth \th} \\ \epm,
\ \ \
M(T)= \bpm 0 & 1  \\ 1 & 0 \epm .
\end{align}
One can check
\begin{align}
  M(T) M(U_\th) & =
\bpm
0 &  \e^{m\imth \th} \\
\e^{n\imth \th} &  0
\epm
 ,
\nonumber\\
M(U_{\th})
  M(T)
& =
\bpm
0 &  \e^{n\imth \th} \\
\e^{m\imth \th} &  0
\epm
 .
\end{align}
 $M(T) M(U_\th)$ and $M(U_\th) M(T)$ differ by a total phase only when $m= n
\in \Z$, in that case $M(T) M(U_\th)=M(U_\th) M(T)$.  Note that $M(T)M(T)=1$.
So we have a trivial projective representation.
If we choose
$M(T)= \bpm 0 & 1  \\ 1 & 0 \epm$,
we will have
 $M(T)M(T)=-1$.
But we still have a trivial projective representations,
since if we add an phase factor
$\t M(T)=\imth M(T)$, we have $\t M(T)\t M(T)=-1$
%If we add a phase factor $\t M(U_\th) = \e^{k \imth \th}
%M(U_\th)$, it will shift $m$ and $n$ by the same $k$. So the projective
%representations with different $n=m$ all belong to the same class.
Thus $U(1)\times Z_2$ has only one trivial class of projective representations.

Second, let us consider the projective representations for
$U(1)\rtimes Z_2$, which may have a form
\begin{align}
M(U_\th)=\bpm \e^{n\imth \th} & 0  \\ 0 & \e^{m\imth \th} \\ \epm,
\ \ \
M(T)= \bpm 0 & 1  \\ 1 & 0 \epm .
\end{align}
One can check
\begin{align}
  M(T) M(U_\th) & =
\bpm
0 &  \e^{m\imth \th} \\
\e^{n\imth \th} &  0
\epm
 ,
\nonumber\\
M(U_{-\th})
  M(T)
& =
\bpm
0 &  \e^{-n\imth \th} \\
\e^{-m\imth \th} &  0
\epm
 .
\end{align}
We have
\begin{align}
 M(T) M(U_\th)
= \e^{-(n+m)\imth \th} M(U_{-\th}) M(T)
\end{align}
Also note that $M(T)M(T)=1$.  So we have a non-trivial
projective representation.
If we add a phase factor $\t M(U_\th) = \e^{k \imth \th}  M(U_\th)$,
we will have
\begin{align}
 M(T) \t M(U_\th)
= \e^{-(n+m-2k)\imth \th} \t M(U_{-\th}) M(T)
\end{align}
So the projective representations with different $n$ and $m$
belong to two classes
$m+n$ = even and $m+n =$ odd.
Thus $U(1)\rtimes Z_2$ has two classes of projective representations
labeled by $\Z_2$.

The projective representation of $U(1)\times Z_2^T$ can have a form
\begin{align}
U_\th & \to  M(U_\th)=\bpm \e^{n\imth \th} & 0  \\ 0 & \e^{m\imth \th} \\ \epm,
\nonumber\\
T & \to  M(T)K= \bpm 0 & 1  \\ 1 & 0 \epm K.
\end{align}
One can check
\begin{align}
  M(T)K M(U_\th) & =
\bpm
0 &  \e^{-m\imth \th} \\
\e^{-n\imth \th} &  0
\epm
K
\nonumber\\
M(U_\th)
  M(T)K
& =
\bpm
0 &  \e^{n\imth \th} \\
\e^{m\imth \th} &  0
\epm
K .
\end{align}
We have
\begin{align}
 M(T)K M(U_\th)
= \e^{-(n+m)\imth \th} M(U_\th) M(T)K
\end{align}
Note that $M(T)KM(T)K=1$.  So we have a projective representation when $m, n
\in \Z$ .  If we add a phase factor $\t M(U_\th) = \e^{k \imth \th}  M(U_\th)$,
then
\begin{align}
 M(T)K \t M(U_\th)
= \e^{-(n+m +2 k)\imth \th} \t M(U_\th) M(T)K .
\end{align}
So the above projective representations belong
to two classes: $m+n=$ even and $m+n=$ odd.

The projective representation of
$Z_2^T\times U(1)$ may also have a form
\begin{align}
M(U_\th)=\bpm \e^{n\imth \th} & 0  \\ 0 & \e^{m\imth \th} \\ \epm,
\ \ \
M(T)K= \bpm 0 & -1  \\ 1 & 0 \epm K.
\end{align}
One can check
\begin{align}
  M(T)K M(U_\th) & =
\bpm
0 & - \e^{-m\imth \th} \\
\e^{-n\imth \th} &  0
\epm
K ,
\nonumber\\
M(U_\th)
  M(T)K
& =
\bpm
0 &  -\e^{n\imth \th} \\
\e^{m\imth \th} &  0
\epm
K .
\end{align}
We have
\begin{align}
 M(T)K M(U_\th)
= \e^{-(n+m)\imth \th} M(U_\th) M(T)K
\end{align}
Note that $M(T)KM(T)K=-1$.  So we also have a projective representation when
$m, n \in \Z$ .  Those projective representations also belong to two classes:
$m+n=$ even and $m+n=$ odd.
So $U(1)\times Z_2^T$ has four classes of projective representations
labeled by $\Z_2\times \Z_2$.

For the $U(1)\rtimes Z_2^T$ group, its
projective representation may have a form
\begin{align}
M(U_\th)=\bpm \e^{n\imth \th} & 0  \\ 0 & \e^{m\imth \th} \\ \epm,
\ \ \
M(T)K= \bpm 0 & -1  \\ 1 & 0 \epm K.
\end{align}
One can check
\begin{align}
  M(T)K M(U_\th) & =
\bpm
0 & - \e^{-m\imth \th} \\
\e^{-n\imth \th} &  0
\epm
K ,
\nonumber\\
M(U_{-\th})
  M(T)K
& =
\bpm
0 &  -\e^{-n\imth \th} \\
\e^{-m\imth \th} &  0
\epm
K .
\end{align}
 $M(T)K M(U_\th)$ and $M(U_\th) M(T)K$ differ by a total phase only when $m= n
\in \Z$.  Note that $M(T)KM(T)K=-1$.
If we add a phase factor $\t M(T)=\e^{\imth \phi}M(T)$,
we still have $\t M(T)K\t M(T)K=-1$
So we have a non-trivial projective representation.
Those projective representations for different $n=m$ all belong to one class.
Thus $U(1)\rtimes Z_2^T$ has two classes of projective representations (the one
discussed above plus the trivial one) labeled by $\Z_2$.

\section{Group cohomology}
\label{Gcoh}

The above discussion on the factor system of a projective representation can be
generalized which give rise to a cohomology theory of group.  In this section,
we will briefly describe the group cohomology theory.\cite{RS}

\subsection{$G$-module}

For a group $G$, let $M$ be a $G$-module, which is an Abelian group (with
multiplication operation) on which $G$ acts compatibly with the multiplication
operation (\ie the Abelian group structure):
\begin{align}
\label{gm}
 g\cdot (ab)=(g\cdot a)(g\cdot b),\ \ \ \ g\in G,\ \ \ \ a,b\in M.
\end{align}

For the most cases studied in this paper, $M$ is simply the $U(1)$ group and
$a$ an $U(1)$ phase.  The multiplication operation $ab$ is the usual
multiplication of the $U(1)$ phases.  The group action is trivial: $g\cdot
a=a$, $g\in G$, $a\in U(1)$.  We will denote such a trivial $G$-module as
$M=U(1)$.

For a group $G$ that contain time-reversal operation, we can define a
non-trivial $G$-module which is denoted as $U_T(1)$.  $U_T(1)$ is also a $U(1)$
group whose elements are the $U(1)$ phases.  The multiplication operation $ab$,
$a,b\in U_T(1)$, is still the usual multiplication of the $U(1)$ phases.
However, the group action is non-trivial now: $g\cdot a=a^{s(g)}$, $g\in G$,
$a\in U_T(1)$,
where $s(g)=1$ if $g$ contains no anti-unitary time reversal transformation $T$
and $s(g)=-1$ if $g$ contains one anti-unitary time reversal transformation
$T$.

The module defined above is actually a model over a ring $\Z$, since we have
the following operation $\Z\times M \to M$:
\begin{align}
 \forall n \in \Z,
 \forall a \in M,\ \ \
a^n \in M.
\end{align}
A model $M$ can be over a more general ring $R$ if  we have the operation $R
\times M \to M$:
\begin{align}
 \forall n \in ,
 \forall a \in M,\ \ \
a^n \in M,
\end{align}
such that
\begin{align}
 a^n b^n=(ab)^n,\
 a^n a^m = a^{m+n}, \
 a^{nm}=(a^m)^n, \
 a^{1_R}=a,
\end{align}
if $R$ has multiplicative identity $1_R$.

Such a general concept of a module over a ring is a generalization of the
notion of vector space, wherein the corresponding scalars are allowed to lie in
an arbitrary ring. As we have seen, modules also generalize the notion of
Abelian groups, which are modules over the ring of integers

\subsection{Algebraic definition of group cohomology}

Let $\om_n(g_1,...,g_n)$ be a function of $n$ group
elements whose value is in the $G$-module $M$. In other words, $\om_n:
G^n\to M$.
Let $\cC^n(G,M)=\{\om_n \}$ be the space of all such
functions.
Note that $\cC^n(G,M)$ is an Abelian group
under the function multiplication
$ \om''_n(g_1,...,g_n)= \om_n(g_1,...,g_n) \om'_n(g_1,...,g_n) $.
We define a map $d_n$ from $\cC^n[G,U_T(1)]$ to $\cC^{n+1}[G,U_T(1)]$:
\begin{align}
&\ \ \ \
(d_n \om_n) (g_1,...,g_{n+1})=
\nonumber\\
&
[g_1\cdot \om_n (g_2,...,g_{n+1})]
\om_n^{(-1)^{n+1}} (g_1,...,g_{n}) \times
\nonumber\\
&\ \ \ \ \
\prod_{i=1}^n
\om_n^{(-1)^i} (g_1,...,g_{i-1},g_ig_{i+1},g_{i+2},...g_{n+1})
\end{align}
Let
\begin{align}
 \cB^n(G,M)=\{ \om_n| \om_n=d_{n-1} \om_{n-1}|  \om_{n-1} \in \cC^{n-1}(G,M) \}
\end{align}
and
\begin{align}
 \cZ^n(G,M)=\{ \om_{n}|d_n \om_n=1,  \om_{n} \in \cC^{n}(G,M) \}
\end{align}
$\cB^n(G,M)$ and $\cZ^n(G,M)$ are also Abelian groups
which satisfy $\cB^n(G,M) \subset \cZ^n(G,M)$ where
$\cB^1(G,M)\equiv \{ 1\}$.
The $n$-cocycle of $G$ is defined as
\begin{align}
 \cH^n(G,M)= \cZ^n(G,M) /\cB^n(G,M)
\end{align}

Let us discuss some examples.  We choose $M=U_T(1)$ and $G$ acts as: $g\cdot
a=a^{s(g)}$, $g\in G$, $a\in U_T(1)$.  In this case $\om_n(g_1,...,g_n)$ is just
a phase factor.
{}From
\begin{align}
 (d_0 \om_0)(g_1)= \om_0^{s(g_1)}/\om_0
\end{align}
we see that
\begin{align}
 \cZ^0[G,U_T(1)]=\{  \om_0| \om_0^{s(g_1)}=\om_0 \} \equiv U_T^G(1)
\end{align}
If $G$ contain time reversal, $U_T^G(1)=\{ 1,-1\}$.
If $G$ does not contain time reversal, $U_T^G(1)=U(1)$.
Since $\cB^0[G,U_T(1)]\equiv \{ 1\}$ is trivial,
we obtain $\cH^0[G,U_T(1)]=U_T^G(1)$.

{}From
\begin{align}
 (d_1 \om_1)(g_1,g_2)= \om_1^{s(g_1)}(g_2)\om_1(g_1)/\om_1(g_1g_2)
\end{align}
we see that
\begin{align}
 \cZ^1[G,U_T(1)]=\{  \om_1| \om_1(g_1)\om_1^{s(g_1)}(g_2)=\om_1(g_1g_2) \} .
\end{align}
Also
\begin{align}
\cB^1[G,U_T(1)]= \{ \om_1|\om_1(g_1)=\om_0^{s(g_1)}/\om_0\}
\end{align}
$\cH^1[G,U_T(1)]=\cZ^1[G,U_T(1)]/\cB^1[G,U_T(1)]$ is the set of all the
inequivalent 1D representations of $G$.

{}From
\begin{align}
&\ \ \ \ (d_2 \om_2)(g_1,g_2,g_3)
\\
&=
\om_2^{s(g_1)}(g_2,g_3) \om_2(g_1,g_2g_3)/\om_2(g_1g_2,g_3)\om_2(g_1,g_2)
\nonumber
\end{align}
we see that
\begin{align}
& \cZ^2[G,U_T(1)]=\{  \om_2|
\\
&\ \ \ \om_2(g_1,g_2g_3)\om_2^{s(g_1)}(g_2,g_3) =\om_2(g_1g_2,g_3)\om_2(g_1,g_2)
 \} .
\nonumber
\end{align}
and
\begin{align}
& \cB^2[G,U_T(1)]=\{ \om_2|\om_2(g_1,g_2)=\om_1^{s(g_1)}(g_2)\om_1(g_1)/\om_1(g_1g_2)
 \} .
\end{align}
The 2-cohomology group
$\cH^2[G,U_T(1)]=\cZ^2[G,U_T(1)]/\cB^2[G,U_T(1)]$ classify the
projective representations discussed in section \ref{prorep}.

{}From
\begin{align}
&\ \ \ \ (d_3 \om_3)(g_1,g_2,g_3,g_4)
\nonumber\\
&= \frac{ \om_3^{s(g_1)}(g_2,g_3,g_4) \om_3(g_1,g_2g_3,g_4)\om_3(g_1,g_2,g_3) }
{\om_3(g_1g_2,g_3,g_4)\om_3(g_1,g_2,g_3g_4)}
\end{align}
we see that
\begin{align}
& \cZ^3[G,U_T(1)]=\{  \om_3|
\\
&\ \ \ \frac{ \om_3^{s(g_1)}(g_2,g_3,g_4) \om_3(g_1,g_2g_3,g_4)\om_3(g_1,g_2,g_3) }
{\om_3(g_1g_2,g_3,g_4)\om_3(g_1,g_2,g_3g_4)}
=1
 \} .
\nonumber
\end{align}
and
\begin{align}
&\ \ \ \
 \cB^3[G,U_T(1)]
\\
&=\{ \om_3| \om_3(g_1,g_2,g_3)=\frac{
\om_2^{s(g_1)}(g_2,g_3) \om_2(g_1,g_2g_3)}{\om_2(g_1g_2,g_3)\om_2(g_1,g_2)}
 \},
\nonumber
\end{align}
which give us the 3-cohomology group
$\cH^3[G,U_T(1)]=\cZ^3[G,U_T(1)]/\cB^3[G,U_T(1)]$.

In this paper, we will show that $\cH^{1+d}[G,U_T(1)]$ can classify
SPT phases in $d$-spatial dimensions with an on-site unitary symmetry group
$G$.  Here the on-site symmetry group $G$ may contain time-reversal operations.

\subsection{Geometric interpretation of group cohomology}

In the following, we will describe a
geometric interpretation of group cohomology.
First, let us introduce
the map $\nu_n: G^{n+1} \to M$ that satisfy
\begin{align}
\label{gnun}
 g\cdot \nu_n(g_0,g_1,...,g_n) =  \nu_n(gg_0,gg_1,...,gg_n),
\end{align}
for any $g\in G$.  We will call such a map $\nu_n$ a $n$-cochain:
\begin{align}
\label{ncyc}
\cC^n(G,M)=
\{\nu_n|
g\cdot \nu_n(g_0,...,g_n) =  \nu_n(gg_0,...,gg_n)
\}
.
\end{align}
$\om_n$ discussed above is one-to-one related to $\nu_n$
through
\begin{align}
\om_n(g_1,...,g_n)
&=
\nu_n(1,g_1,g_1g_2,...,g_1\cdots g_n)
\nonumber\\
&=
\nu_n(1,\t g_1, \t g_2, ...,\t g_n)
\end{align}
where
$ \t g_i =g_1g_2\cdots g_i$.

We can rewrite the $d_n$ map,
$d_n: \om_n \to \om_{n+1}$, as
$d_n: \nu_n \to \nu_{n+1}$:
\begin{widetext}
\begin{align}
&\ \ \ \
(d_n \nu_n) (1, g_1, g_1g_2,...,g_1\cdots g_{n+1})
\nonumber\\
&=g_1\cdot
\nu_n (1, g_2,g_2g_3,...,g_2\cdots g_{n+1})
\nu_n^{(-1)^{n+1}} (1, g_1,g_1g_2,...,g_1\cdots g_{n})
\nu_n^{-1}(1,g_1g_2,g_1g_2g_3,...,g_1\cdots g_{n}) \times
\nonumber\\
&
\ \ \ \ \ \ \ \ \ \ \ \ \ \ \ \ \ \ \ \ \ \
\ \ \ \ \ \ \ \ \ \ \ \ \ \ \ \ \ \ \ \ \ \
\ \ \ \ \ \ \ \ \ \ \ \ \ \ \ \ \ \ \ \ \ \
\ \ \ \ \ \ \ \ \ \ \ \ \ \ \ \ \ \ \ \ \ \
\ \ \ \ \ \ \ \ \ \ \ \ \ \ \ \ \ \ \ \ \ \
\nu_n(1,g_1,g_1g_2g_3,...,g_1\cdots g_{n}) ...
\nonumber\\
&=
\nu_n (g_1, g_1g_2,g_1g_2g_3,...,g_1g_2\cdots g_{n+1})
\nu_n^{(-1)^{n+1}} (1, g_1,g_1g_2,...,g_1\cdots g_{n})
\nu_n^{-1}(1,g_1g_2,g_1g_2g_3,...,g_1\cdots g_{n}) \times
\nonumber\\
&
\ \ \ \ \ \ \ \ \ \ \ \ \ \ \ \ \ \ \ \ \ \
\ \ \ \ \ \ \ \ \ \ \ \ \ \ \ \ \ \ \ \ \ \
\ \ \ \ \ \ \ \ \ \ \ \ \ \ \ \ \ \ \ \ \ \
\ \ \ \ \ \ \ \ \ \ \ \ \ \ \ \ \ \ \ \ \ \
\ \ \ \ \ \ \ \ \ \ \ \ \ \ \ \ \ \ \ \ \ \
\nu_n(1,g_1,g_1g_2g_3,...,g_1\cdots g_{n})  ...
\nonumber\\
&=
\nu_n (\t g_1, \t g_2,\t g_3,...,\t g_{n+1})
\nu_n^{(-1)^{n+1}} (1, \t g_1,\t g_2,...,\t g_{n})
\nu_n^{-1}(1,\t g_2,\t g_3,...,\t g_{n})
\nu_n(1,\t g_1,\t g_3,...,\t g_{n})  ...
\end{align}
\end{widetext}
The above can be rewritten as
(after the renaming $\t g_i \to g_i$)
\begin{align}
\label{dnnun}
&\ \ \ \ (d_n \nu_n) ( g_0, g_1,..., g_{n+1})
\nonumber\\
& = \prod_{i=0}^{n+1}
\nu_n^{(-1)^i}( g_0,.., g_{i-1}, g_{i+1},..., g_{n+1})
\end{align}
which is a more compact and a nicer expression of the $d_n$ operation.

When $n=1$, we have
\begin{align}
\label{d1}
(d_1  \nu_1) ( g_0,  g_1, g_2)
=
 \nu_1 ( g_1,  g_2)  \nu_1 ( g_0,  g_1) /  \nu_1 ( g_0, g_2)
\end{align}
For $n=2$:
\begin{align}
\label{d2}
(d_2  \nu_2) ( g_0,  g_1, g_2, g_3)
&=
\frac{ \nu_2 ( g_1,  g_2, g_3)  \nu_2 ( g_0, g_1, g_3)}{
 \nu_2 ( g_0,  g_2, g_3)  \nu_2 ( g_0, g_1, g_2) }
\end{align}
and for $n=3$:
\begin{align}
\label{d3}
&\ \ \ \
(d_3  \nu_3) ( g_0,  g_1, g_2, g_3,g_4)
\\
&=
\frac{ \nu_3(g_1,g_2,g_3,g_4)\nu_3(g_0,g_1,g_3,g_4)\nu_3(g_0,g_1,g_2,g_3)}{
 \nu_3(g_0,g_2,g_3,g_4)\nu_3( g_0,g_1,g_2,g_4) }
\nonumber
\end{align}

We may represent the 1-cochain, 2-cochain, and 3-cochain graphically by a line,
a triangle, and a tetrahedron with a branching structure respectively (see Fig.
\ref{d1d2}).  We note that, for example, when we use a tetrahedron with a
branching structure to represent a 3-cochain $\nu_3(g_0,g_1,g_2,g_3)$, the last
variable $g_3$ is at the vertex with all the edges point to the vertex (see
Fig.  \ref{d1d2}b).  After removing the $g_3$ vertex and the connected edges,
$g_2$ is at the vertex with all the remaining edges point to the vertex (see
Fig.  \ref{d1d2}b). This can be repeated.  We see that a tetrahedron with a
branching structure gives rise to a natural order $g_0,g_1,g_2,g_3$.  In
general, a $d$-cochain can be represented by a $d$-dimensional simplex with a
branching structure.  We also note that a $d$-dimensional simplex with a
branching structure can have two different chiralities (see Fig. \ref{chiral}).
The  simplex with one chirality correspond to $\nu_d$ and the  simplex with the
other chirality correspond to $\nu_d^{-1}$  (see \eqn{nus}).

In this way, we obtain a graphical representation of eqn.  (\ref{d1}) and
\eqn{d2} as in Fig.  \ref{d1d2}.  In the graphical representation, \eqn{dnu1}
implies that the value of a 1-cocycle $\nu_1$ on the closed loop (such as a
triangle) is 1 and \eqn{dnu2} implies that the value of a 2-cocycle $\nu_2$ on
the closed surface (such as a tetrahedron) is 1.

Let us choose $M=U(1)$ and consider a 1-form $\Om_1$ on the plan in Fig.
\ref{d1d2}a.  Then the differential form expression
\begin{align}
 \int_{(g_0,g_1,g_2)} \dd \Om_1=
\int_{g_0}^{g_1} \Om_1 -\int_{g_0}^{g_2} \Om_1+ \int_{g_1}^{g_2} \Om_1
\end{align}
give us \eqn{d1} if we set
\begin{align}
(d_1
\nu_1) ( g_0,  g_1, g_2) =\text{exp}\Big(\imth  \int_{(g_0,g_1,g_2)} \dd \Om_1\Big)
\end{align}
and
\begin{align}
 \nu_1(g_i,g_j)=\text{exp}\Big(\imth  \int_{g_i}^{g_j} \Om_1\Big).
\end{align}
Here $\int_{(g_0,g_1,g_2)}$ is the integration on the triangle $(g_0,g_1,g_2)$
in Fig. \ref{d1d2}a.  Similarly the differential form expression
\begin{align}
\int_{(g_0,g_1,g_2,g_3)} \dd \Om_2
&=
\int_{(g_1,g_2,g_3)} \Om_2
-\int_{(g_0,g_2,g_3)} \Om_2
\nonumber\\
&
+\int_{(g_0,g_1,g_3)} \Om_2
-\int_{(g_0,g_1,g_2)} \Om_2
\end{align}
give us \eqn{d2} if we set
\begin{align}
(d_2 \nu_2) ( g_0,  g_1, g_2, g_3) =
\text{exp}\Big(\imth  \int_{(g_0,g_1,g_2,g_3)} \dd \Om_2\Big)
\end{align}
and
\begin{align}
 \nu_2(g_i,g_j,g_k)=\text{exp}\Big(\imth  \int_{(g_i,g_j,g_k)} \Om_2\Big).
\end{align}
This leads to a geometric picture of group cohomology.  For example, if $\Om_2$
is a closed form, $\dd \Om_2=0$, the corresponding $\nu_2(g_i,g_j,g_k)$ will be
a cocycle.  If $\Om_2$ is an exact form, $\Om_2=\dd \Om_1$, the corresponding
$\nu_2(g_i,g_j,g_k)$ will be a coboundary.

\subsection{Cohomology on symmetric space}

We would like to mention that cohomology can also be defined on symmetric space $G/H$
where $H$ is a subgroup of $G$.  However, cocycles on the symmetric space $G/H$
can also be viewed as cocycles on the group space $G$
(the maximal symmetric space) and we have
$ \cZ^d(G/H, M) \subset \cZ^d(G, M)$ .
As a result, the SPT phases described by the quantized
topological $\th$-terms on the  symmetric space $G/H$ can all be described by the
quantized  topological $\th$-terms on the maximal symmetric space $G$.  So
classifying quantized  topological $\th$-terms  on the maximal symmetric space
$G$ lead to a classification of all SPT phases.

\section{Branching structure of a complex}
\label{branchapp}

\subsection{Branched simplex and its geometric meaning}

In geometry, a simplex is a generalization of the notion of a
triangle or tetrahedron to arbitrary dimensions. Specifically, an
$n$-simplex is an $n$-dimensional polytope which is the convex hull
of its $n+1$ vertices. It can also be viewed as a complete graph of
its $n+1$ vertices. For example, a $2$-simplex is a triangle, a
$3$-simplex is a tetrahedron, and a $4$-simplex is a pentachoron. An
$n$-simplex is the fundamental unit cell of $n$-manifolds, any
$n$-manifold can be divided into a set of $n$-simplexes through the
standard triangulation procedure. It is obvious that any invariant
under the re-triangulation of $n$-manifolds would automatically be a
topological invariant.

One of such examples is the famous state sum invariants of
$3$-manifolds first proposed by Turaev and Viro\cite{TV9265}. The
basic idea in their construction is associating a special data
set(e.g., $6j$-symbol) with each tetrahedron, and then showing the
states sum invariants under re-triangulations. However, their
construction requires a very high tetrahedral symmetry for the data
set, based on the assumption that all the vertices/edges/faces in a
tetrahedron are indistinguishable. Indeed, in a more general set up,
labeling the vertices/edges/faces is important because they are
actually distinguishable objects.

\begin{figure}[tb]
\begin{center}
\includegraphics[scale=0.3]{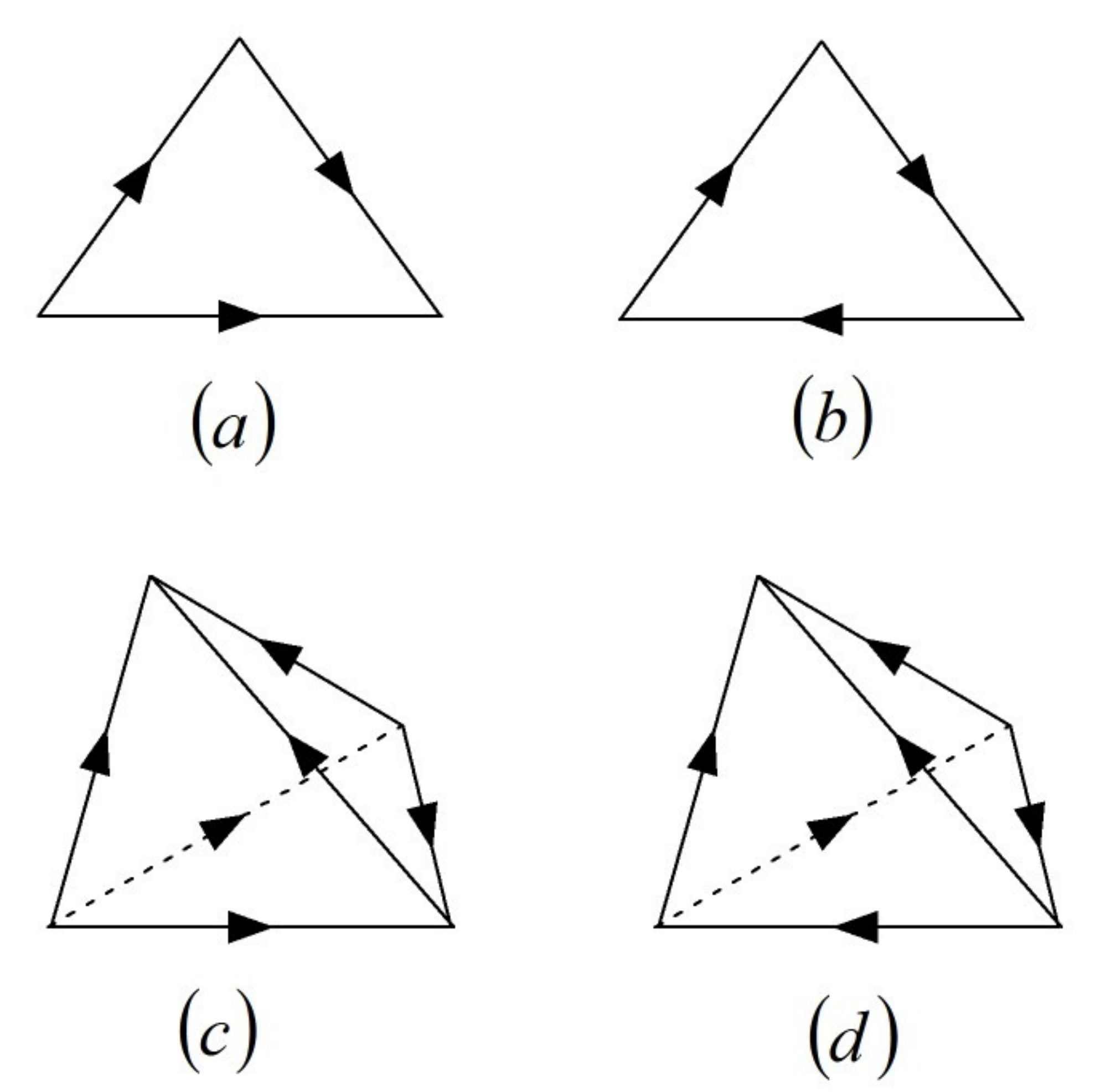}
%Fig. 22
\end{center}
\caption{Examples of allowed ((a),(c)) and unallowed ((b),(d))
branching for a $2$-simplex and a $3$-simplex. \label{branch}}
\end{figure}

A nice local scheme to label an $n$-simplex is given by a branching
structure. A branching is a choice of an orientation of each edge of
an $n$-simplex such that there is no oriented loop on any triangle.
For example, Fig. \ref{branch} (a) is a branched $2$-simplex and (c)
is a branched $3$-simplex. However, (b) is not allowed because all
its three edges contain the same orientations and thus form an
oriented loop. (d) is also not allowed because one of its triangle
contains an oriented loop. Actually, a consistent branched
triangulation can always be induced by a global labeling of the
vertices (We notice any labeling of the vertex
$v^i,i=0,1,2,\cdots,v^n$ will imply a nature ordering $v^i<v^j$ if
$i<j$). This is because any global ordering will induce a consistent
local ordering for all the triangles of an $n$-simplex. If we
associate an orientation from $i$ to $j$ if $v^i<v^j$, it is obvious that
there will be no oriented loop on any triangle.

\begin{figure}[tb]
\begin{center}
\includegraphics[scale=0.35]{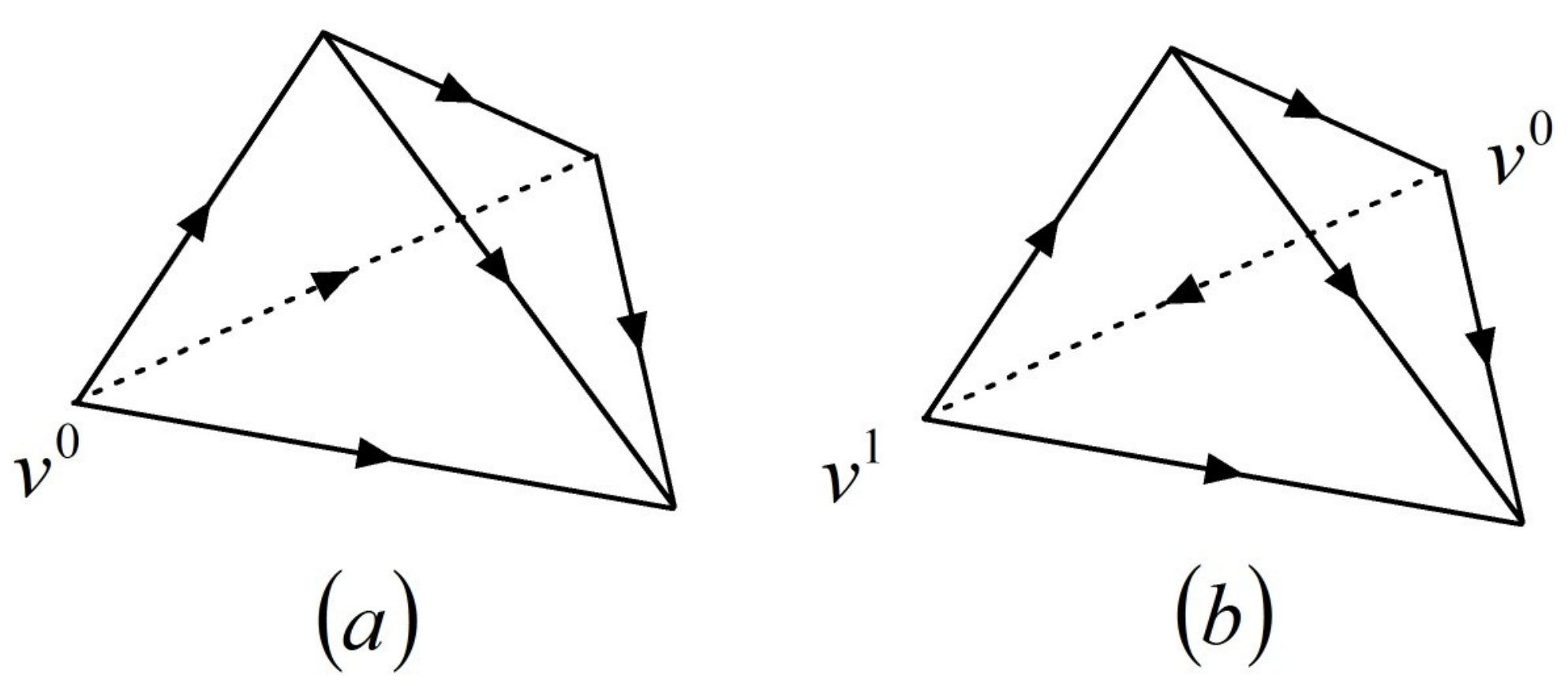}
%Fig. 23
\end{center}
\caption{(a): If a $3$-simplex contains a vertex with no incoming
edge, we can label this vertex as $v^0$ and canonically label the
vertices of the remaining $2$-simplex as $v^1,v^2,v^3$. Such a
scheme can be applied for arbitrary $n$-simplex if $n-1$-simplex has
a canonical label. (b): If a $3$-simplex contains no vertex without
incoming edge, then there must be a vertex with one incoming
edge.(Because canonical ordering is true for $2$-simplex.)If we
label this vertex as $v^1$, the vertex connect to $v^1$ through a
incoming edge must contain no incoming edge, otherwise the branching
rule will be violated. The above argument is true for $n$-simplex if
$n-1$ simplex can be canonically ordered.\label{ordering}}
\end{figure}

\begin{figure}[tb]
\begin{center}
\includegraphics[scale=0.35]{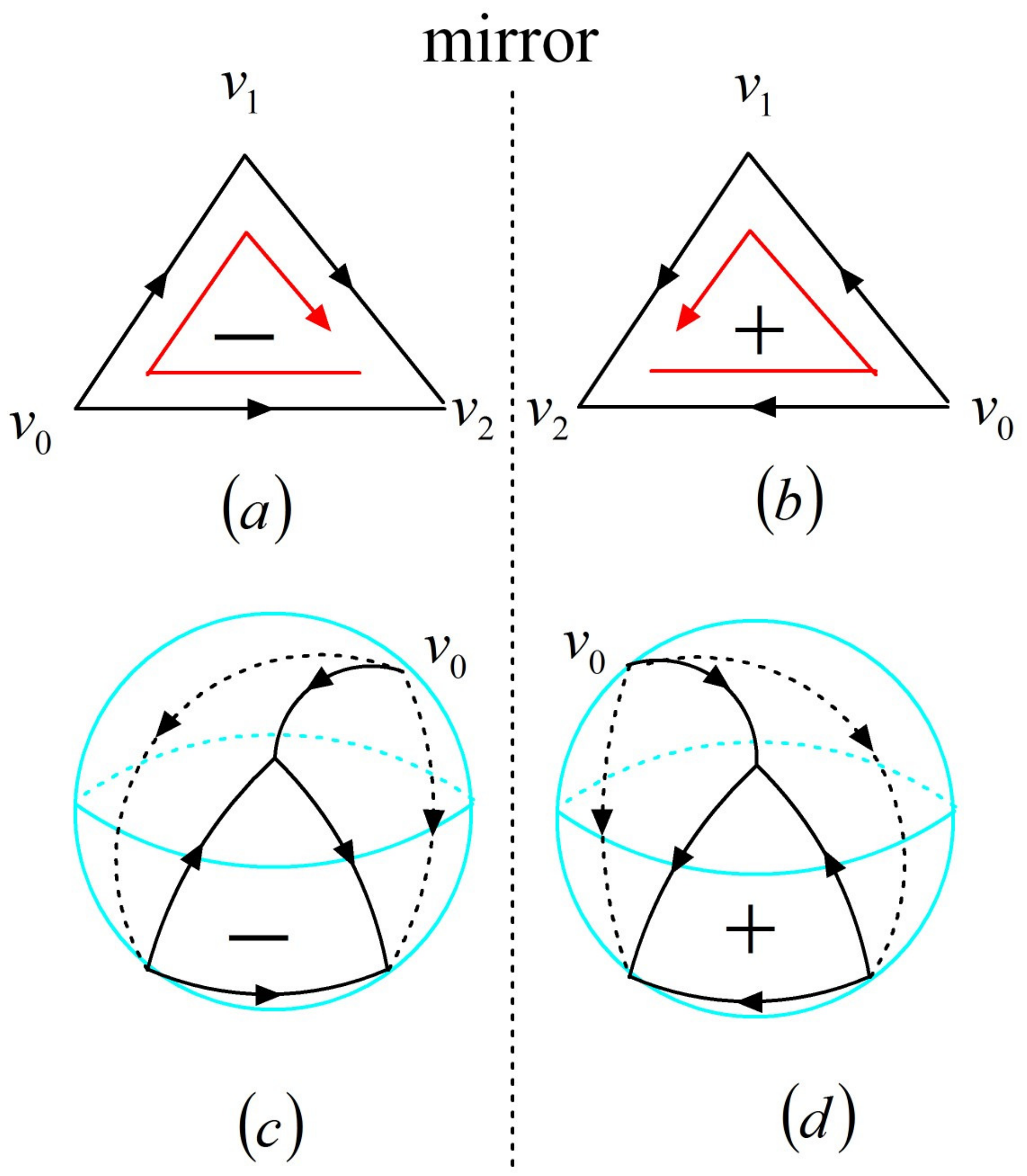}
%Fig. 24
\end{center}
\caption{
(Color online)
(a) and (b): $2$-simplex has two different chiralities,
depending on the clockwise or anticlockwise ordering of the
vertices. (c) and (d): The chirality of the $3$-simplex can be
determined by the chirality of the $2$-simplex which is opposite to
the vertex $v^0$. Similarly, the chirality of $n$-simplex can be
determined by the chirality of $n-1$ simplex which is opposite to
$v^0$. \label{chiral}}
\end{figure}

A branched $n$-simplex will have the following properties:

(a). Any given branching structure for an $n$-simplex will uniquely
determine a canonical ordering of the vertices. For example, Fig.
\ref{branch} (a) is a branched $2$-simplex with three vertices, one
of them contains no incoming edges, one of them contains one
incoming edge and the rest of them contains two incoming edges.
Thus, we can canonically identify the vertex corresponding to each
$v^i$, $i= 0,1,2$. Such a canonical labeling scheme can be applied
to any $n$-simplex, due to the fact that the $n+1$ vertices of any
$n$-simplex will be uniquely associated with $0,1,2,\cdots,n$
incoming edges.
\begin{proof}
Assuming the above statement is true for an $n$-simplex (The
statement is true when $n=2$, see Fig. \ref{branch}.), let us prove
it is also true for $(n+1)$-simplex. See Fig. \ref{ordering}, if the
$(n+1)$-simplex contains a vertex with no incoming edge, we can drop
this vertex and apply the statement for the remaining $n$-simplex.
If we label the $n+1$ vertices of the $n$-simplex as
$1,2,\cdots,n+1$, it is clear the vertex with no incoming edge can
be labeled as $0$. In the following we will prove a branched $(n+1)$-simplex 
must contain a vertex with no incoming edge. If an $(n+1)$-simplex 
does not contain any vertex with no incoming edge, it must
contain a vertex $v^1$ with one incoming edge. This is because if we
remove an arbitrary vertex (denoted as $v_0$) of the $n+1$-simplex,
the statement is true for the rest $n$-simplex. Hence we can always
find a vertex with one incoming edge. Let us denote this vertex as
$v^1$ and it is clear that the orientation of the edge that connects
$v^0,v^1$ must be outgoing towards $v^0$(Otherwise $v^1$ is a vertex
with no incoming edge). However, in this case, the edges that
connect $v^0$ and other vertices must be outgoing from $v^0$, if the
branching rule is not violated. Thus, $v^0$ is the vertex with no
incoming edge.
\end{proof}

(b) Although the branching rule of $n$-simplex uniquely determines
the ordering of the vertices, it could not uniquely
determine an $n$-simplex. This is simply because the mirror image of
a branched $n$-simplex is also a branched $n$-simplex with the same
vertices ordering. Thus, any branched $n$-simplex has a unique
chirality $\pm1$.
\begin{proof}
It is clear that $2$-simplex has two different chiralities (see Fig.
\ref{chiral}). Assuming that an $n-1$-simplex has a unique chirality, let
us deform the boundary of an $n$-simplex (which can be divided into
$n-1$-simplex) into an $n-1$ sphere. Due to the fact that there is
one and only one vertex $v^0$ of an $n$-simplex without incoming
edges, we can make a canonical convention and determine the
chirality of the $n$ simplex by the chirality of the $n-1$ simplex
opposite to $v^0$. Such a definition is sufficient because mirror
reflection will always change the chirality of the boundary of any
$n$-simplex. Indeed, we can define the chirality of the $n$ simplex
by the chirality of the $n-1$ simplex opposite to any $v^i$ up to a
global sign ambiguity (e.g., reversing the chiralities for all
$n$-simplices.).
\end{proof}

\begin{figure}[tb]
\begin{center}
\includegraphics[scale=0.4]{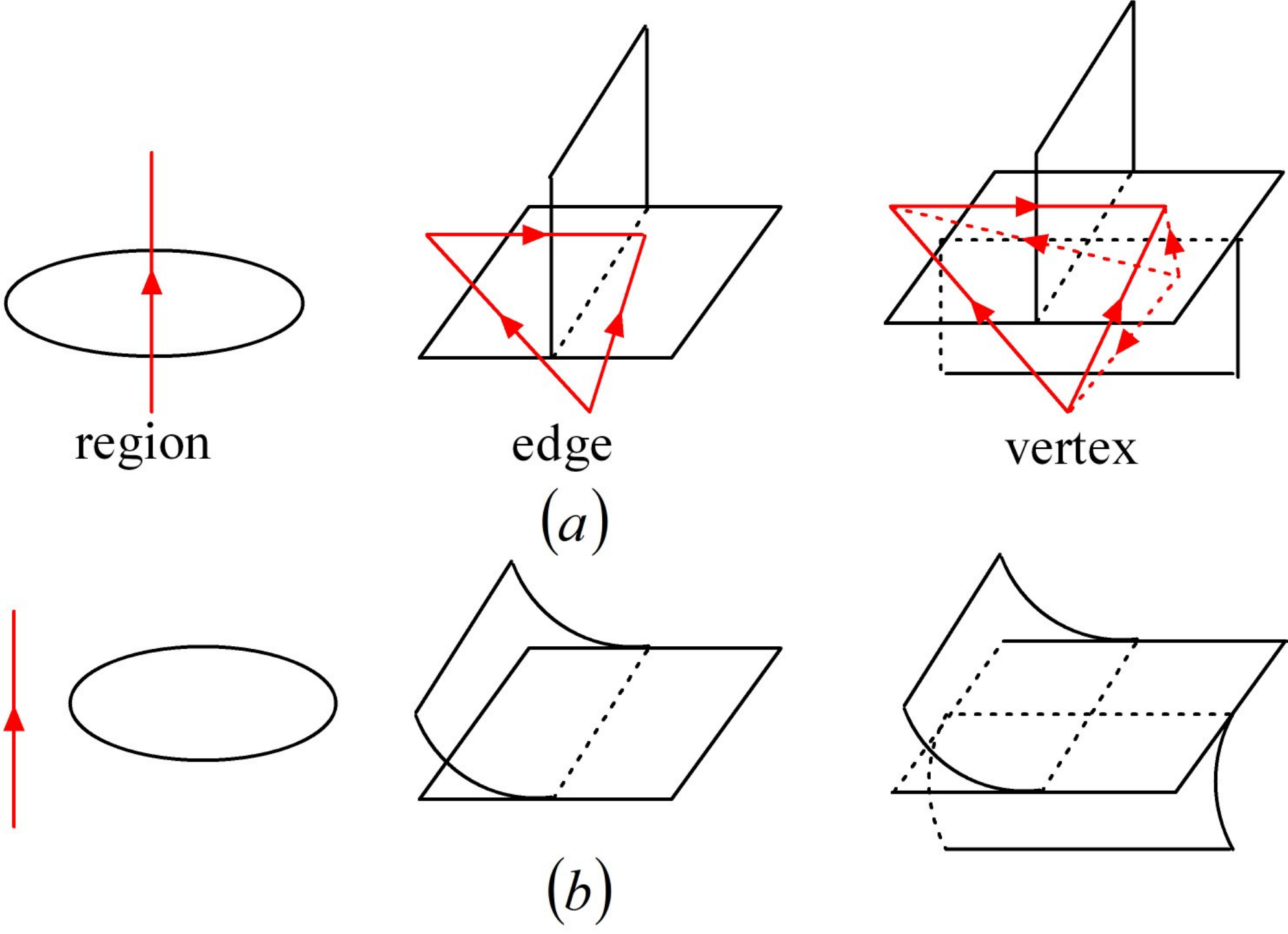}
%Fig. 25
\end{center}
\caption{
(Color online)
(a): Dual representation of branched tetrahedron. (b): We
can always induce a smooth structure on oriented manifold from the
branched polyhedron. The arrow on the left denotes the orientation
of the regions, which is locally identical to their orientations in
(a). \label{smooth}}
\end{figure}

The above two properties allow us to use the branched $n$ simplex to
represent an $n$-cocycle:
\begin{align}
\label{nus}
\nu_{n}^{s_{ij\ldots k}}(g_i,g_j,\cdots,g_k),
\end{align}
where $s_{ij\ldots k}=\pm1$ are determined
by the chirality of the simplex and $g_i,g_j,\cdots,g_k$ are defined
on the canonically ordered vertices $v^0,v^1,\cdots,v^{n}$.

Finally, let us briefly mention the geometric meaning of the
branched tetrahedron in $3$ dimensions.  See in Fig. \ref{smooth}
(a), in a dual picture, the orientations of the edges of tetrahedron
correspond to the orientations of the region of the simple
polyhedron. A branching on a simple polyhedron allows us to smoothen
its singularities and equip it with a smooth structure as shown in
Fig. \ref{smooth} (b). At a more rough level, it can be shown that
branched tetrahedron can be used to represent the Spin$^c$-structures
on the ambient manifolds\cite{C0527}.

\subsection{Basic moves}
To show the topological invariance of the amplitude:
\begin{align}
Z=\frac{\sum_{\{g_i\}}}{|G|^{N_v}}\prod \nu_{n}^{s_{ij\ldots
k}}(g_i,g_j,\cdots,g_k),
\end{align}
we need to generalize the Turaev-Viro moves to their branched
versions in arbitrary dimensions. Because each move will have many
different branched versions, it is not easy to check all the
branched versions case by case. In the following, we will introduce
a simple way to look at the basic moves.

\subsubsection{Graphic representation of $(\dd_n\nu_n)(g_0,g_1,\cdots,g_{n+1})$ and basic moves}

In last section we have shown a branched $n$-simplex can represent
an $n$-cocycle $\nu_{n}(g_i,g_j,\cdots,g_k)$ or its inverse
$v_{n}^{-1}(g_i,g_j,\cdots,g_k)$, depending on the chirality of the
branched $n$-simplex. Here we want to show the boundary of a
branched $n+1$-simplex can represent
$(\dd_n\nu_n)(g_0,g_1,\cdots,g_{n+1})=\prod_{i=0}^n
\nu_n^{(-1)^i}(g_0,\cdots,g_{i-1},g_{i+1},\cdots,g_{n+1})$. Since
any $n+1$-simplex branched simplex has a canonical ordering for its
$n+2$ vertices and its boundary contains $n+1$ $n$-simplices (We can
label these $n+1$ $n$-simplices as $S_n(v_i)$, where $v_i$ is the
vertex opposite to the $n$-simplex.), it is not surprising if we use
the $n$-simplex $S_n(v_i)$ to represent
$\nu_n(g_0,\cdots,g_{i-1},g_{i+1},\cdots,g_{n+1})$ or its inverse.
However, the key difficulty is that we need to show that the chirality of
the $n$-simplex $S_n(v_i)$ is determined by $\pm(-1)^i$, where the
global sign $\pm$ depends on the chirality of the $n+1$-simplex.
\begin{proof}
It is easy to check that the above statement is true for $n=2$. Thus, we
can represent $(\dd_2\nu_2)(g_0,g_1,g_2,g_3)$ as a branched
tetrahedron. Its boundary $2$-simplex $S_2(v^i)$ has opposite
chirality for even and odd $i$. If we assume the above statement is
true for $n-1$, let us proof it is also true for $n$. First let us
remove the vertex $v^0$ from the $n+1$-simplex that represents
$(\dd_n\nu_n)(g_0,g_1,\cdots,g_{n+1})$. By applying the statement to
the rest $n$-simplex, whose boundary contains $n$ $n-1$-simplices
$S_{n-1}(v^i)$ ($i=1,2,\cdots,n$) with chirality $\pm(-1)^i$.
However, according to the definition, the chirality of any
$S_n(v^i)$ ($i=1,2,\cdots,n$) can be defined by the $S_{n-1}(v^i)$
simplex which is opposite to $v^0$, thus we prove
$S_n(v^i)$ ($i=1,2,\cdots,n$) will also have opposite chiralities for
even and odd $i$. To prove that the above statement is also true for
$S_n(v^0)$, we can remove any vertex $j\neq 0$ and apply the same
scheme. Although there can be a global sign ambiguity for the
chirality of any $n$-simplex $S_n(v^i)$ with $i\neq j$, it is
sufficient to show $S_n(v^i)$ ($i=0,1,\cdots,n$) will have opposite
chiralities for even and odd $i$. Thus, $S_n(v^i)$ ($i=0,1,\cdots,n$)
will have chirality $\pm(-1)^i$ with the global sign $\pm$
determined by the chirality of the $n+1$-simplex.
\end{proof}

\begin{figure}[tb]
\begin{center}
\includegraphics[scale=0.4]{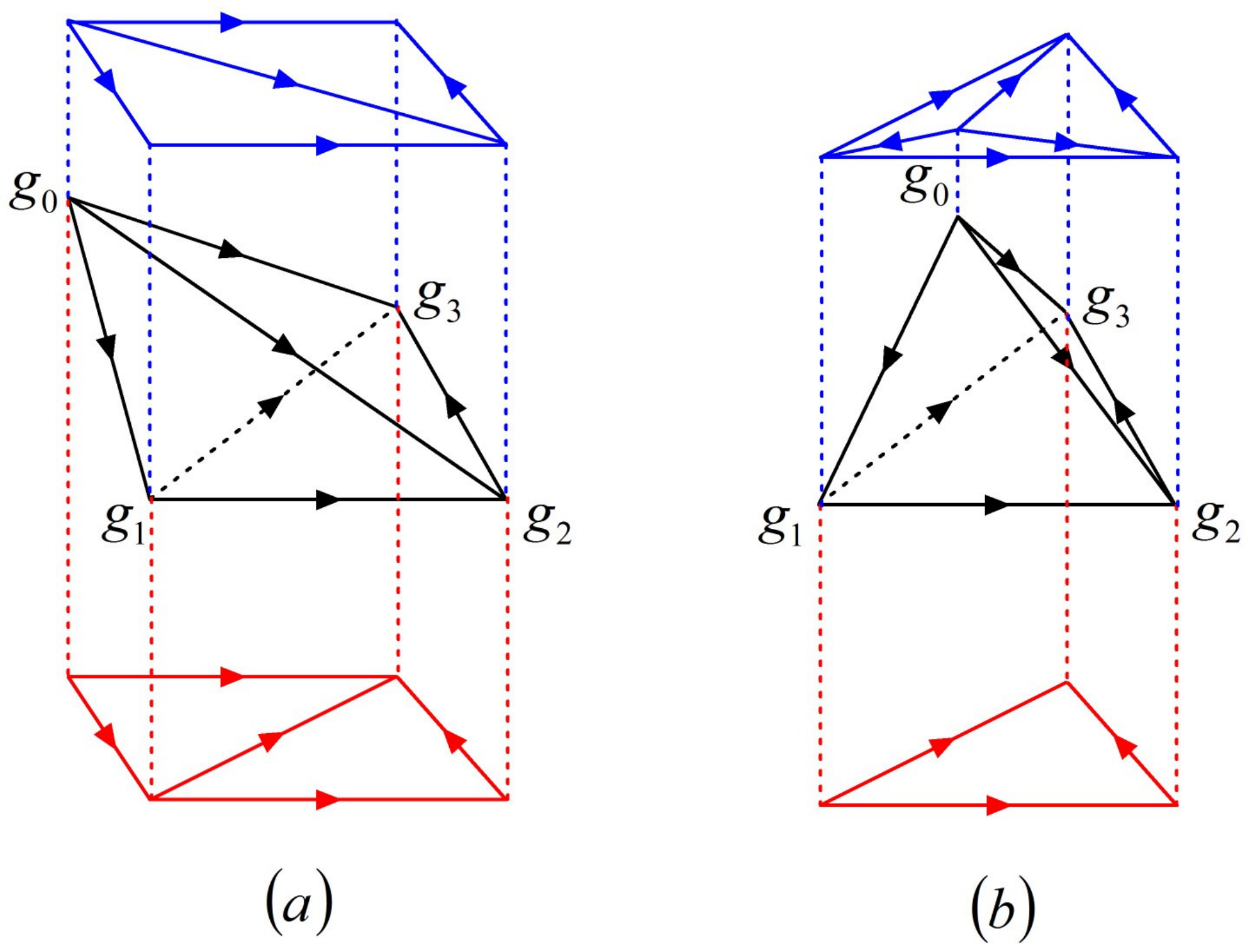}
%Fig. 26
\end{center}
\caption{
(Color online)
$(\dd_2\nu_2)(g_0,g_1,g_2,g_3)$ can be represented as the
boundary of a $3$-simplex. (a) and (b) correspond to two different
basic moves of $2$-simplexes. \label{moves}}
\end{figure}

Based on graphic representation of
$(\dd_n\nu_n)(g_0,g_1,\cdots,g_{n+1})$, it is easy to check that all the
basic moves are actually induced by the identity:
\begin{align}
\prod_{i=0}^n
\nu_n^{(-1)^i}(g_0,\cdots,g_{i-1},g_{i+1},\cdots,g_{n+1})\equiv1
\label{idenity}
\end{align}
Due to this identity, the chirality of the $n+1$-simplex is not
important because if we inverse both sides we will end up with the
same identity. Thus, we can pick up any $n+1$-simplex to represent
$(\dd_n\nu_n)(g_0,g_1,\cdots,g_{n+1})$ and project it into the
$n$-plane from opposite directions. The shadows of these two
projections are $n$-manifold with exact the same vertices. However,
they may correspond to different ways of triangulations. Thus, each
side of the equations of the basic moves will correspond to the two
different ways of projection. For example, Fig. \ref{moves} (a)
represents $2\leftrightarrow2$ moves:
\begin{align}
\nu_2(g_0,g_1,g_3)\nu_2(g_1,g_2,g_3)=\nu_2(g_0,g_1,g_2)\nu_2(g_0,g_2,g_3)
\end{align}
and Fig. \ref{moves} represents $1\leftrightarrow3$:
\begin{align}
\nu_2(g_1,g_2,g_3)=\nu_2(g_0,g_1,g_2)\nu_2(g_0,g_2,g_3)\nu_2^{-1}(g_0,g_1,g_3)
\end{align}
However, all these two equations will be equivalent to the identity:
\begin{align}
\nu_2(g_1,g_2,g_3)\nu_2^{-1}(g_0,g_2,g_3)
\nu_2(g_0,g_1,g_3)^{-1}\nu_2(g_0,g_1,g_2)\equiv1
\end{align}
We also notice the projection from opposite directions will induce
opposite chiralities for the boundary of the $3$-simplex, that's why
we need to change the chiralities of the $2$-simplex in one side of
basic moves. Such a change corresponds to inverse $\nu_2$ when we
move it from left side to right side of Eq. (\ref{idenity}), which
is consistent with multiplication rules of the complex number
$\nu_2$. It is also clear that the above two different moves correspond
to projections in different ways, hence all of them are equivalent
to the above identity. Similar argument is true for arbitrary
dimensions. In conclusion, the identity Eq. (\ref{idenity}) will
induce the correct $2\leftrightarrow n$ and $1\leftrightarrow n+1$
moves in $n$ dimensions.

\subsubsection{Some final details}

\begin{figure}[tb]
\begin{center}
\includegraphics[scale=0.35]{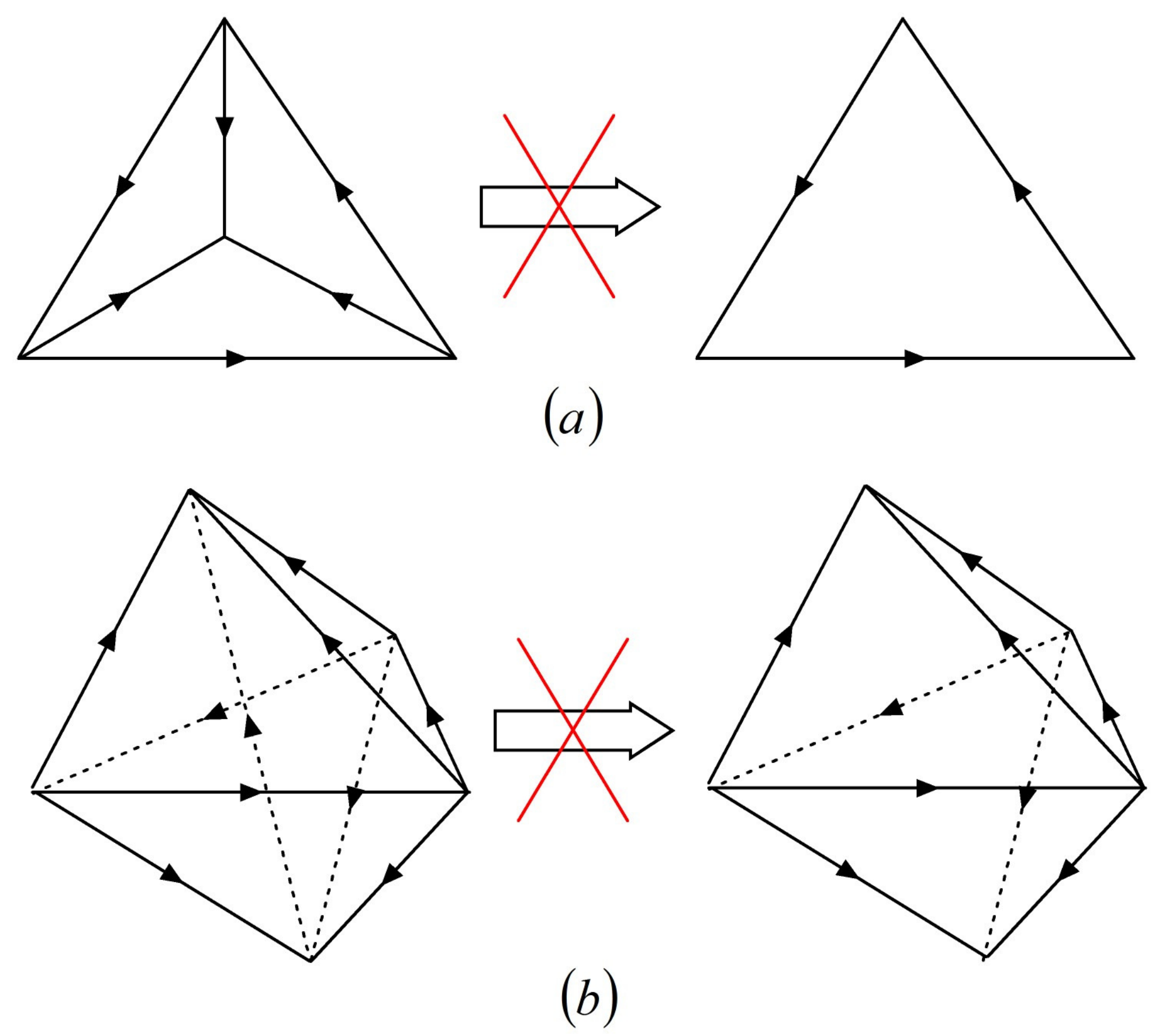}
%Fig. 27
\end{center}
\caption{
(Color online)
Examples of unallowed branched moves in $2$D and $3$D.
\label{unallow}}
\end{figure}
It looks like we have successfully generalized the basic moves to
their branched versions, however, there are still some subtle issues
here, especially in $2$D and $3$D. This is because the $n+1$-simplex
representation of the basic moves relies on the assumption that any basic
moves can be associated with a consistent branching structure, which
is not generically true in $2$D and $3$D. Fig. \ref{unallow} (a) is
an example of unallowed $3\rightarrow 1$ move in $2$D and (b) is an
example of unallowed $3\rightarrow 2$ move in $3$D. Although a
global labeling scheme will not allow triangulations to contain any
local pieces like the left part of Fig. \ref{unallow}, however,
those local configurations can be generated during local moves
because each simplex still satisfies the branching rule. In this
case, we can not directly apply these unallowed local moves to the
local pieces. Fortunately, in $2$D and $3$D it has been
proved\cite{C0527} that any branched triangulations can still be
connected through all the allowed moves. In high dimensions, we can
show that there are no unallowed moves like these. In the following let
us prove this statement.

We notice that the $2\rightarrow n$ move can be realized by adding one
more edge while $1\rightarrow n+1$ move can be realized by adding
one more vertex and $n+1$ edges, let us show that these two moves are
always allowed, by adding proper orientation(s) on the edge(s).
\begin{proof}
It is trivial to show that the $1\rightarrow n+1$ move is always allowed,
by adding vertex without incoming edge. The existence of
$2\rightarrow n$ move can be slightly more complicated. One can
easily check that it is true when $n=2$. Let us assume that it is true for
$n-1$-simplex now. We label the two unconnected vertices as $v^a$,
$v^b$ and label other vertices as $v^i$.(Notice we don't require
$a,b,i$ have an ordering here.) To show there always exist a
proper orientation for the edge $ab$, we only need to show that any
triangles made by $v^a$, $v^b$ and $v^i$ will not violate the
branching rule. If there exists a vertex $v^i$ containing two
incoming edges from $v^a,v^b$ or containing two outgoing edges
towards $v^a,v^b$, we can remove this vertex and apply the statement
to the rest $n-1$ simplex. It does not matter what's the
orientation on $ab$, the triangle $abi$ will not violate branching
rule. If such a vertex does not exist, we can show that there can be only
two cases: either $v^a$ contains no incoming edge, $v^b$ contains no
outgoing edge or the opposite case. In both cases, we can find a
proper orientation for the edge $ab$.
\end{proof}

The inverse of the above moves, namely, the $n\rightarrow 2$ move
can be realized by removing one edge, and the $n+1\rightarrow 1$ move
can be realized by removing one vertex. Now let us show that these moves
are always possible when $n>3$.
\begin{proof}
To prove that these moves are always possible in dimensions $n>3$, let us
understand why sometimes they are impossible in $2$D and $3$D.
Actually, this is simply because three edges of an oriented triangle
which violates the branching rule can belong to three different
simplices before we apply $3\rightarrow 1$ or $3\rightarrow 2$ move.
However, after we apply the move, they belong to the same triangle
and hence violate the branching rule. In high dimensions, when we
apply these inverse move, we always start from a complete graph and
the number of simplices is always larger than $3$. Thus, any
triangle must belong to one of the $n$-simplex and will not violating
the branching rule. If there is no triangle violate the branching
rule in a complete graph, of course there will be no triangle
violating the branching rule by removing edge or vertex.
\end{proof}

\section{(1+1)D solutions of \eqn{UPsi} }
\label{11D}

\subsection{ $U^{i}(g)$ is a linear representation }

\begin{figure}[tb]
\begin{center}
\includegraphics[scale=1.6]{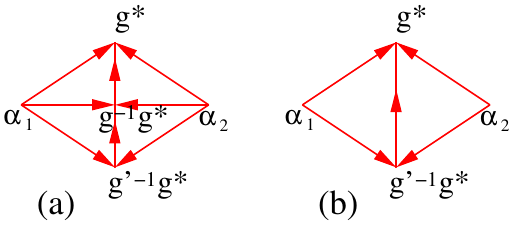}
%Fig. 28
\end{center}
\caption{
(Color online)
The evaluation of the 2-cocycle $\nu_2$ on the above two complexes with
branching structure gives rise to two phase factors in \eqn{f2f2f2a} and
\eqn{f2f2f2b}, which shows that the ratio of the two factors, \eqn{f2f2f2}, is
equal to 1, since the complexes in (a) and (b) overlap.
}
\label{f221}
\end{figure}

To show that $U^{\v i}(g)$ defined in \eqn{fom2om2} is a linear representation
of $G$, let us compare the combined actions of  $U^{\v i}(g)$ and $U^{\v
i}(g'g^{-1})$ with the action of $U^{\v i}(g')$ which are given by (see Fig.  \ref{f221})
\begin{align}
\label{f2f2f2a}
&\ \ \ \
U^{\v i}(g'g^{-1})
U^{\v i}(g)
|\al_1,\al_2\>
\\
&=
f_2(\al_1,\al_2,g,g^*)
f_2(g\al_1,g\al_2,g'g^{-1},g^*)
|g'\al_1,g'\al_2\>
\nonumber
\end{align}
and
\begin{align}
\label{f2f2f2b}
U^{\v i}(g')
|\al_1,\al_2\>
&=
f_2(\al_1,\al_2,g',g^*)
|g'\al_1,g'\al_2\>
\end{align}
We see that
\begin{align}
\label{f2f2f2}
&\ \ \ \
f_2(\al_1,\al_2,g,g^*)
f_2(g\al_1,g\al_2,g'g^{-1},g^*)
f_2^{-1}(\al_1,\al_2,g',g^*)
\nonumber \\
&=
\frac{\nu_2(\al_1,g^{-1}g^*,  g^*)} {\nu_2(\al_2,g^{-1}g^*,  g^* ) }
\frac{\nu_2(g\al_1,gg'^{-1}g^*,  g^*)} {\nu_2(g\al_2,gg'^{-1}g^*,  g^* ) }
\frac{\nu_2(\al_2,g'^{-1}g^*,  g^*)} {\nu_2(\al_1,g'^{-1}g^*,  g^* ) }
\nonumber\\
&=
\frac{\nu_2(\al_1,g^{-1}g^*,  g^*)} {\nu_2(\al_2,g^{-1}g^*,  g^* ) }
\frac{\nu_2(\al_1,g'^{-1}g^*,  g^{-1}g^*)} {\nu_2(\al_2,g'^{-1}g^*,  g^{-1}g^* ) }
\frac{\nu_2(\al_2,g'^{-1}g^*,  g^*)} {\nu_2(\al_1,g'^{-1}g^*,  g^* ) }
\end{align}
The above expression can be represented as Fig. \ref{f221} which indicates that
the expression is equal to 1.
Thus $U^{\v i}(g)$ defined in \eqn{alalU} form a unitary representation of $G$.

\subsection{ $U^{i}(g)$ satisfies \eqn{UPsi} }

\begin{figure}[tb]
\begin{center}
\includegraphics[scale=1.3]{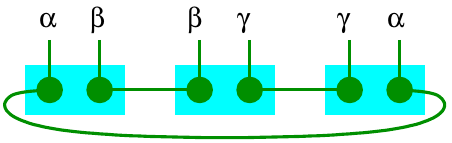}
%Fig. 29
\end{center}
\caption{
(Color online)
The 1D state \eqn{pSRE1D} on a ring.
The degrees of freedom form maximally entangled dimer states.
}
\label{ring}
\end{figure}

The action of $\otimes U^{\v i}(g)$ on the 1D state on a ring in Fig.
\ref{ring} is given by
\begin{align}
&\ \ \
\otimes_{\v i} U^{\v i}(g)
|\al,\bt;\bt,\ga;\ga,\al\>
\\
&=
f_2(\al,\bt,g ,g^* )
f_2(\bt,\ga,g ,g^* )
f_2(\ga,\al,g ,g^* ) \times
\nonumber\\
&\ \ \ \ \
|g\al, g\bt;g\bt, g\ga;g\ga, g\al\>
.
\nonumber
\end{align}
{}From \eq{fom2om2}, we see that
\begin{align}
&\ \ \ \
f_2(\al,\bt,g ,g^* )
f_2(\bt,\ga,g ,g^* )
f_2(\ga,\al,g ,g^* )
\nonumber\\
&=
\frac{\nu_2(\al,g^{-1}g^*,  g^*)} {\nu_2(\bt,g^{-1}g^*,  g^* ) }
\frac{\nu_2(\bt,g^{-1}g^*,  g^*)} {\nu_2(\ga,g^{-1}g^*,  g^* ) }
\frac{\nu_2(\ga,g^{-1}g^*,  g^*)} {\nu_2(\al,g^{-1}g^*,  g^* ) }
\nonumber\\
&=1
\end{align}
We find that
\begin{align}
\otimes_{\v i} U^{\v i}(g)
|\al,\bt;\bt,\ga;\ga,\al\>
=
|g\al, g\bt;g\bt, g\ga;g\ga, g\al\>
.
\nonumber
\end{align}
The state $|\Psi_\text{pSRE}\>$ on a ring is invariant under the symmetry
transformation.  So, $U^{\v i}$ defined in \eqn{alalU} is indeed a solution of
\eqn{UPsi}. We can obtain one solution for every cocycle in $\cH^2(G, U(1) )$
and each solution correspond to a SPT phase in $1$ dimensions.

\subsection{States at the chain end form a projective representation}

\begin{figure}[tb]
\begin{center}
\includegraphics[scale=1.3]{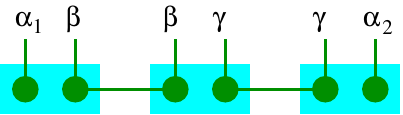}
%Fig.30 
\end{center}
\caption{
(Color online)
A segment of 1D chain with open ends.  The degrees of freedom not on the end
form maximally entangled dimer states.
}
\label{seg}
\end{figure}

Now let us consider the action of on-site symmetry transformation $\otimes_{\v
i} U^{\v i}(g)$ on a segment with boundary (see Fig. \ref{seg}):
\begin{align}
&\ \ \
\otimes_{\v i} U^{\v i}(g)
|\al_1,\bt;\bt,\ga;\ga,\al_2\>
\\
&=
f_2(\al_1,\bt,g ,g^* )
f_2(\bt,\ga,g ,g^* )
f_2(\ga,\al_2,g ,g^* ) \times
\nonumber\\
&\ \ \ \ \
|g\al_1, g\bt;g\bt, g\ga;g\ga, g\al_2\>
\nonumber\\
&=
\frac{\nu_2(\al_1,g^{-1}g^*,  g^*)} {\nu_2(\al_2,g^{-1}g^*,  g^* ) }
.
\nonumber
\end{align}
or
\begin{align}
\label{Unu2nu2}
\otimes_{\v i} U^{\v i}(g)
|\al_1,\al_2\>_0
&=
\frac{\nu_2(\al_1,g^{-1}g^*,  g^*)} {\nu_2(\al_2,g^{-1}g^*,  g^* ) }
|g\al_1, g\al_2\>_0
,
\end{align}
where
\begin{align}
|\al_1,\al_2\>_0=
 \sum_{\bt,\ga}
|\al_1,\bt;\bt,\ga;\ga,\al_2\> .
\end{align}
Eqn. (\ref{Unu2nu2}), is the same as \eqn{alalU} and \eqn{fom2om2}.
Thus, $\otimes_{\v i} U^{\v i}(g)$ form a linear representation of $G$.

Note that $|\al_1,\al_2\>_0$ is the ground state of our fixed-point model on a
segment of chain, where all the internal degrees of freedom form the maximally
entangled dimers (just like the ground state on a ring), while the boundary
degrees of freedom are labeled by $\al_1$ and $\al_2$ on the chain ends.
$\al_1$ and $\al_2$ label the effective low energy degrees of freedom
$|\al_1,\al_2\>_0$.  Those low energy degrees of freedom form a linear
representation of the symmetry transformation as expected.  Eqn.
(\ref{Unu2nu2}) describe how the boundary low energy degrees of freedom
$|\al_1,\al_2\>_0$ transform under the symmetry transformation.

\begin{figure}[tb]
\begin{center}
\includegraphics[scale=0.8]{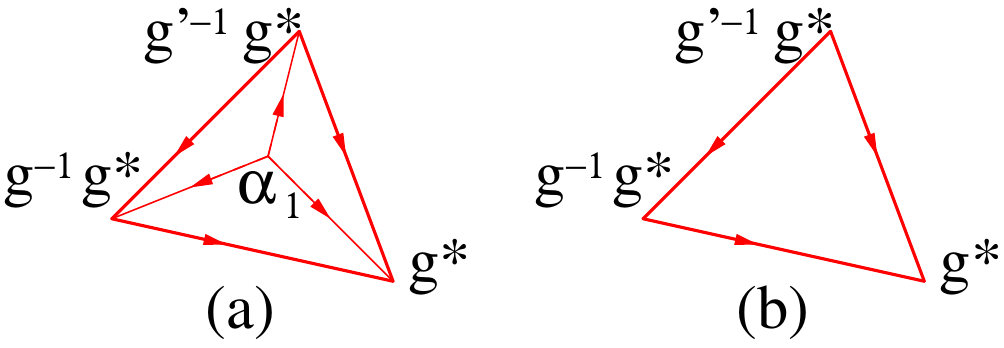}
%Fig. 31
\end{center}
\caption{
(Color online)
(a) The graphic representation of $\frac{ \nu_2(\al_1,g'^{-1}g^*,  g^{-1}g^*)
\nu_2(\al_1,g^{-1}g^*,  g^*) }{ \nu_2(\al_1,g'^{-1}g^*,  g^*) } $.
(b) The graphic representation of $\nu_2(g'^{-1}g^*,g^{-1}g^* , g^*)$
which allows us to show
$\frac{ \nu_2(\al_1,g'^{-1}g^*,  g^{-1}g^*)
\nu_2(\al_1,g^{-1}g^*,  g^*) }{ \nu_2(\al_1,g'^{-1}g^*,  g^*) }
= \nu_2(g'^{-1}g^*,g^{-1}g^* , g^*)
$.
}
\label{f3U3a}
\end{figure}

\begin{figure}[tb]
\begin{center}
\includegraphics[scale=1.4]{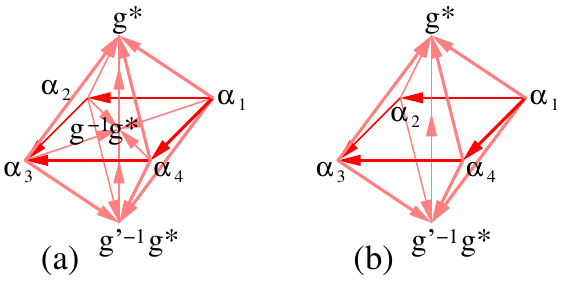}
%Fig. 32
\end{center}
\caption{
(Color online)
(a) The graphic representation of the phase factor \eqn{fom321a1}.
(b) The graphic representation of the phase factor \eqn{fom321a2}.
The graphic representations indicate that the two phases
are the same.
}
\label{fom321a}
\end{figure}

On the other hand, the symmetry transformation $\otimes_{\v i} U^{\v i}(g)$
factorize (see \eqn{Unu2nu2}), also as expected.  This is because the degrees
of freedom labeled by $\al_1$ and $\al_2$ are located far apart and decouple.
We have (on the end whose states are labeled by $\al_1$)
\begin{align}
\label{Unu2}
\otimes_{\v i} U^{\v i}(g)
|\al_1\>_0
&=
\nu_2(\al_1,g^{-1}g^*,  g^*)
|g\al_1\>_0
.
\end{align}
Such transformation satisfies (see Fig. \ref{f3U3a})
\begin{align}
&\ \ \ \
\otimes_{\v i} U^{\v i}(g'g^{-1})
\otimes_{\v i} U^{\v i}(g)
|\al_1\>_0
\nonumber\\
&=
\frac{
\nu_2(g\al_1,gg'^{-1}g^*,  g^*)
\nu_2(\al_1,g^{-1}g^*,  g^*)
}{
\nu_2(\al_1,g'^{-1}g^*,  g^*)
}
\otimes_{\v i} U^{\v i}(g')
|\al_1\>_0
\nonumber\\
&=
\frac{
\nu_2(\al_1,g'^{-1}g^*,  g^{-1}g^*)
\nu_2(\al_1,g^{-1}g^*,  g^*)
}{
\nu_2(\al_1,g'^{-1}g^*,  g^*)
}
\otimes_{\v i} U^{\v i}(g')
|\al_1\>_0
\nonumber\\
&=
\nu_2(g'^{-1}g^*,g^{-1}g^*,  g^*)
\otimes_{\v i} U^{\v i}(g')
|\al_1\>_0
.
\end{align}
We see that the degrees of freedom on one end form a projective
representation labeled by the 2-cocycle $\nu_2$, the same 2-cocycle $\nu_2$
that characterize the symmetry transformation of the SRE state.

\section{(2+1)D solutions of \eqn{UPsi}}
\label{12D}

\subsection{ $U^{i}(g)$ is a linear representation }

\begin{figure}[tb]
\begin{center}
\includegraphics[scale=0.7]{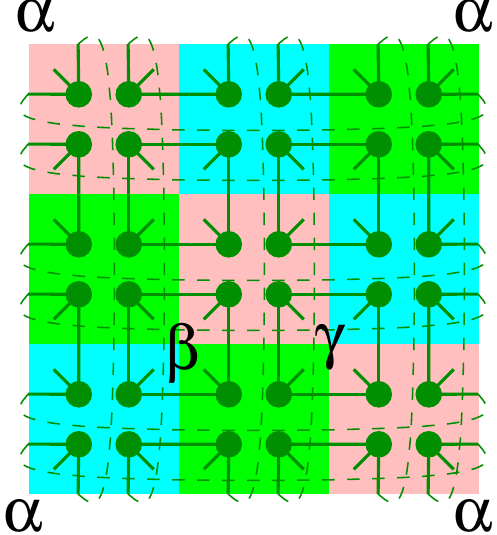}
%Fig. 33
\end{center}
\caption{
(Color online)
A 2D $|\Psi_\text{pSRE}\>$ state on a torus.
In $|\Psi_\text{pSRE}\>$, the linked dots carry the same index
$\al,\bt,\ga,...$
}
\label{tor}
\end{figure}

\begin{figure}[tb]
\begin{center}
\includegraphics[scale=0.7]{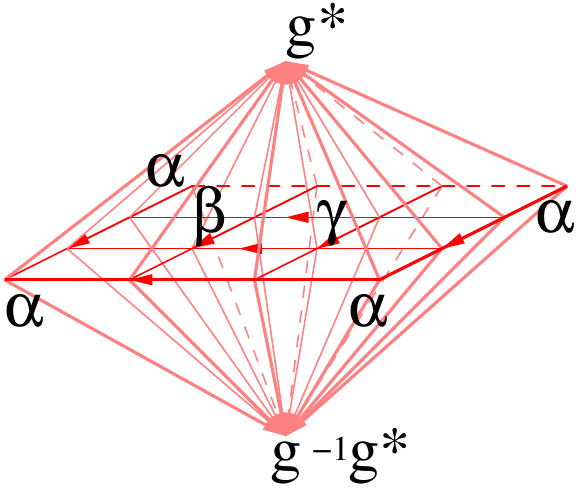}
%Fig. 34
\end{center}
\caption{ (Color online) The graphic representation of the phase
$F_3$ in \eqn{Uf3pSRE}.  $F_3$ is the value of a 3-cocycle $\nu_3$
on the above complex with a branching structure. Note that the top
pyramid and the bottom pyramid each form a solid torus (due to the
periodic boundary condition) and the whole complex is a sphere.  So
$F_3=1$.  Note that the two pyramids on top and blow each small
square represent the phase factor $f_3$ in \eqn{fom3om3}. }
\label{torG}
\end{figure}

To show that $U^{\v i}$ defined in \eqn{fom3om3}
is a linear representation of $G$,
let us compare the action of two symmetry transformations: $ U^{\v
i}(g) U^{\v i}(g^{-1}g')$ with the action of $ U^{\v i}(g^{\prime}) $,
which changes $| \al_1,
\al_2, \al_3, \al_4\>$ to $| g'\al_1, g'\al_2, g'\al_3, g'\al_4|>$.
One has a phase factor
\begin{align}
\label{fom321a1}
&\ \ \ \
\frac{
\nu_3( \al_1,\al_2,g^{-1}g^*,g^*)
\nu_3( \al_2,\al_3,g^{-1}g^*,g^*)}{
\nu_3( \al_4,\al_3,g^{-1}g^*,g^*)
\nu_3( \al_1,\al_4,g^{-1}g^*,g^*)
}\times
\nonumber\\
&\ \ \ \
\frac{
\nu_3( g\al_1,g\al_2,gg'^{-1}g^*,g^*)
\nu_3( g\al_2,g\al_3,gg'^{-1}g^*,g^*)}{
\nu_3( g\al_4,g\al_3,gg'^{-1}g^*,g^*)
\nu_3( g\al_1,g\al_4,gg'^{-1}g^*,g^*)
}
\nonumber\\
&=
\frac{
\nu_3( \al_1,\al_2,g^{-1}g^*,g^*)
\nu_3( \al_2,\al_3,g^{-1}g^*,g^*)}{
\nu_3( \al_4,\al_3,g^{-1}g^*,g^*)
\nu_3( \al_1,\al_4,g^{-1}g^*,g^*)
}\times
\\
&\ \ \ \
\frac{
\nu_3( \al_1,\al_2,g'^{-1}g^*,g^{-1}g^*)
\nu_3( \al_2,\al_3,g'^{-1}g^*,g^{-1}g^*)}{
\nu_3( \al_4,\al_3,g'^{-1}g^*,g^{-1}g^*)
\nu_3( \al_1,\al_4,g'^{-1}g^*,g^{-1}g^*)
}
\nonumber
\end{align}
and the other has a phase factor
\begin{align}
\label{fom321a2}
\frac{
\nu_3( \al_1,\al_2,g'^{-1}g^*,g^*)
\nu_3( \al_2,\al_3,g'^{-1}g^*,g^*)}{
\nu_3( \al_4,\al_3,g'^{-1}g^*,g^*)
\nu_3( \al_1,\al_4,g'^{-1}g^*,g^*)
}
\end{align}
{}From their graphic representations Fig. \ref{fom321a}, we see that the two
phases are the same.  Thus $ U^{\v i}(g)$ form an unitary representation of the
symmetry group $G$.

\subsection{ $U^{i}(g)$ satisfies \eqn{UPsi} }

Following a similar approach as for the (1+1)D case, we can also
show that the state $|\Psi_\text{pSRE}\>$ on a 2D complex (see Fig.
\ref{tor})  that is a boundary of another graph is invariant under
the symmetry transformation $\otimes_{\v i} U^{\v i}$ (see Fig.
\ref{torG}):
\begin{align}
\label{Uf3pSRE}
 \otimes_{\v i} U^{\v
i} |\Psi_\text{pSRE}\>
=F_3 |\Psi_\text{pSRE}\>
=|\Psi_\text{pSRE}\>
\end{align}
So, $U^{\v i}$ defined in \eqn{fom3om3} is indeed a solution of
\eqn{UPsi}. We can obtain one solution for every cocycle in $\cH^3(G, U_T(1) )$
and each solution correspond to a SPT phase in $2$ dimensions.

\subsection{ The action of $\otimes U^{i}(g)$
on $|\Psi_\text{pSRE}\>$ with boundary }

\begin{figure}[tb]
\begin{center}
\includegraphics[scale=0.7]{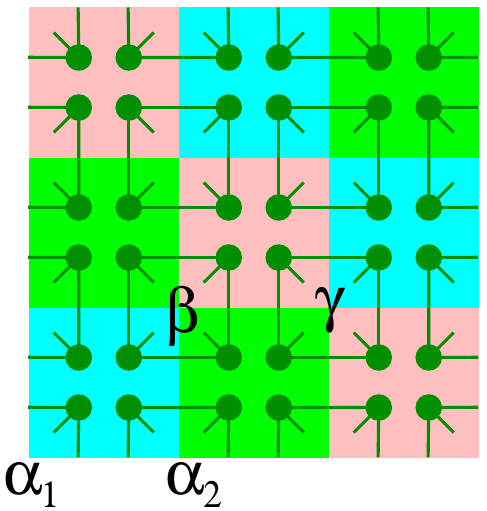}
%Fig. 35
\end{center}
\caption{
(Color online)
A 2D $|\Psi_\text{pSRE}\>$ state on an open square.
In $|\Psi_\text{pSRE}\>$, the linked dots carry the same index
$\al_1,\al_2,\bt,\ga,...$
The indices on the boundary are given by $\al_1,\al_2,...$
The indices inside the square are given by $\bt,\ga,...$
}
\label{sq}
\end{figure}

\begin{figure}[tb]
\begin{center}
\includegraphics[scale=0.7]{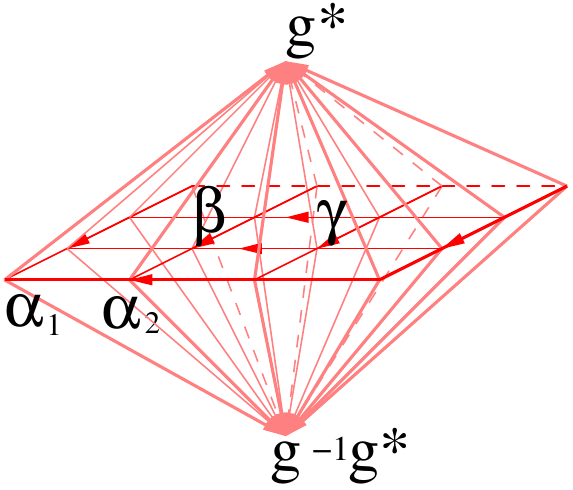}
%Fig. 36
\end{center}
\caption{ (Color online) The graphic representation of the phase $\t
F_3(g,g^*; \al_1,\al_2,\bt,\ga,...)$ in \eqn{Uf3sq}. Compare to the
complex in Fig. \ref{torG}, the above complex do not have the
periodic boundary condition. } \label{sqG}
\end{figure}

Now let us consider the action of  $\otimes_{\v i} U^{\v i}$ on a state in
Fig. \ref{sq} with a boundary (see Fig. \ref{sqG}):
\begin{align}
\label{Uf3sq}
&\ \ \ \
 \otimes_{\v i} U^{\v i}(g) |\al_1,\al_2,\bt,\ga,...\>
\nonumber\\
& = \t F_3(g,g^*; \al_1,\al_2,\bt,\ga,...) |g\al_1,g\al_2,g\bt,g\ga,...\>
\end{align}
{}From the  Fig. \ref{sqG} and the geometric meaning of the cocycles,
we find that
\begin{align}
\label{Ufp3}
 \t F_3(g,g^*; \al_1,\al_2,\bt,\ga,...)=
\prod_{\<ij\>} \nu^{s_{ij}}_3(\al_i,\al_j, g^{-1}g^*,g^*)
\end{align}
where $\prod_{\<ij\>}$ is a product over the nearest neighbor bonds $\{ij\}$,
$|i-j|=1$, around the boundary.  The direction $i\to j$ is the direction of the
bond and $s_{ij}=1$ f $i>j$, $s_{ij}=-1$ f $i<j$.
Since $\t F_3$ is independent of the indices $\bt,\ga,...$ that are
not on the boundary, we find
\begin{align}
\label{Uedge}
&\ \ \ \
  \otimes_{\v i} U^{\v i}(g) |\{\al_i\} \>_0
=
\prod_{\<ij\>} \nu^{s_{ij}}_3(\al_i,\al_j, g^{-1}g^*,g^*)
|\{g\al_i\} \>_0
\end{align}
where $|\{\al_i\} \>_0$ is the SPT state with a boundary which
depends on the indices $\{\al_i\}$ on the boundary:
\begin{align}
 |\{\al_i\} \>_0=\sum_{\bt,\ga,...\in G}
|\al_1,\al_2,\bt,\ga,...\> .
\end{align}

We see that the action of $\otimes_{\v i} U^{\v i}(g)$ on $|\{\al_i\} \>_0$ is
very similar to the action of a single $U^{\v i}(g)$ on a single site (compare
Figs. \ref{fom3} and \ref{sqG}).  Using a similar approach, we can show that
$\otimes_{\v i} U^{\v i}(g)$ indeed form a linear representation (see Fig.
\ref{fom321a}), when viewed as an operator $U_b(g)$ acting
on the boundary state $|\{\al_i\} \>_0$.

To summarize, we discussed the form of on-site symmetry transformations
$\otimes_{\v i} U^{\v i}(g)$ in a basis where the many-body ground state is a
simple product state.  We find that different on-site symmetry transformations
can be constructed from each 3-cocycle $\nu_3$ in $\cH^3[G,U_T(1)]$.

We would like to stress that, in such a simple basis, the symmetry
transformation $\otimes_{\v i} U^{\v i}(g)$ on the boundary \eq{Uedge} has a
very unusual locality property: Due to the non-trivial phase factor
$\prod_{\<ij\>} \nu^{s_{ij}}_3(\al_i,\al_j, g^{-1}g^*,g^*)$, we cannot view
$U_b(g)$ (acting on the boundary state $|\{\al_i\} \>_0$) as a direct product
of local operators acting on each boundary sites $|\al_i\>$.  (Note that we can
view the boundary state $|\{\al_i\} \>_0$ as $|\{\al_i\} \>_0=\otimes_{i \in
\text{boundary}} |\al_i\>$.) Therefore, $U_b(g)$ is \emph{not} a on-site
symmetry transformation on the \emph{boundary}.

In the above, we have viewed $i$ as effective sites on the boundary with
physical states $|\al_i\>$ on each site. We see that the symmetry
transformation is not an on-site symmetry transformation. If we view, instead,
each nearest neighbor bond $\<ij\>$ as an effective site with physical states
$|\al_i\al_j\>$ on each site, then the symmetry transformation will be an
``on-site'' symmetry transformation, but the states on different bounds are not
independent and $|\{\al_i\} \>_0\neq \otimes_{\<ij\> \in \text{boundary}}
|\al_i\al_j\>$.

Thus in a basis where the many-body ground state is a simple product state,
although $\otimes_{\v i} U^{\v i}(g)$ is an on-site symmetry transformation
when acting on the bulk state, it cannot be an on-site symmetry transformation
when viewed as a symmetry transformation acting on the effective low energy
degrees of freedom on the boundary when the 3-cocycle $\nu_3$ is non-trivial.
This is the non-trivial physical properties that characterize a non-trivial SPT
phase in (2+1)D (see appendix \ref{MPUO} and \Ref{CLW1152} for more details).

\section{Two symmetry representations in our fixed-point model} \label{twosymm}

In the old basis in the path integral formalism [see eqn.~(\ref{Psiold})], the
wave function is complicated, but the many-body on-site symmetry
transformation has the following locality structure
\begin{align}
\otimes_i U^i(g)
\end{align}
where $U^i(g)$ is the symmetry transformation on the $i^{th}$ site
\begin{align}
  U^i(g) |g_i\>=|gg_i\> .
\end{align}
This is the definition of the so called on-site symmetry transformation.

%In the new basis [see \eq(\ref{Psinew})], the action of the symmetry transformation becomes
%\begin{align}
%&\ \ \ \
%\otimes_i U^i(g) |\{g_i\}_M\>'
%\nonumber\\
%& =
%\frac{  \prod_{\{ij...*\}} \nu_{1+d}^{s_{ij...*}}(g_i,g_j,...,g^*)}{
%\prod_{\{ij...*\}} \nu_{1+d}^{s_{ij...*}}(gg_i,gg_j,...,g^*)}
%|\{gg_i\}_M\>'
%\nonumber\\
%& =
%\frac{  \prod_{\{ij...*\}} \nu_{1+d}^{s_{ij...*}}(g_i,g_j,...,g^*)}{
%\prod_{\{ij...*\}} \nu_{1+d}^{s_{ij...*}}(g_i,g_j,...,g^{-1}g^*)}
%|\{gg_i\}_M\>'
%\nonumber\\
%& =
%|\{gg_i\}_M\>'
%\end{align}
%provided that the complex $M$ is a boundary of another complex
%$M_\text{ext}$. This appears to have the same form as that in the
%old basis $|\{g_i\}_M\>$. But the locality structure of the symmetry
%transformation is very different in the old and new basis.

In the new basis[see eqn.~(\ref{Psinew})], the wave function is simple but the many-body symmetry transformation
is no longer an on-site symmetry transformation.
It has the following form
\begin{align}
&\ \ \ \
\otimes_i U^i(g) |\{g_i\}_M\>'
\nonumber\\
& =
\frac{  \prod_{\{ij...*\}} \nu_{1+d}^{s_{ij...*}}(g_i,g_j,...,g^*)}{
\prod_{\{ij...*\}} \nu_{1+d}^{s_{ij...*}}(gg_i,gg_j,...,g^*)}
|\{gg_i\}_M\>'
\nonumber\\
& =
\frac{  \prod_{\{ij...*\}} \nu_{1+d}^{s_{ij...*}}(g_i,g_j,...,g^*)}{
\prod_{\{ij...*\}} \nu_{1+d}^{s_{ij...*}}(g_i,g_j,...,g^{-1}g^*)}
|\{gg_i\}_M\>'
\nonumber\\
& =
|\{gg_i\}_M\>'
\end{align}
where $U^i(g)$ is the symmetry transformation that acts on the $i^{th}$ and
$(i\pm 1)^{th}$ sites [in (1+1)D for example]
%\begin{align}
%\otimes_i  U^i(g)
%\end{align}
\begin{align}
 U^i(g) |g_{i-1},g_i,g_{i+1}\>'
=f_2(g_{i-1}, g_i,g_{i+1},g) |g_{i-1},gg_i,g_{i+1}\>' .
\end{align}
Here the phase factor $f_2$ is given by the 2-cocycles
\begin{align}
&\ \ \ \ f_2(g_{i-1}, g_i,g_{i+1},g )
\nonumber\\
&=\frac{
\nu_2(g_{i-1},g_i,g^*)
\nu_2(g_i,g_{i+1},g^*)
}
{
\nu_2(g_{i-1},gg_i,g^*)
\nu_2(gg_i,g_{i+1},g^*)
}
\end{align}
where $g^*$ is an fixed element in $G$.
For example we may choose $g^*=1$.

\begin{figure}[tb]
\begin{center}
\includegraphics[scale=0.8]{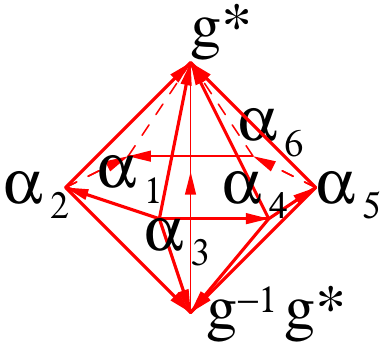}
%Fig. 37
\end{center}
\caption{
(Color online)
The graphic representation of the product of the phase
factor
$\prod_{\<ij\>} \nu^{s_{ij}}_3(\al_i,\al_j, g^{-1}g^*,g^*)$
in \eqn{Uedge}.
}
\label{fp3}

\end{figure}
\begin{figure}[tb]
\begin{center}
\includegraphics[scale=0.8]{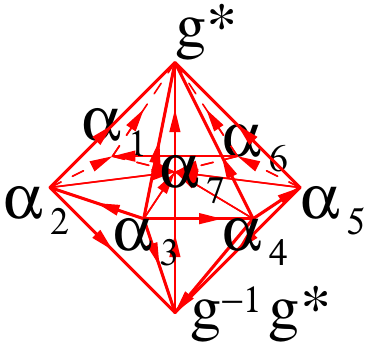}
%Fig. 38
\end{center}
\caption{ (Color online) The graphic representation of the product
of the phase factor $\frac{ \prod_{\{ij...*\}}
\nu_{1+d}^{s_{ij...*}}(\al_i,\al_j,...,g^*)}{ \prod_{\{ij...*\}}
\nu_{1+d}^{s_{ij...*}}(\al_i,\al_j,...,g^{-1}g^*)} $ in \eqn{fp3aeq}
for a 2D complex $(\al_1,...,\al_7)$ with a boundary
$(\al_1,...,\al_6)$. } \label{fp3a}
\end{figure}

We have seen that, in the new basis, we still have $ \otimes_i
U^i(g) |\{g_i\}_M\>' =|\{gg_i\}_M\>' $ if the state $ |\{g_i\}_M\>'$
is defined on a complex $M$ which is the boundary of another complex
$M_\text{ext}$.  It is hard to see the non on-site structure of
$\otimes_i U^i(g)$.  To expose the non on-site structure of
$\otimes_i U^i(g)$ in the new basis, let us consider the action of
$\otimes_i U^i(g)$ on a state defined on a complex that has a
boundary.  In this case, we still have
\begin{align}
\label{fp3aeq}
&\ \ \ \
\otimes_i U^i(g) |\{g_i\}_M\>'
\nonumber\\
& =
\frac{  \prod_{\{ij...*\}} \nu_{1+d}^{s_{ij...*}}(g_i,g_j,...,g^*)}{
\prod_{\{ij...*\}} \nu_{1+d}^{s_{ij...*}}(g_i,g_j,...,g^{-1}g^*)}
|\{gg_i\}_M\>'
\end{align}
But now, the phase factor is not equal to 1.

In Fig. \ref{fp3a}, we give a graphic representation of the above
phase factor $\frac{  \prod_{\{ij...*\}}
\nu_{1+d}^{s_{ij...*}}(\al_i,\al_j,...,g^*)}{ \prod_{\{ij...*\}}
\nu_{1+d}^{s_{ij...*}}(\al_i,\al_j,...,g^{-1}g^*)} $ for a 2D
complex with a boundary.  We see that the complex in Fig. \ref{fp3a}
and Fig. \ref{fp3} have the same surface.  So the phase factor
represented by Fig. \ref{fp3a} equal to that represented by Fig.
\ref{fp3}. So \eqn{Uedge} is the same as \eqn{fp3aeq}.

We have discussed two ways to classify SPT phases.  The first way to
classify SPT phases is to classify symmetry transformations that act
on simple wave function $|\Psi_\text{pSRE}\>$, which lead to  \eqn{Uedge}.  The
second way to  classify SPT phases is to classify fixed-point action-amplitude
(the  topological terms) which lead to  \eqn{fp3aeq}.  The above
analysis indicates that the two ways to classify SPT phases are
equivalent.

The equivalence between the two formalisms \eqn{Uedge} and \eqn{fp3aeq} will become
more clear after a duality transformation.

%\section{Duality between the ground states in the Lagrangian and the Hamiltonian formalisms}\label{app: dual}
In the following we will show that the ground state wave function (\ref{Psinew}) in
the Lagrangian formalism is dual to the ground state wave function (\ref{pSRE1D}) in the Hamiltonian
formalism discussed in Sec. \ref{Cfm} and \ref{clsymm}. Furthermore, after the duality transformation,
the the symmetry representations (\ref{eq:Lgsymm}) are the same as that defined in eqn. (\ref{alalU}) or \eqn{Uedge}.

\begin{figure}[htbp]
\centering
\includegraphics[width=3.in]{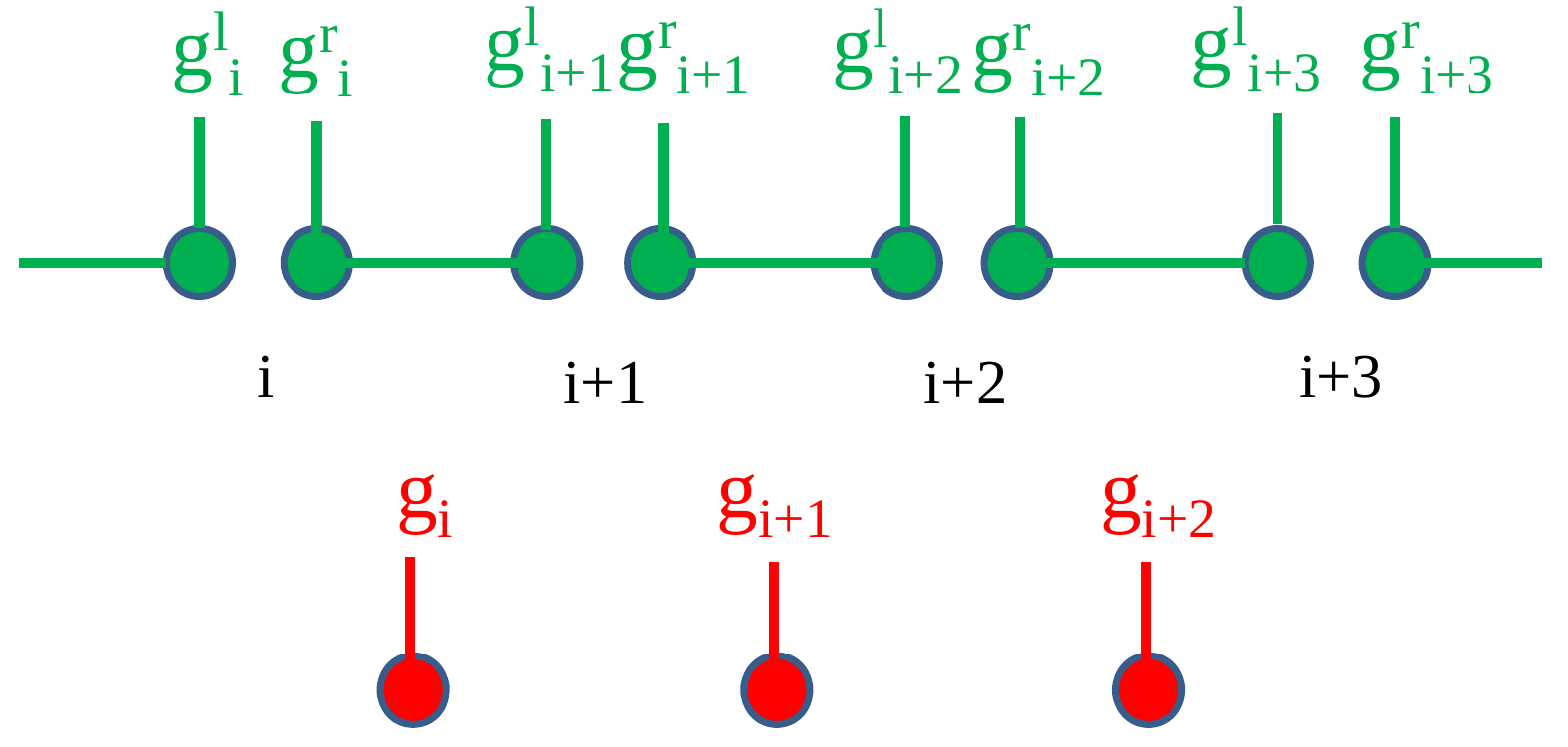}
%Fig. 39
\caption{(Color online) The dual transformation in the new bases in 1D.
} \label{fig:dual1d}
\end{figure}

Let us illustrate the above result in 1D. Firstly, we introduce the dual transformation which maps a state to its dual wave function living on the dual lattice. In the dual transformation, the bases $|g_{i}\rangle$ at site $i$ correspond
to the bond $|g_i^r,g_{i+1}^l\rangle$ in the dual lattice (see Fig.~\ref{fig:dual1d}), where $g_i^r=g_{i+1}^l=g_i$ and the amplitude of the configuration
remains unchanged. In this way, we obtain the dual wave function $\Psi_d(\{g_i^l,g_i^r\})$ of $\Psi(\{g_i\} )$.

Now we introduce the new bases $|\{g_i^l,g_i^r\}\rangle'$ through the LU transformation introduced in eqn.~(\ref{Psinew}),
\begin{eqnarray}
|\{g_i^l,g_i^r\}\rangle' &=& \prod_{i} \nu_{2}(g_i,g_{i+1},g^*)|\{g_i^l,g_{i}^r\}\rangle\nonumber\\
&=&\prod_i \nu_2(g_i^r,g_{i+1}^r,g^*)|\{g_i^l,g_{i}^r\}\rangle\nonumber\\
&=&\prod_i \left[\nu_2(g_{i+1}^l,g_{i+1}^r,g^*)|{g_{i+1}^l,g_{i+1}^r}\rangle\right]
\end{eqnarray}

In the new bases, the fixed point state in the dual lattice becomes a direct product of bonds. Notice that the previous local unitary transformation in eqn.~(\ref{Psinew}) becomes on-site unitary transformation. Furthermore, in the new bases the symmetry representation also becomes on-site and is fractionalized into two `projective' operations:
\begin{eqnarray}
&&\otimes_i U^i(g)|\{g_i^l,g_i^r\}\rangle' \nonumber\\&=& \prod_{i} \nu_{2}(g_{i+1}^l,g_{i+1}^r,g^*)|\{gg_{i+1}^l,gg_{i+1}^r\}\rangle\nonumber\\
%&=&\frac{\prod_{i}\nu_2(g_i,g_{i+1},g^*)}{\prod_{i}\nu_2(gg_i,gg_{i+1},g^*)}|\{gg_i^l,gg_i^r\}\rangle'\nonumber\\
&=&\prod_{i}\frac{\nu_2(g_{i+1}^l,g_{i+1}^r,g^*)}{\nu_2(g_{i+1}^l,g_{i+1}^l,g^{-1}g^*)}|\{gg_{i+1}^l,gg_{i+1}^r\}\rangle'\nonumber\\
%&=&\frac{\prod_{i}\nu_2(g_{i+1}^l,g^{-1}g^*,g^*)}{\prod_{i}\nu_2(g_{i+1}^r,g^{-1}g^*,g^*)}|\{gg_i^l,gg_i^r\}\rangle'\nonumber\\
&=&\prod_{i}\frac{\nu_2(g_{i+1}^l,g^{-1}g^*,g^*)}{\nu_2(g_{i+1}^r,g^{-1}g^*,g^*)}|\{gg_{i+1}^l,gg_{i+1}^r\}\rangle'
\end{eqnarray}
Above formula is the same as eqn.~(\ref{fom2om2}).

\begin{figure}[htbp]
\centering
\includegraphics[width=2.in]{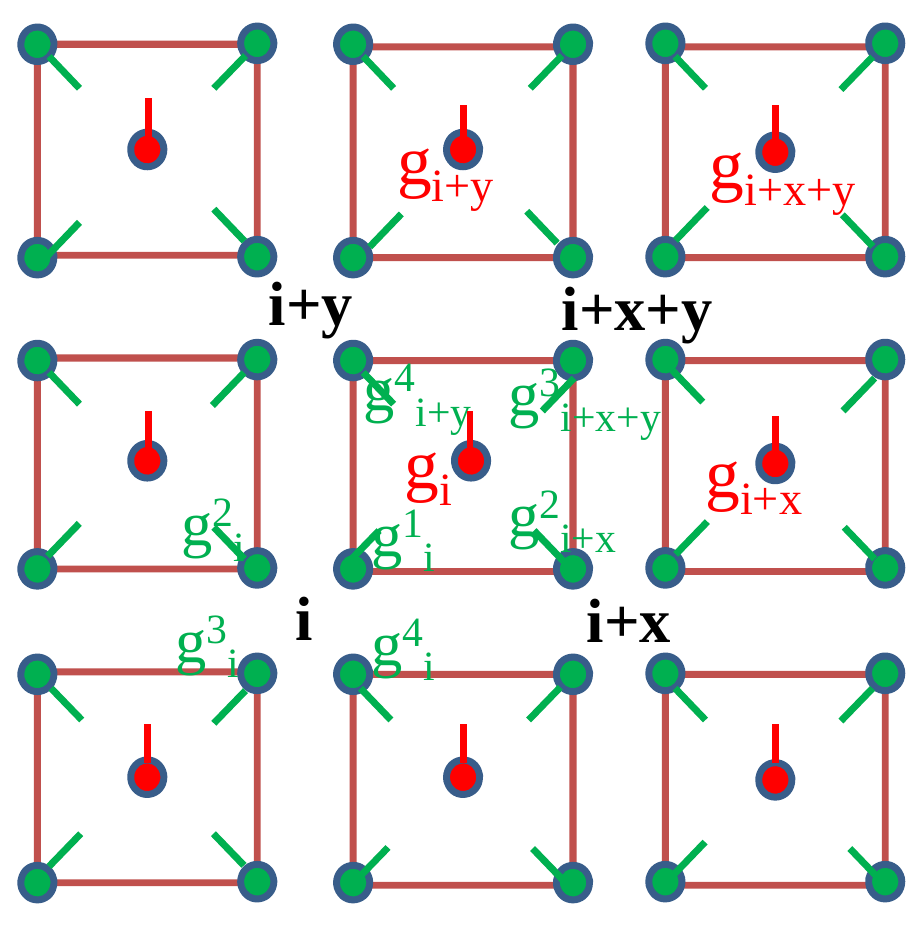}
%Fig. 40
\caption{(Color online) The duality transformation in 2-dimension. The green dots represent the dual lattice of the red dots. In the new bases, the wave function in the green lattice is the same as the one introduced in Sec. III and IV.
} \label{fig:dual2d}
\end{figure}

Similarly, %in 2D, the fixed point wave function from the path integral is dual to the one constructed in Sec.V.
we can illustrate the result in 2D. Now the basis $|g_i\rangle$ correspond to $|g_i^1,g_{i+x}^2,g_{i+x+y}^3,g_{i+y}^4\rangle$ in the dual lattice (see Fig.~\ref{fig:dual2d}). After the dual transformation, the wave function $\Psi(\{g_i\} )$ becomes $\Psi_d(\{g_i^1,g_i^2,g_i^3,g_i^4\})$ (here $g_i^1=g_{i+x}^2=g_{i+x+y}^3=g_{i+y}^4=g_i$). Again, we introduce the LU transformation
\begin{eqnarray}
&&|\{g_i^1,g_i^2,g_i^3,g_i^4\} \rangle'\nonumber\\
&=& \prod_{i} \frac{\nu_{3}(g_i,g_{i+x},g_{i+y},g^*)} {\nu_{3}(g_{i+x},g_{i+y},g_{\tilde i },g^*)}|\{g_i^1,g_i^2,g_i^3,g_i^4\}\rangle\nonumber\\
&=&\prod_{i} \frac{\nu_{3}(g_{\tilde i }^3,g_{\tilde i }^4,g_{\tilde i }^2,g^*)} {\nu_{3}(g_{\tilde i }^4,g_{\tilde i }^2,g_{\tilde i }^1,g^*)}|\{g_{\tilde i }^1,g_{\tilde i }^2,g_{\tilde i }^3,g_{\tilde i }^4\}\rangle%\nonumber\\
\end{eqnarray}
where $\tilde i=i+x+y$. The LU transformation between the old bases and the new ones is an on-site. In the new bases, the fixed point wave function is a direct product of plaquettes. The symmetry operation now becomes
\begin{eqnarray}
&&\otimes_i U^i(g)|\{g_i^1,g_i^2,g_i^3,g_i^4\}\rangle' \nonumber\\
&=& \prod_{i} \frac{\nu_{3}(g_{\tilde i }^3,g_{\tilde i }^4,g_{\tilde i }^2,g^*)} {\nu_{3}(g_{\tilde i }^4,g_{\tilde i }^2,g_{\tilde i }^1,g^*)}|\{gg_{\tilde i }^1,gg_{\tilde i }^2,gg_{\tilde i }^3,gg_{\tilde i }^4\}\rangle\nonumber\\
&=& \prod_{i} \frac{\nu_{3}(g_{\tilde i }^3,g_{\tilde i }^4,g_{\tilde i }^2,g^*)} {\nu_{3}(g_{\tilde i }^4,g_{\tilde i }^2,g_{\tilde i }^1,g^*)} \prod_{i}\frac{\nu_{3}(g_{\tilde i }^4,g_{\tilde i }^2,g_{\tilde i }^1,g^{-1}g^*)} {\nu_{3}(g_{\tilde i }^3,g_{\tilde i }^4,g_{\tilde i }^2,g^{-1}g^*)}\nonumber\\ &&\times|\{gg_{\tilde i }^1,gg_{\tilde i }^2,gg_{\tilde i }^3,gg_{\tilde i }^4\}\rangle'\nonumber\\
&=& \prod_{i} \frac{\nu_{3}(g_{\tilde i }^3,g_{\tilde i }^4,g^{-1}g^*,g^*) \nu_{3}(g_{\tilde i }^4,g_{\tilde i }^1,g^{-1}g^*,g^*)} {\nu_{3}(g_{\tilde i }^3,g_{\tilde i }^2,g^{-1}g^*,g^*) \nu_{3}(g_{\tilde i }^2,g_{\tilde i }^1,g^{-1}g^*,g^*) } \nonumber\\ &&\times |\{gg_{\tilde i }^1,gg_{\tilde i }^2,gg_{\tilde i }^3,gg_{\tilde i }^4\}\rangle'
\end{eqnarray}
Above equation agree with eqn.~(\ref{eq: sym2D}).

From the above examples, we can see that after the `dual transformation' the ground state wave function and its symmetry representation in the Lagrangian formalism are the same as the Hamiltonian formalism as we discussed in Sec.\ref{cnsymm}.

\section{$(2+1)D$ SPT states constructed from 3-cocycles and
matrix product unitary operator}
\label{MPUO}

Based on the 3-cocycles $\nu_3(g_0,g_1,g_2,g_3)$ of group $G$, we can construct
short range entangled models with SPT order as discussed in section
\ref{2+1Dcase} (also discussed in appendix \ref{12D}). In order to assess the
non trivialness of the SPT order of a certain model, in \Ref{CLW1152} we
developed the tool of matrix product unitary operators (MPUO) and used it to
show that the particular model we gave in that paper--the CZX model--has very
special boundary properties and hence nontrivial SPT order. In this section, we
are going to apply the MPUO method to the general models constructed in section
\ref{2+1Dcase} and show that for the model constructed from a 3-cocycle
$\nu_3(g_0,g_1,g_2,g_3)$, the effective MPUO on the boundary transform with the
same 3-cocycle. Therefore, according to the result in \Ref{CLW1152},
models constructed from nontrivial $\nu_3(g_0,g_1,g_2,g_3)$ must either break
the symmetry or have gapless excitations if the system has a boundary.
Moreover, we can show the contrary for models constructed from trivial
$\nu_3(g_0,g_1,g_2,g_3)$. That is, for models constructed from trivial
$\nu_3(g_0,g_1,g_2,g_3)$ we are going to explicitly construct a short range
entangled symmetric state for the effective symmetry on the boundary. For basic
definition and properties of MPUO, see \Ref{CLW1152}.

We consider in this paper models with on-site symmetry of group $G$. SPT order
exist in models whose ground states on a closed manifold are short range
entangled and symmetric under the on-site symmetry. The ground state is unique
and gapped. If the system has boundary, on the other hand, there are low energy
effective degrees of freedom are on the boundary. The effective symmetry on the
boundary however, may no longer take an on-site form. In general, the effective
symmetry on the 1D boundary of a 2D model can be written as a matrix product
unitary operator
\be
U=\sum_{\{i_k\},\{i_k'\}}\Tr(T^{i_1,i'_1}T^{i_2,i'_2}...T^{i_N,i'_N})|i'_1i'_2...i'_N\>\<i_1i_2...i_N|
\ee
where $i$ and $i'$ are input and output physical indices and for fixed $i$ and
$i'$, $T^{i,i'}$ is a matrix.

For the models defined in section \ref{2+1Dcase}, the effective symmetry $\t
U(g)$ on the boundary takes the form(see appendix \ref{12D})
\be
\t U(g) |\{\alpha_i\}\> = \prod_{i,j} \nu_3^{s_{ij}}(\alpha_i,\alpha_j,g^{-1}g^*,g^*)|\{g\alpha_i\}\>
\ee
where $\prod_{ij}$ is a product over the nearest neighbor bonds $\{ij\}$, $|i-j| = 1$, around the boundary. The direction $i \to j$ is the direction of the bond and $s_{ij} =1$ if $i>j$, $s_{ij} = -1$ if $i < j$. This symmetry operator on a 1D chain can be expressed as a MPUO. If the bond goes from $\alpha_i$ to $\alpha_{i+1}$
\be
\begin{array}{l}
T_{i}^{\alpha_i,g\alpha_i}(g)=\sum_{\alpha_{i+1}}\nu_3^{-1}(\alpha_i,\alpha_{i+1},g^{-1}g^*,g^*) |\alpha_i\>\<\alpha_{i+1}| \ ,\forall \alpha_i \\
\text{other terms are zero}
\end{array}
\ee
If the bond goes from $\alpha_{i+1}$ to $\alpha_i$,
\be
\begin{array}{l}
T_{i}^{\alpha_i,g\alpha_i}(g)=\sum_{\alpha_{i+1}} \nu_3(\alpha_{i+1},\alpha_i,g^{-1}g^*,g^*)|\alpha_i\>\<\alpha_{i+1}| \ ,\forall \alpha_i \\
\text{other terms are zero}
\end{array}
\ee
Now we compose multiple MPUOs and find their reduction rule. We will see that
the reduction rule is related to the same $\nu_3$. First, the combination of
$T_i(g_2)$ and $T_i(g_1)$ gives (if the bond goes from $\alpha_i$ to
$\alpha_{i+1}$)
\be
\begin{array}{l}
T_i(g_1,g_2)^{\alpha_i,g_1g_2\alpha_i}=
\sum_{\alpha_{i+1},\alpha'_{i+1}} \nu_3^{-1}(\alpha_i,\alpha_{i+1},g_2^{-1}g^*,g^*)
\\
\nu_3^{*s(g_2)}(g_2\alpha_i,\alpha'_{i+1},g_1^{-1}g^*,g^*)
|\alpha_i,g_2\alpha_i\>\<\alpha_{i+1},\alpha'_{i+1}|
\end{array}
\ee
This can be reduced to
\begin{align}
&\ \ \ \
T_i(g_1g_2)^{\alpha_i,g_1g_2\alpha_i}
\nonumber\\
&= \sum_{\alpha_{i+1}}\nu_3^{-1}(\alpha_i,\alpha_{i+1},g_2^{-1}g_1^{-1}g^*,g^*) |\alpha_i\>\<\alpha_{i+1}|
\end{align}
by applying the following projection operator to the right side of the matrices
\begin{align}
&\ \ \ \
P^r_{g_1,g_2}
\\
&=\sum_{\alpha_{i+1}} \nu_3^{-1}(\alpha_{i+1},g_2^{-1}g_1^{-1}g^*,g_2^{-1}g^*,g^*)|\alpha_{i+1},g_2\alpha_{i+1}\>\<\alpha_{i+1}|
\nonumber
\end{align}
and the hermitian conjugate of
\begin{align}
&\ \ \ \ P^l_{g_1,g_2}
\nonumber\\
&=\sum_{\alpha_i} \nu_3^{-1}(\alpha_i,g_2^{-1}g_1^{-1}g^*,g_2^{-1}g^*,g^*)|\alpha_i,g_2\alpha_i\>\<\alpha_i|
\end{align}
to the left side of the matrices. This is because,
\begin{align}
&\ \ \ \
\nu_3(g_2\alpha_i,g_2\alpha_{i+1},g_1^{-1}g^*,g^*)
\nonumber\\
&=\nu_3^{s(g_2)}(\alpha_i,\alpha_{i+1},g_2^{-1}g_1^{-1}g^*,g_2^{-1}g^*)
\end{align}
and the 3-cocycle condition of $\nu_3$
\be
\begin{array}{l}
\nu_3(\alpha_i,\alpha_{i+1},g_2^{-1}g^*,g^*)\nu_3(\alpha_i,\alpha_{i+1},g_2^{-1}g_1^{-1}g^*,g_2^{-1}g^*) \\
\nu_3^{-1}(\alpha_i,g_2^{-1}g_1^{-1}g^*,g_2^{-1}g^*,g^*)\nu_3(\alpha_{i+1},g_2^{-1}g_1^{-1}g^*,g_2^{-1}g^*,g^*) \\
=\nu_3(\alpha_i,\alpha_{i+1},g_2^{-1}g_1^{-1}g^*,g^*)
\end{array}
\ee
It is easy to check that the same reduction procedure applies when the bond
goes from $\alpha_{i+1}$ to $\alpha_i$. The above definition of $P^l$ and $P^r$
has picked a particular gauge choice of phase for $P^l$ and $P^r$.

Next we consider the combination of three MPUOs and find the corresponding
3-cocycle associated with different ways of combining the three MPUOs into one.
If we combine $T(g_2)$,$T(g_1)$ first and then combine $T(g_1g_2)$ with
$T(g_3)$, the combined operation of $P_{g_1,g_2}$ and $P_{g_1g_2,g_3}$ is (we
omit the site label $i$)
\begin{align}
& \ \ \ \
(P_{g_1,g_2} \otimes I) P_{g_1g_2,g_3}
\\
&=
\sum_{\alpha} \nu_3(\alpha,g_3^{-1}g_2^{-1}g_1^{-1}g^*,g_3^{-1}g_2^{-1}g^*,g_3^{-1}g^*) \times
\nonumber \\
&\ \ \
\nu_3(\alpha,g_3^{-1}g_2^{-1}g_1^{-1}g^*,g_3^{-1}g^*,g^*)|\alpha,g_3\alpha,g_2g_3\alpha\>\<\alpha| .
\nonumber
\end{align}
On the other hand, if we combine $T(g_3)$,$T(g_2)$ first and then combine $T(g_2g_3)$ with $T(g_1)$, the combined operator of $P_{g_2,g_3}$ and $P_{g_1,g_2g_3}$ is
\be
\begin{array}{l}
(I \otimes P_{g_2,g_3})P_{g_1,g_2g_3}  = \\
\sum_{\alpha} \nu_3(\alpha,g_3^{-1}g_2^{-1}g^*,g_3^{-1}g^*,g^*) \times \\
\nu_3(\alpha,g_3^{-1}g_2^{-1}g_1^{-1}g^*,g_3^{-1}g_2^{-1}g^*,g^*)|\alpha,g_3\alpha,g_2g_3\alpha\>\<\alpha|
\end{array}
\ee
These two differ by a phase factor
\be
\nu_3(g_3^{-1}g_2^{-1}g_1^{-1}g^*,g_3^{-1}g_2^{-1}g^*,g_3^{-1}g^*,g^*)
\ee
Hence we see that, the reduction procedure of $T$'s is associative up to phase.
The phase factor is the same 3-cocycle that we used to construct the model.
From the result in \Ref{CLW1152} we know that if $\nu_3$ is nontrivial,
the model we constructed has a nontrivial boundary which cannot have a gapped
symmetric ground state. It must either break the symmetry or be gapless.
Therefore, the model constructed with nontrivial 3-cocycles belong to
nontrivial SPT phases.

On the other hand, if the model is constructed from a trivial 3-cocycle, the
boundary effective symmetry does allow SRE symmetric state. Actually, the SRE
symmetric state on the boundary can be constructed explicitly for the models
discussed here. If $\nu_3$ is trivial, it takes the form of a 3-coboundary
\be
\nu_3(g_0,g_1,g_2,g_3)=\frac{\mu_2(g_1,g_2,g_3)\mu_2(g_0,g_1,g_3)}{\mu_2(g_0,g_2,g_3)\mu_2(g_0,g_1,g_2)}
\ee
where $\mu_2$ is an arbitrary 2-cochain. Note that it is not necessarily a cocycle.
The effective symmetry on the boundary can hence be written as
\be
\begin{array}{l}
\t U(g) |\{\alpha_i\}\> = \\
\prod_{i,j} \left(\frac{\mu_2(\alpha_j,g^{-1}g^*,g^*)\mu_2(\alpha_i,\alpha_j,g^*)}{\mu_2(\alpha_i,g^{-1}g^*,g^*)\mu_2(\alpha_i,\alpha_j,g^{-1}g^*)}\right)^{s_{ij}}|\{g\alpha_i\}\>
\end{array}
\ee
The $\mu_2(\alpha_i,g^{-1}g^*,g^*)$ terms cancel out in the product of phase
factors, and the remaining terms can be grouped into two sets
$\prod_{ij} \mu_2^{s_{ij}}(\alpha_i,\alpha_j,g^*)$
and
$
\prod_{ij} \mu_2^{-s_{ij}}(\alpha_i,\alpha_j,g^{-1}g^*)=\prod_{ij} \mu_2^{-s_{ij}s(g)}(g\alpha_i,g\alpha_j,g^*)
$.
Define $\Theta(g)=\prod_{ij} \sum_{\alpha_i,\alpha_j}\mu_2^{s_{ij}}(\alpha_i,\alpha_j,g^*)|\alpha_i\alpha_j\>\<\alpha_i\alpha_j|$. $\Theta(g)$ is a product of local unitaries. It is easy to see that
\be
\t U(g)= \Theta^{\dag}(g) \left(\sum_{\{\alpha_i\}}|\{g\alpha_i\}\>\<\{\alpha_i\}|\right) \Theta(g)
\ee
(a complex conjugation operation needs to be added if $\t U(g)$ is
anti-unitary). The term in the middle is an on-site operation which permutes
the basis. It has a simple symmetric state which is a product state
$
\otimes_i (\sum_{\alpha_i} |\alpha_i\>)
$.
Therefore
$
 \Theta^{\dag}(g) \otimes_i (\sum_{\alpha_i} |\alpha_i\>)
$
is a symmetric state of $\t U(g)$. Because $\otimes_i (\sum_{\alpha_i}
|\alpha_i\>)$ is a product state and $ \Theta^{\dag}(g)$ is a product of local
unitaries, this is a short range entangled state. Therefore, we have explicitly
constructed a short range entangled symmetric state on the boundary if the
model is constructed from a trivial 3-cocycle $\nu_3$.

\section{Calculations of group cohomology}

In the section, we will calculate group cohomology for some simple groups.  We
will first present a direct calculation from the definition of the group
cohomology.  Then we will present some more advance results.

\subsection{Canonical choice of cocycles}

Let us consider an $n$-cocycles $\nu_n'$ which satisfies the
condition $\dd \nu_n'=1$. By a proper transformation
by a coboundary $b_n$:
$\nu_n=\nu'_nb_n^{-1}$, we can choose a particular cocycle $\nu_n$ in a given cohomology class that satisfies
\begin{subequations}\label{gauge}
\begin{eqnarray}
&\nu_n({\bf{g_0,g_0}},g_1,...,g_{n-2},g_{n-1})=&1\label{gaugea}\\
&\nu_n(g_1,{\bf g_0,g_0},...,g_{n-2},g_{n-1})=&1\label{gaugeb}\\
&...\ \ \ ...\ \ \ ...&\nonumber\\
&\nu_n(g_1,g_2,g_3..., g_{n-1},{\bf g_0,g_{0}})=&1.
\end{eqnarray}
\end{subequations}

To this, let us focus on \eqn{gaugea}. To prove that the choice \eqn{gaugea} is valid in general, it is
equivalent to prove that we can always choose a cocycle
in a cohomology class that satisfies
\begin{eqnarray}\label{General}
\nu_n(\underbrace{\bf{g_0,...,g_0}}_{m\text{ terms}},g_1,...,g_{n+1-m})=1,
\end{eqnarray}
where $m$ is the number of repeating index $g_0$ with $2\leq m\leq n+1$ and $g_1\neq g_0$.

Firstly, we will show that a cocycle can satisfy \eqn{General}
for $m=n+1$, which means
\begin{eqnarray}\label{m=n+1}
\nu_n(\underbrace{g_0,g_0,...,g_0}_{n+1})=1.
\end{eqnarray}
If $n$ is odd,  \eqn{m=n+1} can be easily shown from the cocycle condition
\[
(\dd \nu_n)(\underbrace{g_0,g_0,g_0,...,g_0}_{n+2})=\nu_n(\underbrace{g_0,g_0,...,g_0}_{n+1})=1.
\]
If $n$ is even, then we can introduce a coboundary $b_n(\underbrace{g_0,g_0,...,g_0}_{n+1}
)=(\dd \mu_{n-1})(\underbrace{g_0,g_0,...,g_0}_{n+1})=\mu_{n-1}(\underbrace{g_0,...,g_0}_{n})$,
where $\mu_{n-1}$ is a cochain. If we require that
$\mu_{n-1}(\underbrace{g_0,...,g_0}_n)=\nu_n'(\underbrace{g_0,g_0,...,g_0}_{n+1})$, then after the gauge
transformation $\nu_n=\nu_n'b_n^{-1}$ we have
$\nu_n(\underbrace{g_0,g_0,...,g_0}_{n+1})=1$. Thus, we have proved the validity of
\eqn{General} in the case $m=n+1$.

Now we show that a cocycle can satisfy \eqn{General} for the case $m=2k+1$ ($1\leq k\leq [n/2]$, where $[n/2]$ is the integer part of $n/2$, i.e., $[n/2]=n/2$ if $n$ is even and $[n/2]=(n-1)/2$ if $n$ is odd),
namely, $\nu_n(\underbrace{g_0,...,g_0}_{2k+1},g_1,g_2,...,g_{n-2k})=1$. Here we assume $g_1\neq g_0$.
Again, we introduce the gauge transformation $\nu_n=\nu_n'b_n^{-1}$ with $b_n=(\dd \mu_{n-1})$. We requires that the cochain $\mu_{n-1}$ satisfies
\begin{eqnarray}\label{fixing1}
&&\mu_{n-1}(\underbrace{g_0,...,g_0}_{2k},g_1,g_2,...,g_{n-2k})\nonumber\\&=&
\nu_n'(\underbrace{g_0,g_0,...,g_0}_{2k+1},g_1,g_2,...,g_{n-2k})\nonumber\\&&
\left[\mu_{n-1}^{-1}(\underbrace{g_0,g_0,...,g_0}_{2k+1},g_2,g_3,...,g_{n-2k})\right.\nonumber\\
&&\left.\mu_{n-1}(\underbrace{g_0,g_0,...,g_0}_{2k+1},g_1,g_3,...,g_{n-2k})...\right.\nonumber\\&&
\left.\mu_{n-1}^{(-1)^{n+1}}(\underbrace{g_0,g_0,...,g_0}_{2k+1},g_1,g_2,...,g_{n-2k-1})\right]^{-1},
\end{eqnarray}
for $k=[n/2]$, $k=[n/2]-1$, ..., $k=1$ in sequence.  Notice that there is
one-to-one correspondence between
$\nu_n'(\underbrace{g_0,g_0,...,g_0}_{2k+1},g_1,g_2,...,g_{n-2k})$ and
$\mu_{n-1}(\underbrace{g_0,...,g_0}_{2k},g_1,g_2,...,g_{n-2k})$. In the square
bracket, the terms which have odd number of successive index $g_0$ are free
variables. Equation (\ref{fixing1}) can always be satisfied by letting
$\mu_{n-1}(\underbrace{g_0,...,g_0}_{2k},g_1,g_2,...,g_{n-2k})$ equal to the
right-hand side of \eqn{fixing1} (we will illustrate it by several examples).
In other words, \eqn{fixing1} is a constrain for the components of $\mu_{n-1}$
which have even number of successive index $g_0$, and at the same time the
components of $\mu_{n-1}$ which have odd number of successive index $g_0$ can
be chosen arbitrarily.  Thus, we can always find a $\mu_{n-1}$ that satisfies
\eqn{fixing1}.

From \eqn{fixing1}, we have
\begin{eqnarray*}
&&\ \ \ \ b_n(\underbrace{g_0,g_0,...,g_0}_{2k+1},g_1,g_2,...,g_{n-2k})\\ &&=(\dd \mu_{n-1})(\underbrace{g_0,g_0,...,g_0}_{2k+1},g_1,g_2,...,g_{n-2k})\\&&=\nu_n'(\underbrace{g_0,g_0,...,g_0}_{2k+1},g_1,g_2,...,g_{n-2k}).
\end{eqnarray*}
Consequently, we obtain $\nu_n(\underbrace{g_0,g_0...,g_0}_{2k+1},g_1,g_2,...,g_{n-2k})=1$ after the gauge transformation $\nu_n=\nu_n'b_n^{-1}$. From this result and the cocycle condition
\begin{eqnarray*}
&&(\dd \nu_n)(\underbrace{g_0,g_0,...,g_0}_{2k+1},g_1,g_2,...,g_{n+1-2k})\\&=&
\nu_n(\underbrace{g_0,...,g_0}_{2k},g_1,g_2,...,g_{n+1-2k})\\&&
\nu_n^{-1}(\underbrace{g_0,g_0,...,g_0}_{2k+1},g_2,g_3...,g_{n+1-2k})...\\&&
\nu_n^{(-1)^{n+1}}(\underbrace{g_0,g_0,...,g_0}_{2k+1},g_1,g_2,...,g_{n-2k})\\&=&1,
\end{eqnarray*}
we obtain $\nu_n(\underbrace{g_0,...,g_0}_{2k},g_1,g_2,...,g_{n+1-2k})=1$, which proves \eqn{General} in the case $m=2k$.

Above proof includes the cases of $1\leq k\leq [n/2]$. Together with \eqn{m=n+1}, we have finished the proof of \eqn{General}, or equivalently, \eqn{gaugea}. Notice that the only requirement in the proof is \eqn{fixing1}.

To prove \eqn{gaugeb} in general, it is equivalent to prove the following equations
\begin{eqnarray}\label{General2}
\nu_n(g_1,\underbrace{\bf{g_0,...,g_0}}_{m},g_2,g_3,...,g_{n-2},g_{n+1-m})=1,
\end{eqnarray}
here $m$ is the number of repeating index $g_0$ with $2\leq m\leq n$ and $g_1\neq g_0$ (the case $g_1=g_0$ reduces to \eqn{General} and has been proved already).

Let begin with the case $m=n$, namely,
\begin{eqnarray}\label{m=n}
\nu_n(g_1,\underbrace{g_0,g_0,...,g_0}_{n})=1.
\end{eqnarray}
If $n$ is odd, we introduce $\nu_n=\nu_n'b_n^{-1}$, with $b_n(g_1,\underbrace{g_0,g_0,...,g_0}_{n})=(\dd \mu_{n-1})(g_1,\underbrace{g_0,g_0,...,g_0}_n) =\mu_{n-1}(\underbrace{g_0,g_0,...,g_0}_n)\mu_{n-1}^{-1} (g_1,\underbrace{g_0,...,g_0}_{n-1})$. If we require
\begin{eqnarray*}
&&\mu_{n-1} (g_1,\underbrace{g_0,g_0,...,g_0}_n)
\\&=&\nu_{n}'^{-1}(g_1,\underbrace{g_0,...,g_0}_{n-1})\mu_{n-1}(\underbrace{g_0,g_0,...,g_0}_{n}),
\end{eqnarray*}
then we obtain $b_n(g_1,\underbrace{g_0,g_0,...,g_0}_{n})=\nu_n'(g_1,\underbrace{g_0,g_0,...,g_0}_{n})$ and consequently $\nu_n(g_1,\underbrace{g_0,g_0,...,g_0}_{n})=1$. If $n$ is even, then from the cocycle condition
\begin{eqnarray*}
&&(\dd \nu_n)(g_1,\underbrace{g_0,g_0,g_0,...,g_0}_{n+1})\\&=&\nu_n(\underbrace{g_0,g_0,g_0,...,g_0}_{n+1})
\nu_n^{-1}(g_1,\underbrace{g_0,g_0,...,g_0}_n)\\&=&1,
\end{eqnarray*}
we obtain $\nu_n(g_1,\underbrace{g_0,g_0,...,g_0}_n)=1$, here we have used \eqn{m=n+1}.

Now we prove \eqn{fixing2} for the case $m=2k+1$ ($1\leq k\leq [(n-1)/2]$, where $[(n-1)/2]$ is the integer part of $(n-1)/2$), namely, $\nu_n(g_1,\underbrace{g_0,...,g_0}_{2k+1},g_2,g_3,...,g_{n-2k})=1$ with $g_1,g_2\neq g_0$.
Again, we introduce the gauge transformation $\nu_n=\nu_n'b_n^{-1}$ with $b_n=(\dd \mu_{n-1})$. We requires that the cochain $\mu_{n-1}$ satisfies
\begin{eqnarray}\label{fixing2}
&&\mu_{n-1}^{-1}(g_1,\underbrace{g_0,...,g_0}_{2k},g_2,g_3,...,g_{n-2k})\nonumber\\
&=&\nu_n'(g_1,\underbrace{g_0,g_0,...,g_0}_{2k+1},g_2,g_3,...,g_{n-2k})\nonumber\\&&
\left[\mu_{n-1}(\underbrace{g_0,g_0,...,g_0}_{2k+1},g_2,g_3,...,g_{n-2k})\right.\nonumber\\&&
\left.\mu_{n-1}(g_1,\underbrace{g_0,g_0,...,g_0}_{2k+1},g_3,g_4,...,g_{n-2k})...
\right.\nonumber\\&&\left.\mu_{n-1}^{(-1)^{n+1}}(g_1,\underbrace{g_0,g_0,...,g_0}_{2k+1},g_2,g_3,...,g_{n-2k-1})\right]^{-1},
\end{eqnarray}
for $k=[(n-1)/2]$, $k=[(n-1)/2]-1$, ..., $k=1$ in sequence.
Again, there is one-to-one correspondence between $\nu_n'(g_1,\underbrace{g_0,g_0,...,g_0}_{2k+1},g_2,g_3,...,g_{n-2k})$ and $\mu_{n-1}(g_1,\underbrace{g_0,...,g_0}_{2k},g_2,g_3,...,g_{n-2k})$. Equation (\ref{fixing2}) can always be satisfied by constraining the value of $\mu_{n-1}^{-1}(g_1,\underbrace{g_0,...,g_0}_{2k},g_2,g_3,...,g_{n-2k})$ to equal to the right-hand side (we will illustrate it by several examples).

From \eqn{fixing2}, we have
\begin{eqnarray*}
&&b_n(g_1,\underbrace{g_0,g_0,...,g_0}_{2k+1},g_2,g_3,...,g_{n-2k})\\
&=&(\dd \mu_{n-1})(g_1,\underbrace{g_0,g_0,...,g_0}_{2k+1},g_2,g_3,...,g_{n-2k})\\&=&
\nu_n'(g_1,\underbrace{g_0,g_0,...,g_0}_{2k+1},g_2,g_3,...,g_{n-2k}).
\end{eqnarray*}
Consequently, after the gauge transformation $\nu_n=\nu_n'b_n^{-1}$, we obtain $\nu_n(g_1,g_0,g_0,...,g_0,g_2,g_3,...,g_{n-2k})=1$. From this result and \eqn{General} and the cocycle condition
\begin{eqnarray*}
&&(\dd \nu_n)(g_1,\underbrace{g_0,g_0,...,g_0}_{2k+1},g_2,g_3,...,g_{n+1-2k})\\&=&
\nu_n(\underbrace{g_0,g_0,...,g_0}_{2k+1},g_2,g_3,...,g_{n+1-2k})
\\&&\nu_n^{-1}(g_1,\underbrace{g_0,...,g_0}_{2k},g_2,g_3...,g_{n+1-2k})\\&&
\nu_n(g_1,\underbrace{g_0,g_0,...,g_0}_{2k+1},g_3,g_4,...,g_{n+1-2k})...\\&&
\nu^{(-1)^{n+1}}(g_1,\underbrace{g_0,g_0,...,g_0}_{2k+1},g_2,g_3,...,g_{n-2k})\\&=&1,
\end{eqnarray*}
we obtain $\nu_n(g_1,\underbrace{g_0,...,g_0}_{2k},g_2,g_3,...,g_{n+1-2k})=1$, which proves the case $m=2k$.

Above proof includes the cases of $1\leq k\leq [(n-1)/2]$. Together with \eqn{m=n}, we have finished the proof of \eqn{General2}, or equivalently, \eqn{gaugeb}. Notice that in the proof we have used two conditions \eqn{fixing1} and \eqn{fixing2}. Obviously, they can be satisfied simultaneously.

The remaining part of \eqn{gauge} can be proved by the same
procedure and will not be repeated here. We stress that all of the
equations in \eqn{gauge} can be satisfied simultaneously, because
in proving different equations we are fixing the values of different classes of components of the
$(n-1)$-cochain $\mu_{n-1}$.

As examples, let us illustrate that \eqn{fixing1} and \eqn{fixing2} can be satisfied simultaneously for $n\leq 4$. When $n=2$ ($k=1$), \eqn{fixing1} becomes
\begin{eqnarray*}
\mu_1(g_0,g_0)=\nu_2'(g_0,g_0,g_0),
\end{eqnarray*}
which can be satisfied obviously. We do not need to consider \eqn{fixing2} for $n=2$.

When $n=3$ ($k=1$), \eqn{fixing1} becomes
\begin{eqnarray*}
\mu_2(g_0,g_0,g_1)=\nu_3'(g_0,g_0,g_0,g_1)\mu_2^{-1}(g_0,g_0,g_0),
\end{eqnarray*}
which can be satisfied by constraining the value of $\mu_2(g_0,g_0,g_1)$ to be equal to the right-hand side.
On the other hand, \eqn{fixing2} becomes
\begin{eqnarray*}
\mu_2^{-1}(g_1,g_0,g_0)=\nu_3'(g_1,g_0,g_0,g_0)\mu_2^{-1}(g_0,g_0,g_0),
\end{eqnarray*}
which can also be satisfied by properly choosing the value of
$\mu_2(g_1,g_0,g_0)$.

Finally, when $n=4$, there are two cases $k=2$ and $k=1$.   For $k=2$, \eqn{fixing1} becomes
\begin{eqnarray*}
\mu_3(g_0,g_0,g_0,g_0)=\nu_4'(g_0,g_0,g_0,g_0,g_0),
\end{eqnarray*}
which can be satisfied obviously. For $k=1$, \eqn{fixing1} becomes
\begin{eqnarray*}
\mu_3(g_0,g_0,g_1,g_2)=&&\nu_4'(g_0,g_0,g_0,g_1,g_2)\\
&&\times[\mu_3^{-1}(g_0,g_0,g_0,g_2)\mu_3(g_0,g_0,g_0,g_1)]^{-1},
\end{eqnarray*}
which can be satisfied by constraining the value of $\mu_3(g_0,g_0,g_1,g_2$ to
be equal to the right-hand side.

One the other hand, when $n=4$ and $k=1$, \eqn{fixing2} becomes (We do not need to consider \eqn{fixing2} for $k=2$)
\begin{eqnarray*}
\mu_3^{-1}(g_1,g_0,g_0,g_2)&=&\nu_4'(g_1,g_0,g_0,g_0,g_2)\\
&&[\mu_3(g_0,g_0,g_0,g_2)\mu_3^{-1}(g_0,g_0,g_0,g_1)]^{-1},
\end{eqnarray*}
which can be satisfied by restraining the value of $\mu_3(g_1,g_0,g_0,g_2)$.

Now let us see what happens for the term
$\nu_n(\mathbf{g_0},g_1,g_2,...,g_{n-1},\mathbf{g_0})$ with $g_1,g_{n-1}\neq g_0$. Considering the coboundary
\begin{eqnarray}
&&b_n(g_0,g_1,g_2,...,g_{n-1},g_0)\nonumber\\
&=&(\dd \mu_{n-1})(g_0,g_1,g_2,...,g_{n-1},g_0)\nonumber\\
&=&\mu_{n-1}(g_1,g_2,...,g_{n-1},g_0)\mu_{n-1}^{-1}(g_0,g_2,...,g_{n-1},g_0)\nonumber\\
&&...\mu_{n-1}^{(-1)^{n}}(g_0,g_1,...,g_{n-1}),\nonumber
\end{eqnarray}
Notice that the two cochains $\mu_{n-1}(g_1,g_2,...,g_{n-1},g_0)$
and $\mu_{n-1}^{(-1)^{n}}(g_0,g_1,...,g_{n-1})$ may cancel each
other in some condition. In that case, $b_n(g_0,g_1,g_2,...,
g_{n-1},g_0)$ has less degrees of freedom than $\nu'_n(g_0,g_1,
g_2,...,g_{n-1},g_0)$, so we CAN NOT always set $\nu_n(g_0,g_1,
g_2,...,g_{n-1},g_0)=1$.

\subsection{Group cohomology of $Z_2$}

\label{Z2gc}

Using the properties obtained above, we will show that for the group
$Z_2=\{E,\sigma\}$ (where $E$ is the identity and $\sigma^2=E$) ,
\begin{eqnarray}\label{Z2}
&&\cH^{2m-1}[Z_2,U(1)]=\Z_2,\nonumber\\
&&\cH^{2m}[Z_2,U(1)]=\Z_1,\ \ \ \ m\geq1
\end{eqnarray}
and for the time reversal $Z^T_2=\{E,T\}$ group,
\begin{eqnarray}\label{Z2T}
&&\cH^{2m-1}[Z_2^T,U_T(1)]=\Z_1,\nonumber\\
&&\cH^{2m}[Z_2^T,U_T(1)]=Z_2,\ \ \ \ m\geq1.
\end{eqnarray}
We note that $Z_2$ and $Z_2^T$ are the same group. However, the generator in
$Z_2^T$ has a non-trivial action on the module.  Also $U(1)$ and $U_T(1)$ are
the same as Abelian group. The subscript $T$ in the module $U_T(1)$ is used to
indicate that the group $Z_2^T$ has a non-trivial action on the module.

Let us begin with \eqn{Z2}. Firstly, $\nu'_{2m-1}(g_0,g_1,...,g_{2m+1})$ have
even number of group indices. From \eqn{gauge}, we can set
$\nu_{2m-1}(g_0,g_1,...,g_{2m+1})=1$ if any two neighboring indices are the
same. So the only possible nontrivial one is when the group indices vary
alternatively, namely, the component
$\nu_{2m-1}(E,\sigma,...,E,\sigma)= \nu_{2m-1}(\sigma,E,...,
\sigma,E)$. Considering the cocycle condition
\begin{eqnarray}\label{Z2_2m+1}
&&(\dd \nu_{2m-1})(E,\sigma,...,\sigma,E)\nonumber\\
&=&\nu_{2m-1}(\sigma,E,...,\sigma,E)
\nu_{2m-1}^{-1}(E,E,\sigma,...,\sigma,E)\nonumber\\
&&\nu_{2m-1}(E,\sigma,\sigma,E,...,\sigma,E)...\nu_{2m-1}(E,\sigma,...,E,\sigma)\nonumber\\
&=&[\nu_{2m-1}(\sigma,E,...,\sigma,E)]^2\nonumber\\
&=&1.
\end{eqnarray}
here we have used \eqn{gauge}. So we have
\[
\nu_{2m-1}(E,\sigma,...,E,\sigma)= \nu_{2m-1}(\sigma,E,...,
\sigma,E)=\pm1.
\]
Now we need to show that these two solutions are
not gauge equivalent. Consider the coboundary
\begin{eqnarray}
&&b_{2m-1}(\sigma,E,...,\sigma,E)\nonumber \\
&=&(\dd \mu_{2m-2})(\sigma,E,...,\sigma,E)\nonumber \\
&=&\mu_{2m-2}(E,\sigma,...,\sigma,E)\mu_{2m-2}^{-1}(\sigma,\sigma,E,...,\sigma,E)...\nonumber \\
&&\mu_{2m-2}(\sigma,E,...,\sigma, E,E)\mu_{2m-2}^{-1}(\sigma,E,...,E,\sigma)\nonumber \\
&=&\mu_{2m-2}^{-1}(\sigma,\sigma,E,...,\sigma,E)\mu_{2m-2}(\sigma,E,E,...,\sigma,E)...\nonumber \\
&&...\mu_{2m-2}(\sigma,E,...,\sigma, E,E).
\end{eqnarray}
Notice that $\mu_{2m-2}(E,\sigma,...,\sigma,E)$ is canceled by
$\mu_{2m-2}^{-1} (\sigma,E,...,E,\sigma)$. In all of the remaining
components, a pair of neighboring group indices are the
same. The values of these components have been fixed in the gauge
choice \eqn{gauge}. Consequently, the value of the coboundary
$b_{2m-1} (\sigma,E,...,\sigma,E)$ is also fixed. But there are two cocycles satisfying \eqn{Z2_2m+1}, so they must belong to two different classes.

Secondly, $\nu_{2m}(g_0,g_1,...,g_{2m})$ contains odd number of
group indices. The only possible nontrivial one is
$\nu_{2m}(E,\sigma,...,\sigma,E)= \nu_{2m}(\sigma,E,...,E, \sigma)$.
Considering the coboundary
\begin{eqnarray}\label{Z2_2m}
&&b_{2m}(E,\sigma,...,\sigma,E)\nonumber\\
&=&(\dd \mu_{2m-1})(E,\sigma,...,\sigma,E)\nonumber\\
&=&\mu_{2m-1}(\sigma,E,...,\sigma,E)
\mu_{2m-1}^{-1}(E,E,\sigma,...,\sigma,E)\nonumber\\
&&\mu_{2m-1}(E,\sigma,\sigma,E,...,\sigma,E)...\mu_{2m-1}(E,\sigma,...,E,\sigma)\nonumber\\
&=&[\mu_{2m-1}(\sigma,E,...,\sigma,E)]^2...
\end{eqnarray}
Notice that the component $\mu_{2m-1}(\sigma,E,...,\sigma,E)$ is free
since it is not fixed by the gauge choice \eqn{gauge}. So
the $b_{2m}(E,\sigma,...,\sigma,E)$ has the same
degrees of freedom as $\nu'_{2m}(E,\sigma,...,\sigma,E)$, and we can always set $\nu_{2m}(E,\sigma,...,\sigma,E)=1$ with the gauge transformation $\nu_n=\nu_n'b_n^{-1}$ . Consequently, we have $\cH^{2m}[Z_2,U(1)]=\Z_1.$

Conditions are on the contrary for the time reversal group $Z^T_2$,
because of the relation $\nu_n(Tg_0,Tg_1,...,Tg_n)=\nu_n^{-1}
(g_0,g_1,...,g_n)$. Corresponding to \eqn{Z2_2m+1}, we have
\begin{eqnarray}
&&(\dd \nu_{2m})(E,T,...,E,T)\nonumber\\
&=&\nu_{2m}(T,E,...,E,T)
\nu_{2m}^{-1}(E,E,T,...,E,T)\nonumber\\
&&\nu_{2m}(E,T,T,E,...,E,T)...\nu_{2m}^{-1}(E,T,...,T,E)\nonumber\\
&=&[\nu_{2m}(T,E,...,E,T)]^2\nonumber\\
&=&1,
\end{eqnarray}
which result in 
\[
\nu_{2m}(T,E,...,E,T)=\nu_{2m}^{-1} (E,T,...,T,E)=\pm1.
\] 
Similar to \eqn{Z2_2m+1}, these two solutions are not gauge equivalent. Consequently, $\cH^{2m}[Z_2^T,U_T(1)]=\Z_2.$
Similarly, corresponding to \eqn{Z2_2m}, we have
\begin{eqnarray}
&&b_{2m-1}(E,T,...,E,T)\nonumber\\
&=&(\dd \mu_{2m-2})(E,T,...,E,T)\nonumber\\
&=&\mu_{2m-2}(T,E,...,E,T)
\mu_{2m-2}^{-1}(E,E,T,...,E,T)\nonumber\\
&&\mu_{2m-2}(E,T,T,E,...,E,T)...\mu_{2m-2}^{-1}(E,T,...,T,E)\nonumber\\
&=&[\mu_{2m-2}(T,E,...,E,T)]^2...
\end{eqnarray}
The free component $\mu_{2m-2}(T,E,...,E,T)$ guarantees that the
$b_{2m-1}(E,T,...,E,T)$ has the same degrees of
freedom as $\nu'_{2m-1}(E,T,...,E,T)$, so we can set $\nu_{2m-1}(E,T,...,E,T)=1$ and consequently
$\cH^{2m-1}[Z_2^T,U_T(1)]=\Z_1.$

\subsection{Group cohomology of $Z_n$ over a generic $Z_n$-module}

The cohomology group $\cH^d[Z_n,M]$ has a very simple form.
To describe the simple form in a more general setting,
let us define Tate cohomology groups $\hat\cH^d[G,M]$.

For $d$ to be $0$ or $-1$, we have
\begin{align}
 \hat\cH^0[G,M] &= M^{G}/\text{Img}(N_{G},M),
\nonumber\\
 \hat\cH^{-1}[G,M] &= \text{Ker}(N_{G},M)/I_{G}M.
\end{align}
Here $M^{G}$, $\text{Img}(N_{G},M)$, $\text{Ker}(N_{G},M)$, and
$I_{G}M$ are submodule of $M$.
$M^{G}$ is the maximal submodule that is invariant under the
group action. Let us define a
map $N_{G}: M\to M$ as
\begin{align}
 a\to \prod_{g\in G} g\cdot a, \ \ \ a\in M.
\end{align}
$\text{Img}(N_{G},M)$ is the image of the map
and $\text{Ker}(N_{G},M)$ is the
kernel of the map.
The submodule $I_{G}M$ is given by
\begin{align}
 I_{G}M=
\{ \prod_{g\in G} (g\cdot a)^{n_g} | \sum_{g\in G} n_g=0,\ a\in M\}
\end{align}
In other words, $I_{G}M$ is generated by $(g\cdot a) a^{-1}$,
$\forall\ g\in G,\ a \in M$.

For $d$ other then $0$ and $-1$,
Tate cohomology groups $\hat\cH^d[G,M]$ is given by
\begin{align}
 \hat\cH^d[G,M] &=\cH^d[G,M], \ \text{ for } d>0
\nonumber\\
% \hat\cH^0[G,M] &=\cH^0[G,M],
%\nonumber\\
% \hat\cH^{-1}[G,M] &=\cH^{-1}[G,M],
%\nonumber\\
 \hat\cH^d[G,M] &=\cH_{-d-1}[G,M], \ \text{ for } d<-1 .
\end{align}

For cyclic group $Z_n$, its (Tate) group cohomology over a generic $Z_n$-module $M$
is given by\cite{DJ,RS}
\begin{align}
\label{ZnCoh}
\hat \cH^d[Z_n,M]=
\begin{cases}
\hat\cH^0[Z_n,M] & \text{ if } d=0 \text{ mod } 2,\\
\hat\cH^{-1}[Z_n,M] & \text{ if } d=1 \text{ mod } 2.\\
\end{cases}
\end{align}
where
\begin{align}
 \hat\cH^0[Z_n,M] &= M^{Z_n}/\text{Img}(N_{Z_n},M),
\nonumber\\
 \hat\cH^{-1}[Z_n,M] &= \text{Ker}(N_{Z_n},M)/I_{Z_n}M.
\end{align}

For example, when the group action is trivial,
we have $M^{Z_n}=M$ and $ I_{Z_n}M=\Z_1$.
The map $N_{Z_n}$ becomes $N_{Z_n}: a\to a^n$.
For $M=\Z$, we have
$\text{Img}(N_{Z_n},\Z)=n \Z$
and $\text{Ker}(N_{Z_n},\Z)=\Z_1$.
For $M=U(1)$, we have
$\text{Img}(N_{Z_n},\Z)=U(1)$
and $\text{Ker}(N_{Z_n},\Z)=\Z_n$.
So we have
\begin{align}
\label{ZnZ}
 \cH^d[Z_n,\Z]=
\begin{cases}
\Z & \text{ if } d=0, \\
\Z_n  & \text{ if } d=0 \text{ mod } 2,\ \ d>0\\
\Z_1  & \text{ if } d=1 \text{ mod } 2.
\end{cases}
\end{align}
and
\begin{align}
 \cH^d[Z_n,U(1)]=
\begin{cases}
U(1) & \text{ if } d=0, \\
\Z_1  & \text{ if } d=0 \text{ mod } 2,\ \ d>0\\
\Z_n  & \text{ if } d=1 \text{ mod } 2.
\end{cases}
\end{align}
which reproduces the result mentioned in \Ref{DW9093}, and the result
obtained in the last subsection for $d=2$.

What does a non-trivial cocycle in $\cH^d[Z_n,U(1)]$ looks like?
Since $\cH^1[Z_n,U(1)]$ describes the 1D unitary representation
of $Z_n=\{0,1,...,k,...,n-1\}$, we find that
the $m^\text{th}$ 1-cocycles in $\cH^1[Z_n,U(1)]=\Z_n$ are represented by
complex function
$\om^{(m)}_1(k)=\nu_1^{(m)}(0,k)=\e^{mk\imth 2\pi/n}$, $k \in Z_n$.

If a group operation $T$ acts on
$Z_n$ by inversion: $TkT^{-1}=-k \text{ mod } n$, $k\in Z_n$,
then $T$ act on  an 1-cocycle $\om_m(k)$ in $\cH^1[Z_n,U(1)]=\Z_n$
as $T\cdot \om_m(k) =\om_m(-k) =\om_{-m \text{ mod } n}(k)$.
Since $\cH^1[Z_n,U(1)]=\cH^2[Z_n,\Z]$, we find that
\begin{align}
\label{TalH2Zn}
 T\cdot \al = -\al, \ \ \ \al \in \cH^2[Z_n,\Z].
\end{align}
A similar result can also be obtained for $U(1)$ group:
\begin{align}
\label{TalH2U1}
 T\cdot \al = -\al, \ \ \ \al \in \cH^2[Z_n,\Z].
\end{align}
Such a result will be useful later.

We can also use the above approach to calculate
some other cohomology groups.
To calculate $ \cH^d[Z_2^T,U_T(1)]$, we note that the invariant submodule
$[U_T(1)]^{Z_2^T}=\Z_2$, and the map $N_{Z_2^T}$ becomes $a\to 1$.  So
$\text{Img}[N_{Z_2^T},U_T(1)]=\Z_1$ and
$\text{Ker}[N_{Z_2^T},U_T(1)]=U_T(1)$.  Also $I_{Z_2^T}U_T(1)=U_T(1)$.  Thus
\begin{align}
 \cH^d[Z_2^T,U_T(1)]=
\begin{cases}
\Z_2  & \text{ if } d=0 \text{ mod } 2,\\
\Z_1  & \text{ if } d=1 \text{ mod } 2,
\end{cases}
\end{align}
which reproduces the result
obtained in the last subsection.

Now let us calculate $ \cH^d[Z_2^T,\Z_T]$,
where $Z_2^T=\{E,T\}$ has a non-trivial action on the integer
module $\Z_T$:
\begin{align}
\label{Z2TZT}
 T\cdot n = -n,\ \ \   E\cdot n =n,\ \ \ n\in \Z_T .
\end{align}
We note that the invariant submodule
$[\Z_T]^{Z_2^T}=\Z_1$, and the map $N_{Z_2^T}$ becomes $n\to 0$.  So
$\text{Img}[N_{Z_2^T},\Z_T]=\Z_1$ and
$\text{Ker}[N_{Z_2^T},\Z_T]=Z_T$.  Also $I_{Z_2^T}\Z_T=2\Z_T$.  Thus
\begin{align}
 \cH^d[Z_2^T,\Z_T]=
\begin{cases}
\Z_1 & \text{ if } d=0, \\
\Z_1  & \text{ if } d=0 \text{ mod } 2,\ \ d>0\\
\Z_2  & \text{ if } d=1 \text{ mod } 2.
\end{cases}
\end{align}

Next, let us consider $\cH^d[Z_2\times Z_p,U(1)]$ where $p$ is an odd number.
Notice that $ Z_2\times Z_p= Z_{2p}=\{1,zt,z^2,z^3t,...,z^{p-1},t,z,...\}$
where $t$ generates $Z_2$, $z$ generates $Z_p$, and $zt=tz$ generates $Z_{2p}$.
So we have
\begin{align}
\label{Z2Zp}
 \cH^d[ Z_2\times  Z_p,U(1)]=
\begin{cases}
U(1) & \text{ if } d=0, \\
\Z_1  & \text{ if } d=0 \text{ mod } 2,\ \ d>0\\
\Z_2\times \Z_p  & \text{ if } d=1 \text{ mod } 2.
\end{cases}
\end{align}

%Last, we consider $\cH^d[ Z_2^T\times Z_p,U_T(1)]$ where $p$ is an odd number.
%Here $ Z_2^T$ is generated by time reversal $T$ and acts non-trivially on
%$U_T(1)$: $T: a\to a^{-1}$, $a\in U_T(1)$.  $ Z_p$ acts trivially on $U_T(1)$.
%Notice that $ Z_2^T\times Z_p= Z_{2p}=\{1,zT,z^2,z^3T,...,z^{p-1},T,z,...\}$
%where $T$ generates $Z_2^T$, $z$ generates $Z_p$, and $zT=Tz$ generates
%$Z_{2p}$.  So we can use the formula \eqn{ZnCoh}.

Last, we consider $\cH^d[ Z_{2p}^T,U_T(1)]$.  $Z_{2p}^T=\{1,z,z^2,...\}$ where
$z^{2n-1}$ contains a time-reversal operation and $z^{2n}$ contains no
time-reversal operation.  Thus $s(z^n)=(-)^n$.  $Z_{2p}^T$  acts non-trivially
on $U_T(1)$: $z^n: a\to a^{[(-)^n]}$, $a\in U_T(1)$.  So we have $[U_T(1)]^{
Z_{2p}^T}=\Z_2$, and the map $N_{Z_{2p}^T}$ becomes $a\to 1$.  So
$\text{Img}[N_{ Z_{2p}^T},U_T(1)]=\Z_1$ and $\text{Ker}[N_{
Z_2^T},U_T(1)]=U_T(1)$.  Also $I_{ Z_{2p}^T}U_T(1)=U_T(1)$.  Thus
\begin{align}
\label{Z2pT}
 \cH^d[ Z_{2p}^T,U_T(1)]=
\begin{cases}
\Z_2  & \text{ if } d=0 \text{ mod } 2,\\
\Z_1  & \text{ if } d=1 \text{ mod } 2.
\end{cases}
\end{align}
When $p$ is odd, $Z_{2p}^T=Z_2^T\times  Z_p$, and we have
\begin{align}
\label{Z2TZp}
 \cH^d[ Z_2^T\times  Z_p,U_T(1)]=
\begin{cases}
\Z_2  & \text{ if } d=0 \text{ mod } 2,\\
\Z_1  & \text{ if } d=1 \text{ mod } 2.
\end{cases}
\end{align}

\subsection{Some useful tools in group cohomology}

To calculate more complicated group cohomology, such as
$\cH^d[Z_m\times Z_n,U(1)]$, we would like to
introduce some mathematical tools here.

\subsubsection{Cohomology on continuous groups}

In the above discussion of group cohomology, we have assumed that the symmetry
group $G$ is finite.  For continuous group, one can also define group
cohomology.  One may naively expected that, for continuous group, the cochain
$\nu_d(\{g_i\})$ should be a continuous function of $g_i$'s in $G$.  Such a
choice of cochain indeed give us a definition of group cohomology for
continuous groups, which is denoted as $\cH_c^d[G,U(1)]$.\cite{JS,WW} 

However, continuous cochain is not the right choice and $\cH_c^d[G,U(1)]$ is
not the right type of group cohomology.  Although $\cH_c^1[G,U(1)]$ does
classify all the 1D representations of $G$, $\cH_c^2[G,U(1)]$ only classifies a
subset of projective representations.\cite{JS,WW} In fact,
$\cH_c^2[G,U(1)]$ only classified topologically split group extensions
of $G$ by $U(1)$:
\begin{align}
 1\to U(1) \to E \to G\to 1
\end{align}
such that, as a space, $E=U(1)\times G$.\cite{JS}  However, a generic
projective representation can be viewed as an $U(1)$ extension of $G$ where the
extension, as a space, can be a principal $U(1)$ bundle over $G$.

So we need to come up with a generalized definition of group cohomology, such
that the resulting $\cH^2[G,U(1)]$ classifies the projective representations of
$G$.  In fact, there are many different generalized definitions of group
cohomology for continuous groups.\cite{JS,CS,WW}  What is the right definition?
We note that the cochain $\nu_d(\{g_i\})$ is related to the action-amplitude
$\e^{ - S(\{g_i\}) }$ (see \eqn{AAdis}) that describe our physics system.  In
general, the action-amplitude $\e^{ - S(\{g_i\}) }$ is a continuous function of
$g_i$.  However, the cochain $\nu_d(\{g_i\})$ is actually a fixed-point
action-amplitude which is a limit of the usual continuous action-amplitude away
from the fixed point.  So the fixed-point action-amplitude, and hence the
cochain $\nu_d(\{g_i\})$, may not be continuous function of $g_i$. For example,
as a limit of continuous functions, it can be piecewise continuous function.

We also note that the cochain appears in the path integral.  So only the
integration values of the cochain over sub-regions of $G$ are physical.  Two
cochains are regarded as the same if their integrations over any sub-regions of
$G$ are the same.

The above considerations suggest that the proper choice of the cochain
$\nu_d(\{g_i\})$ is that the cochains should be measurable
functions.\cite{MFun} Measurable functions are more general than continuous
functions, which can be roughly viewed as piecewise continuous functions.  Such
a choice of cochain defines a group cohomology called Borel
cohomology.\cite{JS} We will use $\cH^d_B[G,U(1)]$ to denote such a group
cohomology.  The SPT phases with a continuous symmetry are classified by the
Borel cohomology group $\cH_B^d[G,U(1)]$.  It has be shown that the second
Borel cohomology group $\cH^2[G,U(1)]$ classifies the projective
representations of $G$,\cite{JS} which classifies the 1D SPT phase with an
on-site symmetry $G$.

On page 16 of \Ref{WW}, it is mentioned in
Remark IV.16(3)
that $\cH^d_B(G,\R)=
\cH^d_c(G,\R)=\Z_1$ (there, $\cH^d_B(G,M)$ is denoted as
$\cH^d_\text{Moore}(G,M)$ which is equal to
$\cH^d_\text{SM}(G,M)$. And $\cH^d_c(G,M)$ is denoted as $\cH^d_\text{glob,c}(G,M)$). 
It is also shown in Remark IV.16(1) and in Remark IV.16(3) that
$\cH^d_\text{SM}(G,\Z)=H^{d}(BG,\Z)$
and $\cH^d_\text{SM}(G,U(1))=H^{d+1}(BG,\Z)$,
%On page 14, it is shown that\\ 
%(3) $\cH^d_\text{loc,c}(G,U(1))=\cH^{d}_\text{SM}(G,U(1))$\\
%Thus $\cH^d_B(G,U(1))=\cH^{d+1}_B(G,\Z)=
%H^{d+1}(BG,\Z)$ 
(where $G$ can have a non-trivial
action on $U(1)$ and $\Z$, and $H^{d+1}(BG,\Z)$ is the usual topological
cohomology on the classifying space $BG$ of $G$).
%and $\cH^d_\text{loc,c}(G,U(1))$ is the group cohomology where the cochain
%$\nu_d(g_0,...,g_d)$ is a continuous function near $g_i=1$.)  
Therefore, 
%we can
%use $\cH^d_\text{loc,c}(G,U(1))$ to calculate $\cH^d_B(G,U(1))$.  
we have
\begin{align} \label{HdR} 
& \cH^d_B(G,U(1))=\cH^{d+1}_B(G,\Z)=H^{d+1}(BG,\Z),
\nonumber\\
& \cH^d_B(G,\R)=\Z_1,\ \ d>0.  
\end{align} 
%These results will be useful later.  
These results are valid for both
continuous groups and discrete groups, as well as for $G$ having a non-trivial
action on the modules $U(1)$ and $\Z$.  In this paper, we use $\cH^d(G,M)$ to
denote the Borel group cohomology class $\cH^d_B(G,M)$.

%\subsubsection{$\cH^d(G,\R)=\Z_1$}
%
%Many group cohomology classes with real coefficient are trivial.  Suppose that
%$G$ is finite, then $|G|\hat \cH^d(G,M)=\Z_1$, where $|G|$ is the total number
%elements in $G$.  Using this result, we can show that if $M$ is a $G$-module on
%which multiplication by $|G|$ is an isomorphism , then $\hat
%\cH^d(G,M)=\Z_1$.\cite{RS} Thus $\hat \cH^d(G,\R)=\Z_1$.  For a compact Lie
%group $G$, on page 33 of \Ref{CS}, it is stated that
%$\cH^d(G,\R)=\cH_c^d(G,\R)$ where $\cH_c^d(G,\R)$ is the group cohomology of
%continuous cochains.  In the last page of \Ref{JS}, it is stated that
%$\cH_c^d(G,\R)=\Z_1$.  Thus, suppose that $G$ is a finite group or a compact
%Lie group, then $ \cH^d[G,\R]=\Z_1$ for $d\geq 1$.
%%where $G$ may have a non-trivial action on $\R$.

\subsubsection{Relation between group cohomology classes
with different modules}
\label{cohM}

Let $A$, $B$, $C$ be $G$-modules related by an exact sequence:
\begin{align}
0\to A\to B\to C\to 0
\end{align}
Then there is a long exact sequence in cohomology:
\begin{align}
0\to &
\cH^0(G,A)\to \cH^0(G,B)\to \cH^0(G,C)\to
\nonumber\\
& \cH^1(G,A)\to \cH^1(G,B)\to \cH^1(G,C)\to ...
\end{align}
We also have
\begin{align}
... \to & \hat \cH^{-1}(G,A)\to \hat \cH^{-1}(G,B)\to \hat \cH^{-1}(G,C)\to
\nonumber\\
  & \hat \cH^0(G,A)\to \hat \cH^0(G,B)\to \hat \cH^0(G,C)\to
\nonumber\\
& \hat \cH^1(G,A)\to \hat \cH^1(G,B)\to \hat \cH^1(G,C)\to ...
\end{align}

\begin{figure}[tb]
\begin{center}
\includegraphics[scale=0.8]{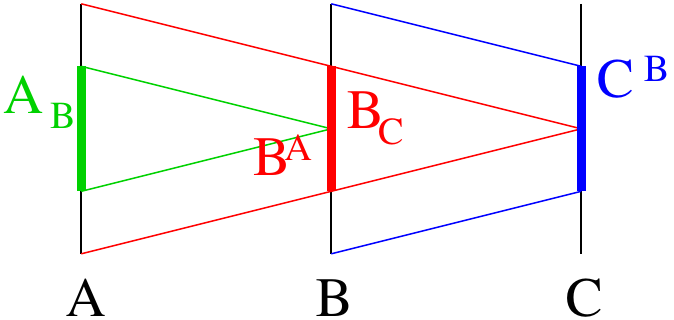}
\end{center}
%Fig. 41
\caption{
(Color online)
The graphic representation of $A\to B$ and $A\to B\to C$.
}
\label{ABC}
\end{figure}

Here 0 represents the trivial module $\Z_1$ with only one elements and
\begin{align}
\cH^0(G,A)=A^G\equiv \{a| g\cdot a=a, g\in G, a\in A\}.
\end{align}
An arrow $A\to B$ means that $A$ maps into a submodule in $B$, $B^A \subset B$,
where a submodule in $A$, $A_B$, maps into the identity $1\in B$.  In other
words, $B^A$ is the image of the map and $A_B$ is the kernel of the map. Those
maps preserve the operations on the modules.  An exact sequence $A\to B\to C$
means that $B^A=B_C$ (see Fig. \ref{ABC}).  So $0\to A\to B\to C\to 0$,
basically, is another way to say $C=B/A$.

Since
\begin{align}
 0\to \Z \to \R \to U(1) \to 0,
\end{align}
we have
\begin{align}
... \to &
\hat \cH^0(G,\Z)\to \hat \cH^0(G,\R)\to \hat \cH^0(G,U(1))\to
\nonumber\\
& \hat \cH^1(G,\Z)\to \hat \cH^1(G,\R)\to \hat \cH^1(G,U(1))\to
\nonumber\\
& \hat \cH^2(G,\Z)\to \hat \cH^2(G,\R)\to \hat \cH^2(G,U(1))\to ...
\end{align}
Since $\hat \cH^d(G,\R)=\Z_1$ [see \eqn{HdR}]
when $G$ is a finite group or a compact Lie group, we have
\begin{align}
 0\to  \hat \cH^d(G,U(1))\to  \hat \cH^{d+1} (G,\Z) \to 0
\end{align}
So [also using \eqn{HdR}]
\begin{align}
 \cH^d(G,U(1)) =  
 \cH^{d+1} (G,\Z)
= H^{d+1} (BG,\Z)
\end{align} 
for any finite group $G$, and for any compact Lie group $G$ (if $d\geq 1$).
This allows us to use topological-space cohomology $H^{d+1} (BG,\Z)$ to
calculate group cohomology $ \cH^d(G,U(1))$.

\begin{figure}[tb]
\begin{center}
\includegraphics[scale=0.8]{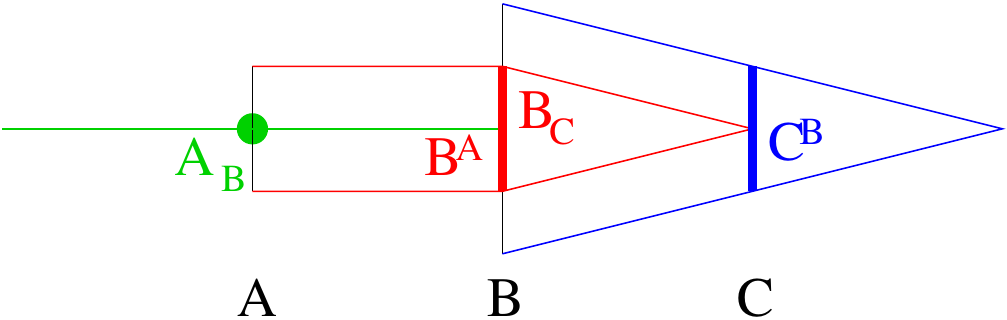}
\end{center}
%Fig.42 
\caption{
(Color online)
The graphic representation of $0\to A\to B\to C\to 0$,
where $A_B=1$, $A\sim B^A=B_C$, and
$C^B=C$.
}
\label{ABCext}
\end{figure}

We may also assume that $G$ has a non-trivial action
on the modules $\R$ and $\Z$
(which are renamed $\R_T$ and $\Z_T$):
\begin{align}
 g\cdot n & = s(g) n,\ \ \ n\in \Z_T ,
\nonumber\\
 g\cdot x & = s(g) x,\ \ \ x\in \R_T ,
\end{align}
where $s(g)=\pm 1$.  We have
\begin{align}
 0\to \Z_T \to \R_T \to \R_T/\Z_T \to 0,
\end{align}
and
\begin{align}
... \to &
\hat \cH^0(G,\Z_T)\to \hat \cH^0(G,\R_T)\to \hat \cH^0(G,U_T(1))\to
\nonumber\\
& \hat \cH^1(G,\Z_T)\to \hat \cH^1(G,\R_T)\to \hat \cH^1(G,U_T(1))\to
\nonumber\\
& \hat \cH^2(G,\Z_T)\to \hat \cH^2(G,\R_T)\to \hat \cH^2(G,U_T(1))\to ...
\end{align}
where we have used $\R_T/\Z_T=U_T(1)$.  Since $\hat \cH^d(G,\R_T)=\Z_1$ when
$G$ is a finite group or a compact Lie group, we have
\begin{align}
 0\to  \hat \cH^d(G,U_T(1))\to  \hat \cH^{d+1} (G,\Z_T) \to 0
\end{align}
So $\hat \cH^d(G,U_T(1)) =  \hat \cH^{d+1} (G,\Z_T)$ for any finite group or
compact Lie group $G$.  (On page 35 of \Ref{CS}, it is stated that $ \cH^{d}
(G,U(1))= \cH^{d+1} (G,\Z)=H^{d+1}(BG,\Z)$ for compact Lie group $G$.)

\subsubsection{Module and $G$-module}

\newcommand{\qplus}{\,\text{\bf \tt +}\,}
\newcommand{\qminus}{\,\text{\bf \tt -}\,}
\newcommand{\qtimes}{\,\text{\bf \tt *}\,}

In the next subsection, we are going to describe K\"unneth formula for group
cohomology.  As a preparation for the description, we will discuss the concepts
of module and $G$-module here.

The concept of module is a generalization of the notion of vector space.
Since two vectors can add, two elements in a module $M$,
$a,b \in M$, also support an additive $\qplus$ operation:
\begin{align}
 a \qplus b \in M.
\end{align}
The $\qplus$ operation commute and has inverse.
So $\qplus$ is an Abelian group multiplication.
The module $M$ equipped with the $\qplus$ operation
is an Abelian group.

Vector spaces also have a scaler product operation, so do modules.  The
coefficients $n$ that we can multiply to elements $a$ in a module form a
ring $R$.  We will use $\qtimes$ to describe the scaler product operation: $n
\qtimes a \in M$, $n \in R$ and $a\in M$.

A ring $R$ is a set equipped with two binary operations: addition $+$ : $R
\times R \to R$ and multiplication $\cdot$: $R \times R \to R$, where $\cdot$
may not have a inverse.  A ring becomes a field if $\cdot$ does have an
inverse, except for the additive identity ``0''.

The scaler product $\qtimes$ satisfies, for $n,m \in R$ and $a,b \in M$,
\begin{align}
 n\qtimes (a\qplus b) &= (n\qtimes a) \qplus (n\qtimes  b),
\nonumber\\
 (n + m) \qtimes a &=  (n\qtimes a) \qplus (m\qtimes  a),
\nonumber\\
 (n \cdot m) \qtimes a &=  n\qtimes (m\qtimes  a),
\\
 1_R \qtimes a &=a, \text{ if $R$ has multiplicative identity }1_R.
\nonumber
\end{align}
We will call the structure defined by $(M,R,\qplus,\qtimes)$
a module $M$ over $R$.

A  module over $R$ is $R$-free if the module has a basis (a linearly
independent generating set): there exist elements $x_1,x_2,...  \in M$, such
that for every element $a \in M$, there is a unique set $n_i\in R$ such that
$a= (n_1\qtimes x_1) \qplus (n_2\qtimes x_2) \qplus ... ...  $.

Here are some example of modules.  A ring $R$ is a module over itself.  Another
simple example is $\Z$ over $\Z$.  The module $\Z$ over $\Z$ is a free module.
The basis set contains only one element $1$.  The third example is the module
$\Z_2$ over $\Z$.  Such a module is not free, since if we choose $1$ as the
basis, the element $0\in \Z_2$ can have several expressions: $0=0 \times
1\text{ mod } 2=2\times 1 \text{ mod } 2$.  However, the module $\Z_2$ over
$\Z_2$ is a free module with basis $1$.

The module that we are going to use in this paper
is formed by pure complex phases $M=U(1)$.
It is a module over $\Z$.
The $\qplus$ and $\qtimes$ operations are defined as
\begin{align}
 a\qplus b=ab,\ \ \ \
 n\qtimes a=a^n,\ \ \ \
a,b\in U(1),\ n\in \Z.
\end{align}

Just like vector spaces, we can define direct sum of two modules, $M_1$ over $R$
and $M_2$ over $R$, which produces a third module $M_3$ over $R$.  As a space,
$M_3$ is given by $M_3=M_1\times M_2$.
The $\qplus$ and $\qtimes$ operations on
$M_3=M_1\times M_2$ are given by
\begin{align}
 (a_1,a_2)\qplus (b_1,b_2) & = (a_1\qplus b_1, a_2\qplus b_2),
\nonumber\\
 n\qtimes (a_1,a_2) &=(n\qtimes a_1,n\qtimes a_2),
\nonumber\\
 n \in R, \ \
a_1,b_1 & \in M_1,\ \
a_2,b_2 \in M_2 .
\end{align}
The resulting module is denoted as $M_1\oplus M_2$.  However, in this paper
(see Table \ref{tb}), we will use $M_1\times M_2$ to denote $M_1\oplus M_2$.
Say, for module $\Z_n$ over $\Z$, $\Z_n\times Z_m = \Z_n\oplus \Z_m$ and
$\Z_n^2=\Z_n\times Z_n = \Z_n\oplus \Z_n$, \etc.

We can also define tensor product $\otimes_R$, which maps two modules, $M_1$
over $R$ and $M_2$ over $R$, to a third module $M_3$ over $R$:
$M_3=M_1\otimes_R M_2$.  If the two modules $M_1$ and $M_2$ are free with basis
$\{x_i\}$ and $\{y_i\}$ respectively, then their tensor product
$M_3=M_1\otimes_R M_2$ is simply a module over $R$ with $\{ x_i\otimes y_j\}$
as basis.
If the modules $M_1$ and/or $M_2$ are not free, then
their tensor product is more complicated:
$M_3=M_1\otimes_R M_2$ is module over $R$ whose elements have the
form
\begin{align}
& [n_1 \qtimes (a_1\otimes_R b_1)] \qplus
 [n_2 \qtimes (a_2\otimes_R b_2)] \qplus ... ...
\nonumber\\
& n_i\in R, \ \ \ \ a_i\in M_1,\ \ \ b_i\in M_2,
\end{align}
subject to the following reduction relation
\begin{align}
& ((a_1\qplus a_2)\otimes_R b)\  \qminus (a_1\otimes_R b)\  \qminus (a_2\otimes_R b) =0,
\nonumber\\
& (a \otimes_R (b_1\qplus b_2))\  \qminus (a \otimes_R b_1)\  \qminus (a \otimes_R b_2)) =0,
\nonumber\\
&
n\qtimes (a\otimes_R b)=
((n\qtimes a)\otimes_R b)=
(a\otimes_R (n\qtimes b)) .
\end{align}
Such a definition allows us to obtain the following result:
\begin{align}
\label{tnprd}
& \Z \otimes_\Z M = M \otimes_\Z \Z =M ,
\nonumber\\
& \Z_n \otimes_\Z M = M \otimes_\Z \Z_n = M/nM ,
\nonumber\\
& \Z_m \otimes_\Z \Z_n  =\Z_{(m,n)} ,
%\nonumber\\
%& U(1) \otimes_\Z U(1)  = U(1) ,
\nonumber\\
&  (A\times B)\otimes_R M = (A \otimes_R M)\times (B \otimes_R M)   ,
\nonumber\\
& M \otimes_R (A\times B) = (M \otimes_R A)\times (M \otimes_R B)   ;
\end{align}
where $(m,n)$ is the greatest common divisor of $m$ and $n$.
In the above $\times$ really represents $\oplus$.

We can also define torsion product $\text{Tor}_1^R(, ) $, which maps two
modules, $M_1$ over $R$ and $M_2$ over $R$, to a third module $M_3$ over $R$:
$M_3=\text{Tor}_1^R(M_1,M_2)$.
We will not discuss the definition of
the  torsion product. We just list some simple results here:
\begin{align}
\label{trprd}
& \text{Tor}_1^R(A,B) \simeq \text{Tor}_1^R(B,A)  ,
\nonumber\\
& \text{Tor}_1^\Z(\Z, M) = \text{Tor}_1^\Z(M, \Z) = 0,
\nonumber\\
& \text{Tor}_1^\Z(\Z_n, M) = \{m\in M| nm=0\},
\nonumber\\
& \text{Tor}_1^\Z(\Z_m, \Z_n) = \Z_{(m,n)} ,
%\nonumber\\
%& \text{Tor}_1^\Z(U(1), U(1)) = 0 ,
\nonumber\\
& \text{Tor}_1^R(A\times B,M) = \text{Tor}_1^R(A, M)\times\text{Tor}_1^R(B, M),
\nonumber\\
& \text{Tor}_1^R(M,A\times B) = \text{Tor}_1^R(M,A)\times\text{Tor}_1^R(M,B)
,
\end{align}
Again $\times$ really represents $\oplus$.

A $G$-module over $R$ is a module over $R$ that also admits a group $G$ action:
$g\cdot a \in M$ for $a\in M$ and $g\in G$.
The group action is compatible with the
$\qplus$ and $\qtimes$ operations:
\begin{align}
 g\cdot(a\qplus b)= (g\cdot a)\qplus (g\cdot b),\ \ \ \
 g\cdot(n\qtimes a)= n\qtimes (g\cdot a).
\end{align}
In the group cohomology $\cH^d(G,M)$, $M$ is a $G$-module over $R$. In fact,
$\cH^d(G,M)$ is also a $G$-module over $R$.

\subsubsection{K\"unneth formula for group cohomology}

Now, we are ready to describe
the K\"unneth formula for group cohomology.
Let $M$ (resp. $M'$) be an arbitrary $G$-module (resp. $G'$-module)
over a principal ideal domain $R$.
We also assume that
either $M$ or $M'$ is $R$-free. Then we have a
K\"unneth formula for group cohomology\cite{G0691,Kunn}
\begin{align}
\label{kunnES}
0 & \rightarrow
\prod_{p=0}^d \cH^p(G,M)\otimes_R \cH^{d-p}(G',M')
\nonumber\\
& \rightarrow
\cH^d(G\times G',M\otimes_R M')
\nonumber\\
& \rightarrow
\prod_{p=0}^{d+1}
\text{Tor}_1^R(\cH^p(G,M),\cH^{d-p+1}(G',M'))
\rightarrow 0
\end{align}
If both  $M$ and $M'$ are $R$-free, then the sequence splits and we have
\begin{align}
\label{kunn}
&\ \ \ \ \cH^d(G\times G',M\otimes_R M')
\nonumber\\
&=\Big[\prod_{p=0}^d \cH^p(G,M)\otimes_R \cH^{d-p}(G',M')\Big]\times
\nonumber\\
&\ \ \ \ \ \
\Big[\prod_{p=0}^{d+1}
\text{Tor}_1^R(\cH^p(G,M),\cH^{d-p+1}(G',M'))\Big]  .
\end{align}
If $R$ is a field $K$, we have
\begin{align}
&\ \ \ \ \cH^d(G\times G',M\otimes_K M')
\nonumber\\
&=\Big[\prod_{p=0}^d \cH^p(G,M)\otimes_K \cH^{d-p}(G',M')\Big] .
\end{align}

For the cases studied in this paper, we have $R=M=\Z$ (\ie $G$ acts trivially on
$\Z$) and $M'=\Z_T$ (\ie $G'$ may act non-trivially on $\Z_T$).  So
$M\otimes_\Z M' = \Z_T$, on which $G$ acts trivially and $G'$ may act
non-trivially. Also the sequence splits.

\subsubsection{Cup product for group cohomology}

Consider two cochains $\nu_{n_1}\in \cC^{n_1}(G,M_1)$ and $\nu_{n_2} \in
\cC^{n_2}(G,M_2)$. From $\nu_{n_1}$ and $\nu_{n_2}$, we can construct a third cochain
$\nu_{n_1+n_2} \in \cC^{n_1+n_2}(G,M_1\otimes_\Z M_2)$:
\begin{align}
&\ \ \ \
 \nu_{n_1+n_2}(g_0,g_1,...,g_{n_1+n_2})
\\
&=
 \nu_{n_1}(g_0,g_1,...,g_{n_1})\otimes_\Z
 \nu_{n_2}(g_{n_1},g_{n_1+1},...,g_{n_1+n_2})
.
\nonumber
\end{align}
The above mapping $ \cC^{n_1}(G,M_1)\times  \cC^{n_2}(G,M_2)
\to  \cC^{n_1+n_2}(G,M_1\otimes_\Z M_2)$ is called the cup product
and is denoted as
\begin{align}
 \nu_{n_1+n_2}=\nu_{n_1}\cup\nu_{n_2}.
\end{align}

The cup product has the following nice property
\begin{align}
 d \nu_{n_1+n_2}= [(d \nu_{n_1})\cup\nu_{n_2}] \qplus [ (-)^{n_1}
\qtimes \nu_{n_1}\cup(d \nu_{n_2}) ] ,
\end{align}
where
\begin{align}
&\ \ \ \
 (d \nu_{n})(g_0,...,g_{n+1})
\nonumber\\
&=
\qplus_{i=0}^{n+1} [(-)^i\qtimes \nu_{n}(g_0,...,\hat g_i,...,g_{n+1})].
\end{align}
where the sequence $g_0,...,\hat g_i,...,g_{n+1}$ is the sequence
$g_0,...,g_{n+1}$ with $g_i$ removed.
So if $\nu_{n_1}$ and $\nu_{n_2}$ are cocycles
$\nu_{n_1}\cup\nu_{n_2}$ is also a cocycle.
Thus the cup product defines the mapping
\begin{align}
  \cH^{n_1}(G,M_1)\times  \cH^{n_2}(G,M_2)
\to  \cH^{n_1+n_2}(G,M_1\otimes_\Z M_2) .
\end{align}
As a map between classes of cocycles, the
 cup product satisfies\cite{RS}
\begin{align}
& \nu_{n_1}\cup\nu_{n_2}=(-)^{n_1n_2}\qtimes (\nu_{n_2}\cup\nu_{n_1}),
\nonumber\\
& (\nu_{n_1}\cup\nu_{n_2})\cup\nu_{n_3}
=\nu_{n_1}\cup(\nu_{n_2}\cup\nu_{n_3}).
\end{align}

Let
\begin{align}
 H^*(G,M)=
H^0(G,M)\oplus
H^1(G,M)\oplus...
\end{align}
$H^*(G,M)$ has an additive operation $\qplus$ inherited from the module $M$.
The cup product provided an multiplicative operation on $ H^*(G,M)$.  So
$H^*(G,M)$ with the  additive operation $\qplus$ and the  multiplicative
operation $\cup$ is a ring which is called the group cohomology ring for the
group $G$.

\subsection{Group cohomology of $Z_m\times Z_n$}
\label{ZmZn}

Let us first calculate $\cH^d[Z_m\times Z_n, \Z]$.
Using \eqn{kunn}, we find that
\begin{align}
& \cH^d(\Z_m \times Z_n; \Z) \cong
\left(\prod_{i= 0}^d \cH^i(Z_m;\Z) \otimes_\Z \cH^{d-i}(Z_n;\Z)\right)
\nonumber\\
& \times
\left(\prod_{p =0}^{n+1} \text{Tor}_1^\Z(\cH^p(Z_m;\Z),\cH^{d+1-p}(Z_n;\Z))\right) .
\end{align}
The above can be calculated using \eqn{ZnZ} and the simple properties of the
tensor product $\otimes_\Z$ and torsion product $\text{Tor}_1^\Z$ in
\eqn{tnprd} and \eqn{trprd}.
For example
\begin{align}
 &\ \ \ \
 \cH^0(\Z_m \times Z_n; \Z)
\nonumber\\
& \cong
\left(\Z \otimes_\Z \Z\right)
\times \text{Tor}_1^\Z(\Z_1,\Z)
\times \text{Tor}_1^\Z(\Z,\Z_1)
\nonumber\\
& = \Z
.
\end{align}
\begin{align}
 &\ \ \ \
 \cH^1(\Z_m \times Z_n; \Z)
\nonumber\\
& \cong
\left(\Z_1 \otimes_\Z \Z\right)
\times \left(\Z \otimes_\Z \Z_1\right)
\times \text{Tor}_1^\Z(\Z,\Z_n)
\nonumber\\
&\ \ \ \ \ \ \ \
\times \text{Tor}_1^\Z(\Z_1,\Z_1)
\times \text{Tor}_1^\Z(\Z_m,\Z)
\nonumber\\
& =\Z_1
.
\end{align}
\begin{align}
 &\ \ \ \
 \cH^2(\Z_m \times Z_n; \Z)
\nonumber\\
& \cong
\left(\Z_m \otimes_\Z \Z\right)
\times \left(\Z \otimes_\Z \Z_n\right)
\times \text{Tor}_1^\Z(\Z,\Z_1)
\times \text{Tor}_1^\Z(\Z_1,\Z_n)
\nonumber\\
&\ \ \ \ \ \ \ \
\times \text{Tor}_1^\Z(\Z_m,\Z_1)
\times \text{Tor}_1^\Z(\Z_1,\Z)
\nonumber\\
& =\Z_m\times \Z_n
.
\end{align}

%
%\begin{align}
% &\ \ \ \
% \cH^3(\Z_m \times Z_n; \Z)
%\nonumber\\
%& \cong
% \text{Tor}_1^\Z(\Z,\Z_n)
%\times \text{Tor}_1^\Z(\Z_m,\Z_n)
%\times \text{Tor}_1^\Z(\Z_n,\Z)
%\nonumber\\
%& =
%\Z_{(m,n)}
%.
%\end{align}
%%
%\begin{align}
% &\ \ \ \
% \cH^4(\Z_m \times Z_n; \Z)
%\nonumber\\
%& \cong
%\left(\Z \otimes_\Z \Z_n\right)
%\times \left(\Z_m \otimes_\Z \Z_n\right)
%\times \left(\Z_m \otimes_\Z \Z\right)
%\nonumber\\
%& =
%\Z_{(m,n)}
%\times \Z_m
%\times \Z_n
%.
%\end{align}
%
%\begin{align}
% &\ \ \ \
% \cH^5(\Z_m \times Z_n; \Z)
%\nonumber\\
%& \cong
%\text{Tor}_1^\Z(\Z_m,\Z_n)
%\times \text{Tor}_1^\Z(\Z_m,\Z_n)
%\nonumber\\
%& =
%\Z_{(m,n)}
%\times \Z_{(m,n)}
%.
%\end{align}
%

Using the relation
$ \cH^d(\Z_m \times Z_n; U(1))= \cH^{d+1}(\Z_m \times Z_n; \Z)$,
$d>0$, we find
\begin{align}
 \cH^d[Z_m\times Z_n,U(1)]=
\begin{cases}
U(1) & d=0, \\
\Z^{d/2}_{(m,n)}  & d= \text{ even} \\
\Z_m\times \Z_n \times \Z^{(d-1)/2}_{(m,n)}  & d= \text{ odd} .
\end{cases}
\end{align}
This agrees with \eqn{Z2Zp}.

\subsection{Group cohomology of $Z_2^T\times Z_n$}

Let us first calculate $\cH^d[Z_2^T\times Z_n, \Z_T]$
where $Z_2^T$ acts on $\Z_T$ non-trivially (see \eqn{Z2TZT}).
Using \eqn{kunn}, we find that
\begin{align}
& \cH^d(\Z_2^T \times Z_n; \Z_T) \cong
\left(\prod_{i= 0}^d \cH^i(Z_2^T;\Z_T) \otimes_\Z \cH^{d-i}(Z_n;\Z)\right)
\nonumber\\
& \times
\left(\prod_{p =0}^{n+1} \text{Tor}_1^\Z(\cH^p(Z_2^T;\Z_T),\cH^{d+1-p}(Z_n;\Z))\right) .
\end{align}
%Thus
%\begin{align}
% \cH^0(\Z_2^T \times Z_n; \Z_T)
% \cong
%\left(\Z_1 \otimes_\Z \Z\right)
% = \Z_1
%.
%\end{align}
%%
%\begin{align}
% \cH^1(\Z_2^T \times Z_n; \Z_T)
% \cong
%\left(\Z_2 \otimes_\Z \Z\right)
% =\Z_2
%.
%\end{align}
%%
%\begin{align}
% \cH^2(\Z_2^T \times Z_n; \Z_T)
% \cong
%\text{Tor}_1^\Z(\Z_2,\Z_n)
% =\Z_{(2,n)}
%.
%\end{align}
%%
%\begin{align}
%&\ \ \ \
% \cH^3(\Z_2^T \times Z_n; \Z_T)
% \cong
%\left(\Z_2 \otimes_\Z \Z_n\right)
%\left(\Z_2 \otimes_\Z \Z\right)
%\nonumber\\
%& =
%\Z_{(2,n)} \times \Z_2
%.
%\end{align}
%%
%\begin{align}
% &\ \ \ \
% \cH^4(\Z_2^T \times Z_n; \Z_T)
% \cong
%\text{Tor}_1^\Z(\Z_2,\Z_n)
%\times \text{Tor}_1^\Z(\Z_2,\Z_n)
%\nonumber\\
%& =
%\Z^2_{(2,n)}
%.
%\end{align}

Using the relation
$ \cH^d(\Z_2^T \times Z_n; U_T(1))= \cH^{d+1}(\Z_2^T \times Z_n; \Z_T)$,
$d>0$, we find
\begin{align}
\label{Z2TZn}
 \cH^d[Z_2^T\times Z_n,U_T(1)]=
\begin{cases}
\Z_2 \times \Z^{d/2}_{(2,n)}  & d= \text{ even} \\
\Z^{(d+1)/2}_{(2,n)}  & d= \text{ odd}
.
\end{cases}
\end{align}
When $n$ is odd, this agrees with \eqn{Z2TZp}.

\subsection{Group cohomology of $D_{2h}$}

The group $D_{2h}$ is the same as $Z_2\times Z_2 \times Z_2^T$. So we can use
the K\"unneth formula and the results in subsection \ref{ZmZn} to calculate
$\cH^d[D_{2h},U_T(1)]$.
Let us first calculate $\cH^d[Z_m\times Z_n \times Z_2^T,\Z_T]$:
\begin{align}
&\ \ \ \
\cH^0[Z_m\times Z_n \times Z_2^T,\Z_T]
\cong \Z_1 .
\end{align}
\begin{align}
&\ \ \ \
\cH^1[Z_m\times Z_n \times Z_2^T,\Z_T]
 \cong
(\Z \otimes_\Z \Z_2)
=\Z_2 .
\end{align}
\begin{align}
&\ \ \ \
\cH^2[Z_m\times Z_n \times Z_2^T,\Z_T]
\nonumber\\ &
 \cong
 \text{Tor}_1^\Z(\Z_m\times \Z_n,\Z_2)
\nonumber\\ &
=
 \text{Tor}_1^\Z(\Z_m,\Z_2) \times
 \text{Tor}_1^\Z(\Z_n,\Z_2)
\nonumber\\ &
=
\Z_{(2,m)} \times \Z_{(2,n)}
 .
\end{align}
\begin{align}
&\ \ \ \
\cH^3[Z_m\times Z_n \times Z_2^T,\Z_T]
\nonumber\\ &
 \cong
(\Z \otimes_\Z \Z_2)
\times [ (\Z_m\times \Z_n) \otimes_\Z \Z_2]
\times \text{Tor}_1^\Z(\Z_{(m,n)},\Z_2)
\nonumber\\ &
=\Z_2\times
\Z_{(2,m)} \times \Z_{(2,n)}
\times \Z_{(m,n)}
 .
\end{align}
\begin{align}
&\ \ \ \
\cH^4[Z_m\times Z_n \times Z_2^T,\Z_T]
\nonumber\\ &
 \cong
(\Z_{(m,n)} \otimes_\Z \Z_2)
\times \text{Tor}_1^\Z(\Z_m\times \Z_n,\Z_2)
\nonumber\\
&\ \ \ \ \ \ \ \ \
\times \text{Tor}_1^\Z(\Z_m\times \Z_n\times \Z_{(m,n)},\Z_2)
\nonumber\\ &
=
\Z^2_{(2,m,n)}
\times \Z^2_{(2,m)}
\times \Z^2_{(2,n)}
 .
\end{align}
\begin{align}
&\ \ \ \
\cH^5[Z_m\times Z_n \times Z_2^T,\Z_T]
\nonumber\\ &
 \cong
(\Z \otimes_\Z \Z_2)
\times [ (\Z_m\times \Z_n) \otimes_\Z \Z_2]
\nonumber\\
&\ \ \ \ \ \ \ \ \
\times [ (\Z_m\times \Z_n\times \Z_{(m,n)}) \otimes_\Z \Z_2]
\nonumber\\
&\ \ \ \ \ \ \ \ \
\times \text{Tor}_1^\Z(\Z_{(m,n)},\Z_2)
\times \text{Tor}_1^\Z(\Z^2_{(m,n)},\Z_2)
\nonumber\\ &
=\Z_2\times
\Z^4_{(2,m,n)}
\times \Z^2_{(2,m)}
\times \Z^2_{(2,n)}
 .
\end{align}
This gives us
\begin{align}
&\ \ \ \
\cH^1[Z_m\times Z_n \times Z_2^T,U_T(1)]
\nonumber\\ &
 \cong
\Z_{(2,m)} \times \Z_{(2,n)}
 .
\end{align}
\begin{align}
&\ \ \ \
\cH^2[Z_m\times Z_n \times Z_2^T,U_T(1)]
\nonumber\\ &
 \cong
\Z_2\times
\Z_{(2,m)} \times \Z_{(2,n)}
\times \Z_{(m,n)}
 .
\end{align}
\begin{align}
&\ \ \ \
\cH^3[Z_m\times Z_n \times Z_2^T,U_T(1)]
\nonumber\\ &
 \cong
\Z^2_{(2,m,n)}
\times \Z^2_{(2,m)}
\times \Z^2_{(2,n)}
 .
\end{align}
\begin{align}
&\ \ \ \
\cH^4[Z_m\times Z_n \times Z_2^T,U_T(1)]
\nonumber\\ &
 \cong
\Z_2\times
\Z^4_{(2,m,n)}
\times \Z^2_{(2,m)}
\times \Z^2_{(2,n)}
 .
\end{align}
The results
$
\cH^1[Z_2\times Z_2 \times Z_2^T,U_T(1)]
=\cH^1[D_{2h},U_T(1)]=\Z_2^2
$
and
$
\cH^2[Z_2\times Z_2 \times Z_2^T,U_T(1)]
=\cH^2[D_{2h},U_T(1)]=\Z_2^4
$
agrees with those in \Ref{CGW1123,LCW1121}
obtained through direct calculations.

\subsection{Group cohomology of $U(1)$}

To calculate $\cH^d[U(1),U(1)]$ (the Borel cohomology) directly from the
algebraic definition is very tricky since $U(1)$ has infinite uncountable many
elements.
Here, we will use a physical argument to calculate it by first calculating
$\cH^d[ Z_n,U(1)]$, and then let $n\to \infty$.  This way, we find
\begin{align}
\label{U1U1}
 \cH^d[U(1),U(1)]=
\begin{cases}
U(1) & \text{ if } d=0, \\
\Z_1  & \text{ if } d=0 \text{ mod } 2,\ \ d>0\\
\Z  & \text{ if } d=1 \text{ mod } 2.
\end{cases}
\end{align}
In \Ref{PM}, it is stated that
\begin{align}
 \cH^d[U(1),\Z]=
\begin{cases}
\Z  & \text{ if } d=0 \text{ mod } 2,\\
\Z_1  & \text{ if } d=1 \text{ mod } 2.
\end{cases}
\end{align}
This is consistent with \eqn{U1U1} since $\cH^d[U(1),U(1)]=\cH^{d+1}[U(1),\Z]$.

We note that the 1D representations of $U(1)$,
$M(U_\th) = \e^{n \imth \th}$, where the $U(1)$ group elements are
denoted as $U_\th$, are labeled by $n \in \Z$.
Also,  $U(1)$ has no non-trivial projective
representation. This is consistent with the above results: $\cH^1[U(1),U_T(1)]=
\Z $ and $\cH^2[U(1),U_T(1)]= \Z_1$.

\subsection{Group cohomology of $Z_2^T \times U(1)$}
As pointed out in the section \ref{1Dproj}, a spin system with time reversal
and $U(1)$ (say generated by $S_z$) symmetries has a symmetry group $U(1)\times
Z_2^T$.  To find the SPT phases of such a bosonic system, we need to calculate
$\cH^d[ U(1)\times Z_2^T,U_T(1)]$.
We simply need to repeat the calculation of $\cH^d[ Z_2^T\times Z_n,\Z_T]$ by
replacing $\Z_n$ with $\Z$.  We note that $\Z_2 \otimes_\Z \Z_n = \Z_{(2,n)}$
and $\text{Tor}_1^\Z (\Z_2 , \Z_n) = \Z_{(2,n)}$, while $\Z_2 \otimes_\Z \Z =
\Z_2$ and $\text{Tor}_1^\Z (\Z_2 , \Z) = \Z_1$.  So we find
\begin{align}
\label{Z2TU1Z}
 \cH^d[Z_2^T\times U(1),\Z_T]=
\begin{cases}
\Z_1 & d=0, \\
\Z_1 & d= \text{ even} \\
\Z^{\frac{d+1}{2}}_2 & d= \text{ odd}
.
\end{cases}
\end{align}
Since $\cH^d[Z_2^T\times U(1),\R_T]=\Z_1$, we have $\cH^d[Z_2^T\times
U(1),U_T(1)]=\cH^{d+1}[Z_2^T\times U(1),\Z_T]$, and

\begin{align}
\label{Z2TU1}
 \cH^d[ Z_2^T\times U(1),U_T(1)]=
\begin{cases}
%\Z_2 &  d=0, \\
\Z^{\frac{d+2}{2}}_2  &  d= \text{ even},\\
\Z_1  &  d= \text{ odd}.
\end{cases}
\end{align}
This can be obtained from \eqn{Z2TZn} by taking $n\to \infty$ and choosing
$\lim_{n\to \infty} \Z_{(2,n)}=\Z_2$ for $d=$ even, and $\lim_{n\to
\infty} \Z_{(2,n)}=\Z_1$ for $d=$ odd.

\subsection{Group cohomology of $U(1) \rtimes Z_2^T$}

As pointed out in the section \ref{1Dproj}, a bosonic system with time reversal
symmetry and boson number conservation has a symmetry group $U(1)\rtimes
Z_2^T$.  To find the SPT phases of such a bosonic system, we need to calculate
$\cH^d[ U(1)\rtimes Z_2^T,U_T(1)]$.  In the last subsection, we calculated
$\cH^d[ U(1)\times Z_2^T,U_T(1)]$.  In those two cases, $Z_2^T$ has a
non-trivial action on the module $M=U_T(1)$.

The other two related cohomology groups are $\cH^d[ U(1)\times Z_2,U(1)]$ and
$\cH^d[ U(1)\rtimes Z_2,U(1)]$, where the $Z_2$ is a usual unitary symmetry.
In those two cases, $Z_2$ has a trivial action on the module $M=U(1)$.

In this section, we will use spectral sequence method to calculate the above
four cohomology groups from the following facts:
\begin{align}
\label{ZnU1fact}
 \cH^{2p}[U(1),\Z] &= \Z,  &
 \cH^{2p+1}[U(1),\Z] &=0,
\nonumber\\
 \cH^{2p}[\Z_n,\Z] &= \Z_n, &
 \cH^{2p+1}[\Z_2,\Z] &= 0,
\nonumber\\
 \cH^{2p}[\Z_2^T,\Z_T] &= 0,  &
 \cH^{2p+1}[\Z_2^T,\Z_T] &= \Z_2.
\end{align}
Let $G$ be one of the $U(1)\times Z_2$, $U(1)\rtimes
Z_2$, $U(1)\times Z_2^T$, and $U(1)\rtimes Z_2^T$.
From the exact sequence of the groups
$ 1\to U(1) \to G \to Z_2 \to 1$,
we have the following
$E^{p,q}_2=\cH^p[Z_2,\cH^q[U(1),M]]$ page:
\begin{widetext}
\begin{align}
\bmm
\cH^0[Z_2,\cH^3[U(1),M]] & \cH^1[Z_2,\cH^3[U(1),M]] & \cH^2[Z_2,\cH^3[U(1),M]] & \cH^3[Z_2,\cH^3[U(1),M]] \\
\cH^0[Z_2,\cH^2[U(1),M]] & \cH^1[Z_2,\cH^2[U(1),M]] & \cH^2[Z_2,\cH^2[U(1),M]] & \cH^3[Z_2,\cH^2[U(1),M]] \\
\cH^0[Z_2,\cH^1[U(1),M]] & \cH^1[Z_2,\cH^1[U(1),M]] & \cH^2[Z_2,\cH^1[U(1),M]] & \cH^3[Z_2,\cH^1[U(1),M]] \\
\cH^0[Z_2,\cH^0[U(1),M]] & \cH^1[Z_2,\cH^0[U(1),M]] & \cH^2[Z_2,\cH^0[U(1),M]] & \cH^3[Z_2,\cH^0[U(1),M]] \\
\emm
\end{align}
\end{widetext}
To calculate $\cH^p[Z_2,\cH^q[U(1),M]]$,
we need to know how $Z_2$ acts on $\cH^q[U(1),M]$ through
how $Z_2$ acts on $U(1)$ group and $M$ module.

First we consider $U(1) \times Z_2$ group and module $M=\Z$.  In this case,
$Z_2$ acts on $U(1)$ group trivially and it acts on $M$ trivially.  As a
result, $Z_2$ acts on $\cH^q[U(1),\Z]$ trivially: $T\cdot \al\to \al$, $\al\in
\cH^q[U(1),\Z_T]$.  Note that $ \cH^d[U(1),\Z]= \Z$, $ \cH^d[\Z_2,\Z]= \Z_2$
for $d=$ even and $ \cH^d[U(1),\Z_T]= \Z_1=0$, $ \cH^d[\Z_2,\Z]= \Z_1=0$ for
$d=$ odd.  We obtain the following $E^{p,q}_2=\cH^p[Z_2,\cH^q[U(1),\Z]]$ page
in the spectral sequence:
\begin{align}
\bmm
0  & 0 &  0   &   0   &   0   &  0 &  0   &   0   \\
\Z & 0 & \Z_2 &   0   & \Z_2  &  0 & \Z_2 &   0   \\
0  & 0 &  0   &   0   &   0   &  0 &  0   &   0   \\
\Z & 0 & \Z_2 &   0   & \Z_2  &  0 & \Z_2 &   0   \\
0  & 0 &  0   &   0   &   0   &  0 &  0   &   0   \\
\Z & 0 & \Z_2 &   0   & \Z_2  &  0 & \Z_2 &   0   \\
0  & 0 &  0   &   0   &   0   &  0 &  0   &   0   \\
\Z & 0 & \Z_2 &   0   & \Z_2  &  0 & \Z_2 &   0   \\
\emm
\end{align}
In the $E_2^{p,q}$ page, we have spectral sequence
\begin{align}
 ... \rightarrow E_2^{p,q} \rightarrow   E_2^{p+2,q-1} \rightarrow ...
\end{align}
generated by
$ d_2^{p,q}: E_2^{p,q} \to E_2^{p+2,q-1} $
such that
$ d_2^{p+2,q-1} d_2^{p,q}=0 $.
So the cohomology of
$d_2^{p,q}$ produces the $E_3^{p,q}$ page:
$ E_3^{p,q}=\text{ker}  d_2^{p,q}/\text{im} d_2^{p-2,q+1} $.
We note that all the non trivial terms in
the $E_2^{p,q}$ page are connected to an incoming zero
and an outing zero.
Thus $d_2^{p,q}=0$, which leads to
$\text{ker}  d_2^{p,q}=E_2^{p,q}$ and
$\text{im}  d_2^{p,q}=0$. So
$E_3^{p,q}=E_2^{p,q}$.
In the $E_3^{p,q}$ page, we have a spectral sequence
\begin{align}
 ... \rightarrow E_3^{p,q} \rightarrow   E_3^{p+3,q-2} \rightarrow ...
\end{align}
and again  all the non trivial terms in the $E_3^{p,q}$ page are connected to
an incoming zero and an outing zero and $d_3^{p,q}=0$.
So
$E_4^{p,q}=E_3^{p,q}(=E_2^{p,q})$. This way, we can show that
the $E_2^{p,q}$ page
stabilizes:  $E_\infty^{p,q}=E_2^{p,q}$.
Using $E_\infty^{p,q}$ with $p+q=n$, we can calculate
$\cH^{n}[U(1)\times Z_2, \Z]$, since
$\cH^{n}[U(1)\times Z_2, \Z]$ has
a filtration
\begin{align}
 0=H^n_{n+1}
\subseteq H^n_n
...
\subseteq H^n_{1}
\subseteq H^n_{0}=\cH^n[U(1)\times Z_2, \Z] ,
\end{align}
such that
$ H^n_p/H^n_{p+1}=E_\infty^{p,n-p} $.
Thus we have
\begin{align}
\label{Z2U1Z}
 \cH^d[ U(1)\times Z_2,\Z] =
\begin{cases}
\Z\times \Z^{d/2}_2,  &  d=0 \text{ mod } 2,\\
\Z_1,  &  d=1 \text{ mod } 2 ,
\end{cases}
\end{align}
which gives
\begin{align}
\label{Z2U1}
 \cH^d[ U(1)\times Z_2,U(1)] =
\begin{cases}
U(1), &  d=0, \\
\Z_1,  &  d=0 \text{ mod } 2, \\
\Z\times \Z^{\frac{d+1}{2}}_2,  &  d=1 \text{ mod } 2.
\end{cases}
\end{align}

Next we consider $U(1) \times Z_2^T$ group and module $M=\Z_T$.  In this case,
$Z_2^T$ acts on $U(1)$ group trivially and it acts on $M$ non-trivially: $T\cdot
n\to -n$, $n\in \Z_T$.  As a result, $Z_2$ acts on $\cH^q[U(1),\Z_T]$
non-trivially: $T\cdot \al\to -\al$, $\al\in \cH^q[U(1),\Z_T]$.  We obtain
the following $E^{p,q}_2=\cH^p[Z_2,\cH^q[U(1),\Z_T]]$
page in the spectral sequence:
\begin{align}
\bmm
0 &  0   &   0   &   0   &  0 &  0   &   0   &   0  \\
0 & \Z_2 &   0   & \Z_2  &  0 & \Z_2 &   0   & \Z_2 \\
0 &  0   &   0   &   0   &  0 &  0   &   0   &   0  \\
0 & \Z_2 &   0   & \Z_2  &  0 & \Z_2 &   0   & \Z_2 \\
0 &  0   &   0   &   0   &  0 &  0   &   0   &   0  \\
0 & \Z_2 &   0   & \Z_2  &  0 & \Z_2 &   0   & \Z_2 \\
0 &  0   &   0   &   0   &  0 &  0   &   0   &   0  \\
0 & \Z_2 &   0   & \Z_2  &  0 & \Z_2 &   0   & \Z_2 \\
\emm
\end{align}
Again, the $E_2^{p,q}$ page stabilizes:  $E_\infty^{p,q}=E_2^{p,q}$.  We have
\begin{align}
\label{Z2TU1Za}
 \cH^d[ U(1)\times Z_2^T,\Z_T] =
\begin{cases}
\Z_1,  &  d=0 \text{ mod } 2,\\
\Z_2^{\frac{d+1}{2}},  &  d=1 \text{ mod } 2 ,
\end{cases}
\end{align}
which gives
\begin{align}
\label{Z2TU1a}
 \cH^d[ U(1)\times Z_2^T,U_T(1)] =
\begin{cases}
\Z_2^{\frac{d+2}{2}},  &  d=0 \text{ mod } 2, \\
\Z_1,  &  d=1 \text{ mod } 2.
\end{cases}
\end{align}
This agrees with \eqn{Z2TU1}.

Third, we consider $U(1) \rtimes Z_2$ group and module $M=\Z$.  In this case,
$Z_2$ acts on $U(1)$ group non-trivially
$TU_\th T=U_{-\th}$
and it acts on $M$ trivially.
As a result, $Z_2$ acts on $\cH^q[U(1),\Z]$
non-trivially: $T\cdot \al\to (-)^{q/2} \al$, $\al\in \cH^q[U(1),\Z]$.
To obtain the above result, we note
$T$ act on the $q$-cocycles in $\cH^q[U(1),\Z]$ in the following way:
\begin{align}
 T: \nu_q(g_0,...,g_q) \to \nu_q(Tg_0T^{-1},...,Tg_qT^{-1}),\ \ \
g_i \in U(1).
\end{align}
Through some explicit calculations, we find that a $T$ transformed 2-cocycle, $
\nu_2(Tg_0T^{-1},Tg_1T^{-1},Tg_2T^{-1}) $, is same as $ -\nu_2(g_0,g_1,g_2) $ up to a
2-coboundary (see \eqn{TalH2U1}).  Thus  $T\cdot \al_2\to - \al_2$ for $\al_2\in
\cH^2[U(1),\Z]$.  The generator $\al_{2p} \in \cH^{2p}[U(1),\Z]$ can be
obtained from the generator $\al_{2} \in \cH^{2}[U(1),\Z]$ by taking the cup
product\cite{RS}
$ \al_{2p}=\al_2\cup \al_2 \cup ...\cup \al_2 $.
For example, the cup product of two 2-cocycles, $\al_2$ and $\al_2$,
gives rise to a 4-cocycle $\al_4$:
$
%\begin{align}
%&\ \ \ \
 \al_4(g_0,g_1,...,g_4)=(\al_2\cup \al_2)(g_0,g_1,...,g_4)
%\nonumber\\
%&
= \al_2(g_0,g_1,g_2) \al_2(g_2,g_3,g_4)
%\end{align}
$.
Therefore $T\cdot \al_{2p} = (-)^p \al_{2p}$.

We obtain the following $E^{p,q}_2=\cH^p[Z_2,\cH^q[U(1),\Z]]$ page in the
spectral sequence:
\begin{align}
\bmm
0  &  0   &   0   &   0   &  0 &  0   &   0   &   0  \\
0  & \Z_2  &  0 & \Z_2 &   0   & \Z_2  &   0  & \Z_2 \\
0  &  0   &   0   &   0   &  0 &  0   &   0   &   0  \\
\Z &  0   & \Z_2  &  0 & \Z_2 &   0   & \Z_2  &   0  \\
0  &  0   &   0   &   0   &  0 &  0   &   0   &   0  \\
0  & \Z_2  &  0 & \Z_2 &   0   & \Z_2  &   0  & \Z_2 \\
0  &  0   &   0   &   0   &  0 &  0   &   0   &   0  \\
\Z &  0   & \Z_2  &  0 & \Z_2 &   0   & \Z_2  &   0  \\
\emm
\end{align}
But now, we can no longer show that the $E_2^{p,q}$ page stabilizes.

So, we can only obtain a weaker result
\begin{align}
\label{Z2rU1Z}
 \cH^d[ U(1)\rtimes Z_2,\Z] \leq
\begin{cases}
\Z\times \Z_2^{\frac d 4},  &  d=0 \text{ mod } 4,\\
\Z_2^{\frac{d-1}{4}},  &  d=1 \text{ mod } 4 ,\\
\Z_2^{\frac{d+2}{4}},  &  d=2 \text{ mod } 4 ,\\
\Z_2^{\frac{d+1}{4}},  &  d=3 \text{ mod } 4 .\\
\end{cases}
\end{align}
(In fact, we can show $E_\infty^{p,q}=E_2^{p,q}$ when $p+q \leq 2$.
So, $\leq$ becomes $=$ for $d=0,1,2$.)
The above gives
\begin{align}
\label{Z2rU1}
 \cH^d[ U(1)\rtimes Z_2,U(1)] \leq
\begin{cases}
U(1),  &  d=0 ,\\
\Z_2^{\frac{d}{4}},  &  d=0 \text{ mod } 4 ,\\
\Z_2^{\frac{d+3}{4}},  &  d=1 \text{ mod } 4 ,\\
\Z_2^{\frac{d+2}{4}},  &  d=2 \text{ mod } 4 ,\\
\Z\times \Z_2^{\frac {d+1} 4},  &  d=3 \text{ mod } 4,\\
\end{cases}
\end{align}

Last we consider $U(1)\rtimes Z_2^T$ group and module $M=\Z_T$.  In this case,
$Z_2^T$ acts on $U(1)$ group non-trivially $TU_\th T=U_{-\th}$ and it acts on
$M$ non-trivially: $T\cdot n\to -n$, $n\in \Z_T$.  As a result, $Z_2^T$ acts on
$\cH^q[U(1),\Z_T]$ non-trivially: $T\cdot \al\to -(-)^{q/2} \al$, $\al\in
\cH^q[U(1),\Z_T]$.  We obtain the following
$E^{p,q}_2=\cH^p[Z_2,\cH^q[U(1),\Z_T]]$ page in the spectral sequence:
\begin{align}
\bmm
0  &  0   &   0   &   0   &  0 &  0   &   0   &   0  \\
\Z &  0   & \Z_2  &  0 & \Z_2 &   0   & \Z_2  &   0  \\
0  &  0   &   0   &   0   &  0 &  0   &   0   &   0  \\
0  & \Z_2  &  0 & \Z_2 &   0   & \Z_2  &   0  & \Z_2 \\
0  &  0   &   0   &   0   &  0 &  0   &   0   &   0  \\
\Z &  0   & \Z_2  &  0 & \Z_2 &   0   & \Z_2  &   0  \\
0  &  0   &   0   &   0   &  0 &  0   &   0   &   0  \\
0  & \Z_2  &  0 & \Z_2 &   0   & \Z_2  &   0  & \Z_2 \\
\emm
\end{align}
Again, the above $E_2$ page may not stabilize.

So we have
\begin{align}
\label{Z2TrU1Z}
 \cH^d[ U(1)\rtimes Z_2^T,\Z_T] \leq
\begin{cases}
\Z_2^{\frac d 4},  &  d=0 \text{ mod } 4,\\
\Z_2^{\frac{d+3}{4}},  &  d=1 \text{ mod } 4 ,\\
\Z\times\Z_2^{\frac{d-2}{4}},  &  d=2 \text{ mod } 4 ,\\
\Z_2^{\frac{d+1}{4}},  &  d=3 \text{ mod } 4 ,\\
\end{cases}
\end{align}
which gives
\begin{align}
\label{Z2TrU1}
 \cH^d[ U(1)\rtimes Z_2^T,U_T(1)] \leq
\begin{cases}
\Z_2^{\frac{d+4}{4}},  &  d=0 \text{ mod } 4 ,\\
\Z\times \Z_2^{\frac{d-1}{4}},  &  d=1 \text{ mod } 4 ,\\
\Z_2^{\frac{d+2}{4}},  &  d=2 \text{ mod } 4 ,\\
\Z_2^{\frac {d+1} 4},  &  d=3 \text{ mod } 4.\\
\end{cases}
\end{align}

We note that $\cH^1[G,U_T(1)]$ classifies the 1D representation and
$\cH^2[G,U_T(1)]$ classifies the projective representation of $G$.  The 1D
representation and the projective representation for groups $U(1)\times Z_2$,
$U(1)\rtimes Z_2$, $U(1)\times Z_2^T$, and $U(1)\rtimes Z_2^T$ are discussed in
section \ref{1Dproj}.  They agree with $\cH^1[G,U_T(1)]$  and $\cH^2[G,U_T(1)]$
calculated here.  In particular, $\leq$ becomes $=$ in \eqn{Z2rU1} and
\eqn{Z2TrU1} for $d=0,1,2$.

\Ref{U1Z2} gives a calculation and obtains a more complete result for
$\cH^d[U(1)\rtimes Z_2, \Z]$ and $\cH^d[U(1)\rtimes Z_2^T, \Z_T]$:
\begin{align}
 \cH^d[ U(1)\rtimes Z_2 , \Z]
=
\begin{cases}
\Z &  d=0 \\
\Z_1 &  d=1, \\
\Z_2 &  d=2,3,5 \\
%\Z_2 &  d=3, \\
\Z_2\times \Z &  d=4, \\
%\Z_2 &  d=5, \\
\Z_2^2 &  d=6. \\
\end{cases}
\end{align}
and
\begin{align}
 \cH^d[ U(1)\rtimes Z_2^T , \Z_T]
=
\begin{cases}
\Z_1 &  d=0 \\
\Z_2 &  d=1,3,4 \\
\Z \text{ or } \Z_2\times \Z &  d=2, \\
%\Z_2 &  d=3, \\
%\Z_2 &  d=4, \\
\Z_2^2 &  d=5. \\
\end{cases}
\end{align}
From that we can obtain
$\cH^d[U(1)\rtimes Z_2, U(1)]$ and $\cH^d[U(1)\rtimes Z_2, U_T(1)]$:
\begin{align}
 \cH^d[ U(1)\rtimes Z_2 , U(1)]
=
\begin{cases}
U(1) &  d=0, \\
\Z_2 &  d=1,2,4 \\
%\Z_2 &  d=3, \\
\Z_2\times \Z &  d=3. \\
%\Z_2 &  d=5, \\
%\Z_2^2 &  d=6. \\
\end{cases}
\end{align}
and
\begin{align}
 \cH^d[ U(1)\rtimes Z_2^T , U_T(1)]
=
\begin{cases}
%\Z_1 &  d=0 \\
\Z_2 &  d=0,2,3 \\
\Z \text{ or } \Z_2\times \Z &  d=1, \\
%\Z_2 &  d=3, \\
%\Z_2 &  d=4, \\
\Z_2^2 &  d=4. \\
\end{cases}
\end{align}
Since $\cH^1[ U(1)\rtimes Z_2^T , U_T(1)]$ classifies the 1D
representation of $ U(1)\rtimes Z_2^T$, from the calculation in subsection
\ref{1Dproj}, we find $\cH^1[ U(1)\rtimes Z_2^T , U_T(1)]=\Z$.
$\leq$ becomes $=$ in \eqn{Z2rU1} and
\eqn{Z2TrU1} for $d=0,1,2,3,4$.

\subsection{Group cohomology of $Z_n \rtimes \Z_2$}

In this section, we are going to use spectral sequence method
to calculate
$\cH^d[ Z_n \times Z_2,U(1)]$,
$\cH^d[ Z_n \times Z_2^T,U_T(1)]$,
$\cH^d[ Z_n \rtimes Z_2,U(1)]$
and
$\cH^d[ Z_n \rtimes Z_2,U_T(1)]$.
The group $Z_n \rtimes Z_2$ contain two subgroups $Z_n=\{U_k, k=0,1,...,n-1\}$
and $Z_2=\{1,T\}$.  We have $TU_k T =U_{-k \text{ mod }n}$ and $T^2=1$.  Just
like the $U(1) \rtimes Z_2$ cases studied in last section, for those four
groups, the $E_2$ page of the spectral sequence do not obviously stabilize.
However, it turn out that the $E_2$ pages do stabilize, can we can calculate
$\cH^d$ directly from the $E_2$ page.

We need to first calculate
$\cH^d(Z_2,\Z_n)$ and $\cH^d(Z_2,\Z_{T,n})$.
To calculate $\cH^d(Z_2,\Z_n)$
using \eqn{ZnCoh}, we note that
$\Z_n^{Z_2}=\Z_n$ and $ I_{Z_2}\Z_n=\Z_1$.
The map $N_{Z_2}$ becomes $N_{Z_2}: a\to 2 a$
for $M=\Z_n$. We have
$\text{Img}(N_{Z_2},\Z_n)= 2\Z_n=\Z_n$
when $n=$ odd and
$\text{Img}(N_{Z_2},\Z_n)= 2\Z_n=\Z_{n/2}$ when $n=$ even.
This gives $\text{Ker}(N_{Z_n},\Z)=\Z_{(2,n)}$.
So we have
\begin{align}
\label{Z2Zn}
 \cH^d[Z_2,\Z_n]=
\begin{cases}
\Z_n & \text{ if } d=0, \\
\Z_{(2,n)}  & \text{ if }  d>0\\
%\Z_{(2,n)}  & \text{ if } d=1 \text{ mod } 2.
\end{cases}
\end{align}

To calculate $\cH^d(Z_2,\Z_{T,n})$
where $Z_2$ acts non-trivially as
$T\cdot a=- a$, $a\in \Z_n$,
we note that
$\Z_{T,n}^{Z_2}=\Z_{(2,n)}$ and $ I_{Z_2}\Z_{T,n}=2\Z_n$.
The map $N_{Z_2}$ becomes $N_{Z_2}: a\to 0$.
So we have
$\text{Img}(N_{Z_2},\Z_n)= \Z_1$.
and $\text{Ker}(N_{Z_n},\Z)=\Z_n$.
So we have
\begin{align}
\label{Z2ZnT}
 \cH^d[Z_2,\Z_{T,n}]= \Z_{(2,n)}
\end{align}

First, we consider $Z_n \times Z_2$ group and module $M=\Z$.  In this case,
$Z_2$ acts on the $Z_n$ subgroup trivially and it acts on $M$ trivially.  As a
result, $Z_2$ acts on $\cH^q[Z_n,\Z]$ trivially.  This allows us to obtain the
following $E^{p,q}_2=\cH^p[Z_2,\cH^q[Z_n,\Z]]$ page in the spectral sequence:
\begin{align}
\bmm
0  &  0   &   0   &   0   &  0 &  0   &   0   &   0  \\
\Z_n &  \Z_{(2,n)}   & \Z_{(2,n)}  &  \Z_{(2,n)} & \Z_{(2,n)} &   \Z_{(2,n)}   & \Z_{(2,n)}  &   \Z_{(2,n)}  \\
0  &  0   &   0   &   0   &  0 &  0   &   0   &   0  \\
\Z_n&  \Z_{(2,n)}   & \Z_{(2,n)}  &  \Z_{(2,n)} & \Z_{(2,n)} &   \Z_{(2,n)}   & \Z_{(2,n)}  &   \Z_{(2,n)}  \\
0  &  0   &   0   &   0   &  0 &  0   &   0   &   0  \\
\Z_n &  \Z_{(2,n)}   & \Z_{(2,n)}  &  \Z_{(2,n)} & \Z_{(2,n)} &   \Z_{(2,n)}   & \Z_{(2,n)}  &   \Z_{(2,n)}  \\
0  &  0   &   0   &   0   &  0 &  0   &   0   &   0  \\
\Z &  0   & \Z_2  &  0 & \Z_2 &   0   & \Z_2  &   0  \\
\emm
\end{align}
We can show $E_\infty^{p,q}=E_2^{p,q}$ when $n=$ odd.  But for $n=$ even,
the above $E_2$ page may not stabilize.
%we can only show $E_\infty^{p,q}=E_2^{p,q}$ when $p+q \leq 1$.
So we only have
\begin{align}
\label{Z2ZnZ}
 \cH^d[ Z_n\times Z_2,\Z] \leq
\begin{cases}
\Z,  &  d=0 ,\\
\Z_2\times \Z_n\times \Z_{(2,n)}^{\frac {d-2} 2},  &  d=0 \text{ mod } 2,\\
\Z_{(2,n)}^{\frac{d-1}{2}},  &  d=1 \text{ mod } 2 ,\\
\end{cases}
\end{align}
which gives
\begin{align}
\label{Z2ZnU1}
 \cH^d[ Z_n\times Z_2,U(1)] \leq
\begin{cases}
U(1),  &  d=0 ,\\
\Z_{(2,n)}^{\frac{d}{2}},  &  d=0 \text{ mod } 2 ,\\
\Z_2\times \Z_n \times \Z_{(2,n)}^{\frac {d-1} 2},  &  d=1 \text{ mod } 2.\\
\end{cases}
\end{align}

Second, we consider $Z_n \times Z_2^T$ group and module $M=\Z_T$.  In this case,
$Z_2$ acts on the $Z_n$ subgroup trivially and it acts on $M$ non-trivially.
As a result, $Z_2^T$ acts on $\cH^q[Z_n,\Z_T]$ non-trivially: $T\cdot \al\to -
\al$, $\al\in \cH^q[Z_n,\Z_T]$.  This allows us obtain the following
$E^{p,q}_2=\cH^p[Z_2^T,\cH^q[Z_n,\Z_T]]$ page in the spectral sequence:
\begin{align}
\bmm
0  &  0   &   0   &   0   &  0 &  0   &   0   &   0  \\
\Z_{(2,n)} &  \Z_{(2,n)}   & \Z_{(2,n)}  &  \Z_{(2,n)} & \Z_{(2,n)} &   \Z_{(2,n)}   & \Z_{(2,n)}  &   \Z_{(2,n)}  \\
0  &  0   &   0   &   0   &  0 &  0   &   0   &   0  \\
\Z_{(2,n)}&  \Z_{(2,n)}   & \Z_{(2,n)}  &  \Z_{(2,n)} & \Z_{(2,n)} &   \Z_{(2,n)}   & \Z_{(2,n)}  &   \Z_{(2,n)}  \\
0  &  0   &   0   &   0   &  0 &  0   &   0   &   0  \\
\Z_{(2,n)} &  \Z_{(2,n)}   & \Z_{(2,n)}  &  \Z_{(2,n)} & \Z_{(2,n)} &   \Z_{(2,n)}   & \Z_{(2,n)}  &   \Z_{(2,n)}  \\
0  &  0   &   0   &   0   &  0 &  0   &   0   &   0  \\
 0   & \Z_2  &  0 & \Z_2 &   0   & \Z_2  &   0  & \Z_2 \\
\emm
\end{align}
Again, we can show $E_\infty^{p,q}=E_2^{p,q}$ for $n=$ odd,
but not for $n=$ even,
%we can only show $E_\infty^{p,q}=E_2^{p,q}$ when $p+q \leq 1$.
So we only have
\begin{align}
\label{Z2TZnZ}
 \cH^d[ Z_n\times Z_2^T,\Z_T] \leq
\begin{cases}
\Z_{(2,n)}^{\frac d 2},  &  d=0 \text{ mod } 2,\\
\Z_2\times \Z_{(2,n)}^{\frac{d-1}{2}},  &  d=1 \text{ mod } 2 ,\\
\end{cases}
\end{align}
which gives
\begin{align}
\label{Z2TZnU1}
 \cH^d[ Z_n\times Z_2^T,U_T(1)] \leq
\begin{cases}
\Z_2\times\Z_{(2,n)}^{\frac{d}{2}},  &  d=0 \text{ mod } 2 ,\\
 \Z_{(2,n)}^{\frac {d+1} 2},  &  d=1 \text{ mod } 2,\\
\end{cases}
\end{align}

Third, we consider $Z_n \rtimes Z_2$ group and module $M=\Z$.  In this case,
$Z_2$ acts on the $Z_n$ subgroup non-trivially
and it acts on $M$ trivially.
As a result, $Z_2$ acts on $\cH^q[Z_n,\Z]$
non-trivially: $T\cdot \al\to (-)^{q/2} \al$, $\al\in \cH^q[Z_n,\Z]$.
To obtain the above result, we note that
$T$ act on the $q$-cocycles in $\cH^q[Z_n,\Z]$ in the following way:
\begin{align}
 T: \nu_q(g_0,...,g_q) \to \nu_q(Tg_0T^{-1},...,Tg_qT^{-1}),\ \ \
g_i \in Z_n.
\end{align}
Through some explicit calculations, we find that a $T$ transformed 2-cocycle, $
\nu_2(Tg_0T^{-1},Tg_1T^{-1},Tg_2T^{-1}) $, is same as $ -\nu_2(g_0,g_1,g_2) $ up to a
2-coboundary.  Thus  $T\cdot \al_2\to - \al_2$ for $\al_2\in
\cH^2[Z_n,\Z]$.  The generator $\al_{2p} \in \cH^{2p}[Z_n,\Z]$ can be
obtained from the generator $\al_{2} \in \cH^{2}[Z_n,\Z]$ by taking the cup
product\cite{RS}
$ \al_{2p}=\al_2\cup \al_2 \cup ...\cup \al_2$.
Therefore $T\cdot \al_{2p} = (-)^p \al_{2p}$.

This allows us
obtain the following $E^{p,q}_2=\cH^p[Z_2,\cH^q[Z_n,\Z]]$ page in the
spectral sequence:
\begin{align}
\bmm
0  &  0   &   0   &   0   &  0 &  0   &   0   &   0  \\
\Z_{(2,n)} &  \Z_{(2,n)}   & \Z_{(2,n)}  &  \Z_{(2,n)} & \Z_{(2,n)} &   \Z_{(2,n)}   & \Z_{(2,n)}  &   \Z_{(2,n)}  \\
0  &  0   &   0   &   0   &  0 &  0   &   0   &   0  \\
\Z_n&  \Z_{(2,n)}   & \Z_{(2,n)}  &  \Z_{(2,n)} & \Z_{(2,n)} &   \Z_{(2,n)}   & \Z_{(2,n)}  &   \Z_{(2,n)}  \\
0  &  0   &   0   &   0   &  0 &  0   &   0   &   0  \\
\Z_{(2,n)} &  \Z_{(2,n)}   & \Z_{(2,n)}  &  \Z_{(2,n)} & \Z_{(2,n)} &   \Z_{(2,n)}   & \Z_{(2,n)}  &   \Z_{(2,n)}  \\
0  &  0   &   0   &   0   &  0 &  0   &   0   &   0  \\
\Z &  0   & \Z_2  &  0 & \Z_2 &   0   & \Z_2  &   0  \\
\emm
\end{align}

We have
\begin{align}
\label{Z2rZnZ}
 \cH^d[ Z_n\rtimes Z_2,\Z] \leq
\begin{cases}
\Z,  &  d=0 ,\\
\Z_2\times \Z_n\times \Z_{(2,n)}^{\frac {d-2} 2},  &  d=0 \text{ mod } 4,\\
\Z_{(2,n)}^{\frac{d-1}{2}},  &  d=1 \text{ mod } 2 ,\\
\Z_2\times \Z_{(2,n)}^{\frac d 2},  &  d=2 \text{ mod } 4,\\
\end{cases}
\end{align}
which agrees with the result obtained
in \Ref{DH} for $n=$ even.
The above
gives
\begin{align}
\label{Z2rZnU1}
 \cH^d[ Z_n\rtimes Z_2,U(1)] \leq
\begin{cases}
U(1),  &  d=0 ,\\
\Z_{(2,n)}^{\frac{d}{2}},  &  d=0 \text{ mod } 2 ,\\
\Z_2\times \Z_{(2,n)}^{\frac {d+1} 2},  &  d=1 \text{ mod } 4,\\
%\Z_{(2,n)}^{\frac{d}{2}},  &  d=2 \text{ mod } 4 ,\\
\Z_2\times \Z_n\times \Z_{(2,n)}^{\frac {d-1} 2},  &  d=3 \text{ mod } 4,\\
\end{cases}
\end{align}
This agrees with a result for the symmetric group on three
elements, $S_3=Z_3\rtimes Z_2$\cite{TL}
%http://math.berkeley.edu/~teichner/Courses/215B/Surveys/TyeLidman.pdf
\begin{align}
 \cH^d[S_3,U(1)]=
\begin{cases}
U(1) & \text{ if } d=0, \\
\Z_1  & \text{ if } d=0 \text{ mod } 2, \ \ d>0\\
\Z_2  & \text{ if } d=1 \text{ mod } 4, \\
\Z_6  & \text{ if } d=3 \text{ mod } 4,
\end{cases}
\end{align}
if we replace $\leq$ by $=$.

Last, we consider $Z_n\rtimes Z_2^T$ group and module $M=\Z_T$.  In this case,
$Z_2^T$ acts on the $Z_n$ subgroup non-trivially and it acts on
$M$ non-trivially: $T\cdot n\to -n$, $n\in \Z_T$.  As a result, $Z_2^T$ acts on
$\cH^q[Z_n,\Z_T]$ non-trivially: $T\cdot \al\to -(-)^{q/2} \al$, $\al\in
\cH^q[Z_n,\Z_T]$.  We obtain the following
$E^{p,q}_2=\cH^p[Z_2,\cH^q[Z_n,\Z_T]]$ page in the spectral sequence:
\begin{align}
\bmm
0  &  0   &   0   &   0   &  0 &  0   &   0   &   0  \\
\Z_n &  \Z_{(2,n)}   & \Z_{(2,n)}  &  \Z_{(2,n)} & \Z_{(2,n)} &   \Z_{(2,n)}   & \Z_{(2,n)}  &   \Z_{(2,n)}  \\
0  &  0   &   0   &   0   &  0 &  0   &   0   &   0  \\
\Z_{(2,n)} &  \Z_{(2,n)}   & \Z_{(2,n)}  &  \Z_{(2,n)} & \Z_{(2,n)} &   \Z_{(2,n)}   & \Z_{(2,n)}  &   \Z_{(2,n)}  \\
0  &  0   &   0   &   0   &  0 &  0   &   0   &   0  \\
\Z_n&  \Z_{(2,n)}   & \Z_{(2,n)}  &  \Z_{(2,n)} & \Z_{(2,n)} &   \Z_{(2,n)}   & \Z_{(2,n)}  &   \Z_{(2,n)}  \\
0  &  0   &   0   &   0   &  0 &  0   &   0   &   0  \\
 0   & \Z_2  &  0 & \Z_2 &   0   & \Z_2  &   0  & \Z_2 \\
\emm
\end{align}

We obtain
\begin{align}
\label{Z2rZnZT}
 \cH^d[ Z_n\rtimes Z_2^T,\Z_T] \leq
\begin{cases}
\Z_{(2,n)}^{\frac d 2},  &  d=0 \text{ mod } 4,\\
\Z_2\times \Z_{(2,n)}^{\frac{d-1}{2}},  &  d=1 \text{ mod } 2 ,\\
\Z_n\times \Z_{(2,n)}^{\frac {d-2} 2},  &  d=2 \text{ mod } 4,\\
\end{cases}
\end{align}
which agrees with the result obtained
in \Ref{DH} for $n=$ even.
The above
gives
\begin{align}
\label{Z2rZnU1T}
 \cH^d[ Z_n\rtimes Z_2^T ,U_T(1)] \leq
\begin{cases}
\Z_2\times \Z_{(2,n)}^{\frac{d}{2}},  &  d=0 \text{ mod } 2 ,\\
\Z_n\times \Z_{(2,n)}^{\frac {d-1} 2},  &  d=1 \text{ mod } 4,\\
\Z_{(2,n)}^{\frac {d+1} 2},  &  d=3 \text{ mod } 4,\\
\end{cases}
\end{align}

Although we cannot prove it, in fact, the $E_2^{p,q}$ pages do stabilize:
$E_\infty^{p,q}=E_2^{p,q}$, for all the four groups discussed here.  For $n=$
odd, we can show that the $E_2^{p,q}$ pages stabilize.  So we can replace
$\leq$ by $=$ in \eqn{Z2ZnU1}, \eqn{Z2TZnU1}, \eqn{Z2rZnU1} and \eqn{Z2rZnU1T}.
For $n=$ even, the results obtained before using the K\"unneth theorem imply
that we can replace $\leq$ by $=$ in \eqn{Z2ZnU1} in \eqn{Z2TZnU1}, and the
results in \Ref{DH} imply that we can replace $\leq$ by $=$ in  \eqn{Z2rZnU1}
and \eqn{Z2rZnU1T}. This suggests that the $E_2^{p,q}$ pages stabilize even when
$n=$ even.

\subsection{Group cohomology of $U(n)$, $SU(n)$, and $Sp(n)$ }

The group cohomology ring of $U(n)$, $SU(n)$, and $Sp(n)$
are given by \cite{SO3}
\begin{align}
 \cH^*[U(n),\Z) & =\Z[c_1,...,c_n],
\nonumber\\
 \cH^*[SU(n),\Z) & =\Z[c_2,...,c_n],
\nonumber\\
 \cH^*[Sp(n),\Z) & =\Z[p_1,...,p_n],
\end{align}
where
$c_i \in \cH^{2i}[U(n),\Z]$ or
$c_i \in \cH^{2i}[SU(n),\Z]$, and
$p_i \in \cH^{4i}[Sp(n),\Z]$.
Here $\Z[x,y,...]$ represents a ring of polynomials
of variables $x$, $y$, ... with integer coefficients.

For example $\cH^*[U(1),\Z]=\Z[c_1]$ mean that the elements in $\cH^*[U(1),\Z]$
has a form $n_0+ n_1 c_1+n_2 c_1^2+n_3 c_1^3 + ...=n_0+ n_1 c_1+n_2 c_1\cup
c_1+n_3 c_1\cup c_1\cup c_1 + ...$.  Note that $c_1$ is a two cocycle and $ n_1
c_1$ is a two cocycle labeled by $n_1\in \Z$.  Also $ n_1 c_1$ is the only 2
cocycle in the expression $n_1 c_1+n_2 c_1\cup c_1+n_3 c_1\cup c_1\cup c_1 +
...$.  Thus $\cH^2[U(1),\Z]=\Z$.  Similarly $n_2 c_1\cup c_1$ is the only
4-cocycle in the expression $n_1 c_1+n_2 c_1\cup c_1+n_3 c_1\cup c_1\cup c_1 +
...$.  Thus $\cH^4[U(1),\Z]=\Z$.  There is no odd cocycles in $n_1 c_1+n_2
c_1\cup c_1+n_3 c_1\cup c_1\cup c_1 + ...$.  Thus $\cH^d[U(1),\Z]=\Z_1$ when
$d=$ odd.

From $\cH^*[SU(2),\Z]=\Z[c_2]$, we find that the elements in $\cH^*[SU(2),\Z]$
has a form $n_0+ n_1 c_2+n_2 c_2\cup c_2+n_3 c_2\cup c_2\cup c_2 + ...$.
Thus
\begin{align}
\label{SU2Z}
 \cH^d[ SU(2) , \Z]=
\begin{cases}
\Z &  d=0 \text{ mod }4 \\
\Z_1 &  d\neq 0  \text{ mod }4 .
\end{cases}
\end{align}
and
\begin{align}
\label{SU2Z}
 \cH^d[ SU(2) , U(1)]=
\begin{cases}
\Z &  d=3 \text{ mod }4 \\
\Z_1 &  \text{otherwise} .
\end{cases}
\end{align}

\subsection{Group cohomology of $SO(3)$ and $SO(3) \times \Z_2^T$}

The group cohomology ring of $SO(3)$ is given by\cite{SO3,G0691}
\begin{align}
 \cH^*[SO(3),\Z] & =\Z[v,c]/(2v),
\end{align}
where $v \in \cH^3[SO(3),\Z]$ and $c \in \cH^4[SO(3),\Z]$.  Here $/(2v)$ means
that the expression $2v$ in the polynomial is regarded as $0$.  The elements in
$\cH^*[SO(3),\Z]$ have a form $\sum n_{i,j} c^iv^j$.  Note that $c^iv^j$ is a
$4i+3j$ cocycle.  The lowest cocycle in the expression $\sum n_{i,j} c^iv^j$ is
a 3 cocycle $n_{01}v$. Since $2v$ is regarded as zero, there are only two 3
cocycles $0$ and $v$ labeled by $n_{01}=0,1$.  Thus $\cH^3[SO(3),\Z]=\Z_2$.
The expression $\sum n_{i,j} c^iv^j$ contains only one 4-cocycle $n_{10}c$
labeled by $n_{10}\in \Z$. Thus $\cH^4[SO(3),\Z]=\Z$.  This way, we find that
\begin{align}
\label{SO3}
 \cH^d[ SO(3) , \Z]=
\begin{cases}
\Z &  d=0,4 \\
\Z_1 &  d=1,2,5 \\
\Z_2 &  d=3,6. \\
\end{cases}
\end{align}

Using \eq{kunn} and $ \cH^d[ SO(3)\times Z_2^T , U_T(1)]= \cH^{d+1}[
SO(3)\times Z_2^T , \Z_T]$, we obtain
\begin{align}
 \cH^d[ SO(3)\times Z_2^T , U_T(1)]
=
\begin{cases}
\Z_2 &  d=0,3 \\
\Z_1 &  d=1, \\
\Z_2^2 &  d=2, \\
%\Z_2 &  d=3, \\
\Z_2^3 &  d=4. \\
\end{cases}
\end{align}

%\bibliography{../../bib/wencross,../../bib/all,../../bib/publst}

~

\end{document}